%% file: Report.tex
\documentclass[12pt, a4paper]{report}
\input{Packages.tex}

\hypersetup{pdftitle = Thesis, pdfauthor = {First Last}, pdfstartview=FitH, pdfkeywords = essay, pdfpagemode=FullScreen, colorlinks, anchorcolor = black, citecolor = black, urlcolor=blue, filecolor=green, linkcolor=black, plainpages=false}
\fancyheadoffset{9pt}
\fancyfootoffset{9pt}
\date{}

\title{Decoding non-invasive brain activity with novel deep-learning approaches}
\author{\\ \Large{Richard Csaky}
\\ Christ Church College
\\
\\
\\
\\ University of Oxford
\\
A thesis presented for the degree of \\ \textit{Doctor of Philosophy}
\\ \\
Michaelmas 2023
}
\begin{document}
% Adjust logo positions here
\AddToShipoutPicture*{\BackgroundPicturea{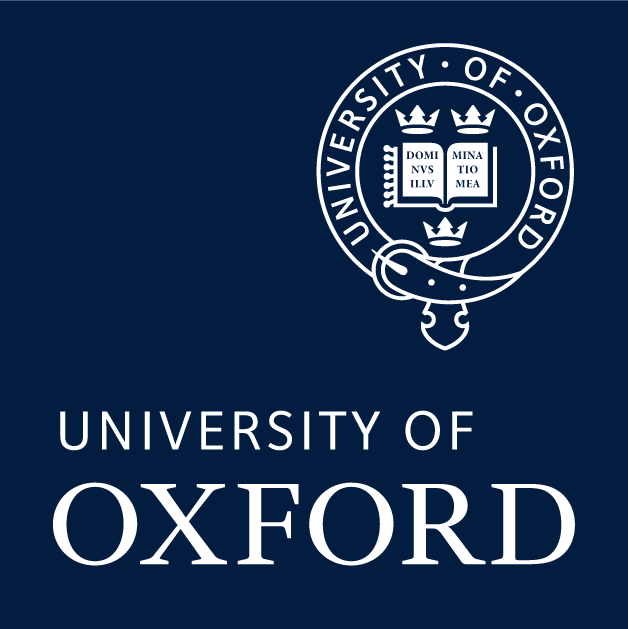}{0.7in}{5.8in}}
\AddToShipoutPicture*{\BackgroundPictureb{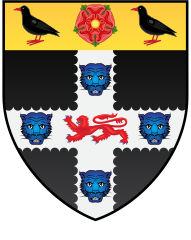}{6.1in}{3.7in}}
\thispagestyle{headings}
	\maketitle
\FloatBarrier
\pagenumbering{roman}

\newpage
\thispagestyle{empty}
\begin{center}
I dedicate this to the memory of my grandma, Zori
\end{center}

\newpage
\thispagestyle{empty}
\vspace*{\fill}
\begin{center}
Copyright \copyright  \thinspace 2023 by Richard Csaky \\ All Rights Reserved
\end{center}
\vspace*{\fill}
\newpage
\thispagestyle{empty}
\epigraph{It may be that our role on this planet is not to worship God, but to create him.}{--- \textup{Arthur C. Clarke}}

\thispagestyle{empty}
\chapter*{Acknowledgements}
I would primarily like to acknowledge my supervisor Mark Woolrich. I had no idea of the kind of journey I would embark during my 3 PhD years, but Mark has given me the research freedom that few can enjoy. While at times I may have needed more guidance, as I learned later, stumbling in the darkness is part of the process. Mark always provided support and an intellectual clarity which I am still in awe of.

I am grateful to my two co-supervisors, Oiwi and Mats, for their support and involvement in my research. Without their sacrifice of long hours to help with experiments, Chapter 6 would have not been possible. Here I would also like to thank Anna Camera, and all OHBA members who helped with my experiments. I am humbled by their selflessness.

I am eternally indebted to Gabor Recski, my Bachelor's and Master's supervisor, who instilled in me fundamental research techniques that I used throughout my PhD. It is under his guidance that I \sout{mastered} was introduced to python, deep learning, latex, and countless other tools, elemental to my research.

Thanks to Arun, Antara, Coby, Arina, Nati, Shu, Ryan, and Clara, for making the UK more homely, and for putting up with my \sout{daily} hourly complaints about the weather. While I lost touch with some of them, I do not forget our time together.

Finally, I would like to thank my parents, and my friends from home, Adam, Janos, Come, and Peter, who have helped me through some dark times during my PhD. It's difficult to put into words the significance of their friendship.

\begin{abstract}

This thesis delves into the world of non-invasive electrophysiological brain signals like electroencephalography (EEG) and magnetoencephalography (MEG), focusing on modelling and decoding such data. The research aims to investigate what happens in the brain when we perceive visual stimuli or engage in covert speech (inner speech) and enhance the decoding performance of such stimuli. The findings have significant implications for the development of brain-computer interfaces (BCIs), leading to assistive communication technologies for paralysed individuals. The thesis is divided into two main sections, methodological and experimental work. A central concern in both sections is the large variability present in electrophysiological recordings, whether it be within-subject or between-subject variability, and to a certain extent between-dataset variability. 

In the methodological sections, we explore the potential of deep learning for brain decoding. The research acknowledges the urgent need for more sophisticated models and larger datasets to improve the decoding and modelling of EEG and MEG signals. We present advancements in decoding visual stimuli using linear models at the individual subject level. We then explore how deep learning techniques can be employed for group decoding, introducing new methods to deal with between-subject variability. Finally, we also explores novel forecasting models of MEG data based on convolutional and Transformer-based architectures. In particular, Transformer-based models demonstrate superior capabilities in generating signals that closely match real brain data, thereby enhancing the accuracy and reliability of modelling the brain’s electrophysiology.

In the experimental section, we present a unique dataset containing high-trial inner speech EEG, MEG, and preliminary optically pumped magnetometer (OPM) data. We highlight the limitations of current BCI systems used for communication, which are either invasive or extremely slow. While inner speech decoding from non-invasive brain signals has great promise, it has been a challenging goal in the field with limited decoding approaches, indicating a significant gap that needs to be addressed. Our aim is to investigate different types of inner speech and push decoding performance by collecting a high number of trials and sessions from a few participants. However, the decoding results are found to be mostly negative, underscoring the difficulty of decoding inner speech.

In conclusion, this thesis provides valuable insight into the challenges and potential solutions in the field of electrophysiology, particularly in the decoding of visual stimuli and inner speech. The findings could pave the way for future research and advancements in the field, ultimately improving communication capabilities for paralysed individuals.

%\keywords{Keyword1 - Keyword2 - Keyword3}
% \vspace{-10mm} %To remove added white space after
\end{abstract}
\tableofcontents
\thispagestyle{plain}
\listoffigures
\listoftables

\chapter*{List of Abbreviations}
\begin{abbreviations}
    \item[BCI] Brain-Computer Interface
    \item[MVPA] Multivariate Pattern Analysis
    \item[RSA] Representational Similarity Analysis
    \item[MEG] Magnetoencephalohraphy
    \item[EEG] Electroencephalography
    \item[OPM] Optically-pumped Magnetometer
    \item[SNR] Signal-to-Noise Ratio
    \item[fMRI] functional Magnetic Resonance Imaging
    \item[PCA] Principal Component Analysis
    \item[ICA] Independent Component Analysis
    \item[LDA] Linear Discriminant Analysis
    \item[HMM] Hidden Markov Model
    \item[CNN] Convolutional Neural Network
    \item[RNN] Recurrent Neural Network
    \item[GPT] Generative Pre-trained Transformer
    \item[AR] Auto-regressive
    \item[PFI] Permutation Feature Importance
    \item[STFT] Short-time Fourier Transform
    \item[LOSO] Leave-one-subject-out
    
\end{abbreviations}

\chapter{Introduction}
\label{Chap1}
\pagenumbering{arabic}

\section{Electrophysiology of the brain}
\subsection{Recording brain activity}
Understanding the intricacies of the human brain remains one of the grand challenges of science, but what does it mean to understand? Does it entail accurately simulating certain functions computationally, as pursued in computational neuroscience \citep{dayan2005theoretical}? Is it synonymous with prediction, as suggested by predictive coding theory \citep{friston2010free}? Does understanding pertain to characterising an individual brain, an average brain, or a collection of distinct brain types \citep{kanai2011structural}? Might understanding enable novel treatments for brain disorders and enhancement of human abilities, aligning with the mission of translational neuroscience \citep{insel2013twenty}?

As with most doctoral theses, this work does not attempt to resolve such expansive questions. Rather, it aims to provide incremental advancements in select research domains, with the aspiration that these innovations may one day contribute to a more comprehensive understanding. For our purposes, understanding may constitute elucidating particular processes in the brain and linking them to cognitive, emotional, or behavioural phenomena \citep{pessoa2022emergent}. Frequently this involves mathematical models that approximate the underlying biology \citep{izhikevich2007dynamical}. We can equate these models themselves with understanding \citep{kriegeskorte2018cognitive}, although the use of deep learning in the models can complicate this notion, causing the model itself to require an additional interpretative effort \citep{arrieta2020explainable}.

The brain contains upwards of 86 billion neurons with quadrillions of synaptic connections \citep{azevedo2009equal}. To achieve tractable levels of understanding, approximations and abstractions are imperative. We can delineate types of understanding by spatial and temporal scale \citep{panahi2021generative}, reflecting the spatiotemporal physical essence of the brain. For example, modelling neurotransmitter dynamics aids comprehension of learning, motivation, and rewards - pivotal constructs in cognitive and behavioural neuroscience \citep{schultz2002getting}. Single neuron models specify input-output characteristics on precise (millisecond) timescales, providing insights at the core of computational neuroscience \citep{izhikevich2004model}. Interconnecting such models allows understanding of local microcircuits. Validating models against experimental data reveals, for instance, which neuronal populations represent different parts of the visual field (retinotopic maps) \citep{wandell2007visual, nasiotis2017high}. Such models also elucidate the rapid temporal propagation of activity across the hierarchical visual system \citep{carlson2013representational}.

Deriving and evaluating models requires actual brain data. For fine spatial scales, this often comes from intracranial electrodes measuring individual neuron spiking or local field potentials (LFPs) \citep{buzsaki2012origin, buzsaki2015tools}. Intracranial electrocorticography (ECoG) provides real-time population-level (10,000s of neurons) activity \citep{miller2009power} by placing electrode grids directly on the brain surface. However, such invasive procedures carry risks associated with surgery and are typically only employed in clinical settings, or in research with patients who already require surgery for medical reasons \citep{waldert2016invasive}. This inevitably constrains data quantity and variety. At larger scales, neural mass models enable whole-brain biophysical simulations \citep{deco2008dynamic, hadida2018bayesian}, while machine learning can model diverse non-invasive recording modalities like electrophysiology or blood-oxygen-level-dependent (BOLD) imaging \citep{friston2005models}.

BOLD techniques including functional magnetic resonance imaging (fMRI) and functional near-infrared spectroscopy (fNIRS) offer the poorest temporal resolution, on the order of seconds. This is because they detect changes in blood flow resulting from variations in local neuronal activity, reflecting the understanding that the body cannot, and does not need to regulate blood flow on the order of milliseconds \citep{buxton2013physics}. However, their spatial resolution is unparalleled, producing dynamic 3D brain images with hundreds of thousands to millions of voxels \citep{buxton2013physics}. Voxel volumes are around $0.5mm^3$, providing localised activity estimates in the form of BOLD changes. On the downside, MRI scanners are sensitive to head motion and place the participant in a constrained noisy space, potentially influencing brain function \citep{van2012influence}.

\subsection{Non-invasive electrophysiology}

This thesis is concerned with non-invasive electrophysiological recording modalities, the modelling of such data, and the kind of understanding and real-world applications that these models might facilitate. The modality inherently limits the spatial and temporal scale of modelling and understanding \citep{kiebel2008dynamic}. Non-invasive electrophysiology like electroencephalography (EEG) and magnetoencephalography (MEG) offers millisecond temporal resolution akin to intracranial recordings \citep{cohen1968magnetoencephalography, berger1929elektroenkephalogramm}. This results from measuring near-instantaneous electromagnetic fields generated by neuronal activity \citep{nunez2006electric}. However, limited spatial resolution remains a key challenge, especially for EEG which measures electrical currents. Signal distortion by the skull and scalp restricts spatial specificity \citep{nunez2006electric}. While MEG is less affected due to measuring magnetic fields, its few hundred sensors still average over millions of neurons \citep{hamalainen1993magnetoencephalography}. MEG also necessitates costly specialised equipment in magnetically shielded rooms \citep{baillet2017magnetoencephalography}. Emerging optically pumped magnetometers (OPMs) may improve MEG sensitivity and flexibility by enabling on-scalp measurements \citep{wens2023exploring}. Their lack of required cooling could expand MEG accessibility \citep{boto2018moving}, potentially enabling brain-computer interface (BCI) applications, although current OPM technology does require the devices to be housed in a magnetically shielded room.

Despite advances, modelling and decoding brain signals from non-invasive electrophysiology remains challenging. Models often fail to accurately decode complex, variable signals across and within individuals \citep{saha2020intra}. While some variability naturally arises from morphological and dynamical diversity \citep{michel2019eeg, wainio2021dynamic}, noise also contributes \citep{faisal2008noise}. Signal-to-noise ratio is thus crucial \citep{nenonen2007total}, with artefacts originating from external sources (e.g. power lines, Earth's magnetic field) or internal ones like breathing, blinks, heartbeats, and muscle activity \citep{uriguen2015eeg}. In later chapters, we discuss artefact reduction via signal processing and machine learning \citep{makeig1995independent}. Attention, fatigue, anatomy, and functional differences also modulate brain recordings \citep{saha2020intra}, complicating generalisation across individuals or sessions. Managing variability in electrophysiological data is an integral theme of this thesis.

One may rightly question how such noisy, spatially-coarse signals can inform understanding or applications. However, the real-time nature, direct neural basis, and non-invasiveness of M/EEG enable massive datasets with exquisite temporal resolution \citep{gifford2022large}. This makes EEG uniquely suited for BCIs in healthy and ill populations \citep{murguialday2011transition}. MEG and OPMs currently remain confined to research due to the shielding requirement, and fNIRS \citep{naseer2015fnirs}, while up and coming in the BCI field is too slow to be used in the kinds of BCI applications we are interested in the experimental work part of this thesis.

As our research is not concerned with the development of new kinds of recording technology, we will have to carefully consider the limitations inherent in non-invasive electrophysiology in our methodological work. These constraints inform the kind of questions we can ask (and hopefully answer), and the kind of understanding we can gain \citep{kiebel2008dynamic}. The whole-brain view of M/EEG facilitates studying dispersed cognitive processes that recruit vastly different brain regions like vision and language, as we will see. We can characterise spatiotemporal dynamics with millisecond precision relative to events \citep{baillet2001electromagnetic}, albeit with limited spatial specificity. Critically, only synchronised activity across tens of thousands of neurons overcomes noise to manifest in M/EEG \citep{singer1999neuronal, buzsaki2006rhythms, buzsaki2004neuronal}. As we will see this consideration is particularly important when studying subtle cognitive processes such as inner speech \citep{alderson2015inner}. To reiterate an important point, one way to deal with the natural variability of ongoing brain signals \citep{wainio2021dynamic} and artefactual contamination is to collect as much data as possible from a single individual \citep{boudewyn2018many}. Repeated measurements allow both better spatial and better temporal specificity, and consequently a deeper level of understanding \citep{hebart2022things}. However, better methods are needed to deal with variability and fully capitalise on the rich information in M/EEG data \citep{quinn2022glm, hebart2022things, van2020post}.

A major opportunity with non-invasive electrophysiology is accumulating large datasets. However, between-subject variability hinders understanding brain activity beyond the individual level \citep{olivetti2014meg}. The success of deep learning with large sets of data motivates the need to unlock the full potential of electrophysiology datasets by developing new methods for dealing with between-participant variability \citep{defossez2022decoding, kostas2021bendr}.

This thesis aims to address the above challenges by exploring the potential of deep learning techniques in the decoding and modelling of electromagnetic brain signals. We delve into these issues through methodological and experimental work seeking to advance electrophysiology. In addition to providing a deeper understanding of the brain (or brains), we hope our findings will contribute to the development of more accurate and reliable BCIs.

\section{Visual and language processing}

Besides ongoing neural activity during rest \citep{raichle2001default}, the brain must also process and react to external stimuli, including complex inputs such as vision and language. Elucidating the neural dynamics underlying these faculties is fundamental in neuroscience and is important for brain-computer interfaces (BCIs) \citep{icscan2018steady, akbari2019towards}. However, decoding brain signals during such tasks poses multiple challenges that this thesis tackles. As discussed previously, these stem principally from various forms of variability.

Visual processing involves a hierarchical cascade that originates in the retina, travelling along the optic nerve to the primary visual cortex (V1) to extract basic features such as orientation and spatial frequency \citep{hubel1962receptive}. This early stage, focused on low-level attributes such as edges and colour, has been thoroughly characterised \citep{livingstone1988segregation}. Information then flows to extrastriate areas including V2, V3, V4, and inferotemporal cortex for processing complex features like object identity \citep{tanaka1996inferotemporal}, motion \citep{albright1984direction}, and spatial location \citep{maunsell1987visual}.

The spatiotemporal dynamics and spectral signatures of visual processing can be examined using electrophysiological techniques like EEG and MEG \citep{baillet2017magnetoencephalography}. For example, visual stimuli can evoke specific amplitude peaks in the signal precisely time-locked to stimulus onset (evoked activity). Due to time-locking, by averaging over individual trials the evoked peaks become more prominent and ongoing oscillations are cancelled out. Evoked responses triggered by changes in the visual field manifest very rapidly, within 100 ms post-stimulus \citep{cichy2014resolving, cichy2016comparison}. Stimuli can also more broadly modulate oscillatory activity in the alpha (8-12 Hz) and gamma (30-80 Hz) frequency bands, reflecting processes like attentional modulation \citep{jensen2010shaping} and object recognition \citep{tallon1999oscillatory, gruber1999selective}, respectively. This induced activity plays a critical role in coordinating neural processing and integration of visual information \citep{klimesch2012alpha, herrmann2010human}. 

In contrast, language processing relies on a distributed network that includes classical perisylvian language areas such as Broca's area for speech production and Wernicke's area for comprehension, among others \citep{friederici2011brain, hickok2007cortical}. These regions are interconnected via white matter tracts, forming an integrated system for linguistic processing. Like visual processing, language tasks also elicit specific oscillatory activity patterns. Theta band (4-8 Hz) oscillations, for instance, have been linked to syllable segmentation and sentence parsing \citep{bastiaansen2006oscillatory}, while gamma oscillations reflect phonetic and semantic processing \citep{obleser2011multiple}. The precise timing of these oscillations is believed to play a critical role in the coordination of neural dynamics during language tasks \citep{peelle2012neural}.

\citep{dikker2020magnetoencephalography} have presented the detailed spatiotemporal dynamics evoked in language processing. Spoken word processing starts in Heschl’s gyrus and the superior temporal gyrus 50-100 ms after stimulus. Written word processing starts in the occipital lobe 100 ms post-stimulus and goes on to the posterior and anterior fusiform gyrus for orthographic and morphological segmentation. Modality-independent processing happens 300-500 ms post-stimulus with lexical access and word meaning in the middle temporal gyrus. Semantic processing takes place 350-500 ms post-stimulus in orbito-frontal areas, and finally syntactic processing at around 600 ms in the inferior frontal gyrus. This late syntactics related activity reflects sentence-level integration.

We could conclude that the spatiotemporal dynamics of language and vision are quite well understood based on hundreds of studies with both invasive and noninvasive recordings. As we are motivated by BCI applications, we wanted to design methods that can improve decoding performance while still offering the kind of spatiotemporal understanding that has been established. A major issue is the limited performance of BCIs that allow a subject to communicate. Current systems are either invasive, posing risks to the user \citep{chaudhary2016brain}, or non-invasive but extremely slow, limiting their practical use. Non-invasive BCIs often rely on slow and effortful control signals, such as the P300 wave or cortical potentials, which can be difficult to control and may require extensive training \citep{birbaumer1999spelling, farwell1988talking, lebedev2006brain, wolpaw2013brain}. To provide improvements in BCI applications we wanted to tackle both the fundamental decoding methods \citep{lotte2018review}, and offer new ways of using BCIs, such as inner speech \citep{martin2018decoding}.

\section{Thesis outline}

One of the main challenges in decoding brain signals is the variability and complexity of these signals \citep{saha2020intra}. Multivariate pattern analysis (MVPA) of MEG and EEG data can be a valuable tool for understanding how the brain represents and discriminates between different stimuli \citep{guggenmos2018multivariate, king2014characterizing}. However, traditional decoding models, such as linear, pairwise, sliding window decoding models, can be computationally intensive and may have limited decoding performance \citep{higgins2022relationship, higgins2022spatiotemporally}. These models typically focus on identifying the spatial and temporal signatures of stimuli, but they may not fully capture the complex patterns of brain activity during visual and language tasks. In contrast, full-epoch decoding models, commonly used for BCI applications, can provide better decoding performance but lack methods to interpret the contributions of spatial and temporal features \citep{haufe2014interpretation,lotte2018review}. To address these challenges, we propose an approach that combines a multiclass, full epoch decoding model with supervised dimensionality reduction. This approach allows us to reveal the contributions of spatiotemporal and spectral features using permutation feature importance, while achieving higher decoding accuracy than traditional sliding window decoders. 

Moving to multi-subject datasets will require decoding methods to be able to deal with high amounts of between-subject variability \citep{varoquaux2017assessing, poldrack2009decoding}. Decoding is typically subject-specific and does not generalise well over subjects \citep{olivetti2014meg, dash2020decoding-imagined}. Naive group modelling approaches have been proposed where a single model is trained on the data from multiple subjects. Due to high amounts of between-subject variability these methods typically perform much worse than subject-dependent modelling \citep{olivetti2014meg, li2021inter, saha2020intra}. Techniques that can overcome this will not only provide richer neuroscientific insights but also make it possible for group-level models to outperform subject-specific models. Here, we propose a method that uses subject embedding, analogous to word embedding in Natural Language Processing (NLP) \citep{Mikolov:2013f}, to learn and exploit the structure in between-subject variability as part of a decoding model. We apply this method to MEG data from a visual task and show that the combination of deep learning and subject embedding can close the performance gap between subject and group-level decoding models.

The final kind of variability lies in the types of datasets collected and experimental conditions used. If we truly want to leverage big data in a brain decoding context, then we will have to deal with this additional layer of variability. Despite the potential of deep learning techniques in decoding brain signals \citep{schirrmeister2017deep}, there is little work on training large unsupervised models on brain data and then fine-tuning these for specific decoding tasks \citep{kostas2021bendr}. This approach has seen massive success in the deep learning field, for various kinds of data, e.g. images, language, audio \citep{Devlin:2019, Krizhevsky:2012, Hinton:2012}. We hypothesise that Wavenet \citep{oord2016wavenet} and Transformer \citep{Vaswani:2017} models can more accurately predict future timesteps than a linear model, and that these models can capture the spectral properties and long-range spatiotemporal dynamics of the data more accurately. Thus, they serve as solid foundation models to deal with between-dataset and between-task variability. We also propose that pre-trained forecasting models can be used to improve downstream decoding performance, much like in NLP \citep{Radford:2018}.

We hope the aforementioned methodological advancements can contribute to improving BCI systems. However, we also wanted to ask whether faster BCI communication can be achieved with previously untapped modalities, such as inner speech \citep{brumberg2010development}. Inner speech refers to the inner voice inside the head that governs our thoughts, specifically when thoughts take the form of language \citep{morin2005possible, alderson2015inner}. Despite the prevalence of inner speech in everyday life, research on this has been limited, particularly when it comes to non-invasive methods \citep{panachakel2021decoding, dash2020decoding-imagined}. Our proof of concept work aims to fill this gap by using EEG and MEG to collect data from three different inner speech paradigms and by conducting an initial decoding analysis. We aim to investigate the decoding performance of inner speech in EEG and MEG with a large number of per-participant trials, the transferability of decoders across sessions and tasks, and the comparison of OPM decoding performance and spatiotemporal dynamics to EEG and MEG.

In conclusion, besides analysing a new modality for BCI communication, inner speech, we address the various types of variability that hinder BCI decoding performance and applicability. We aim to leverage large datasets and deep learning to deal with within-participant, between-participant, and between-dataset variability.

The remainder of this thesis is organised as follows:
\begin{itemize}
    \item[] \textbf{Chapter} \hyperref[Chap2]{\textbf{2}} introduces key concepts in electrophysiological data processing and modelling. This includes signal processing, the use of machine learning in unsupervised modelling, encoding, and decoding, and methods for interpreting such models.
    
    \item[] \textbf{Chapter} \hyperref[Chap3]{\textbf{3}} introduces several solutions for unifying the fields of multivariate pattern analysis and brain computer interface decoding, mainly focusing on linear full-epoch multiclass models and the uncovering of spatiotemporal and spectral information through permutation feature importance. This chapter is part of a published paper \citep{csaky2023interpretable}.
    
    \item[] \textbf{Chapter} \hyperref[Chap4]{\textbf{4}} presents a new method termed subject embedding to deal with between-subject variability in group-level decoding models. It investigates how this subject embedding and deep learning contribute to better group-level modeling. This chapter is part of a published paper \citep{csaky2023group}.
    
    \item[] \textbf{Chapter} \hyperref[Chap5]{\textbf{5}} introduces deep learning methods for unsupervised modelling (forecasting) of MEG data. It is shown that specifically Transformer-based models are capable of accurately generating the spatiotemporal dynamics of real data. We investigate how these capabilities arise through a series of ablations.

    \item[] \textbf{Chapter} \hyperref[Chap6]{\textbf{6}} presents our proof of concept work on inner speech with EEG and MEG. We discuss our experimental pipeline, data collection, data analysis, and decoding results, offering new insights into the decoding of inner speech. Preliminary OPM data is also presented along with a comparison of decoding performance with more standard modalities.

    \item[] \textbf{Chapter} \hyperref[Chap7]{\textbf{7}} discusses the implications of our findings and possible future directions for this research and the field of non-invasive brain decoding.

\end{itemize}

\chapter{Modelling and decoding electrophysiology}
\label{Chap2}

\section{Machine learning}
Machine learning (ML) refers to the automated discovery of models from data. Instead of manually specifying model parameters based on prior knowledge, as with conventional modelling approaches, ML algorithms learn (infer) these parameters directly from data. By learning from data, ML models can capture complex patterns and relationships that may be difficult to a priori specify. Thus, they are well suited to the complexities of high-dimensional whole-brain electrophysiology data. Compared to more biologically inspired modelling, ML is abstract and does not necessarily explain the underlying physical phenomena. For predictive power and abstraction, we are trading interpretability. This has enabled breakthroughs across various domains, from computer vision \citep{Krizhevsky:2012} to natural language processing \citep{Vaswani:2017}.

\subsection{Key components}

At its core, ML comprises four key components: (1) data, (2) model specification, (3) learning objective, and (4) learning algorithm \citep{richards2019deep}. Carefully considering the interplay between these components is crucial for successfully applying ML. We discuss each of these in turn below.

\paragraph{Data} The data used for training is the primary basis for everything an ML model learns. For neuroimaging applications, this typically comprises multivariate time series data reflecting brain activity across multiple spatial locations. Data quality and curation is thus critical. Brain data is often accompanied by synchronised behavioural or task-related data. Electrophysiology data exhibits substantial within-subject variability from noise and artefacts \citep{bigdely2015prep}, as well as between-subject variability in anatomical and functional characteristics \citep{bijsterbosch2018relationship}. Capturing this variability sufficiently during training is key for the model to generalise well across time and subjects. Preprocessing to clean, normalise and extract relevant signals is also important.

\paragraph{Model} The model defines the computational architecture relating inputs to outputs. Choosing an appropriate model class is guided by domain expertise about expected relationships in the data, as well as trade-offs between flexibility, interpretability and trainability. For example, linear models have limited flexibility but are simple and interpretable. Deep neural networks are extremely flexible functional approximators \citep{Hornik:1989} but lack interpretability. Recent work has shown the efficacy of deep networks for learning spatiotemporal relationships from neuroimaging data \citep{gohil2022mixtures}.

\paragraph{Learning objective} The objective quantifies the model's performance on a desired task, providing feedback to drive learning. Objectives may balance various constraints like accuracy, interpretability, and biological plausibility. For example, in regression tasks, a common objective is minimising the mean squared error between predicted and true outputs. In classification, maximising accuracy or minimising cross-entropy loss are typical. Additional objectives can be added to impose further constraints, e.g. regularisation \citep{gohil2022mixtures}.

\paragraph{Learning algorithm} The algorithm optimises model parameters (iteratively) to improve the objective. Algorithms like (stochastic) gradient descent \citep{bottou2010large}, genetic algorithms \citep{goldberg1991comparative}, and reinforcement learning \citep{Sutton:1998} have proven effective for various models and tasks. Factors like scalability, speed of convergence, and avoidance of local optima guide algorithm selection. For deep networks, backpropagation with stochastic gradient descent underpins most state-of-the-art approaches \citep{Goodfellow:2016}.

Together, these components define the ML modelling approach. In the following, we discuss key concepts and strategies for effectively applying ML to brain data.

\subsection{Fundamentals}
ML offers a data-driven approach for uncovering structure in complex, multivariate brain data. However, simply throwing large models and datasets at a problem does not guarantee success. Thoughtfully considering the bias-variance trade-off, model capacity, regularisation, and cross-validation strategies is important for robust, generalisable models \citep{hastie2009elements}.

Perhaps the most important component is the learning objective as this prescribes our goals in mathematical form. Learning objectives can often also be used as a direct metric of goodness, i.e. how well the model accomplishes the prescribed task. An important concept in arriving at such an assessment is the bias-variance trade-off, which provides us with a statistical framework \citep{vapnik1999nature}. From a probabilistic point of view, data represents some probability distribution. Thus, machine learning can be framed as coming up with a model that best captures the training data distribution. Often, we want this model either to be as simple as possible, or to generalise to examples which are not part of the training data. These examples can be either within or outside of the training distribution, prescribing different levels of generalisation \citep{geirhos2018generalisation}.

\begin{figure}[!t]
  \centering
  \includegraphics[width=1.0\linewidth]{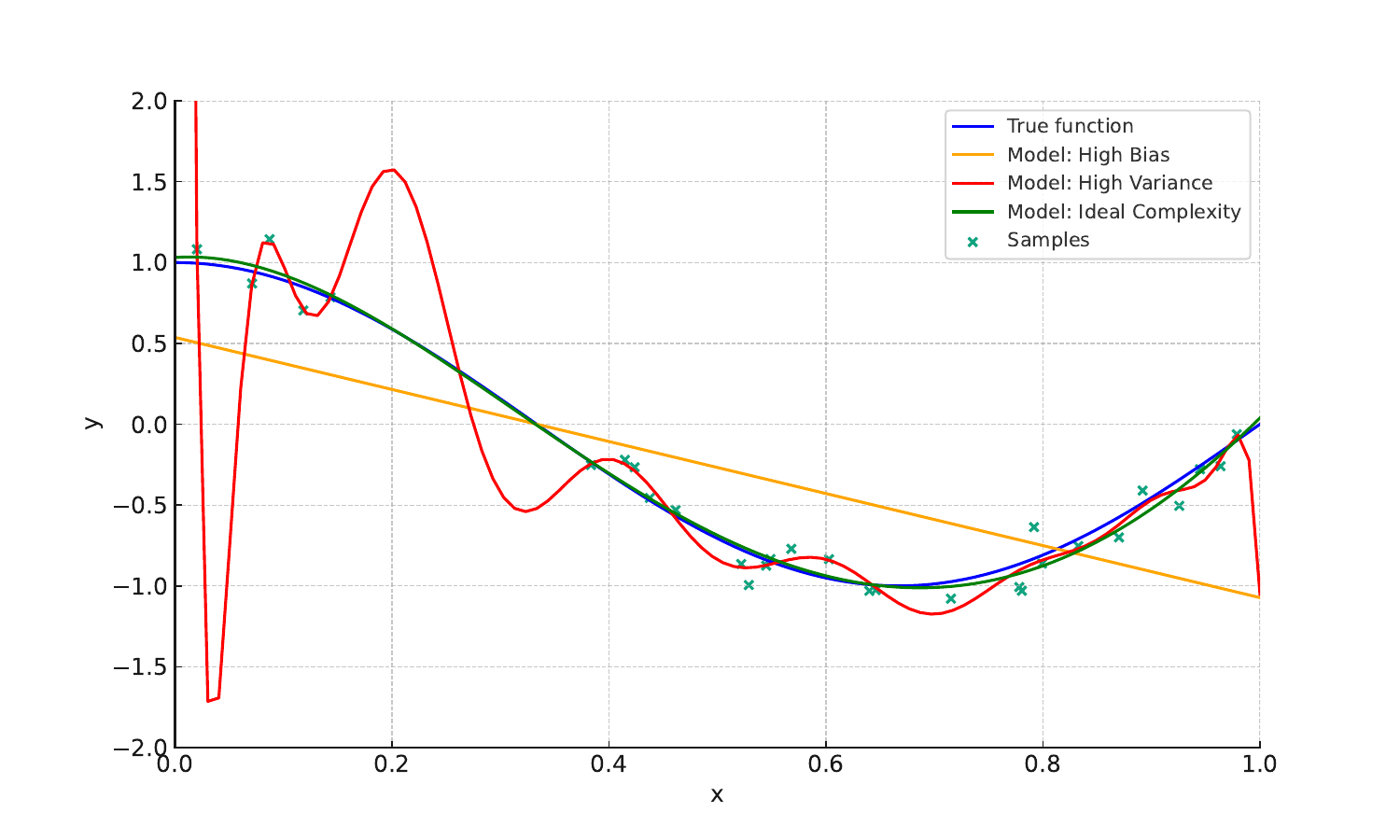}
    \caption{Visualisation of the bias-variance trade-off through a polynomial regression problem. The high-degree polynomial (red) fits the noise in our training samples (green cross) and thus has high variance. In contrast the high-bias linear model (orange) cannot capture the underlying distribution. The optimal model (green) achieves a trade-off between bias and variance and fits the true function (blue) well, however some differences may remain due to the irreducible noise in our samples.}
    \label{fig:bias_variance}
\end{figure}

\paragraph{Bias-variance trade-off} The bias-variance trade-off represents the opposing goals of fitting the training data well (variance error) while being able to generalise to new examples (bias error). Models with insufficient capacity are prone to high bias. Overly complex models overfit the noise in the training data, causing high variance. The ideal model balances bias and variance while also minimising irreducible error from noise \citep{geman1992neural}. This trade-off is visualised through a simple regression problem in Figure~\ref{fig:bias_variance}.

\paragraph{Overfitting} Overfitting happens when the model complexity is so high that it can model the noise as well as the signal in the training data \citep{ying2019overview}. Noise means that the samples in our training data are not fully representative of the true underlying distribution we are trying to model. This is sometimes a consequence of not having the right data, i.e. EEG signals are affected by various noise sources, and always a consequence of not having enough data. Fully defining the underlying distribution would require an infinite set of examples, but our training data is always a subsample of this.

\begin{figure}[!t]
  \centering
  \includegraphics[width=1.0\linewidth]{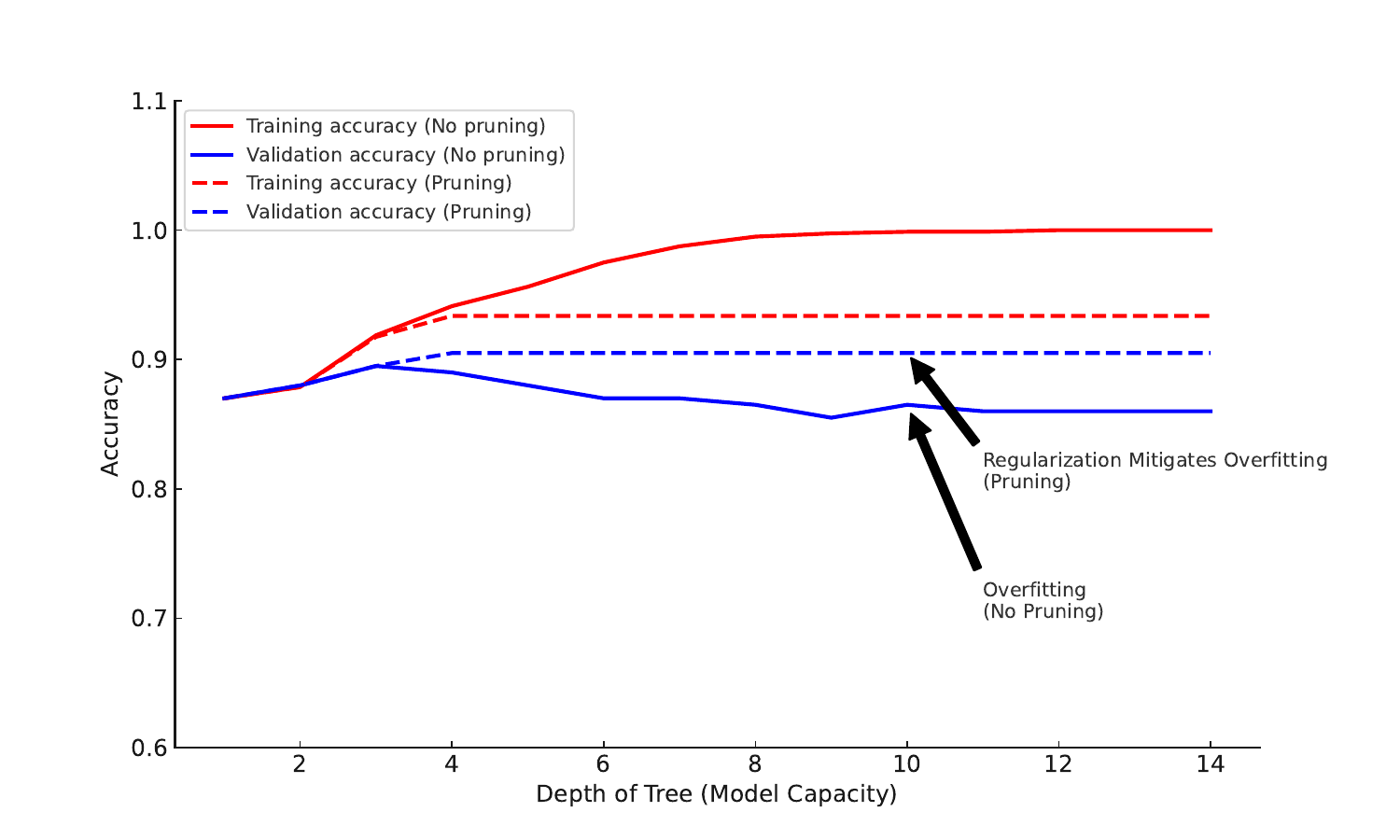}
    \caption{Visualisation of generalisation, overfitting, model capacity, and regularisation on a toy decision tree classification problem. As model capacity increases (depth of tree), both train and validation (generalisation) accuracy improve, up to a certain point after which overfitting occurs, and generalisation performance worsens. When adding regularisation (pruning) to the models (dashed lines), overfitting can be avoided even with larger models.}
    \label{fig:overfitting}
\end{figure}

\paragraph{Model capacity} To deal with noise, overfitting and generalisation, in practise we must employ certain assumptions about the underlying distributions, or collect better and more data. Capacity reflects the model's ability to fit diverse functions. High capacity can improve the fit to training data, but risks overfitting. Capacity depends on factors such as number of parameters and nonlinearity \citep{belkin2019reconciling}. Deep neural networks derive immense capacity through multiple nonlinear layers \citep{raghu2017expressive}. But this needs to be carefully controlled, often requiring large training sets. Simpler linear models have limited capacity. Model selection aims to find the optimal capacity for a given dataset size. The effects of model capacity, overfitting, and regularisation are shown in Figure~\ref{fig:overfitting}.

\paragraph{Regularisation} Constraining model complexity through regularisation or early stopping helps to control variance. Simplicity can be achieved by limiting the number of parameters a model employs, putting mathematical constraints such as linearity, or other forms of regularisation. The latter discourages overfitting by penalising model complexity \citep{Goodfellow:2016}. $\ell_2$ regularisation adds a penalty proportional to the sum of squared parameters to the objective. This shrinks parameters toward zero, effectively constraining the capacity. Dropout randomly omits subsets of activations in neural networks during training as a form of implicit regularisation \citep{Srivastava:2014a}. Other approaches limit parameter ranges or enforce smoothness. The degree of regularisation is tuned to balance under and overfitting.

\paragraph{Cross-validation} Rigorously testing ML models on held-out data is key to ensuring that they generalise beyond the training set. Cross-validation divides the data into training and test sets multiple times to assess performance across different partitions \citep{arlot2010survey}. More sophisticated validation schemes include nested cross-validation for hyperparameter tuning and leaving out entire subjects for evaluating generalisability across individuals.

These are important considerations in our quest for models that can deal with the various types of variability inherent to electrophysiology data. Variability within an individual arises from noise contamination, and we would like to design models capable of generalising to similar events across time. Variability between individuals arises from subtle differences in data distributions, necessitating the need for capturing the underlying (more complex) distribution over multiple brains. Variability over datasets and tasks again expands the distribution of electrophysiological data in non-obvious ways.

\subsection{Model categories}

Various categorisations of modelling approaches can be made based on the learning paradigm, model flexibility, model form, and nature of predictions. These categorisations provide a useful framework for selecting appropriate techniques for modelling brain data.

In terms of learning paradigm, models can be \textit{supervised}, \textit{unsupervised}, or \textit{self-supervised} \citep{Goodfellow:2016}. In supervised learning, the model is trained on a set of input-output pairs with the goal of learning a mapping from inputs to outputs. In contrast, unsupervised learning involves only inputs, with the aim of uncovering latent structure such as clusters or factors \citep{ghahramani2003unsupervised}. Self-supervised learning employs inputs to generate proxy labels that are then used for doing supervised learning. For instance, autoregressive modelling uses past brain activity to predict future activity \citep{oord2018representation}.

Another categorisation considers model \textit{flexibility}: \textit{linear} versus \textit{nonlinear} models. Linear models assume a linear relationship between inputs and outputs, while nonlinear models make no such assumption. Although nonlinear models are more flexible and can capture complex relationships, they risk overfitting and reduced interpretability.

\textit{Generative} models learn the joint distribution $p(x,y)$ over data $x$ and target labels $y$, and can synthesise new data samples. Linear discriminant analysis (LDA) is a simple linear generative model. \textit{Discriminative} models learn the conditional distribution $p(y|x)$ to predict outputs from inputs. Generative models (e.g., LDA) can also be employed for prediction in a classification task, by applying Bayes' theorem to the learned joint probability distribution \citep{murphy2012machine}. Neural networks are most often formulated as discriminative models. Discriminative modelling is advantageous when the relationship between inputs and outputs is well-defined and there is abundant amount of data to avoid overfitting. However, with increasing amounts of noise, variability, and data scarcity, modelling the (generative) joint probability may provide useful assumptions and constraints.

Together, these categorisations delineate the extensive range of modelling techniques applicable to brain data. Careful consideration of the properties of the data and scientific question can guide selection of suitable approaches. In the following sections we detail common machine learning architectures and their application to M/EEG data.

\section{Electrophysiological data analysis}

\subsection{Data characteristics}

Electroencephalography (EEG) and magnetoencephalography (MEG) are two of the most prevalent non-invasive techniques for measuring brain activity with high temporal resolution. EEG electrodes placed on the scalp detect microvolt-level electrical potentials generated by synchronised postsynaptic potentials in neural populations \citep{teplan2002fundamentals}. Standard clinical EEG uses the 10–20 system with 19 electrodes, while high-density EEG utilises up to 256 electrodes to achieve higher spatial resolution \citep{fiedler2022high}. Typical EEG frequency bands include delta (1-4 Hz), theta (4-8 Hz), alpha (8-12 Hz), beta (12-30 Hz), and gamma ($>$30 Hz). Different bands have been linked to various cognitive and behavioural states. For instance, alpha waves reflect relaxed or idle cortical states whereas gamma activity is involved in active processing \citep{wang2010neurophysiological}.

MEG detects femtotesla-level magnetic fields induced by postsynaptic currents, providing millisecond temporal resolution. It uses superconducting quantum interference devices (SQUIDs) to detect the minute neuromagnetic fields emanating from the head \citep{baillet2017magnetoencephalography}. MEG sensor arrays typically contain around 300 sensors housed in a liquid helium Dewar. Magnetically shielded rooms are required to attenuate environmental magnetic noise and enable MEG systems to detect the brain's weak magnetic signals \citep{hamalainen1993magnetoencephalography}. EEG and MEG offer complementary information, with MEG more sensitive to tangential sources and EEG to radial sources \citep{baillet2017magnetoencephalography}. Both modalities provide millisecond temporal resolution critical for tracking rapid neural dynamics.

Recently, optically pumped magnetometers (OPMs) have emerged as a novel MEG sensor technology. OPM-MEG systems utilise an array of OPM sensors that can be placed directly on the scalp, providing higher sensitivity and spatial resolution compared to traditional SQUID sensors \citep{boto2018moving}.

\subsection{Typical preprocessing steps}
\label{ssec:prepoc_pipeline}

Standard preprocessing of electrophysiological data removes noise and artefacts while retaining brain signals. Bandpass filtering eliminates slow drifts below 0.1 Hz from skin potentials and high frequencies above 100 Hz containing muscle noise. Downsampling then reduces the data rate after lowpass filtering. This reduces computation time and the number of features when applying machine learning models to the timeseries. Notch filtering targets removal of 50/60 Hz power line noise and harmonics that can obscure lower amplitude brain signals \citep{widmann2015digital}. Narrow stopbands centered on the noise frequencies are applied, e.g. 59-61 Hz to remove 60 Hz.

Noisy channels and segments are identified and repaired or discarded. Bad channel detection utilises statistical thresholds (e.g., too much or too little variance) to find excessively noisy channels for interpolation from surrounding good channels or removal. In EEG a potential cause for bad channels can be high impedance or bad contact with the scalp. Eye blinks, muscle activity, and motion yield large artefacts detected via amplitude, gradient, variance or visual inspection \citep{fatourechi2007emg}. Thresholds should retain valid data while excluding only clear artefacts.

\begin{figure}[!t]
\begin{subfigure}{0.3\textwidth}
  \centering
  \includegraphics[width=1.0\linewidth]{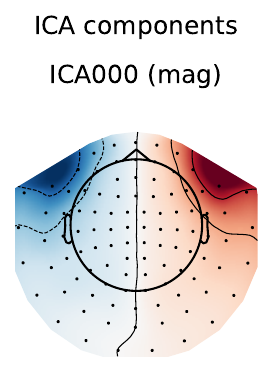}
  \caption{Topographic map of eye blink component}
  \label{fig:eog_topo_example}
\end{subfigure}%
\begin{subfigure}{0.7\textwidth}
  \centering
  \includegraphics[width=1.0\linewidth]{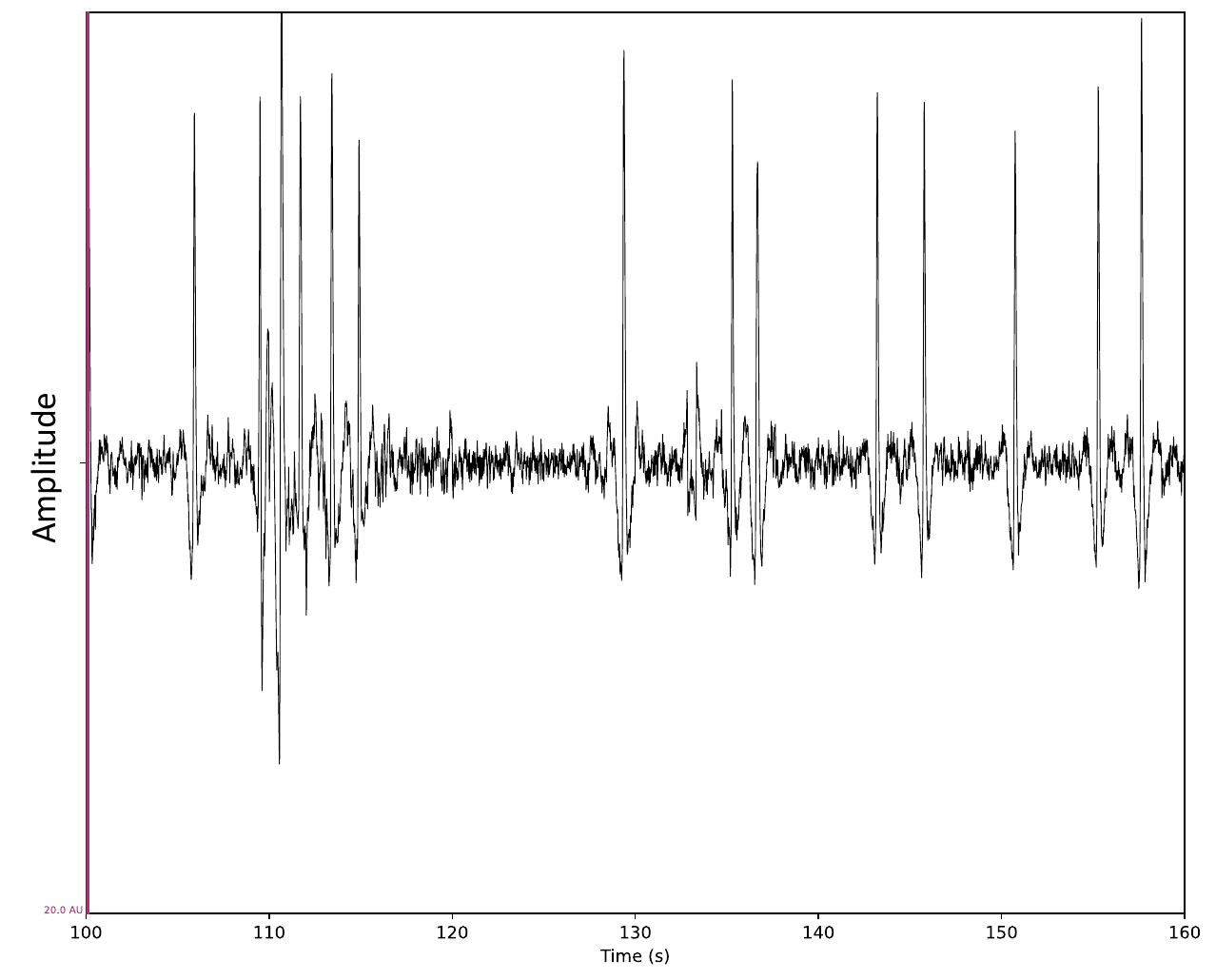}
  \caption{Timeseries of eye blink component}
  \label{fig:eog_timeseries_example}
\end{subfigure}
\caption{An independent component matching eye blink signatures from the ICA decomposition of a MEG recording. Eyeblinks materialise in frontal lateral channels (left) as short large amplitude deviations (right).}
\label{fig:eog_example}
\end{figure}

\begin{figure}[!t]
\begin{subfigure}{0.3\textwidth}
  \centering
  \includegraphics[width=1.0\linewidth]{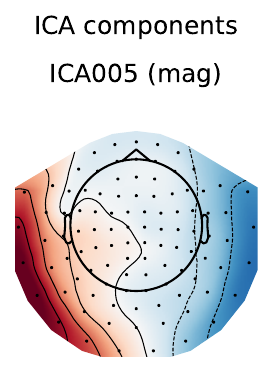}
  \caption{Topographic map of heartbeat component}
  \label{fig:ecg_topo_example}
\end{subfigure}%
\begin{subfigure}{0.7\textwidth}
  \centering
  \includegraphics[width=1.0\linewidth]{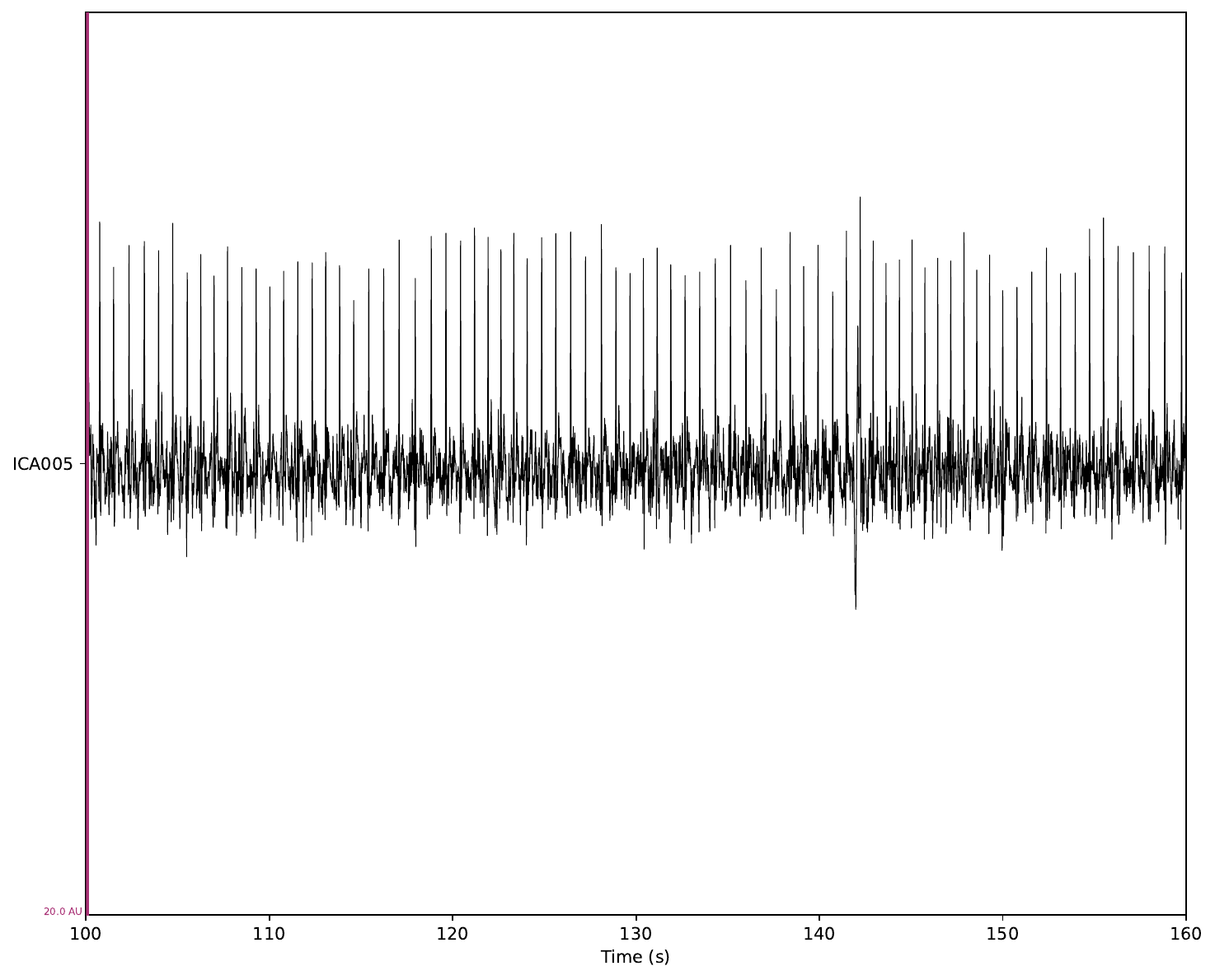}
  \caption{Timeseries of heartbeat component}
  \label{fig:ecg_timeseries_example}
\end{subfigure}
\caption{An independent component matching heartbeat signatures from the ICA decomposition of a MEG recording. Heartbeats show lateral spatial activity (left) and have a consistently repeating high-amplitude pulse-like timeseries. (right)}
\label{fig:ecg_example}
\end{figure}

Independent component analysis (ICA) decomposes the sensor data $X$ into a set of maximally independent components $S$, assuming a linear relationship between the components and the observed signal $X$ \citep{jung2000removing}. For further details see Section~\ref{ssec:pca_ica}. Components corresponding to artefacts like eye blinks or heart beats can be identified from their spatial, temporal, and spectral signatures and removed before reconstructing the data. This cleans artefacts while preserving brain signals. Specifically, components matching eye blinks have corresponding spatial maps with activity focused in frontal lateral channels (Figure~\ref{fig:eog_example}), while heartbeat components exhibit a characteristic repeating temporal profile around 1Hz (Figure~\ref{fig:ecg_example}).

While the preprocessing steps mentioned so far are sufficient for this thesis, often an additional step is mapping sensor-space data to brain sources. Volume conductor modelling creates a head model specifying conductivities of the brain, skull, and surrounding layers for accurate EEG/MEG source localisation \citep{vorwerk2014guideline}. Realistic models built from subject MRIs can optimise localisation accuracy and spatial interpretation. Source reconstruction enables better alignment with underlying brain geometry and is an active research area (see \cite{timms2022time} for an in-depth review).

\begin{figure}[!t]
\begin{subfigure}{0.5\textwidth}
  \centering
  \includegraphics[width=1.0\linewidth]{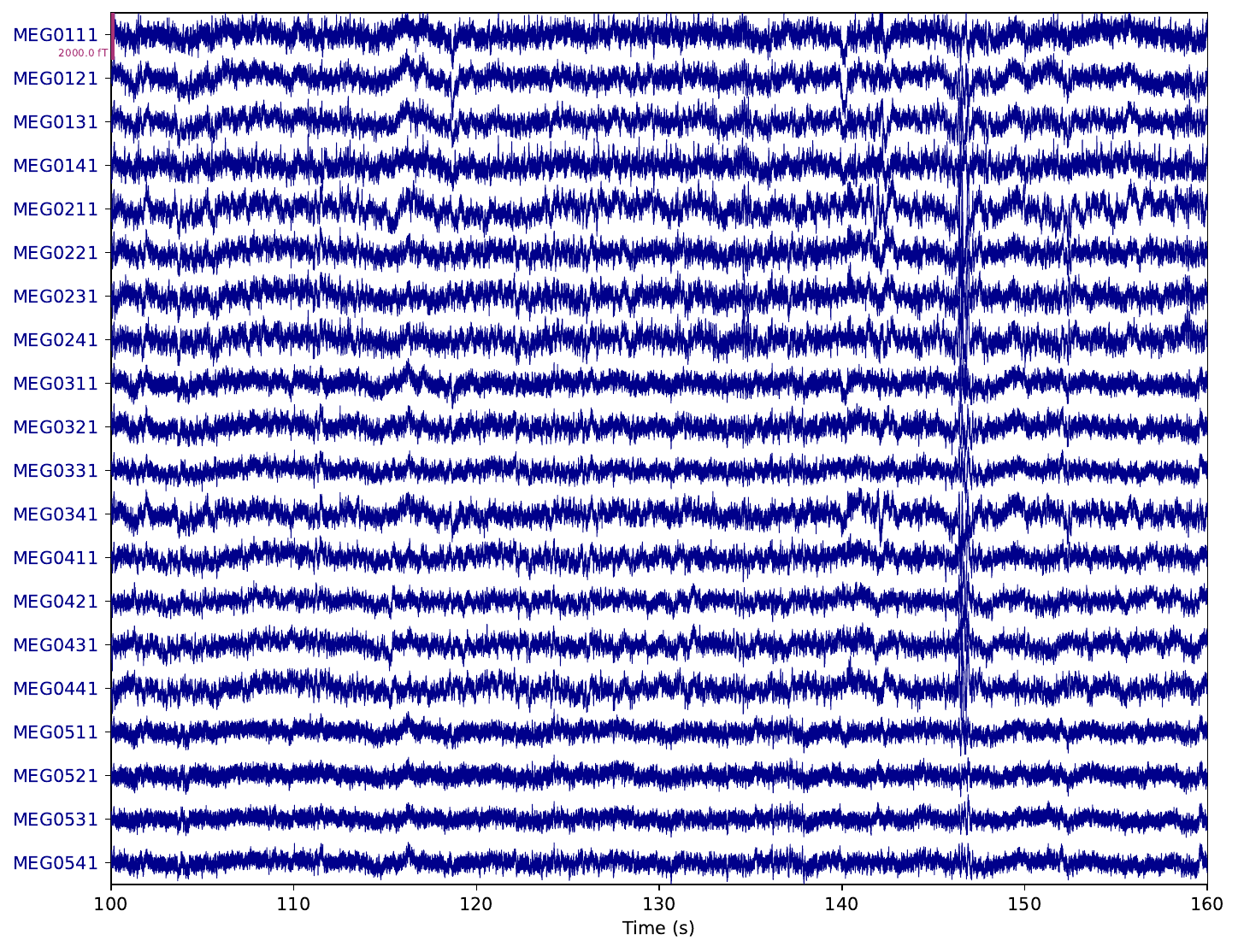}
  \caption{raw MEG timeseries}
  \label{fig:before_clean_example}
\end{subfigure}%
\begin{subfigure}{0.5\textwidth}
  \centering
  \includegraphics[width=1.0\linewidth]{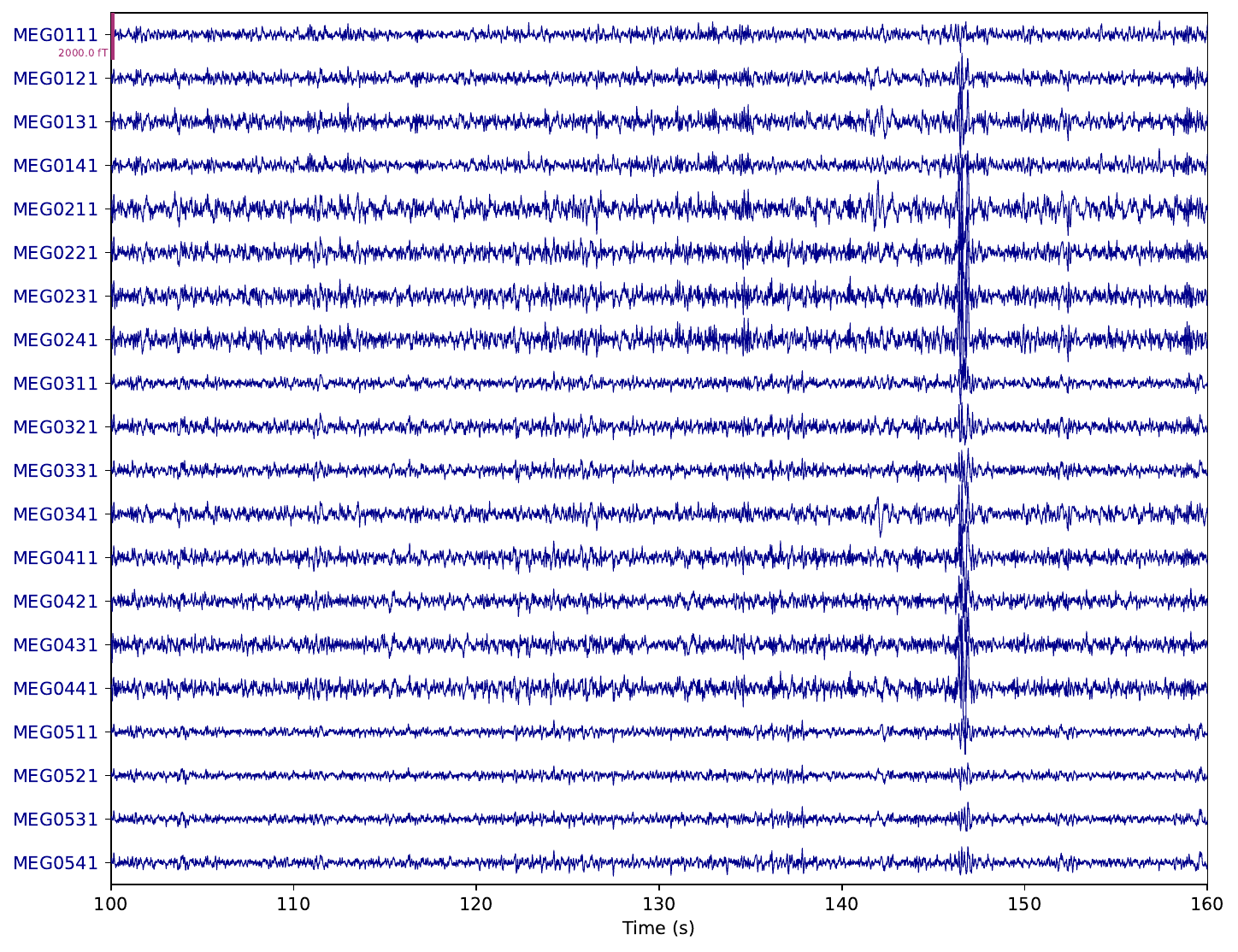}
  \caption{preprocessed MEG timeseries}
  \label{fig:after_clean_example}
\end{subfigure}
\caption{Comparing 20 MEG channels before (left) and after (right) running a typical preprocessing pipeline. The data is less noisy, and low and high-frequency activity has been removed. Some artefacts may remain.}
\label{fig:before_after_example}
\end{figure}

In summary, these preprocessing steps clean electrophysiological data, remove artefacts, and prepare multichannel timeseries for further analysis. Careful preprocessing improves signal quality and enhances the fidelity of subsequent analytic approaches. However, most of these steps can only be applied offline, and thus have limited use in BCI systems. A visual comparison of the MEG timeseries before and after applying the aforementioned preprocessing steps is presented in Figure~\ref{fig:before_after_example}.

\subsection{Analysis methods}
\label{ssec:analysis_methods}

Analyzing electrophysiological data presents unique challenges due to the high-dimensional, multi-channel time series nature of EEG and MEG recordings. In MEG and EEG, a clear 10 Hz oscillation appears that can be observed in real-time recordings when a subject closes their eyes, and gross sleep phases can also be identified in the signal by eye \citep{schulz2008rethinking}. In more general cases the neuronal signals are stochastic and unintelligible to human eyes. This section reviews common methods to interrogate M/EEG data by examining spatial, temporal, or spectral properties. These techniques provide an initial foothold when confronted with new electrophysiological recordings.

Power spectral analysis decomposes the signal into constituent frequencies using Fourier-based methods \citep{cohen2014analyzing}. The power spectral density (PSD) quantifies power within frequency bands of interest. Specifically, the power $P$ at frequency $f$ is computed from the Fourier transform $X(f)$ as:

\begin{equation}
P(f) = \lvert X(f) \rvert^2
\end{equation}

\begin{figure}[!t]
  \centering
  \includegraphics[width=1.0\linewidth]{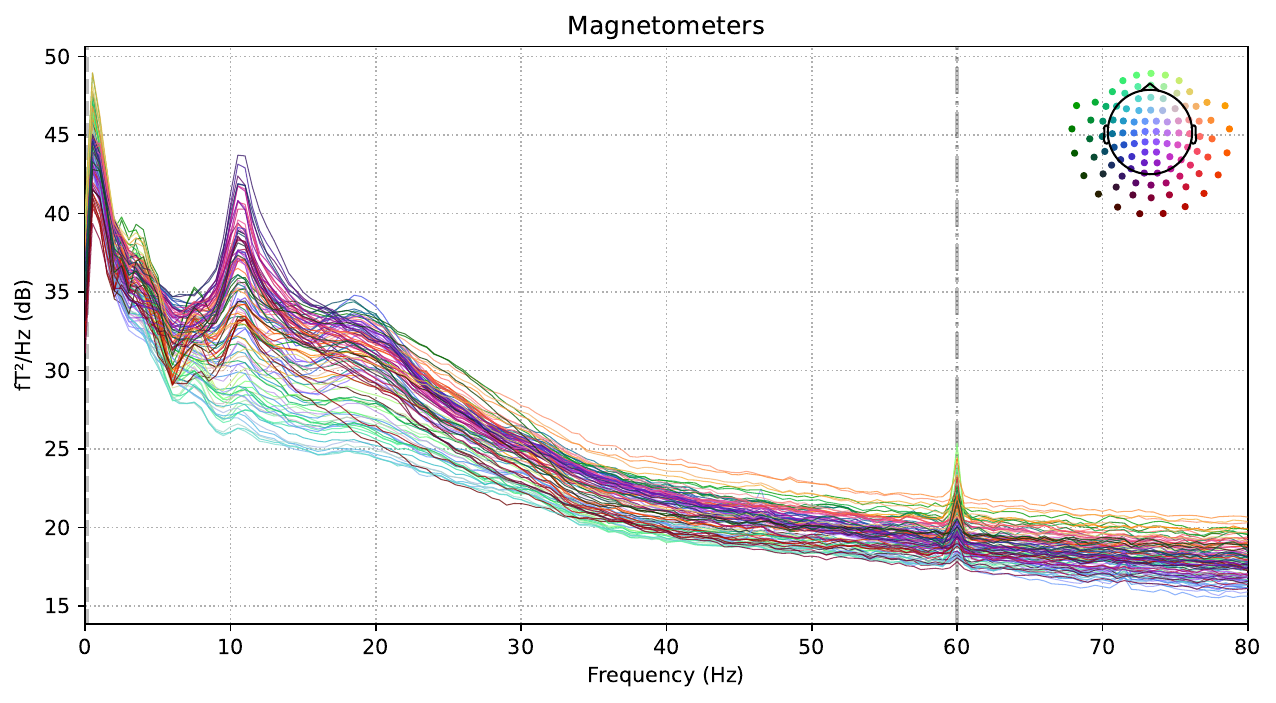}
    \caption{PSD of a typical MEG recording. Each line represents a separate channel. The $1/f$ shape is apparent, as well as a prominent peak around 10 Hz, and a sharp peak at 60 Hz (power line noise).}
    \label{fig:spectra_example}
\end{figure}

Band power in canonical delta, theta, alpha, beta, and gamma ranges can be compared across experimental conditions \citep{buzsaki2006rhythms}. Typical M/EEG spectra follow a $1/f$ distribution with peaks at \textasciitilde10 and \textasciitilde20 Hz, reflecting dominant alpha and beta rhythms in the awake brain \citep{nunez2000toward}. A typical MEG power spectra is shown in Figure~\ref{fig:spectra_example}.

Time-frequency analysis such as wavelet and Hilbert transforms reveal spectral dynamics over time \citep{link2002meg}. This elucidates task-related changes in oscillatory activity, like suppression or enhancement of alpha during visual processing \citep{klimesch2007eeg}. Topographic mapping of the sensor-level PSD highlights the spatial signature of different rhythms. For instance, alpha power localised to visual cortex decreases during visual tasks \citep{bruers2018alpha}. Intrinsic brain networks, like the default mode, also exhibit characteristic spatial spectral patterns \citep{mantini2007electrophysiological}.

For stimulus-driven experiments, evoked responses are obtained by averaging short data segments time-locked to each event \citep{luck2014introduction}. Assuming similar brain responses across trials, this averaging cancels non-stimulus-dependent ongoing brain activity and noise while retaining time-locked signals. Studying peak latencies and amplitudes of visual, auditory, or cognitive components (e.g. P100, N200, P300) provides insights into perceptual and cognitive processes.

\begin{figure}[!t]
  \centering
  \includegraphics[width=1.0\linewidth]{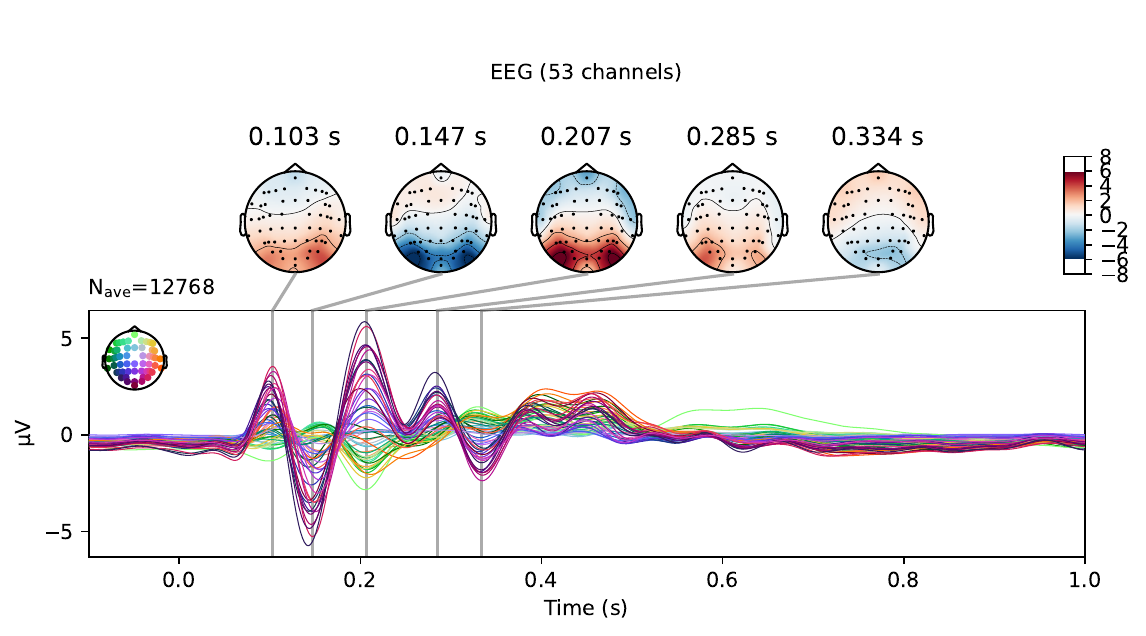}
    \caption{Spatiotemporal evoked activity from a visual EEG dataset. The evoked response can be observed as soon as 100 ms after stimulus presentation, followed by several peaks in the timeseries. Topographic maps show that the evoked response appears in channels over the visual area of the brain.}
    \label{fig:eeg_evoked_ex}
\end{figure}

Topographic mapping visualises the spatial distribution of brain activity on the scalp \citep{tzovara2012tutorial}. Combined with evoked analysis, this reveals spatiotemporal dynamics (Figure~\ref{fig:eeg_evoked_ex}). For better spatial localisation, inverse solutions like minimum norm estimation and beamforming combine sensor data with head models \citep{baillet2001electromagnetic}. This allows sensor data to be linked to the brain regions from which signals originate.

To summarise, non-invasive brain activity can be investigated in terms of temporal, spatial and spectral signatures. We can observe each aspect in isolation, e.g. the average PSD across all channels at timepoints, the average evoked response across all channels, or the mean topographic map across all timepoints. Alternatively, the data can be decomposed into pairs such as temporal-spatial, temporal-spectral, and spatial-spectral, or even across all three dimensions. This latter view would correspond to plotting the wavelet transform as a function of time and space, a topographic PSD map evolving in time. These analyses and views allow various neuroscientific investigations, in both resting-state and task-related brain activity.

Functional connectivity (FC) analyses statistical relationships between the activity in different brain regions, such as coherence \citep{sakkalis2011review}, and is therefore most straightforwardly carried out in source space. A simpler metric is channel covariance, which reveals spatial relationships in the data. Examining time-varying covariance provides insights into ongoing and stimulus-driven whole-brain dynamics \citep{vidaurre2018spontaneous, gohil2022mixtures}. Discussed next, modelling covariance structure using generative models provides useful ways to understand M/EEG data.

In conclusion, electrophysiology provides millisecond insights into human brain function. Typical preprocessing removes noise and artefacts enabling further analysis like spectral decomposition, source localisation, functional connectivity, and evoked response characterisation for understanding brain dynamics and evaluating data quality.

\section{Unsupervised modelling}

EEG and MEG provide rich temporal information about brain dynamics. However, these signals represent an aggregate measure of the activity from many neurons and are challenging to interpret. Unsupervised machine learning techniques, such as Principal Component Analysis (PCA), Independent Component Analysis (ICA), and autoregressive (AR) modelling, have proven useful for extracting meaningful representations of neural dynamics from M/EEG \citep{makeig_applications_1999}.

The main challenge these techniques address is the reduction of the high-dimensional (channels × time points) and noisy raw data into a lower-dimensional, human-interpretable latent space. This section and the next will focus on common machine learning techniques used for analysing electrophysiological data. The last section of this chapter will then describe how these models can be leveraged to understand brain activity in more complex ways than the basic signal processing and statistical methods discussed in Section~\ref{ssec:analysis_methods}.

\subsection{PCA and ICA}
\label{ssec:pca_ica}

As mentioned in Section~\ref{ssec:prepoc_pipeline}, ICA can be used to remove non-brain artefacts from M/EEG data. To better understand ICA, it is useful to first introduce Principal Component Analysis (PCA). PCA and ICA are commonly used for dimensionality reduction and source separation of M/EEG recordings \citep{hyvarinen_independent_2001, delorme_eeglab:_2004}.

Let us assume we have M/EEG data in the form of a matrix $\mathbf{X} \in \mathbb{R}^{C \times T}$, with $C$ channels and $T$ time points. PCA finds a set of orthogonal basis vectors that capture directions of maximum variance in the data. Let $\mathbf{U} = [\mathbf{u}_1,...,\mathbf{u}_D] \in \mathbb{R}^{C \times D}$ be the PCA basis vectors, with $D \leq C$. The PCA decomposition is given by:

\begin{equation}
\mathbf{X} = \mathbf{UDV}
\end{equation}

where the rows of $\mathbf{V} \in \mathbb{R}^{D \times T}$ give the PCA component time courses and $\mathbf{D} \in \mathbb{R}^{D \times D}$ is a diagonal matrix containing the eigenvalues. PCA provides a compressed representation of the data by retaining only the top $K < D$ components that explain the most variance. The choice of $K$ is a trade-off between capturing as much variability in the data as possible while maximally reducing the feature space dimension. Using the matrix $\mathbf{U}$ we can map the decomposed data back to the original channel space. Metrics for evaluating a PCA decomposition include reconstruction error and percentage of variance explained in the original channel space data.

Unlike PCA, ICA aims to find statistically independent latent sources $\mathbf{S} \in \mathbb{R}^{N \times T}$ underlying the observed data:

\begin{equation}
\mathbf{X} = \mathbf{AS} 
\end{equation}

where $\mathbf{A} \in \mathbb{R}^{C \times N}$ is the mixing matrix. Common ICA algorithms include infomax \citep{bell_information-maximization_1995} and FastICA \citep{hyvarinen_fast_1999}.

ICA can better isolate neural and artefact components compared to PCA. The estimated ICA sources do not always have a clear mapping to underlying cortical generators \citep{makeig_applications_1999}. Source localisation, which maps sensor data to cortical sources, requires more sophisticated techniques than PCA or ICA. Thus, these methods all provide complementary information. Due to its data-driven nature, PCA is often used for feature reduction before training machine learning models like Hidden Markov Models or linear classifiers for decoding tasks. By retaining key components explaining most variance, PCA helps mitigate the \textit{curse of dimensionality} when working with limited electrophysiological data.

Typically, PCA and ICA are used to reduce spatial dimensionality while retaining the full temporal resolution. However, we may also want to reduce the temporal (potentially together with spatial) dimensionality, and have a coarser representation of the temporal dynamics. Let us explore such models in the next section.

\subsection{Hidden Markov Models}

To achieve temporal dimensionality reduction, a simple solution is downsampling. However, this can remove useful high-frequency signal. Another option for temporal dimensionality reduction is to classify time points into a limited set of classes, or states. The simplest way to achieve this is by using sliding windows. First, a window of length $T$ (usually 100-200 ms) is slid along the multivariate time series data with a step size of 1 time step. Within each window, statistics like the mean, power spectra, or local (in time) covariance can be derived. Finally, K-means clustering \citep{hartigan1979algorithm} can then be applied on the window statistics (e.g. covariance) to arrive at a discrete set of high-level states governing the dynamics.

States are useful for temporal reduction since empirically their average lifetime is around 100-200 ms \citep{baker2014fast}. Using a small set of states (e.g. 10-20) to explain long recordings acts as a bottleneck, summarising the salient recurring patterns. Studies show state statistics like covariance, power spectral density, and connectivity correspond to identifiable networks and activation patterns during both rest and tasks \citep{vidaurre_discovering_2018}.

\begin{figure}[!t]
\centering
\includegraphics[width=1.0\linewidth]{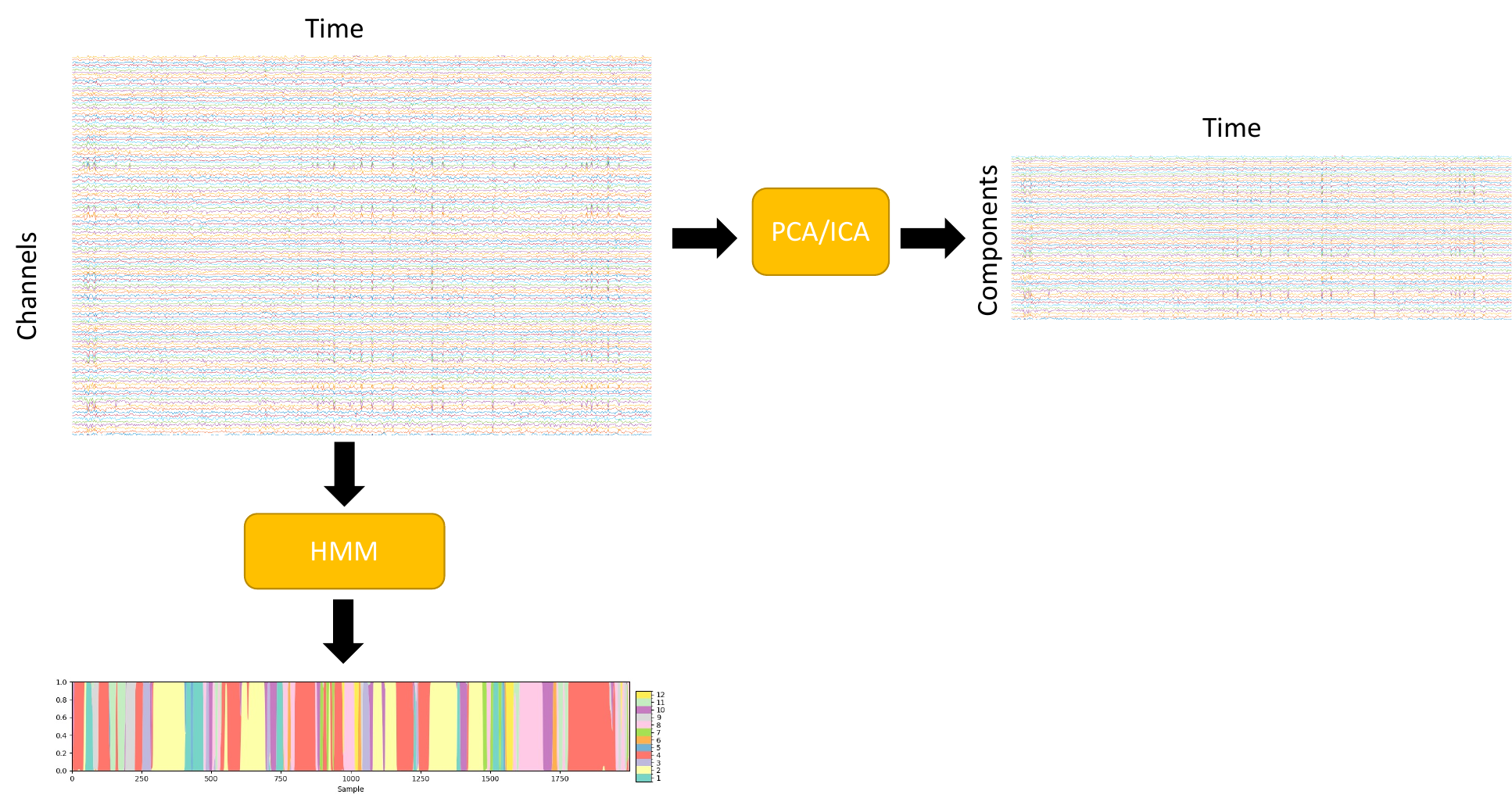}
    \caption{The original high-dimensional (channels $\times$ time) M/EEG data can be transformed with PCA/ICA to reduce spatial dimensionality. The HMM can be used to infer a set of states governing brain dynamics. The state time course provides both a spatial and temporal dimensionality reduction.}
\label{fig:hmm_ica_pca}
\end{figure}

However, using sliding windows to find states has limitations. An alternative is to directly learn the states and model state switching dynamics in a data-driven manner. Hidden Markov models (HMMs) are well-suited for modelling sequential data and discovering recurring patterns \citep{rabiner_tutorial_1989}. A conceptual visualisation of HMM, and PCA/ICA for temporal and spatial reduction is given in Figure~\ref{fig:hmm_ica_pca}.

In an HMM, the observed multivariate time series $\mathbf{x}_1,...,\mathbf{x}_T$ is assumed to be generated by an underlying sequence of hidden states $z_1,...,z_T$ with Markov dynamics:

\begin{align}
\label{eq:markov}
p(z_t | z_{t-1}, ..., z_1) &= p(z_t | z_{t-1}) \\  
p(\mathbf{x}_t | z_t, \mathbf{x}_{t-1}, ..., \mathbf{x}_1) &= p(\mathbf{x}_t | z_t)
\end{align}

The HMM is parameterized by initial state distribution $\pi$, transition matrix $\mathbf{A}$, and observation model parameters $\theta = \mathbf{B}, \boldsymbol{\mu}, \mathbf{\Sigma}$:

\begin{align}
\pi_i &= p(z_1 = i) \\
\mathbf{A}_{ij} &= p(z_t = j | z_{t-1} = i) \\
\mathbf{B}_i &= p(\mathbf{x}_t | z_t = i) \sim \mathcal{N}(\boldsymbol{\mu}_i, \mathbf{\Sigma}_i)
\end{align}

\begin{figure}[!t]
\centering
\begin{tikzpicture}[
    node distance=20mm,
    every node/.style={align=center},
    box/.style={rectangle,draw,minimum width=20mm,minimum height=10mm},
    arrow/.style={->, >=latex, thick}
]

\node[box] (zt_1) at (0,0) {$z_{t-1}$};
\node[box, right=of zt_1] (zt) {$z_{t}$};
\node[box, right=of zt] (zt1) {$z_{t+1}$};

\node[below=of zt_1] (xt_1) {$\mathbf{x}_{t-1}$};
\node[below=of zt] (xt) {$\mathbf{x}_{t}$};
\node[below=of zt1] (xt1) {$\mathbf{x}_{t+1}$};

\draw[arrow] (zt_1) -- node[midway,above] {$\mathbf{A}$} (zt);
\draw[arrow] (zt) -- node[midway,above] {$\mathbf{A}$} (zt1);

\draw[arrow] (zt_1) -- node[midway,right] {$\boldsymbol{\mu}_{z_{t-1}},\mathbf{\Sigma}_{z_{t-1}}$} (xt_1);
\draw[arrow] (zt) -- node[midway,right] {$\boldsymbol{\mu}_{z_{t}},\mathbf{\Sigma}_{z_{t}}$} (xt);
\draw[arrow] (zt1) -- node[midway,right] {$\boldsymbol{\mu}_{z_{t+1}},\mathbf{\Sigma}_{z_{t+1}}$} (xt1);

\draw[arrow] ([xshift=-15mm]zt_1.west) -- node[midway,above] {$\pi$} (zt_1.west);
\draw[arrow] (zt1.east) -- ([xshift=15mm]zt1.east);

\end{tikzpicture}

\caption{A graphical illustration of the HMM. See text for parameters.}
\label{fig:hmm_graph}
\end{figure}
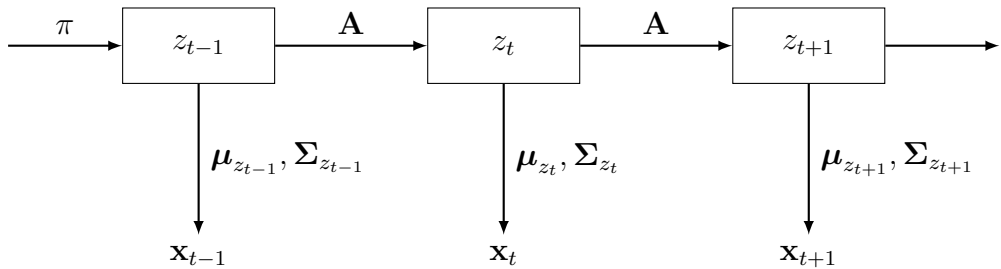

The transition matrix gives state switching probabilities. The observation model is usually a Gaussian with learned mean and covariance per state. Importantly, the spatial dimension is retained as $\mathbf{\Sigma}_i \in \mathbb{R}^{C \times C}$, where $C$ is number of channels. Thus, the HMM provides a high-level state description while able to reproduce the data through the observation model (Figure~\ref{fig:hmm_graph}). To be clear the state time course is both a spatial and temporal reduction of the original high-dimensional data. HMM parameters can be estimated from the data using the Baum-Welch algorithm \citep{baum1970maximization}. The Viterbi path gives the optimal hidden state sequence \citep{forney1973viterbi}. Note that the HMM is a linear model, and the Gaussian observation model can be quite limiting. More realistic observation models like mixtures of Gaussians have been proposed in the literature \citep{bilmes1998gentle}, but these also usually complicate the inference process.

To aid interpretation, state covariances $\mathbf{\Sigma}_i$ can be visualised to characterise recurring brain patterns \citep{vidaurre_discovering_2018}. Section \ref{sec:interpretability_methods} provides more details on interpreting HMMs. More powerful deep generative models can also replace HMMs, as will be discussed in Chapter \ref{Chap5}.

\subsection{Linear autoregressive models}

As we have seen the HMM is an unsupervised generative sequential model well-suited for M/EEG data. Another method of self-supervision is sequential (timeseries) forecasting, where the aim is to model the conditional probability of the future given the past. Autoregressive (AR) models are a powerful class of self-supervised generative models well-suited for forecasting. AR models aim to directly model the conditional probability distribution of future timesteps given the past, i.e. $p(\mathbf{x}_t|\mathbf{x}_{t-1}, \mathbf{x}_{t-2},...,\mathbf{x}_{t-P})$ where $\mathbf{x}_t \in \mathbb{R}^C$ is a multivariate time series with $C$ channels at time $t$. This formulation allows AR models to learn complex temporal dynamics from the data in a fully unsupervised manner.

The AR modelling approach shares similarities with Hidden Markov Models (HMMs), with a couple key distinctions. First, AR models do not enforce a low-dimensional latent state space, instead operating directly on the observed data. Second, AR models do not in general make the Markov assumption, allowing dependence on multiple past timesteps rather than just the previous state (Equation \ref{eq:markov}).

More formally, the log-likelihood \(\log p(\mathbf{x}_t|\mathbf{x}_{t-1},...,\mathbf{x}_{t-p};\theta)\) is maximised with respect to model parameters $\theta$. Using Gaussian noise assumptions, this is equivalent to minimising the squared error between predictions and targets:

\begin{equation}
    \underset{\theta}{\argmin}(\mathbf{x}_t - f(\mathbf{x}_{t-1},\mathbf{x}_{t-2}...,\mathbf{x}_{t-P}; \theta))^2
\end{equation}

The simplest AR model specification uses a linear function for $f$:

\begin{equation}
\mathbf{x}_t = \sum_{i=1}^P \mathbf{A}_i \mathbf{x}_{t-i} + \boldsymbol{\epsilon}_t  
\end{equation}

where $\mathbf{A}_i \in \mathbb{R}^{C \times C}$ are the autoregressive coefficients at time lag $i$ and $\boldsymbol{\epsilon}_t \sim \mathcal{N}(0, \mathbf{\Sigma})$ is Gaussian noise. $P$ controls the model order, which determines the length of temporal memory or receptive field. $C$ is the number of input channels. To be precise, this is known as a multivariate AR (MAR) model, as it captures cross-channel interactions \citep{schlogl2006analyzing}. A simpler (univariate) approach is to avoid modelling the channel interactions in $\mathbf{A}_i$ by letting it be a scalar for each channel (\( \mathbf{A}_i \in \mathbb{R}^{C}\)). This decouples the conditional dependence between channels, as the conditional probability $ p(x_{t,c}|x_{t-1,c},...,x_{t-P,c})$ is modelled with a separate univariate model for each channel $c$. AR models can be fit via ordinary least squares.

MAR models are able to capture linear temporal autocorrelations as well as cross-channel relationships in M/EEG data. They can be interpreted from a signal processing lens as infinite impulse response filters applied to the input \citep{takalo2005tutorial}. This enables analysing model dynamics in the frequency domain using tools from spectral analysis.

A key advantage of MAR models is the ability to generate new data recursively for any length by feeding back previous outputs. However, linearity remains a limitation, which motivates exploring nonlinear AR models in Chapter \ref{Chap5}. We conclude this section by briefly introducing basic nonlinear models.

\subsection{Neural network autoregressive models}
\label{ssec:nn_ar}

Neural networks are powerful function approximators that have proven effective for time series modelling and forecasting. They comprise multiple layers of nonlinear transformations with learned parameters that enable extracting hierarchical features from the input data. Fully-connected neural networks with sufficiently wide hidden layers can approximate any continuous function, as established by the universal approximation theorem \citep{hornik1989multilayer}. However, architectural constraints like weight sharing and recurrence are often incorporated to exploit structure in sequential data. Fully connected networks consist of affine transformations followed by nonlinear functions, such as the sigmoid or ReLU function \citep{nair2010rectified}, stacked on top of each other. To learn the parameters of such models, the backpropagation algorithm \citep{Rumelhart:1985} is used to differentiate an objective function (e.g. sum of squared errors), and an optimisation algorithm such as stochastic gradient descent is used to minimise this function.

Convolutional neural networks (CNNs) leverage local spatial or temporal correlations through weight sharing across space/time \citep{fukushima1982neocognitron}. They comprise convolutional layers that convolve input representations with learned kernels followed by nonlinearities. This equates to extracting finite impulse response (FIR) filters whose coefficients are optimised via gradient descent. The learned kernels act similarly to the autoregressive coefficients in modelling temporal dependencies. Through hierarchical feature extraction, deep CNNs can construct complex spatiotemporal representations \citep{Krizhevsky:2012,chambon_deep_2018}. The weights of a MAR model are equivalent to a CNN with a single layer without any nonlinearities.

For an M/EEG input $\mathbf{X} \in \mathbb{R}^{C \times T}$, a CNN with $L$ layers produces an output representation $\mathbf{H}^{(L)} \in \mathbb{R}^{M \times T}$:

\begin{align}
\mathbf{H}^{(l+1)} &= [\mathbf{h}_1^{(l+1)}; \mathbf{h}_2^{(l+1)}; ... ; \mathbf{h}_{M^{(l+1)}}^{(l
+1)}] \\
\mathbf{h}_{m}^{(l+1)} &= f^{(l)}(\mathbf{b}_{m}^{(l)} +\sum_{i=1}^{M^{(l)}} \mathbf{w}_{m, i}^{(l)}*\mathbf{h}_{ i}^{(l)}) \\
\mathbf{H}^{(0)} &= \mathbf{X}
\end{align}

where $[;]$ denotes concatenation across the channel dimension, $*$ is convolution, $\mathbf{W}_{m}^{(l)}$, $\mathbf{b}_{m}^{(l)}$ are the learned weights/biases corresponding to output channel $m$ in layer $l$. Each layer can have a different nonlinearity $f^{(l)}$, such as ReLU. $M^{(l)}$ and $M^{(l+1)}$ are the number of input/output channels in layer $l$.

By concatenating across channels we can concisely denote the full bias and weight tensors of layer $l$ by $\mathbf{B}^{(l)} \in \mathbb{R}^{M^{(l+1)} \times M^{(l)}}$ and $\mathbf{W}^{(l)} \in \mathbb{R}^{M^{(l+1)} \times M^{(l)} \times K^{(l)}}$ where $K^{(l)}$ is the kernel size in layer $l$. For forecasting, the CNN can be trained to minimise error between model outputs $\mathbf{H}^{(L)}$ and future timesteps. Thus, the target labels are the inputs shifted by 1 timestep.

Stacking layers enables hierarchical feature learning. Temporal downsampling can be achieved via strided convolution or pooling layers like max pooling. This simply takes a time window as input and outputs the maximum value from this window, sliding over the timeseries. Architectural innovations like dilated convolutions \citep{oord2016wavenet} also enable expanding the receptive field to capture longer-range dependencies. CNNs are widely used for end-to-end M/EEG modelling \citep{schirrmeister_deep_2017,chambon_deep_2018}.

Recurrent neural networks (RNNs) are the nonlinear counterpart to HMMs. They comprise recurrent connections that enable maintaining a state over time and modelling long-term temporal contexts \citep{Rumelhart:1986a}. An RNN takes inputs $\mathbf{X} \in \mathbb{R}^{C \times T}$ and outputs a hidden state $\mathbf{h}_t$ and optional outputs $\mathbf{y}_t$ per timestep:

\begin{align}
\mathbf{h}_t &= f(\mathbf{U} \mathbf{x}_t + \mathbf{W} \mathbf{h}_{t-1} + \mathbf{b}) \\
\mathbf{y}_t &= g(\mathbf{S} \mathbf{x}_t + \mathbf{V} \mathbf{h}_t + \mathbf{c}) 
\end{align}

Here, $f$ and $g$ are nonlinearities like Gated Recurrent Unit (GRU) or Long Short-Term Memory (LSTM) cells \citep{hochreiter_long_1997}. $\mathbf{U}$, $\mathbf{W}$, $\mathbf{S}$, $\mathbf{V}$ are learned projections and $\mathbf{b}$, $\mathbf{c}$ are biases. The fixed weights at each timestep coupled with recurrent state enables modelling complex dynamics. The matrix $\mathbf{S}$ corresponds to skip connections as the hidden state is bypassed in the information flow from inputs to outputs. In standard RNN formulations $\mathbf{S}$ is not used. By removing the nonlinearities and bias terms one can recognise a linear state-space system in these equations. Indeed RNN dynamics may be studied in terms of a dynamical state-space system when certain assumptions are met.

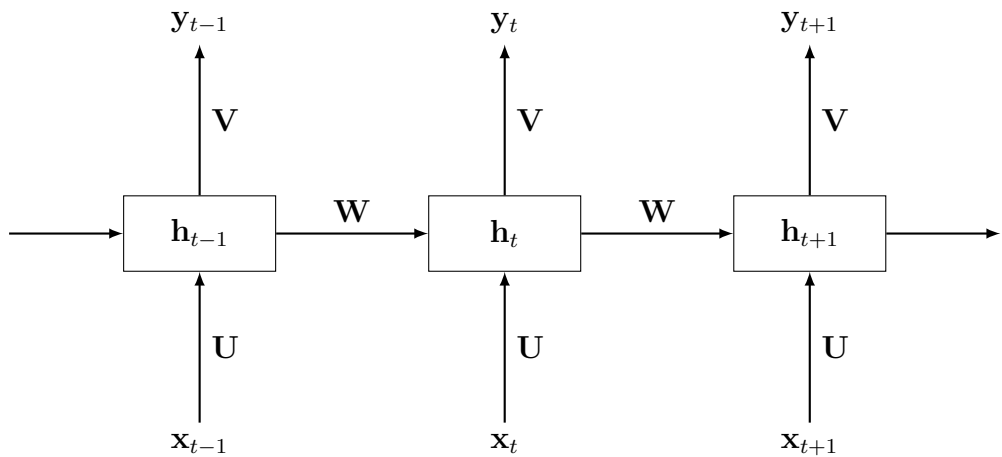
\begin{figure}[!t]
\centering

\begin{tikzpicture}[
    node distance=20mm,
    every node/.style={align=center},
    box/.style={rectangle,draw,minimum width=20mm,minimum height=10mm},
    arrow/.style={->, >=latex, thick}
]

\node[box] (ht_1) at (0,0) {$\mathbf{h}_{t-1}$};
\node[box, right=of ht_1] (ht) {$\mathbf{h}_{t}$};
\node[box, right=of ht] (ht1) {$\mathbf{h}_{t+1}$};

\node[below=of ht_1] (xt_1) {$\mathbf{x}_{t-1}$};
\node[below=of ht] (xt) {$\mathbf{x}_{t}$};
\node[below=of ht1] (xt1) {$\mathbf{x}_{t+1}$};

\node[above=of ht_1] (yt_1) {$\mathbf{y}_{t-1}$};
\node[above=of ht] (yt) {$\mathbf{y}_{t}$};
\node[above=of ht1] (yt1) {$\mathbf{y}_{t+1}$};

\draw[arrow] (ht_1) -- node[midway,above] {$\mathbf{W}$} (ht);
\draw[arrow] (ht) -- node[midway,above] {$\mathbf{W}$} (ht1);

\draw[arrow] (xt_1) -- node[midway,right] {$\mathbf{U}$} (ht_1);
\draw[arrow] (xt) -- node[midway,right] {$\mathbf{U}$} (ht);
\draw[arrow] (xt1) -- node[midway,right] {$\mathbf{U}$} (ht1);

\draw[arrow] (ht_1) -- node[midway,right] {$\mathbf{V}$} (yt_1);
\draw[arrow] (ht) -- node[midway,right] {$\mathbf{V}$} (yt);
\draw[arrow] (ht1) -- node[midway,right] {$\mathbf{V}$} (yt1);

\draw[arrow] ([xshift=-15mm]ht_1.west) -- (ht_1.west);
\draw[arrow] (ht1.east) -- ([xshift=15mm]ht1.east);

\end{tikzpicture}

\caption{Graphical illustration of the recurrence in an RNN layer. Nonlinearities, biases, and the projection $\mathbf{S}$ are omitted.}
\label{fig:rnn_graph}
\end{figure}

RNNs can model multivariate AR dynamics by predicting future values of the timeseries. We can enforce this through the objective function (e.g. mean-squared error) applied to output $\mathbf{y}_t$:
\begin{align}
\mathrm{MSE}(\mathbf{y}_t,\mathbf{x}_{t+1}) &= \frac{1}{C} \sum_{i=1}^C (y_{t,i} - {x}_{t+1,i})^2
\end{align}

While the above description is accurate for a 1-layer RNN, these operations can be stacked on top of each other by feeding the output representations $\mathbf{y}_t$ as input to the next layer. Multiple layers allow deeper feature extraction/representation capabilities. CNNs and RNNs may look somewhat similar, however a CNN layer is state-less and exploits weight-sharing across time, whereas an RNN layer applies the same operation at each timestep while carrying a continually updating hidden state (Figure~\ref{fig:rnn_graph}). In practice several extensions have been proposed to both types of architectures, even combining them into a single model \citep{bashivan2015learning}. These and other, more recent modelling approaches (such as Transformers) will be explored in Chapter~\ref{Chap5}.

\section{Encoding and decoding}

In the previous section, we have seen how unsupervised modelling of M/EEG data can uncover intrinsic brain dynamics. While these models elucidate spontaneous neural processes, they do not account for external stimuli and behaviour. By incorporating such external events, we can study the relationship between brain dynamics and the outside world. This is also necessary for applying our models in brain-computer interfaces (BCIs).

\begin{figure}[!t]
\centering
\includegraphics[width=1.0\linewidth]{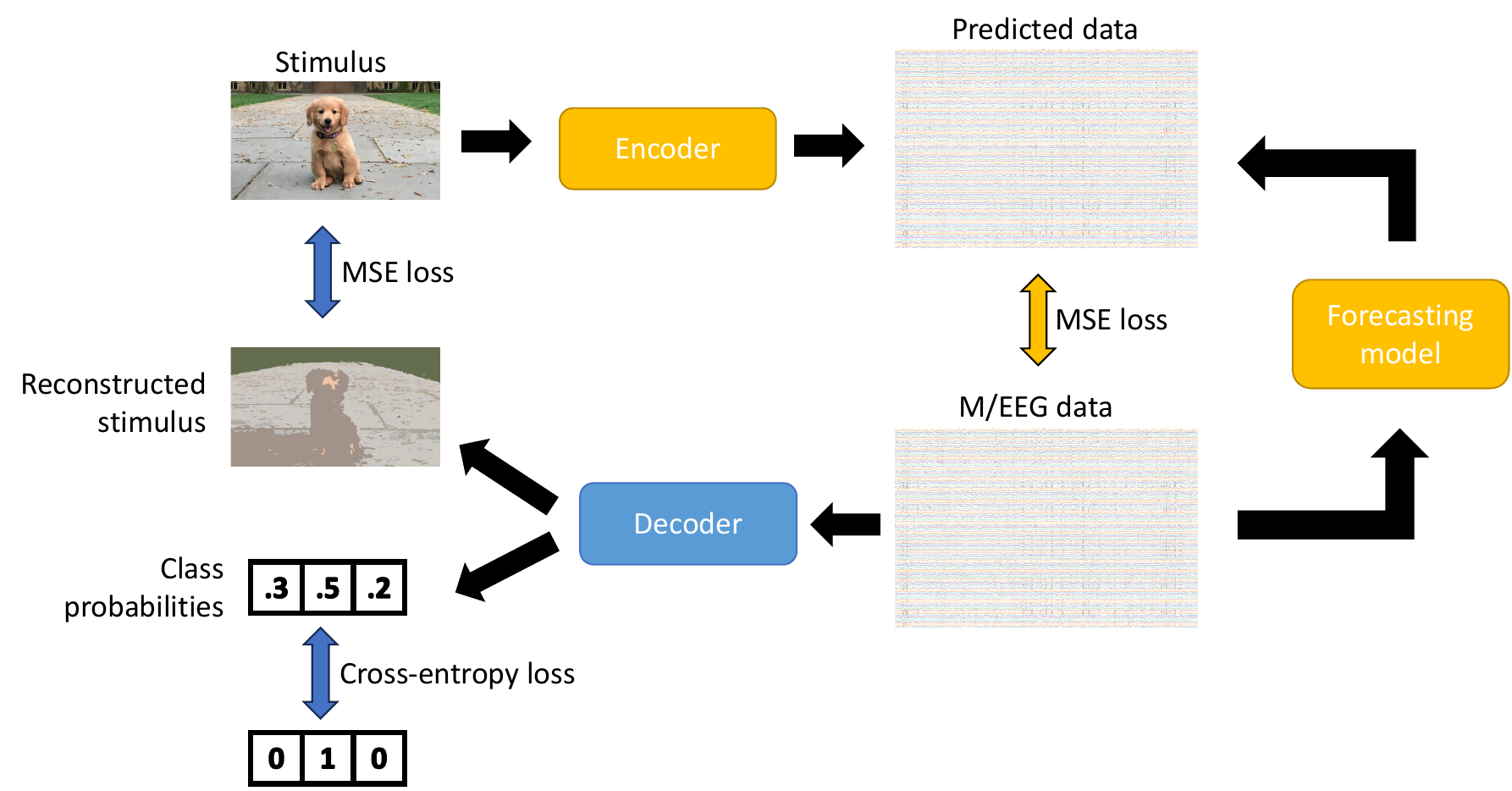}
    \caption{Conceptual comparison of forecasting, encoding, and decoding. A typical forecasting model is fed some brain data to predict future timesteps and trained through the MSE loss. A typical encoder is similar to the forecasting model except that it is fed the stimulus. A typical decoder is similar to the forecasting model except that it has to either reconstruct the stimulus (trained with MSE loss), or predict the stimulus class, trained with the cross-entropy loss. Note that each modelling approach may involve standard feature extraction steps, and thus map features to outputs, instead of raw stimuli or brain data.}
\label{fig:encoder_decoder}
\end{figure}

If we conceptualise the brain as a dynamic system that receives inputs (e.g., visual, auditory) from the environment and generates outputs (e.g., movement, emotion, speech) there are several approaches for investigating input-output relationships. In this thesis, we specifically focus on links between external inputs and resultant brain dynamics. This relationship can be studied bidirectionally, termed encoding and decoding. For a high-level visualization comparing forecasting, encoding, and decoding see Figure~\ref{fig:encoder_decoder}.

Encoding refers to predictive modelling of brain activity evoked by stimuli. Such models elucidate how external inputs are processed and represented in the brain \citep{naselaris2011encoding}. Decoding models predict, classify, or reconstruct stimuli based on elicited brain activity, providing insights into neural representations and importantly enabling BCI applications. Key considerations in encoding models include accurately mapping temporally evolving, stimulus-evoked responses and their spatial propagation across regions. Decoders can also infer these representational dynamics. For BCIs, harnessing knowledge of evoked spatiotemporal dynamics can inform input constraints.

An important aspect of both encoding and decoding is the level of generalisation. As discussed, M/EEG variabilities necessitate particular forms of generalisation. External stimuli pose an additional dimension. The simplest case is predicting responses to identical stimuli in new trials. Increased difficulty arises in generalising to unseen stimuli from the training distribution (e.g., novel dog images). Further generalisation may predict responses to any within-distribution sample (any image). The generalisation levels can be similarly formulated for decoding.

For a thorough introduction to electrophysiological encoding and decoding, \citet{holdgraf2017encoding} provides an excellent resource to interested readers.

\subsection{Encoding}

Linear encoding characterises relationships between stimuli or task conditions $\mathbf{s}_t \in \mathbb{R}^{K}$ and resultant M/EEG response $\mathbf{x}_t \in \mathbb{R}^{C}$ at timepoint $t$:

\begin{equation}
\mathbf{x}_t = \mathbf{W}_t\mathbf{s}_t + \boldsymbol{\epsilon}_t
\end{equation}

where $\mathbf{W}_t$ is a weight matrix and $\boldsymbol{\epsilon}_t \sim \mathcal{N}(0,\mathbf{\Sigma})$ is zero-mean Gaussian noise with covariance $\mathbf{\Sigma}$. Noise can be static or time-varying. $K$ is stimulus feature dimensionality. Weights can also be static or dynamic over time. For static weights, the overall M/EEG response $\mathbf{X} \in \mathbb{R}^{C \times T}$ can be predicted by:

\begin{equation}
\label{eq:linear_encoding}
\mathbf{X} = \mathbf{W}\mathbf{S} + \boldsymbol{\epsilon}
\end{equation}

where $\mathbf{S} \in \mathbb{R}^{K \times T}$ contains the stimulus features at each timepoint. This models the conditional distribution $p(\mathbf{X}|\mathbf{S})$. Given $N$ labelled trials $(\mathbf{S}_n, \mathbf{X}_n)$, linear encoding is fit by least squares if the error is assumed to have a symmetric distribution:

\begin{equation}
\hat{\mathbf{W}} = \underset{\mathbf{\mathbf{W}}}{\mathrm{argmin}} \sum_{n=1}^N | \mathbf{X}_n - \mathbf{WS}_n|_2^2
\end{equation}

minimising squared error between predicted and actual responses across trials.

Once fit, encoding models can be analysed to study neural representations. Weights $\mathbf{w}_c = \mathbf{W}[c,:]$ for channel $c$ indicate its selectivity for stimulus features in $\mathbf{S}$. Spatial mapping of weight magnitudes localises brain areas selective for different features. Comparing weights across channels and stimuli reveals distributed sensory representations.

Linear encoding has limited flexibility in capturing complex relationships between stimuli and responses. Simplicity also restricts insights into nonlinear neural computations/dynamics. Nonlinear approaches, such as neural networks, provide greater modelling flexibility for encoding. A multi-layer fully-connected neural network can be formulated as:

\begin{align}
\mathbf{h}^{(1)} &= f^{(1)}(\mathbf{W}^{(1)}\mathbf{s} + \mathbf{b}^{(1)}) \\
\mathbf{h}^{(l+1)} &= f^{(l)}(\mathbf{W}^{(l+1)} \mathbf{h}^{(l)} + \mathbf{b}^{(l+1)}) \\
\mathbf{\hat{x}} &= g(\mathbf{W}^{(L)} \mathbf{h}^{(L-1)} + \mathbf{b}^{(L)})
\end{align}

where $f(\cdot)$ and $g(\cdot)$ are nonlinear activations, $\mathbf{W}^{(l)}$ and $\mathbf{b}^{(l)}$ are the learned weights/biases of layer $l$, and $\mathbf{\hat{x}}$ is the predicted response. This can be applied per-timestep with $\mathbf{s}=\mathbf{s}_t$, $\mathbf{\hat{x}}=\mathbf{\hat{x}}_t$, and static or time-dependent $\mathbf{W}$. Alternatively, the model can be applied to the whole stimulus $\mathbf{S} \in \mathbb{R}^{K \times T}$ in which case $\mathbf{s} \in \mathbb{R}^{KT}$ is the flattened vector form of $\mathbf{S}$, and similarly for $\mathbf{\hat{x}}$. In this latter case $\mathbf{W}$ is by default time-dependent, since the time dimension is present in the input. 

Stacking layers enables learning hierarchical nonlinear feature dynamics, capturing complex stimuli-response relationships. Note that $T$ can be freely chosen and indeed it can range from 1 timestep up to a larger stimulus window, e.g., several seconds. When setting it to a small number of timesteps (1-20) and prescribing $\mathbf{W}$ to be time-dependent we call this approach sliding window encoding. Training these models is achieved similarly to forecasting models, i.e., using the MSE loss.

The fully-connected network can be replaced by both CNNs or RNNs to provide constraints and introduce more complexity in feature extraction and modelling. Compared to the equations described in Section~\ref{ssec:nn_ar} one simply needs to replace $\mathbf{X} \in \mathbb{R}^{C \times T}$ with $\mathbf{S} \in \mathbb{R}^{K \times T}$ to define an encoding model. Often we can exploit the structure in the stimulus with specialized models, such as CNNs for images, or RNNs for audio or language stimuli.

As shown in Section~\ref{ssec:nn_ar}, CNNs and RNNs models are also well suited for forecasting timeseries. As the goal of encoding is also to predict brain data, it comes as no surprise that providing both past timesteps and stimulus features as input to CNNs and RNNs leads to better encoders. This is also called conditioning the forecasting model on stimulus information. This improves encoding/forecasting performance as instead of predicting open-ended brain activity relying solely on past brain activity, the prediction is constrained/conditioned on the respective external stimulus. A straightforward approach would be to concatenate the flattened stimulus vector $\mathbf{s}$ to the timeseries data $\mathbf{x}_t$ at each timepoint. This also allows for modelling variability in the evoked response by taking into account the past state of the brain. Simple, deterministic encoders predict identical responses to repeated stimuli, unlike real variable data. The high-dimensional output space of the encoder is also better constrained by past brain activity.

\citet{chehab2021deep} combined CNNs and RNNs to predict MEG data from pre-stimulus activity and word features, finding timing differences in feature importance. Deep encoders enable studying neural representations and computations underlying M/EEG responses to complex stimuli. Learned features can be visualised and analysed to reveal encoding transformations \citep{kriegeskorte2015deep}.

\subsection{Decoding}
\label{ssec:background_decoding}

Decoding refers to the inverse (non-causal) process of encoding, in which task conditions or stimuli are inferred from recorded brain activity data. In contrast to encoding models which predict brain activity from stimuli, decoding models estimate stimuli or behaviours from brain activity patterns. Decoding analyses have become increasingly prevalent in neuroimaging research \citep{kay2008identifying, guggenmos2018multivariate,cichy2016comparison}, with applications ranging from brain-computer interfaces \citep{lotte2018review,willett2020high} to basic neuroscience inquiries \citep{haynes2006decoding,kay2008identifying}. Decoding approaches can be applied to diverse experimental paradigms and neural measurement modalities, including the decoding of visual stimuli \citep{cichy2016comparison}, phonemes and words \citep{mugler2014direct,hulten2018cracking, cooney2019classification}, imagined speech \citep{dash2020decoding-imagined}, and motor movements \citep{willett2020high,dash2020decoding,elango2018sequence}.

Stimulus decoding may refer to reconstructing the full sensory stimulus \citep{anumanchipalli2019speech, shen2019deep}, estimating stimulus features or categories \citep{cichy2016comparison,kay2008identifying}, or simply classifying among a predefined set of stimuli \citep{mugler2014direct,hulten2018cracking}. While reconstructing arbitrary novel stimuli poses considerable challenges, classification provides a straightforward decoding objective when experiments utilise fixed stimulus sets. Hence, in this dissertation decoding is carried out primarily through supervised classification.

Conceptually, decoding can be framed as inverting the encoding model from brain activity to stimuli (Equation \ref{eq:linear_encoding}):

\begin{equation}
\mathbf{S} = \mathbf{W}\mathbf{X} + \boldsymbol{\epsilon}
\end{equation}

where $\mathbf{S}$ represents the stimulus, $\mathbf{X}$ is measured brain activity, $\mathbf{W}$ is a weight matrix, and $\boldsymbol{\epsilon} \sim \mathcal{N}(0,\mathbf{\Sigma})$ is zero-mean Gaussian noise. This expresses the conditional distribution $p(\mathbf{S}|\mathbf{X})$ that enables inferring stimuli from brain patterns. While this model can be trained via regression, classification models are more commonly employed for decoding tasks.

For example, consider an MEG recording $\mathbf{X} \in \mathbb{R}^{C \times T}$ comprising $C$ channels and $T$ timepoints. Let $\mathbf{x}_t \in \mathbb{R}^C$ denote the spatial topography across channels at time $t$. Given class labels $y \in {1, \ldots, K}$, where $K$ is the number of classes, linear discriminant analysis (LDA) models the class conditional densities $p(\mathbf{x}_t|y_k)$ as Gaussians $\mathcal{N}(\boldsymbol{\mu}_k,\mathbf{\Sigma})$ and applies Bayes' rule to predict labels \citep{lemm2011introduction}:

\begin{align}
p(y_k|\mathbf{x}_t) &= \frac{p(\mathbf{x}_t|y_k)p(y_k)}{\sum_{l=1}^{K} p(\mathbf{x}_t|y_l)p(y_l)} \\
\mathrm{log}(p(y_k|\mathbf{x}_t))&= \mathbf{\Sigma}^{-1}\boldsymbol{\mu}_k\mathbf{x}_t - \frac{1}{2}\boldsymbol{\mu}_k^T\mathbf{\Sigma}^{-1}\boldsymbol{\mu}_k+\mathrm{log}(p(y_k))
\end{align}

where the class covariance matrix $\mathbf{\Sigma}_k=\mathbf{\Sigma}$ is assumed equal. LDA provides interpretable discriminative spatial patterns through the model weights $\mathbf{W} = \mathbf{\Sigma}^{-1}\boldsymbol{\mu}$. However, LDA's linear decision boundaries limit flexibility for complex data.

For decoding time-varying signals, sliding window LDA trains classifiers on short time segments $\mathbf{x}_{t:t+P}$, where $P$ is the window length. This reveals how decodable information evolves over time \citep{grootswagers2020decoding}, but overlooks long-range dependencies beneficial for decoding. Similar to encoding, the entire trial $\mathbf{X}$ can be flattened into a vector $\mathbf{x} \in \mathbb{R}^{CT}$ for LDA, though high dimensionality may hinder learning. For all of the following models we will denote the inputs with $\mathbf{x}$, which can either refer to 1 timestep, a sliding window, or the full flattened trial, depending on the problem at hand.

Beyond LDA, logistic regression and support vector machines comprise other widely used linear decoding models \citep{lotte2018review}. Logistic regression models the class conditional logits using an affine projection, and then applies the softmax function to obtain probabilities for prediction:

\begin{align}
\mathbf{l} &= \mathbf{W} \mathbf{x} + \mathbf{b} \\
p(\mathbf{y} | \mathbf{x}) &= \frac{e^{\mathbf{l}}}{\sum_{i=1}^K e^{l_i}}
\end{align}

where $\mathbf{l}$ is the logit vector. The softmax function is the extension of the logistic function to more than 2 classes. Logistic regression is probabilistic like LDA but does not assume Gaussian densities, and it is discriminative instead of generative.

Linear models provide interpretable mappings from features to predictions. However, their simplicity can limit decoding complex spatiotemporal brain patterns. As in encoding, multi-layer fully-connected neural networks extend logistic regression via nonlinear hidden layers. The input $\mathbf{x}$ propagates through fully-connected layers to produce class predictions:

\begin{align}
\mathbf{h}^{(1)} &= f^{(1)}(\mathbf{W}^{(1)}\mathbf{x} + \mathbf{b}^{(1)}) \\
\mathbf{h}^{(l+1)} &= f^{(l)}(\mathbf{W}^{(l+1)}\mathbf{h}^{(l)} + \mathbf{b}^{(l+1)}) \\
\hat{\mathbf{y}} &= \mathrm{softmax}(\mathbf{W}^{(L)}\mathbf{h}^{(L-1)} + \mathbf{b}^{(L)})
\end{align}

where $f^{(l)}$ denotes a nonlinearity like ReLU, and $\hat{\mathbf{y}}$ gives predicted class probabilities. Cross-entropy loss trains the network for classification:

\begin{align}
\mathcal{L}(\mathbf{y}, \hat{\mathbf{y}}) = -\mathbf{y}\log{\hat{\mathbf{y}}}
\end{align}

where $\mathbf{y}$ is the true class label distribution. This is also called a one-hot vector since it consists of a 1 at the target class index and zeros everywhere else.

Classification objectives are common for decoding, although regression losses can enable reconstructing richer stimulus representations and provide wider generalisation. However, this comes with considerable challenges as the output dimensionality is much higher than in the case of classification.

Similarly, CNNs and RNNs can be adapted for decoding by changing the targets and loss function compared to forecasting. At a high-level, CNNs insert temporal convolutions before a classifier:

\begin{align}
\mathbf{H} &= \mathrm{CNN}(\mathbf{X}) \\
\hat{\mathbf{y}} &= \mathrm{softmax}(\mathbf{W}\mathrm{flatten}(\mathbf{H}) + \mathbf{b})
\end{align}

While RNNs insert temporal recurrences:

\begin{align}
\mathbf{h}_T &= \mathrm{RNN}(\mathbf{X}) \\
\hat{\mathbf{y}} &= \mathrm{softmax}(\mathbf{W}\mathbf{h}_T + \mathbf{b})
\end{align}

Early work showed fully-connected and convolutional neural networks can decode motor intentions from EEG better than linear SVMs \citep{tabar2016novel}. CNNs have proven effective for motor decoding \citep{schirrmeister_deep_2017} and robust generalization across subjects \citep{lawhern2018eegnet}. The learned filters provide insight into discriminative patterns. Some variants use time-convolutional layers before recurrent layers to extract local temporal features \citep{bashivan_learning_2015}. RNNs directly model the temporal hierarchy of neural processes \citep{kubilius2019brain}.

It is important to mention that an alternative and more traditional way to improve on linear models (instead of using nonlinear models), is to first apply some nonlinear transformation to the input data, e.g. the wavelet transform, and then train the linear decoder \citep{higgins2022relationship, hu2011feature}. In general any encoding or decoding model can operate on a transformed feature space instead of the raw stimuli or brain data.

Next we will provide an introduction to leveraging unsupervised, encoding, and decoding models for understanding the brain.

\section{Interpretability methods}
\label{sec:interpretability_methods}

While the analysis methods presented in Section~\ref{ssec:analysis_methods} focused on utilising signal processing and statistical techniques to characterise brain activity, we can also leverage some of the more advanced modelling approaches described in the previous sections in our quest to understand noninvasive electrophysiology.

\subsection{HMM statistics}
\label{ssec:hmm_stats}

Hidden Markov models (HMMs) have proven to be a powerful tool for modelling sequential M/EEG data and uncovering recurring brain states \citep{vidaurre2018discovering}. Once an HMM is fit to M/EEG observations, the estimated model parameters and hidden state sequences can be analysed in various ways to elucidate the spatiotemporal dynamics of brain activity. This section outlines common techniques for interpreting trained HMMs on M/EEG data.

The HMM transition matrix $\mathbf{A} \in \mathbb{R}^{K\times K}$ describes the Markovian dynamics between states, where $K$ is the number of states:

\begin{equation}
A_{ij} = p(z_t = j | z_{t-1} = i) 
\end{equation}

The overall transition structure characterises dynamic reconfigurations of large-scale brain networks \citep{vidaurre2018discovering}. The spatial covariance patterns $\mathbf{\Sigma}_i$ characterise functional connectivity networks associated with each state \citep{vidaurre2018discovering}.

The power spectral density (PSD) of each state reveals its oscillatory profile. Importantly, we only look at PSD when using the HMM method in conjunction with time-delay embeddings (TDE). This method augments the input to the HMM with multiple lagged versions of the time series. The PSD for state $i$ can be estimated from data segments assigned to that state based on the Viterbi path. Different states often exhibit distinct spectral signatures related to underlying cognitive processes. For example, alpha desynchronisation may indicate active sensory processing \citep{klimesch2012alpha}. Visualising PSD topographically links specific large-scale networks to spectral features like alpha/beta desynchronisation or theta/gamma synchronisation \citep{vidaurre2018discovering}.

The Viterbi path $z^*_1, \ldots, z^*_T$ provides the optimal hidden state sequence explaining the observations. With task data, evoked responses can be computed over the state timecourse. This reveals which states are activated by certain events, and their temporal evolution within the trial.

General statistics can also be calculated from the state timecourses:

\begin{itemize}
\item Fractional occupancy: fraction of time spent in each state.
\item Mean lifetime: average duration of each state visit.

\item Mean interval: average time between consecutive state visits.
\item Switching rate: rate of transitions into each state.
\end{itemize}

Comparing these metrics across states, task conditions, subjects, datasets, or models can reveal insights into differences in brain dynamics. The distribution of the statistics within and across states can also be quantified to characterise variability. Thus HMMs provide a rich set of analysis tools beyond standard spectral analysis, functional connectivity, and evoked responses estimated from brain data.

\subsection{AR generation}
\label{ssec:ar_generation}

Autoregressive (AR) models provide a powerful framework for modelling the dynamics of multivariate M/EEG time series. Once an AR model is fit to M/EEG data, the estimated model parameters and generated data samples can be analysed in various ways to elucidate both local and global spatiotemporal characteristics of brain activity. This section outlines techniques for interpreting trained AR models on M/EEG data.

A multivariate autoregressive (MAR) model characterises how a multivariate time series $\mathbf{x}_t \in \mathbb{R}^C$ evolves linearly over time. The AR coefficients $\mathbf{A}_p \in \mathbb{R}^{C \times C}$ directly encode the autoregressive dynamics, with elements $A_p[i,j]$ relating channel $j$ at lag $p$ to channel $i$ at the current time step. Non-zero off-diagonal elements in $\mathbf{A}_p$ indicate cross-channel interactions, while a diagonal structure reflects independence \citep{chiarion2023connectivity}. The overall cross-channel coefficient structure provides insight into functional brain networks.

The power spectral density (PSD) describes the distribution of power across frequencies. For a linear multivariate AR process, the PSD matrix is \citep{schlogl2006analyzing}:

\begin{equation}
\mathbf{S}(f) = \frac{1}{2\pi} \mathbf{\Sigma} |\mathbf{I} - \sum_{p=1}^P \mathbf{A}_p e^{-i2\pi fp}|^{-2}
\end{equation}

where $\mathbf{I}$ is the identity matrix, and $\mathbf{\Sigma}$ is the noise covariance. Diagonal PSD elements $\mathbf{S}_{ii}(f)$ give the spectrum for each channel $i$. The PSD decompositions provide insight into the oscillatory characteristics captured by the model, including peak frequencies, spectral shape (e.g. $1/f$), and spatial topographies of different rhythms.

A key advantage of AR models is their ability to generate new data points by feeding back their own predictions:

\begin{align}
\hat{\mathbf{x}}_t &= \sum_{p=1}^{P} \mathbf{A}_p \hat{\mathbf{x}}_{t-p} + \boldsymbol{\epsilon}_t \\
\hat{\mathbf{x}}_{t-p} &= \text{past generated data}
\end{align}

This allows synthesising multidimensional time series with similar dynamics as the empirical data. Comparing real and simulated data based on spectral, spatial, and temporal properties helps validate that the model accurately captures the characteristics of interest. Note that recursive generation is straightforward for any of the neural network models described in Section~\ref{ssec:nn_ar}, not only linear models.

Long generative runs enable assessing model stability through metrics like divergence from the empirical covariance matrix \citep{schlogl2006analyzing}. If outputs diverge over time, the model may be underconstrained. Fitting to more data until predictions remain accurate ensures proper characterisation of the system dynamics.

Key empirical measures for comparing real and generated data include:

\begin{itemize}
\item \textbf{Spectral content}: The PSD reveals whether the model accurately captures oscillations in specific frequency bands. 
\item \textbf{Covariance}: The overall covariance matrix measures global correlation structure.
\item \textbf{Evoked responses}: Generating data related to stimuli, e.g. by initialising the generation with the start of the evoked response tests modelled event-related dynamics. 
\item \textbf{HMM statistics}: Comparing the statistics of HMMs (discussed in Section~\ref{ssec:hmm_stats}) trained on real and generated data reveals how well the AR model captures higher-level metrics of brain dynamics.
\end{itemize}

In summary, AR modelling provides a concise framework for capturing the spatiotemporal dynamics in multivariate M/EEG recordings. Analysing model coefficients, spectral characteristics, and generated samples gives insights into oscillatory content and event-related attributes of brain activity. Comparisons against real data validate how accurately the model reproduces empirical dynamics, supporting further analysis of both spontaneous and task-related neural processes.

\subsection{Multivariate pattern analysis}

Multivariate pattern analysis (MVPA) refers to a set of techniques that apply machine learning algorithms to neural data (e.g. M/EEG, fMRI) to uncover distributed neural representations and decode mental states \citep{haxby2001distributed, haynes2006decoding}. In contrast to univariate methods that examine individual sensors or voxels, MVPA examines multivariate patterns of activity across multiple channels or voxels to discriminate between experimental conditions. This provides insights into information encoding in the brain that are not accessible with univariate approaches.

Encoding models can be analysed to study model weights $\mathbf{W}$ reflecting channel tuning, and feature importance for model prediction. This reveals how stimuli are transformed and represented in brain activity.

Once fit, both encoding and decoding models can be analysed to reveal discriminative spatial, temporal, and spectral patterns. This provides insights into neural coding underlying perceptions and behaviours. Similarly to evoked analyses we are interested in the temporal, spatial, and spectral signatures of evoked activity. Leveraging multivariate decoding models for this is the central aim of MVPA.

Sliding-window analysis divides continuous data into short overlapping segments and extracts features from each window separately to characterise temporal dynamics \citep{higgins2022relationship}. Windows may be from 10 ms up to 200 ms long and slide in small steps (10-20 ms). Compared with evoked response analysis, machine learning decoding models trained on time windows provide a complementary view of the temporal activity, by plotting cross-validated accuracy (across trials) for each time window. Such a sliding window decoding approach may reveal the temporal evolution of stimulus-related information and thus discriminability in the brain. Shorter windows provide finer temporal resolution of representative dynamics at the expense of less discriminative power. Longer windows allow achieving higher decoding accuracy and a smoother temporal profile.

For linear sliding window models like LDA, the model weights $\mathbf{W}$ reflect channel contributions. While in encoding models these weights directly quantify the importance of features to brain data, this is not the case for decoding models. Instead, the Haufe transform can be used to quantify these contributions \citep{haufe2014interpretation}.

The Haufe transform works as follows. Let $\mathbf{W}$ be the $C \times K$ matrix of extraction filters from the linear decoding model, where $C$ is the number of channels and $K$ is the number of latent factors. Let $\mathbf{\Sigma_x}$ be the $C \times C$ data covariance matrix and $\mathbf{\Sigma_{\hat{s}}}$ the $K \times K$ covariance matrix of the estimated latent factors $\hat{\mathbf{s}}$.

Then the $M \times K$ matrix $\mathbf{A}$ of activation patterns is given by:

\begin{equation}
\mathbf{A} = \mathbf{\Sigma_x} \mathbf{W} \mathbf{\Sigma_{\hat{s}}}^{-1}
\end{equation}

If the estimated latent factors are uncorrelated, this simplifies to:

\begin{equation}
    \mathbf{A} \propto \mathbf{\Sigma_x} \mathbf{W}
\end{equation}

where $\propto$ denotes proportionality. Plotting $\mathbf{A}$ as a sensor-space topography reveals spatial patterns related to stimulus discriminability. By plotting the activation maps for each sliding window decoding model within a trial, the spatiotemporal stimulus-discriminability can be jointly characterised.

An alternative method of finding spatial patterns of stimulus-discriminability would be to train a separate decoding model on the full trial of each channel, and plotting the accuracies as scalp topographies. This is similar to the temporal sliding window method but across space.

Finally, for spectral investigations one can fit separate decoding models on individual frequency bands from the wavelet transform of the data. Once trained, cross-validation accuracies can be computed to assess which frequency bands contain the most information related to stimuli. Alternatively one can create separate datasets by filtering around bands of interest and training separate models on each band-filtered dataset. These methods can also be combined with sliding window analysis or the Haufe transform to jointly characterise spectro-temporal, and spectro-spatial patterns of stimulus-discriminability.

Another useful concept is temporal generalisability \citep{king2014characterizing}, where a decoding model is trained on one time window and tested on all other time windows (within the trial). This provides a matrix of size timepoints $\times$ timepoints, elucidating which parts of the evoked response contain similar/shared information. This can be similarly extended for spatial and spectral generalisability.

\subsection{Permutation feature importance}

Permutation feature importance (PFI) is a model-agnostic approach that can quantify the contribution of input features to model performance for any black-box model \citep{fisher_all_2019}. By systematically disrupting features in the input and measuring the resultant change in model performance, PFI reveals how much the model relies on each part of the input. This provides an alternative to the multivariate pattern analysis (MVPA) methods described earlier, with the advantage that PFI can be applied to nonlinear models without needing to constrain the input domain or dimensionality. Ablation approaches such as MVPA involve completely removing a feature from the model and re-evaluating performance. While this gives a more direct measure of the impact of that feature, it requires retraining models multiple times. PFI and ablation methods methods can be viewed as complementary and useful in their own right.

For a trained encoder model $g$ with parameters $\theta$, the PFI approach quantifies each feature's contribution $\Delta p_f$ by measuring the change in mean squared error when feature $f$ is randomly permuted across trials:

\begin{equation}
    \Delta p_f = \mathbb{E}_{\mathbf{X},\mathbf{s}}\left[|\mathbf{X} - g(\mathbf{s}_{\perp f};\theta)|_2^2\right] - \mathbb{E}_{\mathbf{X},\mathbf{s}}\left[|\mathbf{X} - g(\mathbf{s};\theta)|_2^2\right]
\end{equation}

where $\mathbf{s}_{\perp f}$ denotes random permutation of feature $f$ in input $\mathbf{s}$. A higher $\Delta p_f$ indicates feature $f$ is more important for encoding model performance.

PFI has been applied to M/EEG data to reveal how linguistic properties such as word length and frequency affect the encoding of language responses at different latencies and sensors \citep{chehab2021deep}. It demonstrated that word length impacts early encoding, while word frequency affects later encoding, matching known stages of linguistic processing.

For a trained decoder model $g$, PFI can localise which parts of the M/EEG input are most relevant for predicting specific evoked responses. It is model-agnostic and therefore can be applied to nonlinear deep neural networks that may better capture complex stimulus-response mappings.

The importance $\Delta p_j$ of each M/EEG feature $j$ is revealed by randomly shuffling it across M/EEG trials:

\begin{equation}
\Delta p_j = \mathbb{E}_{\mathbf{X},\mathbf{y}}\left[\mathcal{M}(\mathbf{y}, g(\mathbf{X}_{\perp j};\theta))\right] - \mathbb{E}_{\mathbf{X},\mathbf{y}}\left[\mathcal{M}(\mathbf{y}, g(\mathbf{X};\theta))\right]
\end{equation}

where $\mathcal{M}$ is the evaluation metric (e.g. accuracy for classification), and $\mathbf{X}_{\perp j}$ denotes shuffling of feature $j$ in $\mathbf{X}$. Here, $j$ can represent either a timepoint vector $\mathbf{x}_t \in \mathbb{R}^{C}$ or a channel vector $\mathbf{x}_c \in \mathbb{R}^{T}$ within the full input $\mathbf{X} \in \mathbb{R}^{C \times T}$.

Higher values of $\Delta p_j$ indicate that feature $j$ is more relevant for the decoding objective. Applying PFI to $\mathbf{x}_t$ across all timepoints $t \in {1, \ldots, T}$ yields an accuracy loss timecourse, revealing discriminative temporal dynamics. Similarly, applying it to $\mathbf{x}_c$  across all channels $c \in {1, \ldots, C}$ creates channel accuracy loss map, localising spatial importance.

Spatiotemporal PFI jointly shuffles feature windows spanning both time and space, $\mathbf{j}=\mathbf{X}[c:c+k, t:t+l]$. This is repeated across all windows by indexing across channels $c$, with spatial window length $k$, and timepoints $t$, with temporal window length $l$. Since the channel ordering in $\mathbf{X}$ may not follow a locality-sensitive layout, methods that incorporate 3D sensor locations could improve the spatial windowing. Overall, spatiotemporal PFI reveals the joint relevance of time and space for stimulus discriminability.

To be clear PFI does not solve the issues raised by \citet{haufe2014interpretation}, as the absence of influence of a feature on decoding performance does not necessarily imply that that feature does not contain stimulus-related information. Furthermore if two features are correlated the model might use both to drive decoding, when in reality only one of the features may reflect stimulus-related activity. PFI is often used throughout the thesis and it is further detailed in Chapter~\ref{Chap3}.

\subsection{Interpreting neural networks}
\label{ssec:interpret_deep}

Interpreting deep learning models becomes more challenging because of the nonlinear functions and high-dimensional nature of the parameter space. Their complexity poses challenges for interpreting what is learned by the models and relating this to neuroscientific principles. This section describes techniques to extract insights from trained deep encoder and decoder models into the representations and computations captured by the networks. Gaining such understanding helps link the models to underlying neural mechanisms. As mentioned, PFI can be used to investigate the input feature importances in any nonlinear model.

A common approach for understanding model decisions is to backpropagate from the output to the input layer to identify salient input patterns via gradients \citep{simonyan2013deep}. Consider a deep neural network decoder $f(\mathbf{X}; \theta)$ that maps an input MEG trial $\mathbf{X} \in \mathbb{R}^{C \times T}$ to predicted stimulus features $\hat{\mathbf{y}} \in \mathbb{R}^D$, where $\theta$ are learned model parameters, and $D$ is the dimensionality of predicted stimulus, or the number of classes in case of classification.
 
The gradient of the loss $\mathcal{L}(\hat{\mathbf{y}}, \mathbf{y})$ (e.g., cross-entropy) between the predicted ($\hat{\mathbf{y}}$) and true stimulus ($\mathbf{y}$) quantifies how small changes in the model input affect the loss. By backpropagating these gradients to the input layer \citep{yosinski2015understanding}, we can compute the input gradient image:

\begin{equation}
\mathbf{G} = \dfrac{\partial \mathcal{L}}{\partial \mathbf{X}} = \dfrac{\partial \mathcal{L}}{\partial \hat{\mathbf{y}}} \dfrac{\partial \hat{\mathbf{y}}}{\partial \mathbf{X}}
\end{equation}

where $\mathbf{G} \in \mathbb{R}^{C \times T}$. This gradient image highlights channels and timepoints most relevant for the model's predictions \citep{sturm2016interpretable}. Alternatively, one can compute saliency maps by taking the absolute gradient magnitude $\mathbf{S} = |\mathbf{G}|$ \citep{simonyan2013deep}. Visualising the gradient image or saliency map reveals spatiotemporal patterns the model relies on for stimulus decoding.

For CNN decoders or forecasting models, visualising the learned filters reveals localised temporal patterns the network detects in the input. Consider a 1D temporal CNN with the following convolutional layer:

\begin{equation}
\mathbf{h}_m^{(l+1)} = f^{(l)}\left(\mathbf{b}_m^{(l)} + \sum_{i=1}^{M^{(l)}} \mathbf{w}_{m, i}^{(l)} * \mathbf{h}_i^{(l)}\right) 
\end{equation}

where $\mathbf{h}_i^{(l)} \in \mathbb{R}^{T}$ is the $i^\text{th}$ input channel, $\mathbf{w}_{m, i}^{(l)} \in \mathbb{R}^{K}$ is a learned 1D kernel of length $K$, $*$ denotes convolution, $\mathbf{b}_m^{(l)}$ is a bias term, and $f^{(l)}$ is a nonlinearity.

Visualising the kernel weights $\mathbf{w}_{m, i}^{(l)}$ shows what patterns the network extracts from each input channel $i$ to form output feature channel $m$. Analysing filter activations on input examples highlights what specific features are detected \citep{zeiler2014visualizing}. Comparing filters across layers reveals hierarchical feature learning that transforms raw signals to higher-level stimulus representations \citep{yosinski2015understanding}.

The internal representations learned within deep neural networks can be analysed by examining activations $\mathbf{H}^{(l)} \in \mathbb{R}^{M^{(l)} \times T^{(l)}}$ at each layer $l$. Dimensionality reduction techniques like PCA can visualise the geometry of high-dimensional activations \citep{rauber2016visualizing}.

Representational similarity analysis characterises the geometry of learned neural representations using pairwise distances between activity patterns \citep{kriegeskorte2008representational}. Consider a decoder $f(\mathbf{X}) = \hat{\mathbf{y}}$ predicting stimulus class $\hat{\mathbf{y}} \in {1, \ldots, K}$ from M/EEG input $\mathbf{X}$. Its layer-wise representations can be analysed by first computing the mean representation pattern $\bar{\mathbf{h}}^{(l)}_k$ for each layer $l$ for each class $k$ by averaging over trials. Then the representational dissimilarity matrix between mean patterns ($\forall k,m \in {1, \ldots, K}$) can be constructed:

\begin{equation}
\mathrm{RDM}^{(l)}(k,m) = d(\bar{\mathbf{h}}^{(l)}_k, \bar{\mathbf{h}}^{(l)}_m)
\end{equation}

where $d$ is a distance metric like Euclidean distance. Visualising and comparing RDMs across layers provides insight into the decoding model's discriminative space \citep{kriegeskorte2008representational}. Comparing to brain RDMs (constructed from evoked responses for example) tests representational alignment between models and neural data \citep{kietzmann2019recurrence}.

The frequency sensitivity of neural networks can be examined by analysing their spectral characteristics. For a 1D temporal convolutional layer, its filters $\mathbf{w}_{m, i}^{(l)} \in \mathbb{R}^K$ act as finite impulse response (FIR) filters on the input. Taking the discrete Fourier transform gives the frequency response:

\begin{equation}
W_{m,i}^{(l)}(f) = \mathcal{F}\{\mathbf{w}_{m,i}^{(l)}\}
\end{equation}

Comparing spectral profiles across layers and kernels reveals what oscillations the model captures. Power spectral analysis can also be applied to network activations $\mathbf{H}^{(l)}$ to examine what oscillations emerge internally, providing insights into the model's spectral representations.

Interpreting encoding and decoding models is crucial for relating predictions to underlying neural processes. Techniques like input backpropagation, layer analysis, representational geometry, and spectral analysis enable understanding hierarchical computations and representations learned by deep neural networks applied to M/EEG data. This links data-driven models with neurophysiological principles to advance mechanistic understanding of the brain. Ongoing research in interpretable deep learning will further bridge predictive and explanatory modelling of brain function \citep{samek2019explainable}.

\chapter{Interpretable full-epoch decoding}
\label{Chap3}

As described in the introduction the aim of the thesis is to deal with the various variability issues in M/EEG data. In this first content chapter we aim to tackle within-subject variability, and push the performance of within-subject decoding while also providing methods for neuroscientific interpretability. In later chapters we will build on these methods and investigate the use of deep learning to deal with more challenging types of variabilities.

Multivariate pattern analysis (MVPA) of MEG and EEG is a valuable tool for understanding how the brain represents and discriminates between different stimuli.  Identifying the spatial and temporal signatures of stimuli is typically a crucial output of these analyses. Such analyses are mainly performed using linear, pairwise, sliding-window decoding models. These allow for relative ease of interpretation, e.g. by estimating a time-course of decoding accuracy, but are computationally intensive and can have limited decoding performance. On the other hand, full epoch decoding models, commonly used for brain-computer interface (BCI) applications, can provide better decoding performance. However, they lack methods for interpreting the contributions of spatial and temporal features.

In this chapter, we propose an approach that combines a multiclass, full epoch decoding model with supervised dimensionality reduction, while still being able to reveal the contributions of spatiotemporal and spectral features using permutation feature importance. Crucially, we introduce a way of doing supervised dimensionality reduction of input features within a neural network optimised for the classification task, improving performance substantially. We demonstrate the approach on 3 different task MEG datasets using image presentations. Our approach consistently achieves higher accuracy than the peak accuracy of a sliding window decoder while estimating the relevant spatiotemporal features in the MEG signal. Finally, we show that our multiclass model can also be used for pairwise decoding (in Appendix~\ref{ssec:multiclass_vs_pairwise}), eliminating the computational burden of training separate models for each pairwise combination of stimuli.

\textit{Note: } Most of this chapter is part of a published paper \citep{csaky2023interpretable}. All of the work has been carried out by the thesis author. Most experiments in this chapter can be reproduced using the associated GitHub repository\footnote{\url{https://github.com/ricsinaruto/MEG-transfer-decoding}}.

\section{Introduction}
Decoding studies tend to prioritise increasing the discriminatory power (accuracy) between stimuli, e.g., in brain-computer interface (BCI) applications \citep{koizumi2018development, cooney2019optimizing, defossez2022decoding}, or gaining interpretable insights as to where and when stimuli are represented in the brain \citep{cichy2014resolving, cichy2016comparison}. These latter approaches are often referred to as multivariate pattern analysis (MVPA), and typically make use of linear, sliding-window decoders. This allows for the extraction of the interpretable spatiotemporal features that drive the decoding; for example, allowing for the estimation of a decoding accuracy time course \citep{cichy2014resolving, cichy2016comparison, cichy2017multivariate, lappe2013beamformer, higgins2022relationship, higgins2022spatiotemporally}. However, it has been demonstrated that, as one would expect, discriminatory power is also important for the effectiveness of MVPA \citep{guggenmos2018multivariate}. Hence, there is a need in MVPA for decoding methods that improve decoding performance, while maintaining the ability to reveal the spatiotemporal features that underlie the decoding.

One possibility for increasing decoding performance is to abandon the use of sliding window approaches and instead use full epoch decoding. Here, we refer to the 500ms following stimulus presentation as the full-epoch. While it is generally good to increase the time window for decoding, as we will later show in the results, using a longer window than 500ms might actually be detrimental. Decoding full-epoch trials has been explored most typically within the context of potential (asynchronous) brain-computer interface (BCI) applications, for example in language tasks \citep{koizumi2018development, cooney2019optimizing, cooney2019classification, hulten2018cracking, dash2020decoding-imagined, defossez2022decoding} and motor tasks \citep{schirrmeister2017deep, dash2020decoding, elango2018sequence}. It should be noted that in synchronous BCI paradigms sliding window decoding might be preferable to minimise prediction lag.

In contrast with the decoding employed in MVPA, BCI applications often use nonlinear multiclass models \citep{lawhern2018eegnet}. These will generally have good discriminatory power (accuracy), but this comes at the expense of poor interpretability, and are thus not directly useful for MVPA.

Within BCI research, dimensionality reduction is often done with established supervised methods such as Common Spatial Patterns (CSP) \citep{blankertz2007optimizing}, or Riemannian classifiers \citep{barachant2014meg}. Supervised variants of PCA have also been introduced, but not for MEG data \citep{kobak2016demixed}. A gold standard approach to designing BCI decoders is the use of a Riemannian classifier that also performs a supervised class separation \citep{barachant2014meg}. Importantly, these methods rely on a separate feature extraction step before applying the classifier. We wanted to include both steps in a single neural network to allow end-to-end optimisation for the classification task. We have found that the features learned by the neural network can be used to also train a standard LDA model, increasing performance substantially over either unsupervised feature reduction or the supervised Riemannian method. 

Some promising approaches have been investigated recently to make full-epoch models more interpretable, such as the linear forward transform \citep{haufe2014interpretation}. However, this approach can only be applied to linear models. Another option is to apply full-epoch and sliding window decoding on the same data in order to get both perspectives, e.g. in \citep{ling2019visual}. Nonetheless, it would be hugely beneficial if a single decoding approach could be used without a loss in performance on both BCI and MVPA.

An additional consideration is computational efficiency. MVPA of MEG data is commonly performed using pairwise decoding methods, i.e. they decode between just two classes at a time \citep{cichy2014resolving, cichy2016comparison, cichy2017multivariate, higgins2022relationship, higgins2022spatiotemporally}. When the number of classes gets large, this becomes computationally burdensome. Here, we propose to overcome this through the use of multi-class decoding.

Taking together the aforementioned issues, we propose an approach that can improve decoding accuracy through the use of full-epoch multi-class decoding, while still being able to reveal the underlying spatiotemporal features that drive the decoding. This allows us to consider and investigate the use of neural network decoding models, and we also show the benefit of using supervised feature reduction. We limit our investigations to linear models, leaving nonlinear models for future work. Importantly, to allow access to interpretable features, we make use of permutation feature importance (PFI).

We assess the proposed approach by systematically comparing it with sliding window decoding on three MEG datasets with visual tasks, finding that our full-epoch decoding outperforms sliding window decoding in terms of accuracy. We then compare PFI with standard alternatives and find that PFI is able to extract the same kind of dynamic temporal, spatial, and spectral information. In addition, we show that pairwise accuracies can easily be gained from a single multiclass model and that these accuracies are on-par with a direct pairwise classification approach. Please see Appendix~\ref{ssec:multiclass_vs_pairwise} for these results.

In short, the aforementioned contributions achieve the best of both worlds: a single decoding model trained on full epochs, empirically good performance, and clear interpretability from an MVPA viewpoint. This approach promises to be useful for both the BCI researcher and the neuroscientist trying to gain insight into the underlying brain activity in a particular task and external stimuli set.

\section{Methods}
\label{sec:methods}
\subsection{Data}

Here, we used three visual MEG datasets: two similar datasets from \cite{cichy2016comparison} and one additional dataset from \cite{liu2019human}. The datasets have been collected with appropriate consent from participants and ethical review by \cite{cichy2016comparison} and \cite{liu2019human}, and do  not contain any personal information. 15 subjects view 118 and 92 different images, respectively in the first two datasets, with 30 repetitions for each image. The third dataset is part of a larger replay study, and we only use the portion of the data where images are presented in random order for 900ms. Here, 22 subjects view 8 different images, with 20-30 repetitions for each image (depending on the subject). The image sets used in the three datasets are different.

We obtained the raw MEG data directly from the authors to run our preprocessing pipeline with MNE-Python \citep{gramfort2013meg}. The 118-image and 92-image data are also available publicly in epoched form\footnote{\url{http://userpage.fu-berlin.de/rmcichy/fusion_project_page/main.html}}. We bandpass filtered raw data between 0.1 and 25Hz and downsampled to 100Hz. As recommended by prior work the sampling rate is 4 times higher than the lowpass filer \citep{higgins2022relationship}. This is done so that representational alias artefacts are eliminated from the sliding window decoding time courses. We also applied whitening, which involved transforming the data with PCA to remove covariance between channels while retaining all components. The PCA was fit on the training set only but applied to both training and test sets.

Many papers have shown that visual information processing in the brain primarily operates in lower frequency ranges. Specifically, theta (4-7Hz), alpha (8-12Hz), and beta (13-30Hz) bands have been implicated in various aspects of visual processing, including object recognition, visual attention, and perceptual decision-making \citep{klimesch1999eeg, engel2010beta, zoefel2017oscillatory}. Therefore, a lowpass filter of 25Hz captures these important frequency bands while reducing the influence of higher frequency signals that are less likely to be related to visual processing.

MEG data, like all bioelectrical signals, are often contaminated by various sources of noise. High-frequency noise, particularly above 30Hz, often originates from sources outside the brain, such as muscle activity or environmental electromagnetic fields \citep{gross2013good}. By using a 25Hz lowpass filter, we can significantly reduce these non-brain noise contributions, thereby improving the signal-to-noise ratio and enhancing the detectability of the brain's visual responses.

While there are meaningful neuronal signals at frequencies above 30Hz (e.g., gamma-band activity), decoding these high-frequency signals from MEG data can be challenging due to lower signal-to-noise ratios. Therefore, unless the specific research question involves high-frequency bands, applying a 25Hz lowpass filter simplifies the data and focuses the analysis on the most relevant and easily interpreted signals. It also allows reducing the sampling rate, and thus the dimensionality of the data which is an important factor for achieving good classification performance with machine learning. In Chapter~\ref{Chap4} we also train decoding models on this dataset using a higher lowpass filter and do not observe improvement in classification accuracy.

In the first two datasets, image presentation lasted for 500ms with an average inter-trial interval of 0.95 seconds. In order to analyse the data using machine learning models, we created two versions of each dataset. The first version consisted of full epochs, with input examples having a shape of [50, 306] (or [90, 273] for the 8-image dataset), where 306 and 273 correspond to the number of MEG channels and 50 and 90 correspond to the number of time points during image presentation. The second version consisted of sliding windows, with input examples having a shape of [10, 306] (or [10, 273] for the 8-image dataset). In this case, we partitioned each trial into overlapping 100ms time windows between 0 and 1000ms post-stimulus and trained separate models on each time window partition as is normally done in the MVPA literature. The difference between consecutive windows was 1 timestep (10ms). As a result, 90 independent sliding window models were trained for each dataset. In the rest of this chapter we use the term \textit{raw} to refer to the pre-processed time domain signal, as opposed to other non-time domain input features.

As opposed to some previous work using a wavelet transform of the trial as features for sliding window decoding \citep{higgins2022relationship}, here we use the raw set of timepoints within the respective 100ms window. This means that we rely more on the decoder to extract relevant frequency information rather than directly providing such information in the input. A more recent approach, termed superlets transform \citep{moca2021time, jorntell2023singular} has been shown to improve classification results by mitigating the time vs. frequency resolution problem \citep{barzan2022time}. However,  our main comparison between sliding-window and full-epoch decoding is performed at the raw data level. Training decoding models on raw data is also beneficial for our goals of using deep learning later in the thesis. By supplying raw data we do not make any assumptions about the types of features that should be used, but rather delegate this task to the model. 

\subsection{Neural network}
\begin{figure}[!t]
  \centering
  \includegraphics[width=1.0\linewidth]{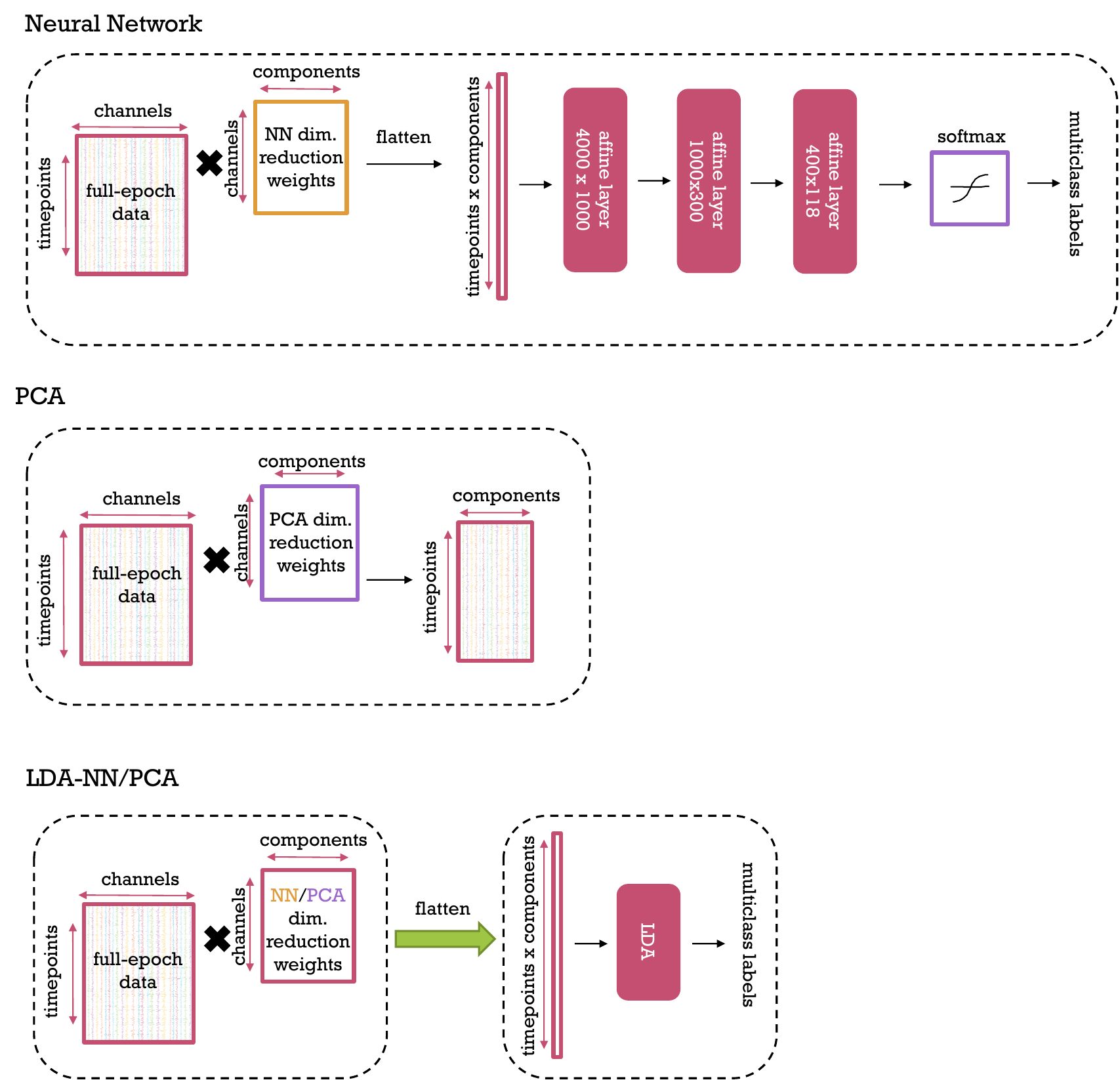}
    \caption{Our Neural Network, PCA, and LDA-NN/PCA methods from top to bottom. Dashed boxes represent separate processing steps, i.e. in the case of LDA-NN and LDA-PCA the respective dimensionality reduction is first used to compute the input features, which are then used to train the LDA model.}
    \label{fig:methods}
\end{figure}

The Neural network (NN) model in this chapter is a four-layer, fully-connected linear neural network which is only run on the full-epoch dataset (Figure~\ref{fig:methods}). The first layer performed a learnable dimensionality reduction, where the full epoch data $\mathbf{X} \in \mathbb{R}^{T \times C}$ was multiplied by a weight matrix $\mathbf{W} \in \mathbb{R}^{C \times K}$, where $K$ was set to 80. This is similar to the projection in principal component analysis, but in this case, the projection and the decoding model are trained simultaneously; therefore, the dimensionality reduction is optimised for the classification objective. To be clear, the input size to the first layer, and thus the dimensionality of this layer, depends on the time window size and number of channels which can be different for each dataset. After the first layer, the data was flattened and three affine transformations were applied in sequence (see Figure~\ref{fig:methods} for dimensionalities). The final layer had an output dimension equal to the number of classes, and the logits from this layer were passed through a softmax function for classification. We chose the intermediate hidden sizes (1000 and 300) to be roughly equally distanced (multiplicatively) between the input and output dimensions of the network (4000 and 118). This rationale was employed for the 118-image dataset primarily and we did not change the hidden sizes for the other two datasets.

The model was trained using cross-entropy loss \citep{good1952rational} for multiclass classification and included dropout between layers during training \citep{Srivastava:2014a}.  It is worth noting that, as no nonlinearities were used, the model could be replaced with a single affine transformation during evaluation. However, deep linear neural networks are known to have nonlinear gradient descent dynamics that change with each additional layer \citep{saxe2013exact}; both the learnable dimensionality-reduction layer and the use of dropout impose additional constraints on the weight matrix during learning.

\subsection{LDA-PCA}
The LDA-PCA approach has two variants: one that is full-epoch, and one that uses a sliding window. In the full-epoch version, PCA is used to do unsupervised dimensionality reduction on the channel dimension of the full-epoch data as an initial, separate step (Figure~\ref{fig:methods}). The resulting PCA-reduced data matrix $\mathbf{H} \in \mathbb{R}^{T \times K}$ is flattened and then used to train a multiclass classifier using LDA.

In the sliding window version, the $\mathbf{H}$ matrix is separated into sliding window matrices $\mathbf{H}_t = \mathbf{H}[t:t+d, :]$, where $d$ was set to 100 ms, and $t \in {1 \dots T-d}$. The data within each window is then flattened in the same manner as in the full-epoch version and fed into separate LDAs that are distinct to each window. The window size of 100 ms was selected due to its previous use in the MVPA literature \citep{higgins2022relationship}, and because it provides a good trade-off between resolution and accuracy. However, we also explore different window sizes in the results.

\subsection{LDA-NN}

In the LDA-NN method, the dimensionality-reducing weight matrix from PCA is replaced with the use of the dimensionality-reducing weight matrix extracted from the pre-trained NN approach (Figure~\ref{fig:methods}). As in LDA-PCA, this projection is then applied to the input data $\mathbf{X}$. The LDA model is applied to the resultant (flattened) features $\mathbf{H} \in \mathbb{R}^{T \times K}$. In the same manner, as LDA-PCA, LDA-NN also has full-epoch and sliding window versions.

\subsection{Permutation feature importance}
To investigate the temporal dynamics of visual information processing, we utilised permutation feature importance (PFI) on our trained models. Specifically, we applied PFI to a trained full-epoch LDA-NN by using sliding windows of 100ms with 1 time point shift for each trial. The information in each window was disrupted by permuting the data across the channel dimension separately for each time window. For instance, if the window was centred around 50ms post-stimulus, the information within that window would be disrupted from 0 to 100ms post-stimulus compared to the original trial, while the rest of the timepoints in the trial remained unchanged. We then evaluated the trained LDA-NN on each of these disrupted trials and compared the accuracy to the original accuracy obtained with the original trials. The greater the accuracy decrease for a trial with disrupted information in a specific time window, the more crucial that time window is to the model's performance and, therefore, the more information it contains relevant to the model's objective of discriminating between images. By repeating this analysis for all time windows, we obtain a temporal profile of the information content, similar to the method of training separate models on individual time windows.

In terms of assessing spatial information content, we followed a similar methodology, albeit with modifications. Here, the disruption involved permuting the data across time points within each channel individually. The outcome of this operation is a sensor space map detailing the decrease in accuracy, which serves as a metric for the visual information content. This map was then compared with others generated by evaluating the per-channel accuracy of individual LDA models trained on the full epoch of each respective channel. Conceptually, this method can be seen as sliding a window (or “search-light”) across the spatial domain, similar to the previous time-based approach. In practice, we ran spatial PFI across sensors (2 gradiometers and 1 magnetometer in the same position) instead of channels, thus permuting these 3 channels together and obtaining a single metric for them. This allows for more robust results.

An alternative would be to permute the gradiometers and magnetometers separately but using a spatial neighbourhood of nearby sensors for smoothing. Due to differences in spatial sensitivity between magnetometers and gradiometers this approach could result in spatial maps with higher resolution. Magnetometers measure the magnitude of the magnetic field and are most sensitive to sources directly below the sensor. Gradiometers on the other hand measure the difference (spatial gradient) of the field and thus are more sensitive to sources slightly to the sides of the sensor. Because they measure a difference, the signal is less affected by magnetic fields coming from outside the head (if spatially homogeneous), and thus decoding models trained on gradiometers usually perform better than those trained on magnetometers.

Additionally, we illustrated the extraction of spatiotemporal information by utilising PFI. The method involved choosing a window that spanned both space (across 4 sensors with 2 gradiometers and 1 magnetometer each, totalling 12 channels) and time (a 100ms window) simultaneously. The spatial window contains the 2 gradiometers and 1 magnetometer on three sides of the sensor in question. By sliding the window and performing the shuffling across all timepoints and channels we obtained a spatiotemporal discriminative information content profile. This comprehensive profile allowed us to understand how the disruption of specific spatiotemporal windows impacts the performance of the trained model, therefore highlighting the importance of those windows in discriminating between visual stimuli.

Finally, we introduce \textit{spectral PFI} to assess the effect of different frequency bands on the visual discrimination objective. Let $\mathbf{X} \in \mathbb{R}^{C \times T}$ be the input trial in the time domain. First, we apply the Fourier transform to each channel:

\begin{equation}
\mathbf{Z} = \mathcal{F}(\mathbf{X})
\end{equation}

to get the frequency domain representation $\mathbf{Z} \in \mathbb{C}^{C \times F}$, with $F$ frequency bands. Then each frequency band $f$ is randomly shuffled across the channel dimension, and the disrupted trial is transformed back to the time domain:

\begin{equation}
\hat{\mathbf{X}} = \mathcal{F}^{-1}(\mathbf{Z}_{\perp f})
\end{equation}

where $\mathbf{Z}_{\perp f}$ denotes random shuffling of feature (frequency band) $f$ in $\mathbf{Z}$. Note that the frequency band can either refer to a single value in $\mathbf{Z}$ or to a window ($l$) of frequencies $\mathbf{Z}[:, f:f+l]$. The window length provides a trade-off between smoothness, power, and specificity of the frequency profile we obtain. The importance $\Delta p_f$ of each frequency band/window $f$ is revealed by comparing the accuracy of disrupted trials with original trials:

\begin{equation}
\label{eq:spectralPFI}
\Delta p_f = \mathbb{E}_{\hat{\mathbf{X}},\mathbf{y}}\left[\mathcal{M}(\mathbf{y}, g(\hat{\mathbf{X}};\theta))\right] - \mathbb{E}_{\mathbf{X},\mathbf{y}}\left[\mathcal{M}(\mathbf{y}, g(\mathbf{X};\theta))\right]
\end{equation}

where $\mathcal{M}$ is our metric of goodness, accuracy, $g$ is the trained model with $\theta$ parameters, and $\mathbf{y}$ are the target classes. 

By applying this method to all frequency bands, we obtained a spectral information content profile, similar to the method of training separate LDA models on features from individual frequency bands \citep{higgins2022relationship}. Similar to spatiotemporal PFI we can combine spatial and spectral PFI, by running spectral PFI on a neighbourhood of 4 sensors at a time (spatial window) to assess the spectral information content of individual MEG channels. Thus the shuffled feature is $\mathbf{j} = \mathbf{Z}[c:c+k, f:f+l]$, where $c$ is the channel index, $k$ is the spatial window size, $f$ is the frequency band index, and $l$ is the frequency window width. We call this spatio-spectral PFI.

Previous work applied sliding window decoding in combination with spectral decoding (i.e., training separate models on individual frequency bands), thus assessing the temporo-spectral information content \citep{higgins2022relationship}. In order to make comparisons with this work, we developed temporo-spectral PFI.

Let $\mathbf{X} \in \mathbb{R}^{C \times T}$ be the input trial in the time domain. We frst apply the short-term Fourier transform with a (Hamming) window size $w=100~ms$ and hop size $h = 1~timestep$ to get the time-frequency representation matching parameters used in \cite{higgins2022relationship}:

\begin{equation}
\mathbf{Z} = \operatorname{STFT}(\mathbf{X}, w, h)
\end{equation}

where $\mathbf{Z} \in \mathbb{C}^{C \times N \times F}$, with $N$ time windows and $F$ frequency bins. Then each feature window $\mathbf{Z}[:, n:n+k, f:f+l]$ is randomly shuffled, were $n$ is the timepoint index, $k$ is the temporal window length, $f$ is the frequency band index, and $l$ is the frequency window width. Note that both the temporal and spectral windows can be set to 1, but usually a small window is better to balance specificity and smoothness. The disrupted trial is transformed back to the time domain:

\begin{equation}
\hat{\mathbf{X}} = \operatorname{iSTFT}(\mathbf{Z}_{\perp n, f})
\end{equation}

where $\mathbf{Z}_{\perp n, f}$ denotes random shuffling of the feature indexed by $(n, f)$ in $\mathbf{Z}$. The importance $\Delta p_{n,f}$ of each time-frequency window is revealed by comparing with the original trials as in Equation~\ref{eq:spectralPFI}:

\begin{equation}
\Delta p_{n,f} = \mathbb{E}_{\hat{\mathbf{X}},\mathbf{y}}\left[\mathcal{M}(\mathbf{y}, g(\hat{\mathbf{X}};\theta))\right] - \mathbb{E}_{\mathbf{X},\mathbf{y}}\left[\mathcal{M}(\mathbf{y}, g(\mathbf{X};\theta))\right]
\end{equation}

By repeating this over all frequency bands and time windows we obtain the temporo-spectral PFI profile.

\subsection{Experimental details}
The primary evaluation metric for the three datasets is classification accuracy across the respective number of classes (118, 92, or 8). The main focus of our analysis was on the 118 and 92-image datasets, with the 8-image dataset, included to demonstrate the effects of a much smaller sample size. Note that such a high number of classes are not commonly decodable in BCI and stimulation settings. All of the main results using our decoding methods (NN, LDA-NN, LDA-PCA) are multiclass. For all analyses, separate models were fit to separate subjects. Training and validation splits were created in a 4:1 ratio for each subject and class, with classes balanced across the splits. The NN approach was trained for 2000 epochs (full passes of the training data as opposed to epochs in the sense of MEG trials) using the Adam optimiser \citep{Kingma:2014}. The high number of epochs was selected as this allowed the training accuracy to converge to almost 100\%, while the validation accuracy also converged to a stable value for most participants. The output layer’s dimensionality was equal to the number of classes in the corresponding dataset. Dropout was set to 0.7 and applied before each of the three hidden layers.

The dimensionality reduction layer and PCA were both set to 80 components, as it is slightly higher than the inherent dimensionality reduction of MaxFilter (of rank 64) which is applied to the MEG data, and thus contains more than 99\% of variance. We briefly tried our pipeline with 60 components as well on 1 subject and found similar results. Thus, as long as the dimensionality reduction is at least the rank of the data there should be no loss of information and similar results should be observed. Projecting to more dimensions than the rank of the data can introduce issues such as linear dependence in the input components. However, we observed that our decoding models are robust and probably due to the large amounts of regularisation used are not affected by such issues.

Validation data was not used for early stopping, and the trained NN dimensionality reduction weight matrix (used in LDA-NN) was extracted after the full 2000 epochs of training on the training data. For the LDA models, the shrinkage parameter was set to \textit{auto} using the sklearn package. Comparisons of interest over methods were evaluated using Wilcoxon signed rank tests, with within-subject pairing and subject-level mean accuracies over validation examples as the samples. We used Bonferroni correction to correct for multiple comparisons. The PyTorch package was used for training \citep{pytorch2019}, and several other packages were utilised for analysis and visualisation \citep{scikit-learn, 2020SciPy-NMeth, harris2020array, pandas:2010, Waskom2021, Hunter:2007}.

\section{Results}
\label{sec:results}

\subsection{Full-epoch models better than sliding-window decoding}

\begin{figure}[!ht]
  \centering
  \includegraphics[width=0.9\linewidth]{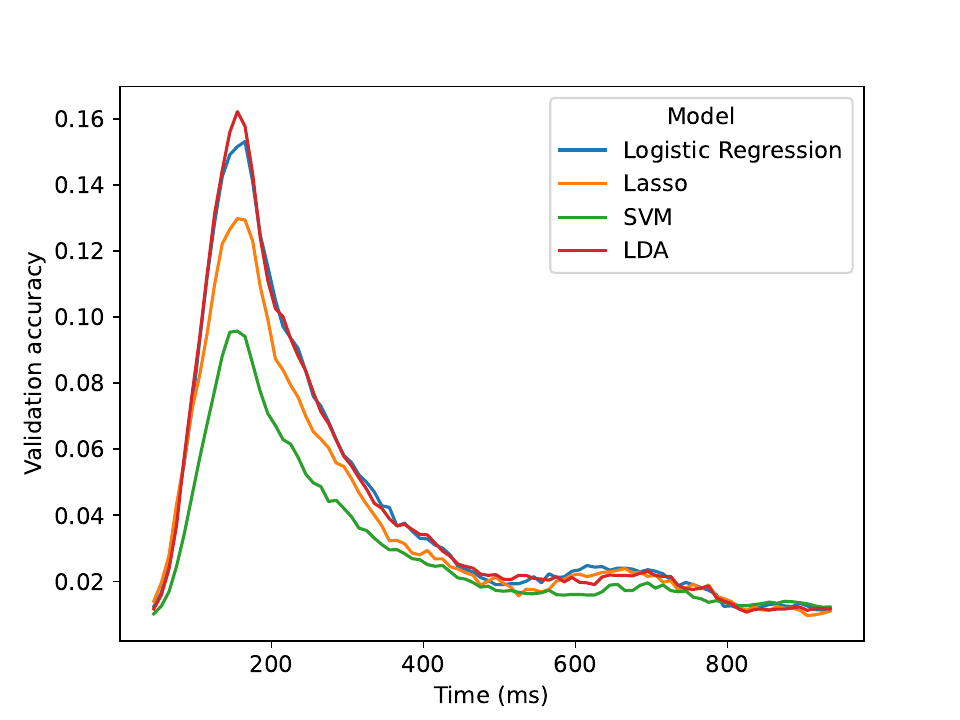}
    \caption{Comparing different sliding window models trained on PCA features on the 118-image dataset for multiclass decoding. The sliding window size is 100ms. Results are averaged across subjects.}
    \label{fig:models}
\end{figure}

We set out to test whether full-epoch decoding is better than timepoint-by-timepoint and sliding-window decoding, which are common practices in the M/EEG literature \citep{carlson2011high, carlson2013representational, su2012spatiotemporal, ramkumar2013feature, cichy2017dynamics, grootswagers2017decoding, kurth2016fast, liu2019human, higgins2022relationship}. We wanted to make sure that our classifier of choice, LDA is at least as good as other commonly used models for multiclass decoding, including support vector machines (SVM), linear discriminant analysis (LDA), logistic regression, and Lasso. The results, depicted in Figure~\ref{fig:models}, indicate that LDA and logistic regression exhibited comparable performance (no statistical difference) and performed better than the other 2 examined models. For this reason, and as described in the methods, we used LDA in all further analyses for comparing different classification strategies.

\begin{figure}[!t]
  \centering
  \includegraphics[width=0.6\linewidth]{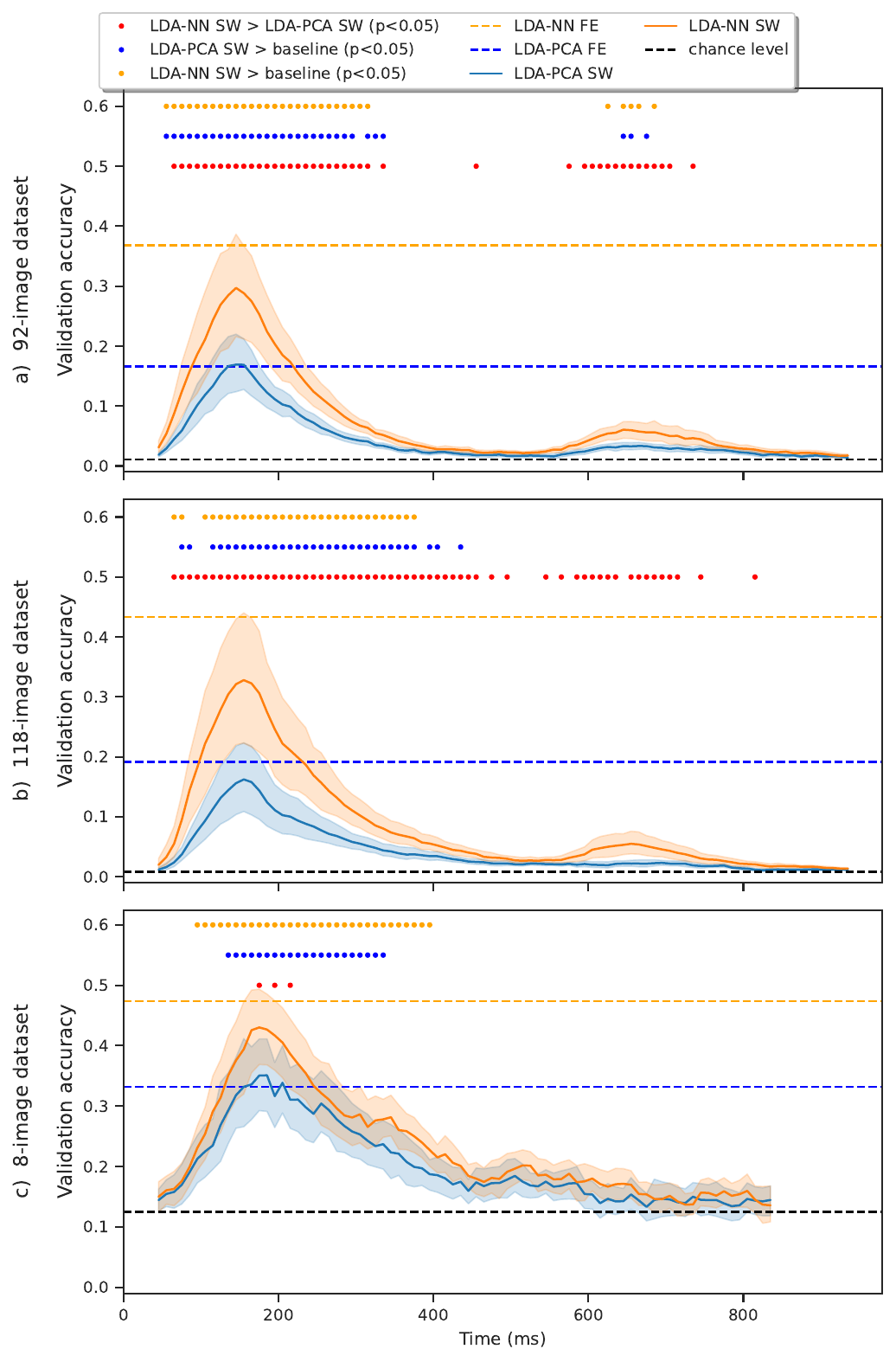}
    \caption{Models trained on the sliding-window versions of the 92-class dataset (top), 118-class dataset (middle) and 8-class dataset (bottom) for multiclass decoding. Wilcoxon signed-rank tests are reported between sliding window LDA-NN and LDA-PCA. We also ran Wilcoxon signed-rank tests between the first timepoint of LDA-NN and LDA-PCA and all other timepoints. This shows statistical significance compared to a “baseline” level. FE stands for full-epoch models, and SW stands for sliding window models. The blue and orange dotted lines are placed at the average performance of full-epoch LDA-NN and LDA-PCA, respectively. All statistical tests are Bonferroni corrected for multiple comparisons across all time points (i.e. p-values are multiplied by 90). Shading indicates the 95\% confidence interval across subjects. For the full-epoch results, please see Figure~\ref{fig:fullepoch_comparison} for distributions across subjects. LDA-NN is better across almost all time points than LDA-PCA, and full-epoch accuracy is higher than peak sliding window accuracy for both LDA-NN and LDA-PCA (except in the 92-class and 8-class datasets).}
    \label{fig:sliding_comparison}
\end{figure}

\begin{figure}[!t]
  \centering
  \includegraphics[width=1.0\linewidth]{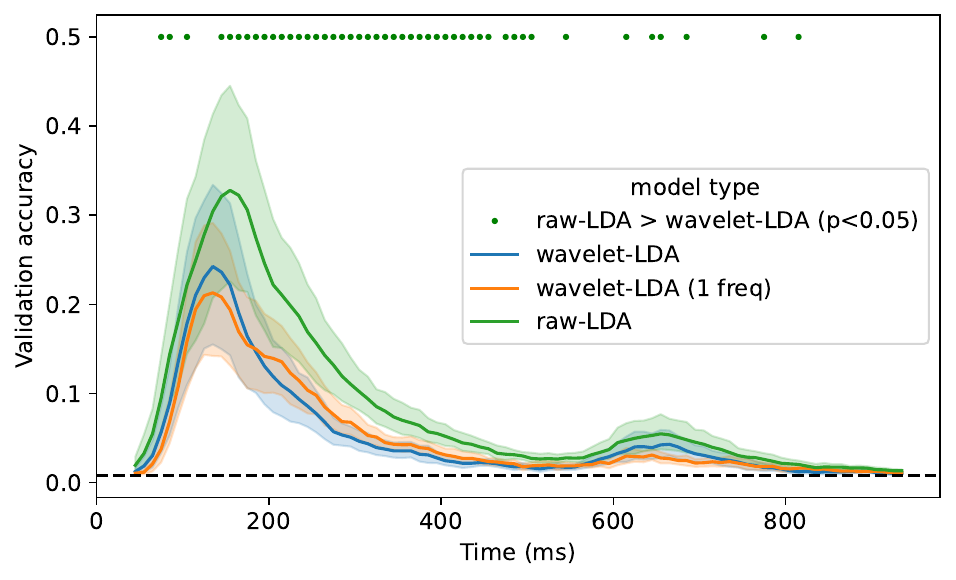}
    \caption{Comparison of our sliding window LDA-NN approach with LDA-NN using wavelet features on the 118-image dataset. The wavelet features are computed after the dimensionality reduction, with the same settings as in \cite{higgins2022relationship}. A hamming window of 10 timesteps was used with an overlap of 9 timesteps. wavelet-LDA corresponds to using a concatenation of all frequency bands for training the LDA model, and wavelet-LDA (1 freq) uses a single frequency band (10Hz). We selected this band based on previous results in \cite{higgins2022relationship}, achieving the best decoding performance using this band only. Results are averaged across subjects, and shading indicates the 95\% confidence interval across subjects.}
    \label{fig:wavelet_comp}
\end{figure}

We also wanted to make sure that the choice of using raw time-domain data is not limiting decding performance. Thus we compared our raw sliding window decoding performance with the wavelet approach employed in \citet{higgins2022relationship} and found the latter to be significantly worse across most timesteps (Figure~\ref{fig:wavelet_comp}).

The performance of multiclass full-epoch models was compared to that of sliding-window decoding for both LDA-PCA and LDA-NN on the three datasets in Figure~\ref{fig:sliding_comparison}. The peak performance of sliding-window decoding was observed at 150-160 ms post-stimulus for the 92 and 118-image datasets, and at 200 ms post-stimulus for the 8-image dataset. These findings are broadly consistent with previous research on the temporal dynamics of visual information processing in MEG \citep{cichy2014resolving, cichy2016comparison, cichy2017multivariate, higgins2022relationship, liu2019human, guggenmos2018multivariate}. For the 92 and 118-image datasets a second smaller peak was observed around 650-660 ms post-stimulus. As the image presentation is switched off at exactly 500ms, we reason that the second peak is due to the brain reacting to this event. The first peak is observed 150-160 ms post-stimulus onset, while the second peak occurs 150-160 ms post-stimulus offset.

Across subjects, the full-epoch LDA-PCA approach demonstrated significantly higher accuracy than the best sliding-window LDA-PCA approach on the 118-class dataset (3.1\% increase, p < 1e-4). On the 92-class dataset, no significant difference was observed between these models, though full-epoch LDA-PCA still outperformed the sliding-window version at most time windows. A similar comparison between full-epoch LDA-NN and peak sliding-window performance showed that full-epoch models had higher accuracy on both the 92- and 118-class datasets (7.1\% and 10.5\% increase, respectively, p < 1e-4). The tests were corrected for multiple comparisons across time points. These results indicate that training a model on the full epoch generally leads to better performance than using the best sliding-window model, except for the LDA-PCA approach on the 92-image dataset. However, as noted in the following section, it is advisable to use an LDA-NN model in any case.

On the 8-image dataset, the full-epoch model had higher accuracy than the peak sliding-window model, though this difference was not significant. It should be noted that the reduced effectiveness of the full-epoch model on this dataset may be due to both the longer epoch of 900 ms and the smaller amount of data. This can lead to overfitting due to a larger number of features and fewer examples.

\begin{figure}[!t]
  \centering
  \includegraphics[width=1.0\linewidth]{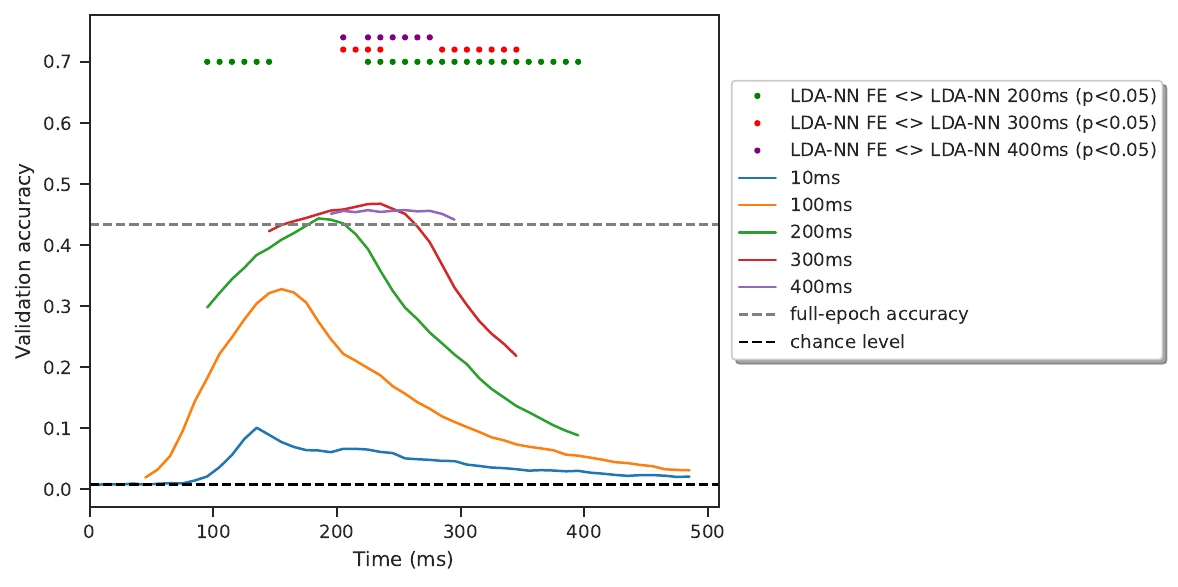}
    \caption{Comparing sliding window LDA-NN with different window sizes on the 118-image dataset. Results are averaged across subjects. Wilcoxon signed-rank tests are reported between the sliding window models and the full-epoch model, Bonferroni corrected for all comparisons in the figure.}
    \label{fig:sliding_window_comp}
\end{figure}

Our results could be affected by the choice of window size for the sliding window LDA (100 ms). Thus, we repeated the sliding window LDA for different window sizes, including a window of 1 sample (i.e., timepoint-by-timepoint decoding), and the results are presented in Figure~\ref{fig:sliding_window_comp}. We trained models using sliding window sizes of 10ms, 100ms, 200ms, 300ms, and 400ms. As expected, using a single time point (10ms) resulted in lower accuracy compared to a 100ms window. As the window size increased, we observed two trends. First, accuracy improved and the peak accuracy of a 200ms window already reached the full-epoch level. Second, the accuracy profile became more distorted and the peak shifted compared to the results obtained with a single time point. In some cases, full-epoch performance was even exceeded by a few percentage points with a 300ms window. This may not be surprising, as a larger window that focuses on the most significant part of the input results in fewer features compared to using the full epoch. However, it is advisable to avoid using a window larger than 100ms in sliding window analysis due to its distortion and lower temporal resolution. One potential solution could be to combine the sliding window models with PFI analysis, but this would be inefficient. We can therefore conclude that using a full-epoch model is the optimal solution, even if it results in slightly lower accuracy. Additionally, we expect that with larger datasets, full-epoch models would outperform sliding window models regardless of window size, as the ratio of features to examples would be reduced.

\subsection{Supervised dimensionality reduction is better than PCA}

\begin{figure}[!ht]
  \centering
  \includegraphics[width=1.0\linewidth]{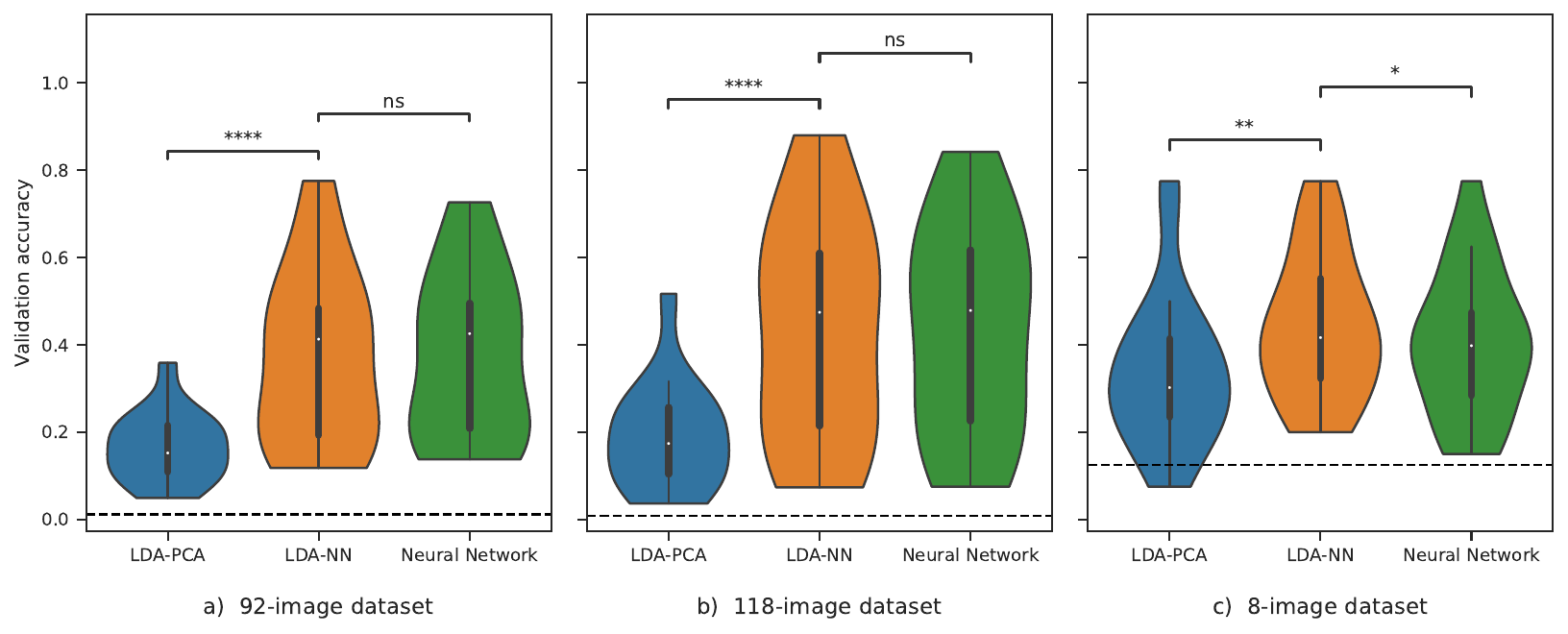}
    \caption{Models trained on the full-epoch versions of the 92-class (left), 118-class (middle), and 8-class (right) datasets for multiclass decoding. The violin plot distributions are shown over the mean individual subject performances. The dashed black line represents the chance level. Wilcoxon signed-rank tests are shown where 4 stars mean p < 1e-4, and 3 stars mean p < 1e-3. “ns” means that the p-value is higher than 0.05.}
    \label{fig:fullepoch_comparison}
\end{figure}

We next investigated the effect of incorporating a learned, supervised dimensionality reduction layer in our models, i.e. a dimensionality reduction optimised to aid a downstream classification task. We, therefore, modified the LDA-PCA approach by replacing the unsupervised dimensionality reduction performed by PCA with the supervised dimensionality reduction (of equal dimensionality) from the Neural Network (NN) approach, as described in Section~\ref{sec:methods}. We refer to this modified approach as LDA-NN. As shown in Figure~\ref{fig:fullepoch_comparison}, this simple change resulted in a significant improvement in performance (20.2\% for the 92-class dataset and 24.2\% for the 118-class dataset, p < 1e-4). We also assessed the performance of the pure NN model and found that it has a similar performance to LDA-NN. In other words, the supervised dimensionality reduction effectively eliminated the performance gap between the LDA and the Neural Network (NN) approach.

The sliding window versions of LDA-PCA and LDA-NN are also compared in Figure~\ref{fig:sliding_comparison}. Across most time points (and all time points around the 2 peaks), LDA-NN is significantly better than LDA-PCA, when Bonferroni corrected for multiple comparisons across time points. Similar conclusions can be drawn on the 8-image dataset, although LDA-NN is better than the NN approach, possibly due to the reduced performance of neural networks on small datasets in general. In summary, our results suggest that using a full-epoch LDA-NN or a simple linear Neural Network results in the best performance across all datasets and that the feature reduction should be learned in a supervised manner for both the LDA and Neural Network models.

\subsection{Temporal PFI}

\begin{figure}[t]
\begin{subfigure}{0.33\textwidth}
  \centering
  \includegraphics[width=1.0\linewidth]{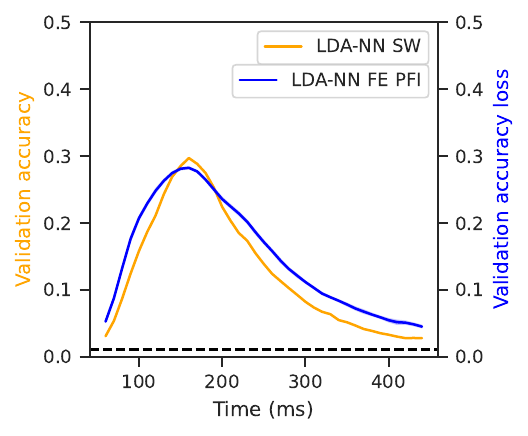}
  \caption{92-image dataset}
  \label{fig:cichy92_temporalPFI}
\end{subfigure}%
\begin{subfigure}{0.33\textwidth}
  \centering
  \includegraphics[width=1.0\linewidth]{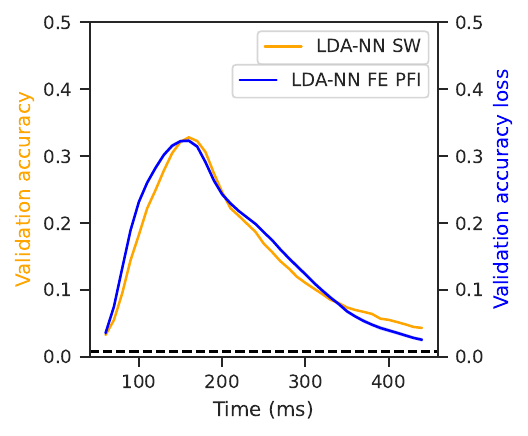}
  \caption{118-image dataset}
  \label{fig:cichy118_temporalPFI}
\end{subfigure}%
\begin{subfigure}{0.33\textwidth}
  \centering
  \includegraphics[width=1.0\linewidth]{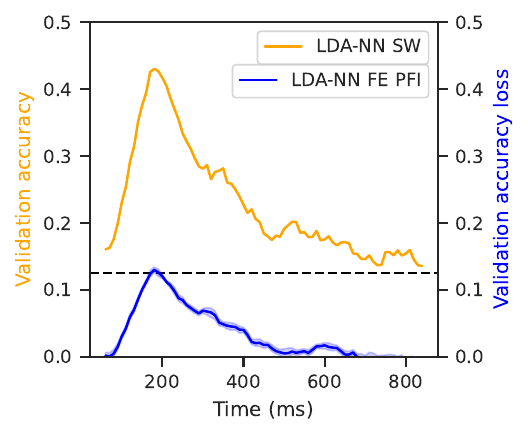}
  \caption{8-image dataset}
  \label{fig:replay_temporalPFI}
\end{subfigure}
\caption{Comparison of multiclass sliding window LDA-NN (orange) and the temporal PFI of multiclass full-epoch LDA-NN (blue) across the three datasets. Results are averaged across all subjects in the respective datasets, and shading indicates 95\% confidence interval across permutations for PFI. Chance level for LDA-NN SW is indicated with a dashed line.}
\label{fig:temporalPFI}
\end{figure}

One of the benefits of sliding window or time-point-by-time-point decoding is that it is straightforward to obtain a time course of decoding accuracy (e.g., Figure~\ref{fig:sliding_comparison}), allowing for interpretation of the temporal dynamics of neural representations. Here we show that full epoch decoding in combination with permutation feature importance (PFI) can give the same qualitative information. The results presented in Figure~\ref{fig:temporalPFI} indicate that temporal PFI applied to a full-epoch LDA-NN model produces temporal profiles similar to those obtained using sliding window LDA-NN models with a window size of 100ms across all three datasets. The peak sliding window performance also aligns well with the peak accuracy loss for PFI.

\begin{figure}[!t]
\begin{subfigure}{0.5\textwidth}
  \centering
  \includegraphics[width=1.0\linewidth]{mvpa_figures/cichy118_temporalPFI.pdf}
  \caption{Standard PFI}
  \label{fig:cichy118_temporalPFI2}
\end{subfigure}%
\begin{subfigure}{0.5\textwidth}
  \centering
  \includegraphics[width=1.0\linewidth]{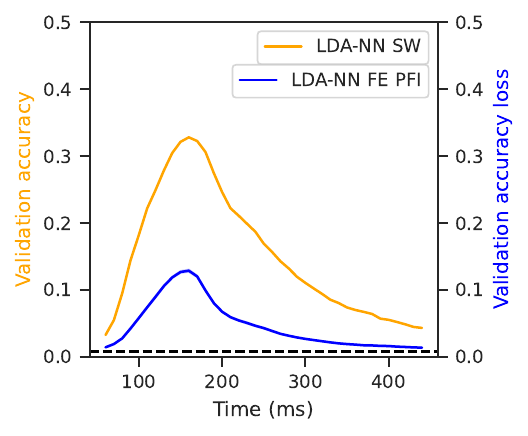}
  \caption{Inverse PFI}
  \label{fig:cichy118_invtemporalPFI}
\end{subfigure}
\caption{Comparison of multiclass sliding window LDA-NN (orange) with standard temporal PFI (a) and inversed temporal PFI (b) using a trained LDA-NN model on the 118-image dataset. Results are averaged across subjects, and shading indicates the 95\% confidence interval across permutations for PFI. Chance level is indicated by the dashed line.}
\label{fig:invtemporalPFI}
\end{figure}

We investigated an alternative method of performing PFI, referred to as inverse PFI. This method is not common in the literature, but could be interesting from an MVPA viewpoint. Inverse PFI differs from standard PFI in that it shuffles values outside a specified time window, rather than within it. Standard PFI assesses the impact of disrupting information within a specific window on performance and therefore reveals the importance of that window for discriminating between images. In contrast, inverse PFI investigates performance when all information outside a specified window is disrupted, thereby providing insight into the performance that can be achieved using only the information contained within the time window. The temporal PFI results for both standard and inverse PFI are presented in Figure~\ref{fig:invtemporalPFI}. While both approaches are similar to the standard sliding window LDA profile, there are some differences as well.

\subsection{Spatial PFI}

\begin{figure}[!t]
\begin{subfigure}{0.5\textwidth}
  \centering
  \includegraphics[width=1.0\linewidth]{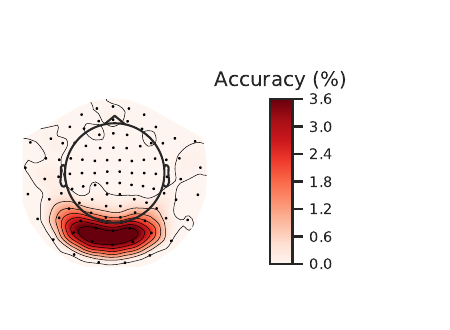}
  \caption{Spatial PFI}
  \label{fig:cichy118_spatialPFI}
\end{subfigure}%
\begin{subfigure}{0.5\textwidth}
  \centering
  \includegraphics[width=1.0\linewidth]{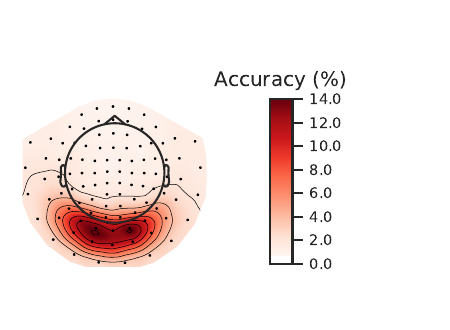}
  \caption{Channel-wise LDA}
  \label{fig:cichy118_channelLDA}
\end{subfigure}
\caption{Comparison of multiclass channel-wise LDA model (b) with the spatial PFI of multiclass full-epoch LDA-NN (a). Spatial maps are averaged across all 15 subjects on the 118-image dataset. Both PFI and the channel-wise LDA model are run on 3-channels in the same location at a time (1 magnetometer and 2 gradiometers).}
\label{fig:spatialPFI}
\end{figure}

We investigated the ability of PFI to accurately capture spatial information by applying it to a full-epoch LDA-NN model on the 118-image dataset. To do this, we permuted time points from the gradiometers and magnetometers located at the same position in the MEG data simultaneously to obtain a single sensor space map. We compared these to the maps obtained by training separate LDA models on the full epoch of the same three sensors (2 gradiometers and 1 magnetometer). This approach can be viewed as a sliding window across space. All PFI results are averaged over the accuracy losses of individual subjects, which can somewhat smear both spatial and temporal profiles. The results, shown in Figure~\ref{fig:spatialPFI}, demonstrate good alignment between the accuracy loss of spatial PFI and per-sensor accuracy of LDA-NN, indicating that PFI can effectively recover spatial information content.

\begin{figure}[!t]
\begin{subfigure}{0.33\textwidth}
  \centering
  \includegraphics[width=1.0\linewidth]{mvpa_figures/cichy118_spatialPFI.pdf}
  \caption{Standard PFI}
  \label{fig:cichy118_spatialPFI2}
\end{subfigure}%
\begin{subfigure}{0.33\textwidth}
  \centering
  \includegraphics[width=1.0\linewidth]{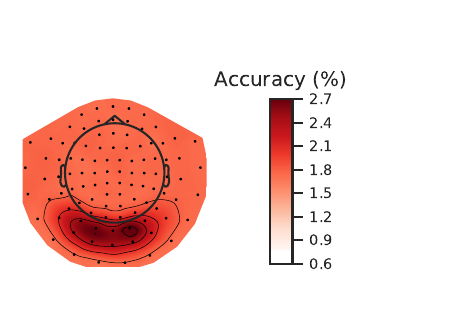}
  \caption{Inverse PFI}
  \label{fig:cichy118_invspatialPFI}
\end{subfigure}%
\begin{subfigure}{0.33\textwidth}
  \centering
  \includegraphics[width=1.0\linewidth]{mvpa_figures/cichy118_channelLDA.pdf}
  \caption{Channel-wise LDA}
  \label{fig:cichy118_channelLDA2}
\end{subfigure}
\caption{Comparison of channel-wise LDA model (c) with the standard spatial PFI (a) and inverse spatial PFI (b) of full-epoch multiclass LDA-NN. Results are averaged across all 15 subjects on the 118-image dataset. Both PFI and the channel-wise LDA model are run on 3-channels in the same location at a time (1 magnetometer and 2 gradiometers).}
\label{fig:invspatialPFI}
\end{figure}

We also conducted the inverse PFI analysis in the spatial domain, the results of which are shown in Figure~\ref{fig:invspatialPFI}. In this domain, the inverse PFI approach exhibits less contrast between visual channels and other channels but appears to distinguish between visual channels more similarly to channel-wise LDA than standard PFI. It is not the aim of this study to determine which approach is superior, as both seem to have their merits.

\subsection{Spatiotemporal PFI}

\begin{figure}[!t]
\begin{subfigure}{0.77\textwidth}
  \centering
  \includegraphics[width=1.0\linewidth]{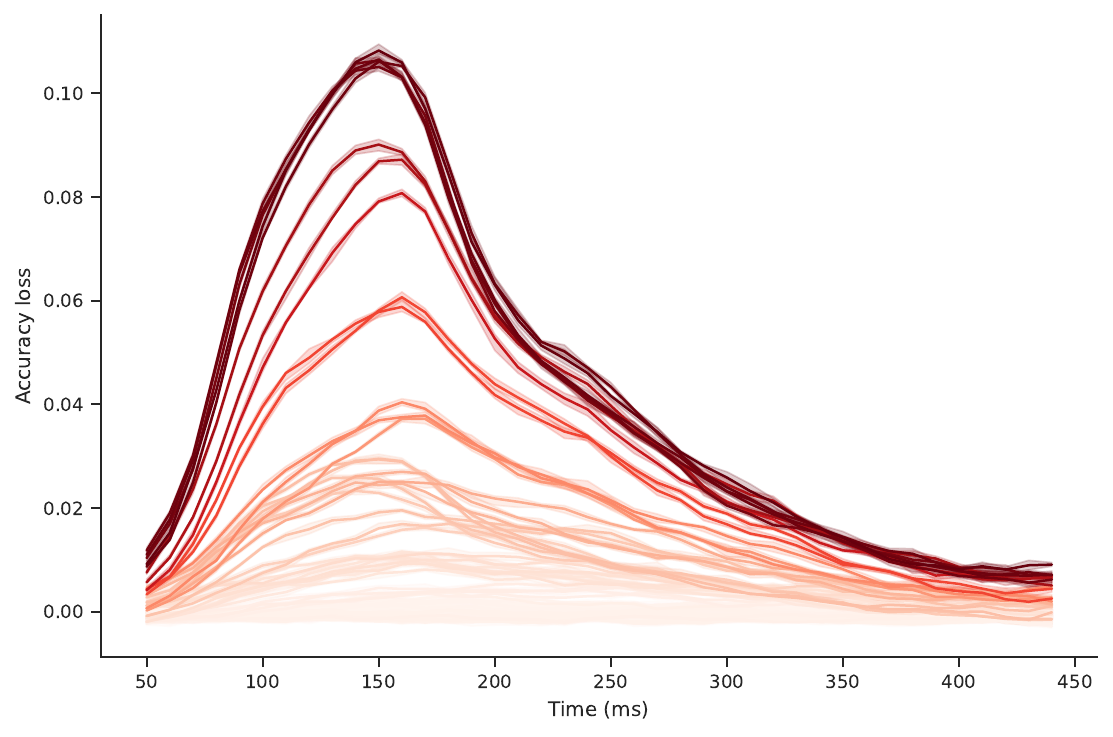}
  \label{fig:cichy118_spatiotemporalPFI}
\end{subfigure}
\begin{subfigure}{1.0\textwidth}
  \centering
  \includegraphics[width=1.0\linewidth]{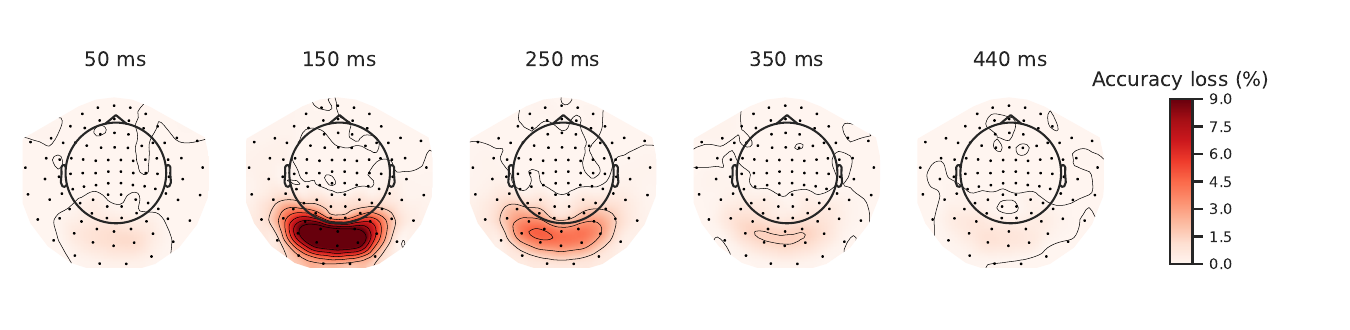}
  \label{fig:cichy118_spatiotemporal_maps}
\end{subfigure}
\caption{Spatiotemporal PFI of multiclass full-epoch LDA-NN on the 118-image dataset. Blocks of 4-channel neighbourhoods and 100ms time windows are shuffled to obtain a spatial and temporal profile jointly. Each line in the temporal profile corresponds to a sensor, and each sensor space map is obtained with a time window centred around the respective time point. The color map of the upper plot is based on the coloring of sensors at 150ms in the lower plot. The shading in the upper plot is across the 10 permutations used for PFI and indicates the 95\% confidence interval. Both temporal and spatial profiles are averaged over subjects.}
\label{fig:spatiotemporalPFI}
\end{figure}

We also employed PFI to extract spatiotemporal information jointly from a trained full-epoch LDA-NN model on the 118-image dataset. Specifically, we used a 100 ms time window and a 4-channel spatial window (i.e., the 2 gradiometers and 1 magnetometer on three sides of the sensors in question) for each time point and channel, shuffling the values within these blocks. This allowed us to unravel the temporal and spatial information simultaneously, showing that only channels located in the visual area exhibited the characteristic temporal profile and that there was a gradient with channels further from the visual area displaying progressively lower peak accuracy loss (Figure~\ref{fig:spatiotemporalPFI}).

We observed that the temporal evolution of the sensor space maps showed the visual area sensors to be consistently the most important for the decoding objective across all time points. In theory, the sliding window LDA and the per-channel LDA approach could be combined to get a similar spatiotemporal profile, where each LDA model is trained on the sliding window of 4 channels at a time. However, in practice accuracy might suffer substantially with so few input features, and it would be computationally taxing considering the amount of LDA models required to train. Overall, PFI proved to be a useful technique for investigating full-epoch data and obtaining spatiotemporal information similar to what can be obtained from individual sliding window models.

\subsection{Spectral PFI}

\begin{figure}[!t]
  \centering
  \includegraphics[width=0.7\linewidth]{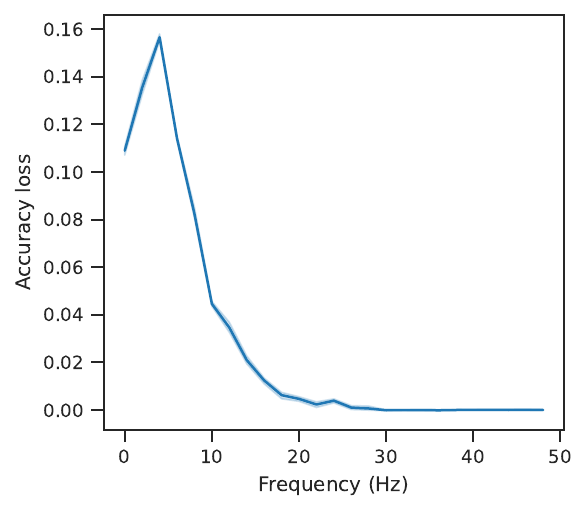}
  \caption{Spectral PFI of multiclass full-epoch LDA-NN on the 118-image dataset. Shading indicates 95\% confidence interval across permutations. Results are averaged across subjects.}
  \label{fig:cichy118_spectralPFI}
\end{figure}

Figure~\ref{fig:cichy118_spectralPFI} presents our spectral PFI results averaged over subjects. This shows a clear peak of spectral information content at 4Hz, after which the power  rapidly declines with increasing frequency. However, it should be noted that, because of the sampling rate of the data and the size of the epochs, the frequency resolution is only 2Hz. This means that the apparent 4Hz peak is due to the 1Hz highpass used for preprocessing the data, and so in actuality there is simply a 1/f characteristic, as is expected in MEG data. We have confirmed this by plotting the psd of the raw (bandpassed) data with a matched frequency resolution, and found the same peak at 4Hz, which shows that this is an artefact of the frequency resolution.

\begin{figure}
  \centering
  \includegraphics[width=0.7\linewidth]{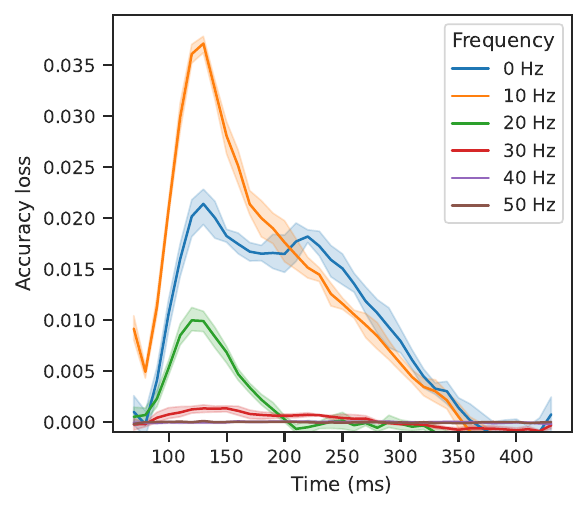}
  \caption{Temporospectral PFI of multiclass full-epoch LDA-NN on the 118-image dataset. Shading indicates 95\% confidence interval across permutations. Results are averaged across subjects.}
  \label{fig:cichy118_temporospectralPFI}
\end{figure}

We present temporospectral PFI in Figure~\ref{fig:cichy118_temporospectralPFI}, which reveals temporal information content within individual frequency bands, in an alternative manner to using separate LDA models trained on wavelet features \citep{higgins2022relationship}. All frequency values represent bands centred around the respective frequencies, except the 0Hz band which represents the true 0Hz signal, i.e. the average over the time window. For computing the STFT we followed the same setup as in \cite{higgins2022relationship}. Because we are using a 100ms window (10 timesteps) for computing the STFT the frequency resolution is 10Hz. When permuting a specific time window, we also permuted the frequency content of the time window right before and after, to obtain a smoother temporal profile.

As expected from the standard temporal PFI, the temporal peak is between 100 and 150ms. Spectrally, higher frequency bands tend to be less and less useful to the decoding objective, confirming the observations of \cite{higgins2022relationship}. However, we think both the figure in \cite{higgins2022relationship}, and the temporospectral PFI analysis are slightly misleading, as they could be interpreted as having a peak in information content in the 10Hz band. As observed in Figure~\ref{fig:cichy118_spectralPFI} this effect is explained simply by the $1/f$ characteristic. Because of the poor frequency resolution, both lower and higher frequencies are represented in the 10Hz band, thus all it shows is the $1/f$ characteristic, and the reason why it is higher than the “0Hz” band is because the 0Hz band contains solely the true 0Hz content. A potentially better approach to disentangling time-frequency information content would be to bandpass the data first into specific frequency bands, then train our decoding model and compute the temporal PFI on each bandpassed data version.

\begin{figure}[!t]
  \centering
  \includegraphics[width=1.0\linewidth]{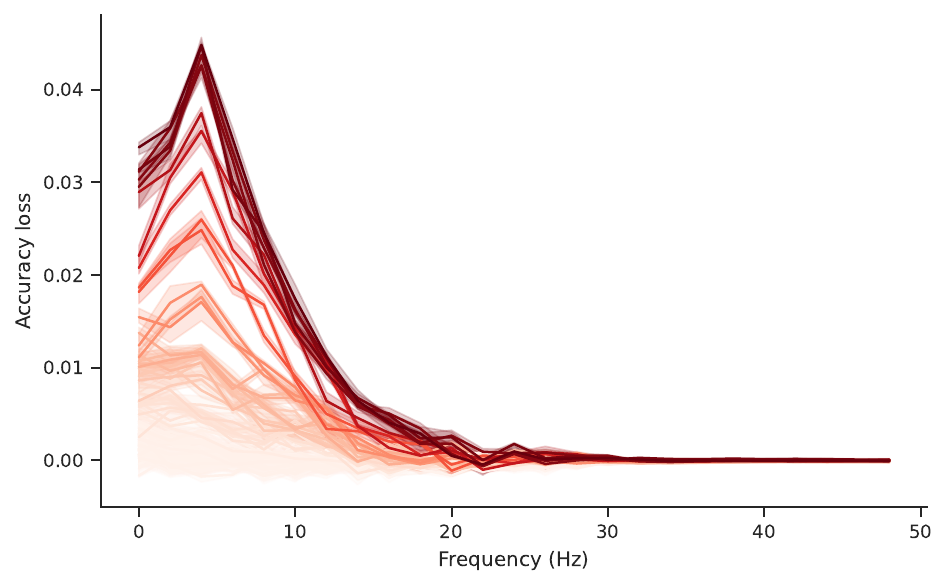}
    \caption{Spatiospectral PFI of multiclass full-epoch LDA-NN on the 118-image dataset, averaged over subjects. Blocks of 4-channel neighbourhoods are shuffled in each frequency to obtain the per-channel frequency profile. Each line corresponds to a sensor. The color map of the upper plot is based on the overall spatial PFI of each sensor, i.e. sensors with high spatial PFI accuracy loss are shown as darker red. The shading is across the permutations used for PFI and indicates the 95\% confidence interval.}
    \label{fig:spatiospectralPFI}
\end{figure}

We note that it is expected that we would find little to no signal above 25Hz, because of the lowpass filter we have employed. In later timepoints (>200ms) the 0Hz band seems to be slightly more important than the 10Hz band, potentially meaning that the classifier relies more on average rather than oscillatory activity after the visual peak. Similar to spatiotemporal PFI we can combine spatial and spectral PFI to assess the spectral information content of individual MEG channels (see Figure~\ref{fig:spatiospectralPFI}).

\section{Discussion}
We made the following contributions in this chapter. We showed empirically that full-epoch models achieve higher accuracy than sliding window decoding models. We showed how temporal, spatial, and spectral brain activity patterns related to stimulus discrimination can be extracted for any black-box full-epoch model. We have shown that learning the dimensionality reduction of input features in a supervised way, within a neural network optimised for the classification task, improves performance substantially. Next, we discuss each result in more detail.

We have found that training a single full-epoch model for multiclass decoding is effective in improving decoding performance. We have shown how this can be used while still providing neuroscientific insights by using PFI to learn which features are contributing to the decoding accuracy. Our results show that a full-epoch model generally performs better than individual sliding window models for visual decoding tasks, and the magnitude of this effect increases with the size of the dataset. The time-efficiency benefits of using a full-epoch model are significant, as training sliding window models takes roughly 10 times longer than a single full-epoch model for a 100 ms time window with a 100 Hz sampling rate.

Our analysis of different window sizes showed that while larger window sizes may improve performance, they are not effective in accurately capturing the temporal profile of information content. It has also been suggested that using equal-length time windows for all trials does not account for trial-by-trial variability, and \cite{vidaurre2018spontaneous} proposed time-resolved decoding using a Hidden Markov Model to segment trials along the time dimension. This approach still involves training multiple models on multiple time windows. We, therefore, recommend using full-epoch models, as they only need to be trained once and contain information from all potentially useful time windows. After training any desired window size can be selected for temporal or spatial investigations through PFI, providing good decoding performance and dynamic spatiotemporal resolution without the need for retraining.

Both dataset size and trial length are important considerations in whether full-epoch decoding is beneficial over sliding-window decoding. Due to having to deal with more input features in full-epoch decoding overfitting becomes an issue with small datasets. Similarly if the experimental setup involves trials of e.g., multiple seconds this may lead to overfitting due to more input features. Thus, full-epoch decoding has considerable limitations in the case of small datasets and long trial windows. The length of the full-epoch window used for decoding should depend on the task at hand (i.e. what is the length of the time period which contains information) and the amount of data available.

There are additional considerations when adapting our methods to BCI settings. In real-time decoding it may be unacceptable to wait for 500ms before making a prediction to the user. In this case it is advisable to use smaller windows for good temporal resolution at the loss of classification performance. In asynchronous BCI paradigms it is possible to use larger windows (full epochs).

We also found that incorporating a supervised dimensionality reduction layer is essential for good decoding performance when using linear neural networks and LDA models. This can be used as a drop-in replacement over standard unsupervised dimensionality reduction typically done with PCA. One limitation of our supervised dimensionality reduction is the increased computation time of training a neural network versus computing a simple PCA. However, the neural network approach provides both the dimensionality reduction and the final decoding model end-to-end. If optimised well, running times are comparable to doing a PCA and then training an LDA model.

To further solidify our results we compared our approach with a Riemannian classifier on the 118-image dataset. We used the pyriemann library \citep{barachant2022pyriemann}, specifically \textit{XdawnCovariances} for covariance computation followed by a tangent space projection. The LDA model was then applied to the features in the tangent space. The average validation accuracy over participants was 0.16, which is lower than the LDA-PCA approach. Both LDA-NN and the neural network are statistically significantly (\(p<1e-3\)) better than this. We note that because of the high number of classes and channels, the dimensionality of the tangent space features was much higher than the features obtained from our neural network approach. We tried using the best possible settings for the Riemannian classifier, i.e. specifying the number of filters (\(n_{filter}\)) to be 1, to reduce the number of features generated by \textit{XdawnCovariances}. However, the main issue is the number of classes, since the number of output features is equal to \(2 \cdot n_{classes} \cdot min(n_{channels}, n_{filter}\)). In standard EEG-BCI applications, the number of classes is much lower than ours (118), which is why Riemannian classifiers may be better suited for EEG-BCI. To fully leverage the Riemannian classifier for this kind of MEG data, an additional feature reduction or selection step may be needed.

We compared PFI results from a full-epoch model with those from individual models trained on either separate time or spatial windows. This demonstrated that PFI can effectively extract both temporal and spatial information, and can also be used to investigate the interaction between these two dimensions. We also introduced a new technique whereby PFI can be used to extract spectral discriminatory information content and confirmed that this matches previous work training individual models on separate frequency band features. PFI is a particularly flexible technique, as it can be applied to nonlinear models and temporal or spatial resolution can be chosen post-hoc without the need for retraining. The performance of full-epoch nonlinear decoding and corresponding PFI analysis is explored in the next chapter. PFI can also be applied to individual conditions or single trials by rerunning with different permutations, enabling the investigation of various neuroscientific questions. Other methods for obtaining temporal and spatial information from trained models, such as the Haufe transform, are limited to linear models and do not provide trial-level patterns \citep{haufe2014interpretation}. As opposed to the statistical nature of PFI, the Haufe transform directly maps the weights of a linear decoding model to input patterns, thus showing which parts of the input are the most important for the decoding objective. One limitatin of PFI compared to the Haufe transform is that the absence of influence on the output does not necessarily mean that those parts of the input (channels or time windows) do not contain information about the target.\\

To conclude, we recommend using a full-epoch multiclass model equipped with a supervised dimensionality reduction in order to achieve the best possible decoding performance while also allowing for flexibility in conducting neuroscientific investigations post-hoc such as MVPA or RSA. Our methods and recommendations scale well with data size and can be readily applied to deep learning models as well, thus bringing the applications of decoding to brain-computer interfaces and representational brain dynamics under a joint approach.

\chapter{Group-level decoding}
\label{Chap4}

In the previous chapter we have seen how we can improve decoding performance at the individual subject level, while providing useful neuroscientific insights. In this chapter we move on from within-subject variability and explore how nonlinear deep learning methods may be used to deal with between-subject variability. As we will see, while linear methods work well at the subject level, tackling group data requires more complex models.

Decoding is typically subject-specific and does not generalise well over subjects, due to high amounts of between subject variability. Techniques that overcome this will not only provide richer neuroscientific insights but also make it possible for group-level models to outperform subject-specific models. In this chapter, we propose a method that uses subject embedding, analogous to word embedding in Natural Language Processing, to learn and exploit the structure in between-subject variability as part of a decoding model, our adaptation of the WaveNet architecture for classification. We apply this to magnetoencephalography data, where 15 subjects viewed 118 different images, with 30 examples per image; to classify images using the entire 1s window following image presentation.

We show that the combination of deep learning and subject embedding is crucial to closing the performance gap between subject- and group-level decoding models. Importantly, group models outperform subject models on low-accuracy subjects (although slightly impair high-accuracy subjects) and can be helpful for initialising subject models. While we have not generally found group-level models to perform better than subject-level models, the performance of group modelling is expected to be even higher with bigger datasets. In order to provide physiological interpretation at the group level, we make use of permutation feature importance (PFI). This provides insights into the spatiotemporal and spectral information encoded in the models. We show that PFI works similarly well with nonlinear full-epoch models as with the linear models in the previous chapter.

\textit{Note: } Most of this chapter is part of a published paper \citep{csaky2023group}. All of the work has been carried out by the thesis author. Most experiments in this chapter can be reproduced using the associated GitHub repository\footnote{\url{https://github.com/ricsinaruto/MEG-group-decode}}.

\section{Introduction}
\label{sec:introduction}

\begin{figure}[!t]
\centering
\begin{subfigure}{0.5\textwidth}
  \includegraphics[width=1.0\linewidth]{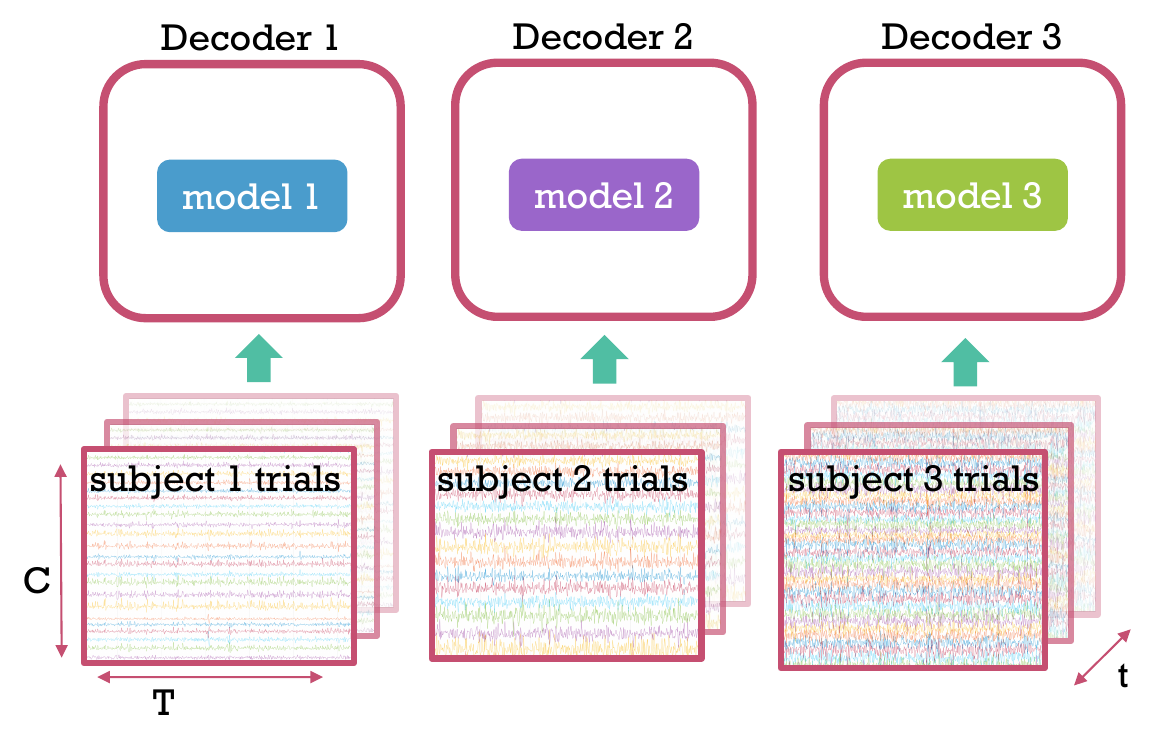}
  \caption{Subject-level models}
  \label{fig:subject_conept}
\end{subfigure}%
\begin{subfigure}{0.5\textwidth}
  \includegraphics[width=1.0\linewidth]{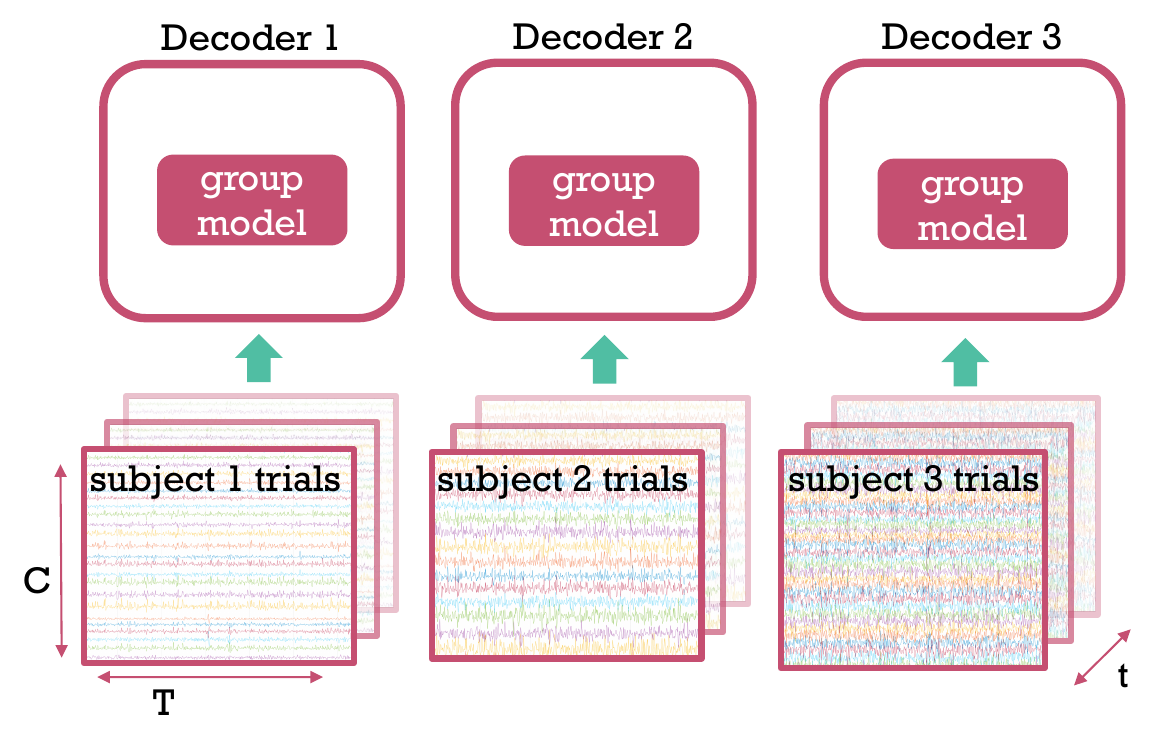}
  \caption{Naïve group-level model}
  \label{fig:group_concept}
\end{subfigure}

\begin{subfigure}{0.6\textwidth}
  \includegraphics[width=1.0\linewidth]{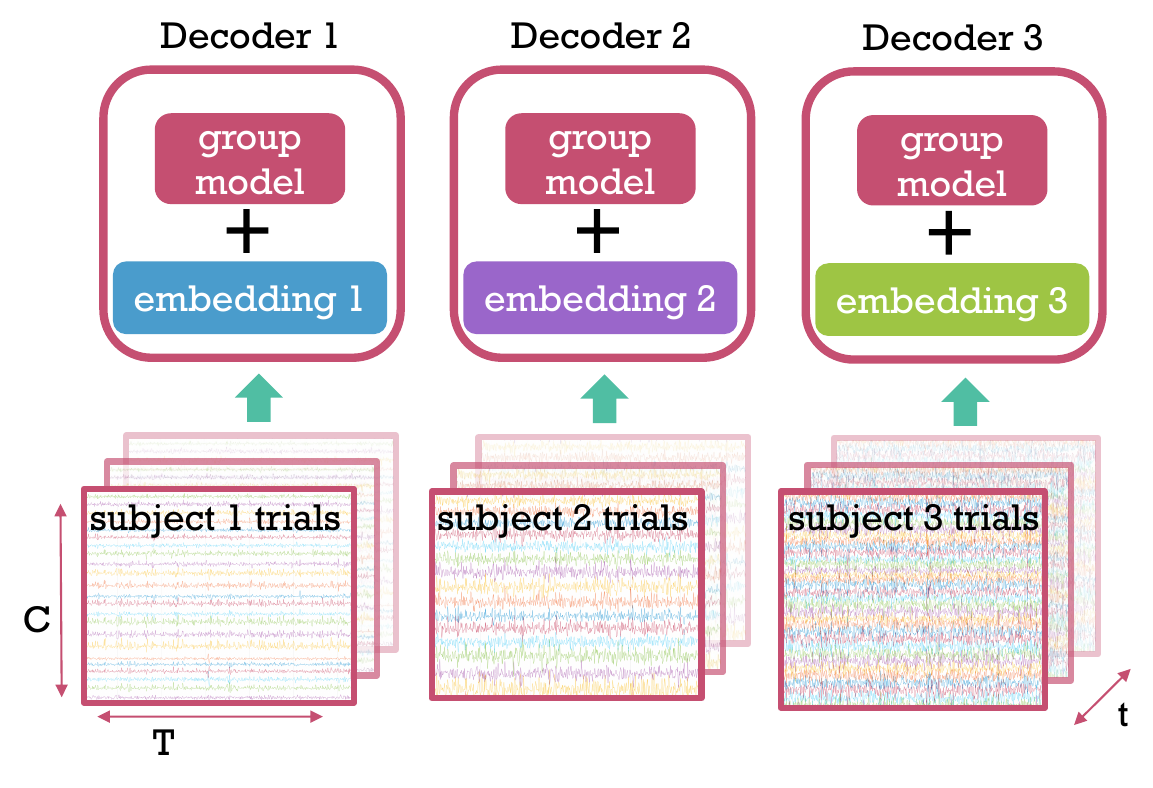}
  \caption{Proposed group-level model}
  \label{fig:groupemb_concept}
\end{subfigure}
\caption{Comparison of subject-level (a), naive group-level (b), the proposed group-level (c) modelling. (a) A separate model is trained on the trials (examples) of each subject. (b) A single, shared model is trained on the trials of all subjects without capturing between-subject variability. (c) A single, shared model is trained on the trials of all subjects with an additional embedding component that is subject-specific. Each trial is $\mathbb{R}^{C \times T}$. Each of the $S$ subjects has $T$ trials.}
\label{fig:subject_group}
\end{figure}

Applications of decoding to brain recordings typically fit separate (often linear) models per dataset, per subject \citep{guggenmos2018multivariate, dash2020decoding-imagined, csaky2023interpretable}. This has the benefit that the decoding is tuned to the dataset/subject, but has the drawback that it is unable to leverage knowledge that could be transferred across datasets/subjects. This is especially desirable for the field of neuroimaging because gathering more data is expensive and often impossible (e.g. in clinical populations). More practical drawbacks of subject-specific (subject-level) models include increased computational load, a higher chance of overfitting, and the inability to adapt to new subjects. We aim to leverage data from multiple subjects and train a shared model that can generalise across subjects (group-level). A conceptual visualisation of subject-level and group-level models is given in Figure~\ref{fig:subject_group}.

Subject-specific modelling is often preferred due to the high levels of between-subject variability in neuroimaging data. An alternative approach would be to train and use the same decoding model across multiple subjects \citep{olivetti2014meg, li2021inter}. We will refer to an approach that does this, while not explicitly modelling any of the between-subject variability, as “naïve group modelling”. Such naïve approaches, effectively pretend that all data comes from the same subject (see Figure~\ref{fig:group_concept}), but due to high amounts of between-subject variability typically perform very badly \citep{saha2020intra, olivetti2014meg, li2021inter}.  The work in this paper is motivated by a need to improve on these methods. If group modelling could be advanced to account for the high amounts of between-subject variability, then this would allow relevant information to be pooled across subjects, resulting in two key benefits. First, we would be able to obtain neuroscientific insights from the decoding models directly at the group level instead of pooling over subjects. Second, with appropriately large multi-subject datasets, group models would be able to outperform subject-level models.

In this chapter we propose a general architecture capable of jointly decoding multiple subjects with the help of subject embeddings (Figure~\ref{fig:groupemb_concept} and Figure~\ref{fig:architecture}). Our main aim is to improve subject-level models by using a single group decoding model that generalises across (and within) subjects. We refer to this as \textit{across-subject} decoding, in which models are trained on part of the data from all subjects and then tested on left-out data from all subjects. This is motivated by the fact that group-level models that perform well in this manner can be useful for gaining neuroscientific insights that are relevant at the group level, as we will show in Sections~\ref{ssec:globalPFI} and \ref{ssec:kernelPFI}. An alternative approach, \textit{Leave-one-subject-out (LOSO) analysis} is also presented in Section~\ref{ssec:generalization}. In LOSO analyses, group-level models are trained on data from multiple subjects and tested on a new, unseen subject \citep{zubarev2019adaptive}, which can be especially useful in zero-shot BCI applications.

Recently, different transfer learning approaches have been proposed to deal with the problem of variability between subjects. \cite{kostas2020thinker} have proposed two distinct methods. First, there is Euclidean alignment, which is very similar to a spatial whitening of the data. We tried this in conjunction with our group model, and found it to lower performance, and thus opted for a simpler channel-wise standardisation. Second, there is mixup regularisation, which is entirely complementary to our approach and can be used in conjunction with it. It is a general regularisation/data augmentation technique and does not specifically deal with inter-subject variability.

Most transfer learning frameworks consist of applying a model trained on one subject to a different (target) subject \citep{elango2018sequence, dash2019towards, cooney2019optimizing, olivetti2014meg, halme2018across, li2021inter}. Some approaches use learnable affine transformations between subjects \citep{elango2018sequence}, while others finetune the whole model on target subjects \citep{cooney2019optimizing, dash2019towards}.

Hyperalignment has been successful for fMRI data to align different subjects to a common cortical space, and some applications have been explored in MEG data as well. For example, \cite{benzhyperalignment} used hyperalignment on MEG data via procrustes matrix transformation to a common sensor space and showed improvement in evoked fields. Similar methods have been explored in recent studies aiming to deal with between-subject variability \citep{ravishankar2021single, michalke2023inter, zhou2020multi}. However, to our knowledge, no prior work has applied it successfully to MEG decoding.

One key consideration is that hyperalignment is a linear method, constraining the transformation between subjects. While this is a sensible assumption, we think that in order to fully leverage data from multiple subjects a nonlinear method is required. Our subject embedding method is fully data-driven without any constraints on the nature of variability between subjects. It may be that this flexibility becomes truly useful when dealing with a large number of subjects, and for a few subjects, the linear assumptions of hyperalignment could work better. Our method also directly optimises the subject embeddings for the decoding objective. It is not clear, whether an unsupervised method, such as hyperalignment would result in better decoding accuracy. We leave it for future work to provide a full comparison between hyperalignment and subject embedding for MEG decoding.

Transfer learning is also popular in the wider machine learning field. Parallels can be drawn with domain adaptation \citep{long2015learning}, or transferring knowledge from large to small datasets within the same domain \citep{wang2019pay, zhuang2020comprehensive}. Natural language processing (NLP) datasets often contain data from widely different sources \citep{radford2022robust}, but due to the sheer size of the dataset and model complexity, training on joint data achieves good results \citep{Radford:2019, Devlin:2018}. Modern approaches to NLP often use Transformer-based \citep{Vaswani:2017} architectures. We believe convolutional architectures are an attractive approach for multi-channel timeseries data, and explore Transformers in the next chapter. There are several issues to overcome when applying Transformers to multi-channel timeseries, similar but perhaps even more challenging than their application to computer vision \citep{parmar2018image, dosovitskiy2020image}. In this work we wanted to keep the model relatively simple, as our dataset size is also limited, and analyse the effect of the subject embedding.

As discussed before, a naive concatenation of subjects does not work well on small neuroimaging datasets. Perhaps the most relevant parallels can be drawn with dialogue and speech modelling work, where inter-speaker differences are modelled using speaker embeddings \citep{Li:2016a, Zhang:2018, saito2019dnn, mridha2021u}. \cite{chehab2021deep} have similarly found that subject embeddings provide a small but significant improvement in encoding MEG data from a language task. They used a combination of recurrent and convolutional neural networks for encoding MEG data. However, limited information is provided on how subject embedding helps, and their results cannot be directly generalised to MEG \textit{decoding}. Our results expand this work to the task of decoding images from MEG data and provide additional insight into how deep learning and subject embeddings help group-level decoding models. In concurrent work, \cite{defossez2022decoding} have also shown the effectiveness of subject embeddings in group-level speech decoding. They have also compared it to subject-specific layers as a way of dealing with between-subject performance and found this latter approach slightly better. The advantage of subject embeddings is that they use less parameters to deal with the between-subject variability and the structure in the learned representations can be readily interpretable.

We make the following contributions using a 15-subject MEG dataset with a visual task \citep{cichy2016comparison}. First, we introduce a group-level model with subject embeddings, substantially improving over naive group modelling and showing the potential improvements in decoding that can be provided over subject-specific decoding models. Second, we provide insight into how non-linearity and subject embedding helps group modelling. Third, we show that we can gain neuroscientific insights from the deep learning-based decoding model, using permutation feature importance \citep{altmann2010permutation} to reveal how meaningful spatiotemporal and spectral information is encoded.

\section{Methods}
\subsection{Data}
\label{ssec:data}
In this work, a task-MEG dataset is used where 15 subjects view 118 different images, with each image viewed 30 times \citep{cichy2016comparison}. The raw epoched data is publicly available\footnote{\url{http://userpage.fu-berlin.de/rmcichy/fusion_project_page/main.html}}, however, we obtained the continuous raw MEG data directly from the authors to be able to run our preprocessing pipeline using MNE-Python \citep{gramfort2013meg}. This is the same dataset as the 118-image dataset used in the previous chapter.

Raw data is bandpass filtered between 0.1 and 125 Hz and line noise is removed with notch filters. Whitening is used to remove covariance between channels for subject-level models. Removal of cross-channel covariance (whitening), or in other words multivariate noise normalisation has been previously found to improve the performance of linear decoding models \citep{guggenmos2018multivariate}. The whitening is simply done by performing a PCA projection over the channels keeping all components. For group-level models no whitening is performed, instead, each channel is individually standardised by removing the mean and dividing by the variance. The reason for not using whitening in the case of group-level models is that it would destroy the alignment of the channels between subjects, as each PCA decomposition projects into a different space. Alternatively, we can run PCA at the group-level on the data concatenated over subjects, however, we did not see an improvement in performance when doing this.

After whitening, we downsample to 250 Hz and 1.024-second epochs are extracted, starting 100 ms before stimulus presentation. This resulted in $\mathbb{R}^{C \times T}$ dimensional trials with $C=306$ and $T=256$. We do multiclass decoding, predicting a separate probability for each of the 118 classes (images). For a summary of the epoched data see Table~\ref{table:epoched_data}.

\begin{table}[ht!]
        \begin{tabular}{m{2cm}m{2cm}m{3cm}m{5cm}}
        %\toprule
           \bf Number of subjects  &\bf Number of classes & \bf Number of samples per class & \bf Dimensions of one sample  \\ \midrule
            
            15 & 118 & 30 & 306 channels $\times$ 256 timesteps  \\

            %\bottomrule
        \end{tabular}
    \caption{\label{table:epoched_data} Dimensons of the epoched dataset.}
\end{table}

\subsection{Models}
Our choice of core decoding model was based on a desire to assess the extent to which group decoding models might allow for the use of more complex, nonlinear networks when compared to subject-specific decoding. In addition, we did not aim to design a new kind of architecture for decoding MEG data, but rather build our model based on CNN-based architectures that have already been proven to be effective on time series data. As such, we used a decoding model based on WaveNet for classification, which has been used successfully in the audio domain \citep{oord2016wavenet, zhang2020mdcnn}, and which we refer to as the Wavenet Classifier. An important aspect of adapting this model is the high dimensionality of MEG data compared to audio. Whereas audio consists of a single channel, MEG has over 300. We simply set the input channels of the first convolutional layer to this number and thus the model processes the channel dimension in a fully-connected way.

The dilated convolutions in WaveNet are effective for modelling time series data, as successive layers extract complementary frequency content of the input \citep{borovykh2018dilated}. While CNN-based architectures have been used successfully on M/EEG data \citep{lawhern2018eegnet}, there is no prior work specifically applying Wavenet to neural decoding. To be clear when we refer to \textit{model} in this section we mean the general (untrained) architecture and not a trained model on some dataset. For all training instances in this paper, we used a randomly initialised model instead of using the pretrained weights from the audio-WaveNet.

\begin{figure}[!t]
  \centering
  \includegraphics[width=1.0\linewidth]{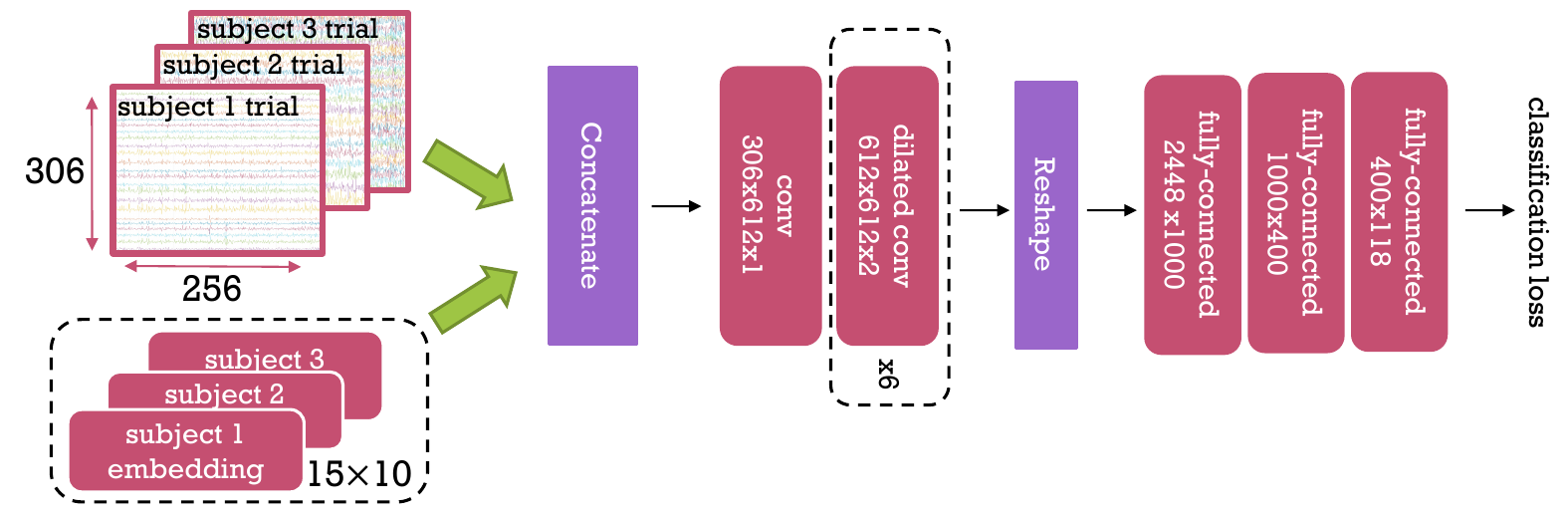}
    \caption{Group-level WaveNet Classifier with subject embeddings. Dashed boxes represent parts of the model which differ between subject-level and group-level versions of our architecture. Red boxes represent learnable parameters. For convolutional layers, the numbers represent \textit{input channels} x \textit{output channels} \(\times\) \textit{kernel size}. For fully-connected layers, the numbers represent \textit{input neurons} \(\times\) \textit{output neurons}. The embedding layer dimensionality is given as $S \times E$, where $S=15$ is the number of subjects, and $E=10$ is the embedding size. Embeddings are concatenated with input trials to provide information about which trial is coming from which subject. The classification loss is cross-entropy.}
    \label{fig:architecture}
\end{figure}

Our Wavenet Classifier model consists of 2 parts: the (temporal) convolutional block, intended to act as a feature extractor; and the fully-connected block, which is designed for classification (Figure~\ref{fig:architecture}). The convolutional block uses a stack of 1D dilated convolutional layers, which include dropout and the inverse hyperbolic sine activation function ($\mathrm{asinh}$). One layer is defined as

\begin{equation}
\mathbf{H}^{(l+1)} = \mathrm{asinh} \left(  \mathrm{Dropout} \left( \mathrm{DilatedConv}^{(l)} \left(  \mathbf{H}^{(l)} \right) \right) \right)
\end{equation}

where $\mathbf{H}^{(l)} \in \mathbb{R}^{M^{(l)} \times T^{(l)}}$ is the input representation at layer $l$ with $M^{(l)}$ channels and $T^{(l)}$ time points, and $\mathbf{H}^{(l+1)}$ is the output representation. $\mathrm{DilatedConv}^{(l)}$ is the dilated convolutional operation at layer $l$. For a single channel it is defined as:

\begin{equation}
y[n] = \sum_{k=0}^{K-1} x[n-d \cdot k] \cdot w[k]
\end{equation}

where $x[n]$ and $y[n]$ is the input and output at time $n$, $w[k]$ is the kernel weight at index $k$, and $d$ is the dilation factor.
The dilation factor is doubled in successive layers, i.e., $d \in {1, 2, 4, \dots}$. In our case the dilated convolution kernel always has 2 learnable values ($K=2$). This, together with the increasing dilations allows for rapid increase of the receptive field with few parameters and layers. In terms of number of channels we have an initial convolutional layer with a kernel size of 1 which simply projects the input channels $C$ to the channel dimensionality used throughout the rest of the convolutional layers: $M^{(l)} = 2*C, \forall{l} \in L$.

For subject-level modelling, we use 3 convolutional layers. For group-level modelling, we use 6 convolutional layers. We arrived at these numbers empirically by training both 3-layer and 6-layer subject-level and group-level models and selecting the best model version in each case, thus providing a fair comparison between subject- and group-level settings. Given there is no pooling and a convolution stride of 1, the output of each layer preserves the temporal dimensionality, except the amount that gets chopped off because of the kernel size itself. When using a convolutional layer with a kernel size $K$, the output representation's temporal dimension is

\begin{equation}
    T^{(l+1)} = T^{(l)} - K - d + 2
\end{equation}

Since the dilation factor is doubled in successive layers, the receptive field of the convolutional block is $2^L$ where $L$ is the number of layers. At the end of the convolutional block, we downsample temporally by the size of the receptive field. For example, the a convolutional block with 6 layers has a receptive field of 64 and thus its output representation has a length of $T-64+1 = 193$,  where $T=256$ is the input trial's length. This is downsampled by a factor of 64, resulting in 4 values per channel.

Next, this downsampled output is flattened and fed into a fully-connected block. The final output is a logit vector corresponding to the 118 classes. The model is trained with the cross-entropy loss for classification, which includes a softmax function that maps the logit vector to a probability distribution over classes.

We assess two versions of each model, one with a Wavenet Classifier that is linear and one that is nonlinear. This allows us to see how nonlinearities (a bedrock of deep learning) interact with group modelling. The linear versions simply correspond to Wavenet Classifier where the activation function is set to be the identity function.

Finally, we divide the group-level modelling into two approaches. First, we have a naive group model, which is our standard 6-layer Wavenet Classifier. Second, we have our proposed group model, which improves on the naive group model through the inclusion of subject embeddings. A high-level mathematical description of subject-level (Equation~\ref{eq:subject}), naïve group-level (Equation~\ref{eq:group}), and the embedding-aided group-level (Equation~\ref{eq:groupemb}) models is given below (corresponding to the 3 panels in Figure~\ref{fig:architecture}).

\begin{equation}
    \forall{s} \in S: \mathbf{y}_s = f_s(\mathbf{X}_s)
    \label{eq:subject}
\end{equation}
\begin{equation}
    \forall{s} \in S: \mathbf{y}_s = f_g(\mathbf{X}_s)
    \label{eq:group}
\end{equation}
\begin{equation}
    \forall{s} \in S: \mathbf{y}_s = f_g(\mathbf{X}_s,\mathbf{e}_s)
    \label{eq:groupemb}
\end{equation}

Where \(s\) denotes a single subject and \(S\) is the set of all subjects. \(\mathbf{y}_s\) and \(\mathbf{X}_s\) are the target variables and input trials of subject \(s\), \(f_s\) is the subject-specific model, and \(f_g\) is the shared group-level model across subjects. \(\mathbf{e}_s\) is the subject-specific learned embedding.

Subject embeddings are introduced as a way of dealing with between-subject variability, similarly to \cite{chehab2021deep}. Like word embeddings in NLP, each subject has a corresponding dense vector \citep{Mikolov:2013d}. This same vector is concatenated with the channel dimension of the input trial across all time points (in each trial). This operation is given in programming notation below.

\begin{equation}
    \mathbf{H}_s = concatenate((\mathbf{X}_s,\mathbf{E}_s), dim=0)
    \label{eq:subjectemb}
\end{equation}

Where \(\mathbf{X}_s \in R^{C\times T}\)  is the input trial consisting of \(C\) channels and \(T\) time points, \(\mathbf{E}_s \in R^{E\times T}\) is the subject embedding of size \(E\) repeated across the $T$ timepoints. $\mathbf{H}_s \in R^{(C+E)\times T}$ is the input that gets fed into the model $f_g$. Embedding size was set to 10 a priori, and the effect of different values is explored in Section~\ref{ssec:embeddings}. Subject embeddings are learnt together with other model weights using backpropagation. We reasoned that an embedding-aided model can learn general features across subjects, with the capability of adapting its internal representations for each subject.

\subsection{Model analysis}
\label{ssec:analysis}
In this section, we describe several approaches to uncovering the information encoded in the WaveNet Classifier. In the previous chapter we have validated the use of PFI in MEG decoding against more traditional methods, such as sliding-window decoding. We leverage PFI here due to its flexibility in obtaining spatiotemporal information from trained models, and its applicability to nonlinear models. The application of temporal and spatial PFI to the Wavenet Classifier follows the same methods described in the previous chapter.

We also extended the PFI method to individual kernels of the Wavenet model. In this case, the feature importance measure is the absolute difference between the kernel output using the original and permuted inputs. We reason that a more important feature will cause a higher output deviation.  The model receives the same permuted inputs as in model-level PFI, the difference is that we look at the output of individual kernels instead of the whole model. Specifically, for a kernel in layer $l$, applied to input channel $i$, contributing to output channel $o$, the feature importance $\Delta p_j$ is

\begin{equation}
\Delta p_j = \mathbb{E}_{\mathbf{X}}\left[f(\mathbf{X}; \theta)_{l, i, o} - f(\mathbf{X}_{\perp j}; \theta)_{l, i, o}\right]
\end{equation}

where $f$ is the trained model with parameters $\theta$. $\mathbf{X}_{\perp j}$ denotes shuffling of feature $j$ in $\mathbf{X}$. This feature can be either temporal, spatial, spectral, or joint across multiple dimensions as described in the previous chapter. For group models we add the learned subject embedding to $\mathbf{X}$ as usual.

\subsection{Experimental details}
Our main evaluation metric is the classification accuracy of the across-subject decoding across the 118 classes. Recall that in across-subject decoding, each subject has a train and test split, and the aim is to see if a single group decoding model generalises across (and within) subjects. Train and validation splits with a 4:1 ratio were constructed for each subject and class. This means that classes are balanced (i.e., contain the same number of examples) across subjects and splits. Subject-level and group-level models are trained and evaluated on the same splits. Note that for each model, an extra training is conducted wherein the (linear) identity function is used as an activation function to assess the influence of nonlinearity. Linear and non-linear models are trained for 500 and 2000 epochs (full passes of the training data), respectively, with the Adam optimiser \citep{Kingma:2014}. Table~\ref{table:model_combinations} lists all of the model and training combinations that are presented in Figure~\ref{fig:group_acc}. In this section when we refer to \textit{model} we mean specific trained models on the respective datasets given in Table~\ref{table:model_combinations}.

Dropout was set to 0.4 and 0.7, and a batch size of 590 and 59 was used for group-level and subject-level models, respectively. The learning rate was set to 0.0001 for group-level, and 0.00005 for subject-level models. Training of a single subject-level and group-level model took 5-15 minutes and 4 hours on an NVIDIA A100 GPU, respectively. For linear models, validation losses (cross-entropy) and accuracies were negatively correlated, i.e. loss decreases while accuracy increases, and eventually both suggested overfitting. Since non-linear models are more expressive, they overfitted sooner according to the loss, but accuracy kept improving until it reached a plateau, never overfitting. Analysing the loss distribution across validation examples (for non-linear models) shows that even during overfitting most examples’ loss keeps decreasing with a few high-loss outliers disproportionately influencing the mean. Since accuracy is binary, outliers are diminished, explaining the apparent difference in learning behaviour. For linear models, this unintuitive behaviour was not observed probably due to inherent model simplicity.

We compute Wilcoxon signed-rank tests for comparisons of interest over trained models, where the pairing is within-subject, and samples are the subject-level mean accuracies over validation trials. We used PyTorch for training \citep{pytorch2019} and several other packages for analysis and visualisation \citep{scikit-learn, 2020SciPy-NMeth, harris2020array, pandas:2010, Waskom2021, Hunter:2007}.

\begin{table}[t!]
\newcolumntype{R}{ >{\centering\arraybackslash}m }
            \begin{tabular}{m{3.3cm}|R{1.2cm}R{1.2cm}R{1.5cm}R{2cm}R{1.6cm}}
            %\toprule
               \bf Method name & \bf Linear/ nonlinear  &\bf No. of conv layers & \bf Subject embeddings & \bf Trained on \textit{subject} or \textit{group} data & \bf Finetuned on \textit{subject} data  \\ \midrule
                
                Linear subject	& linear&3	& no	& subject&no  \\ 
                 Nonlinear subject	& nonlinear&3	& no	& subject&no  \\ 
                 Linear group	& linear&6	& no	& group&no  \\
                Nonlinear group&	nonlinear&	6	&no&	group&	no\\
                Linear group-emb&	linear&	6&	yes&	group&	no\\
                Nonlinear group-emb	&nonlinear&	6	&yes&	group&	no\\
                Nonlinear group finetuned&	nonlinear&	6	&no&	group&	yes\\
                Nonlinear group-emb finetuned&	nonlinear&	6&	yes&	group&	yes\\
                
                \bottomrule
            \end{tabular}
        \caption{\label{table:model_combinations} Model and training combinations and their corresponding naming}
\end{table}

\section{Results}
\label{sec:results_group}

\subsection{Subject embedding aided group models}
\label{ssec:group_models}

Validation accuracies for all trained models are shown in Figure 3. Interestingly, at the subject level, linear models performed slightly better than non-linear (4\% increase, \(p=5.7e-4\)). We think that both the limit in data size and noise levels in the data contribute to the subpar performance of non-linear models when trained/validated within-subject. The large between-subject variability common to MEG datasets is apparent, with individual subjects’ accuracy ranging from 5\% to 88\%. As expected, training naive group models, i.e. a naive application of either the linear or non-linear WaveNet Classifier to the group modelling problem (orange violin plots), results in much worse performance than training subject-level models, i.e. 30\% decrease compared to \texttt{linear subject}. Inferring such high variability implicitly between so few subjects is not trivial.

\begin{figure}[!t]
  \centering
  \includegraphics[width=1.0\linewidth]{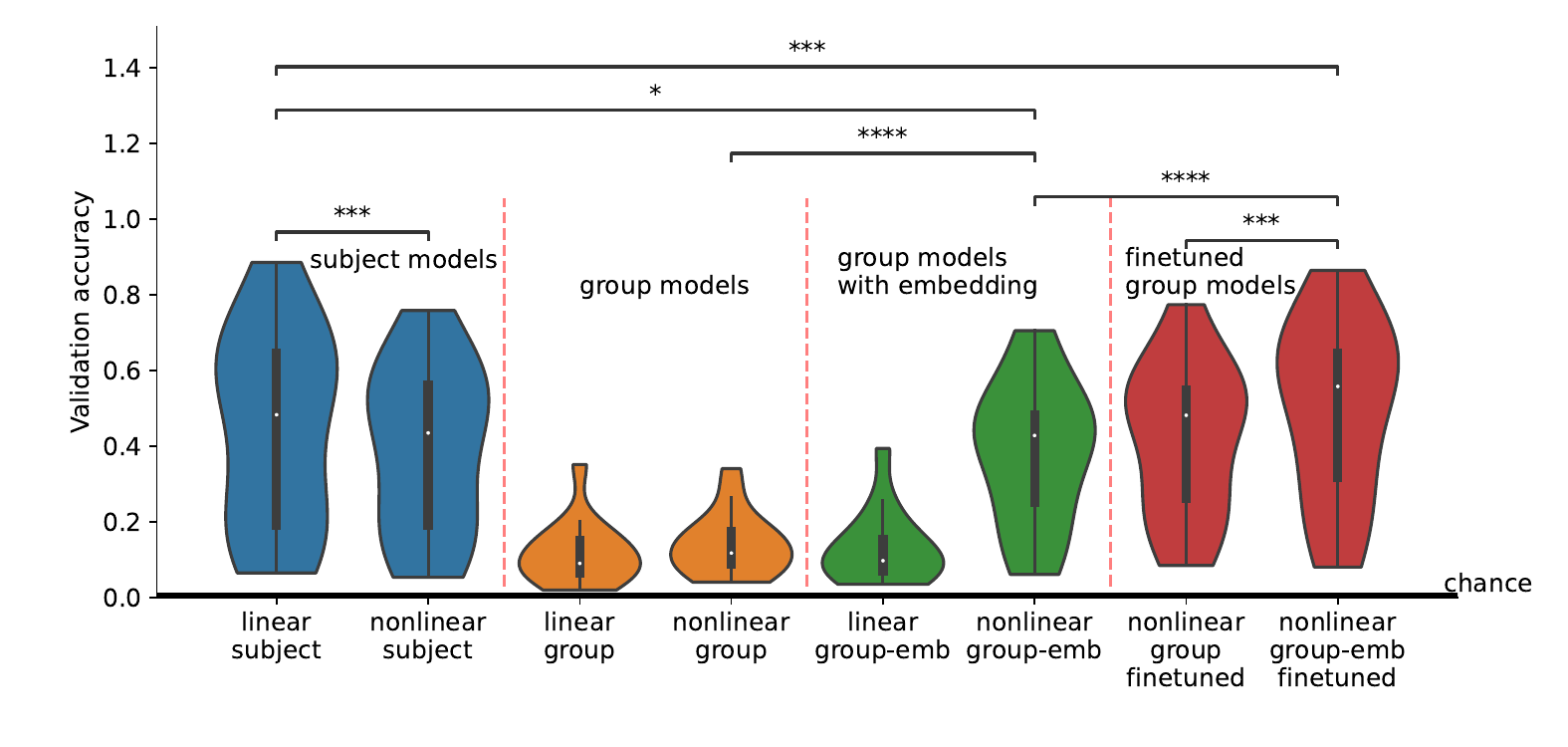}
    \caption{Trained subject-level and group-level models evaluated on the validation set of each subject. Wilcoxon signed-rank tests are shown for comparisons of interest  (\(* = p<5e-2, ** = p<1e-2, *** = p<1e-3, **** = p<1e-4\)). The \texttt{non-linear group-emb finetuned} model is finetuned separately on each subject, initialized with the \texttt{non-linear group-emb} model. Chance level is \(1/118\).}
    \label{fig:group_acc}
\end{figure}

Adding subject embeddings to the non-linear model (\texttt{non-linear group-emb}) improves performance by 24\% (\(p = 1.9e - 6\)), with no increase for the linear model (\texttt{linear group-emb}). This shows that leveraging subject embeddings in conjunction with non-linear activations can narrow the gap with subject-level models (6\% difference with \texttt{linear subject}, \(p = 1.3e - 2\)). Limiting the non-linearity to the first layer resulted in a subpar performance, similar to that of a linear model. This indicates that non-linearity is needed within multiple layers to benefit from subject embeddings. The impact of subject embeddings is further investigated in Section~\ref{ssec:embeddings}.

We also finetuned the embedding-aided group-level model on the training data of each subject separately (\texttt{non-linear group-emb finetuned}) for 500 epochs. We effectively use the group-level model as an initialisation for subject-level models, improving over subject-level models trained from scratch (\texttt{linear subject}), achieving 50\% accuracy (5\% increase, \(p=1e-3\)). This shows that representations learned at the group level are useful for subject-level modelling. In contrast, finetuning a naive group model (\texttt{non-linear group finetuned}) only achieved 42\% accuracy showing that the best performance is reached when finetuning is combined with the best group-model. Thus, in addition to closing the gap between subject-level and group-level modelling, finetuning our embedding-aided model provides the best overall accuracy for subject-level modelling. The variance of \texttt{non-linear group-emb} (19\%) and \texttt{non-linear group-emb finetuned} (24\%) is lower than \texttt{linear subject} (26\%). Generally, group models reduce between-subject variability.

\begin{figure}[!ht]
\begin{subfigure}{0.33\textwidth}
  \centering
  \includegraphics[width=1.0\linewidth]{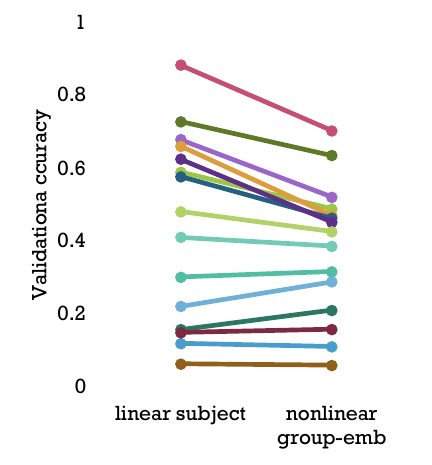}
  \caption{Subject- to group-level}
  \label{fig:subject_to_group}
\end{subfigure}%
\begin{subfigure}{0.33\textwidth}
  \centering
  \includegraphics[width=1.0\linewidth]{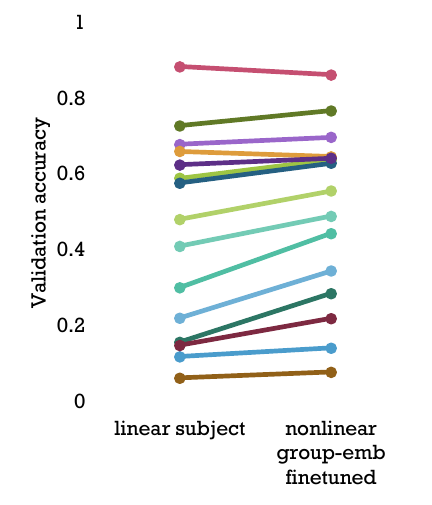}
  \caption{Subject to group finetuned}
  \label{fig:subject_to_finetuned}
\end{subfigure}%
\begin{subfigure}{0.33\textwidth}
  \centering
  \includegraphics[width=1.0\linewidth]{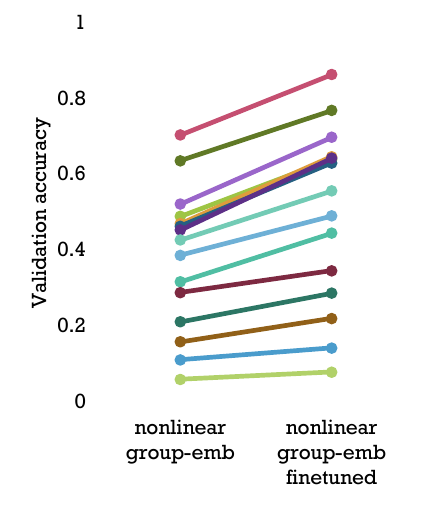}
  \caption{Group to group finetuned}
  \label{fig:group_to_finetuned}
\end{subfigure}
\caption{Accuracy changes across all 15 subjects (individual colours), when comparing trained \texttt{linear subject}, \texttt{non-linear group-emb}, and \texttt{non-linear group-emb finetuned} models. Both \texttt{non-linear group-emb} and the finetuned version clearly reduce the variability of accuracies across subjects and are especially helpful for low-accuracy subjects. When finetuning \texttt{non-linear group-emb} on individual subjects (c), we can see that accuracy increases for all subjects, and especially for high-accuracy subjects. This is unsurprising because these subjects have good enough data on their own for subject-level models to be able to learn well. As seen in (a) and (b) these high-accuracy subjects are usually impaired by group-level models, exactly for the aforementioned reason.}
\label{fig:accuracy_changes}
\end{figure}

In neural decoding, group models are widely understood to perform worse than individual models \citep{guggenmos2018multivariate, dash2020decoding-imagined} But why is this? By plotting per-subject performance in both kinds of models (Figure~\ref{fig:accuracy_changes}), we see something revealing. In the case of \texttt{non-linear group-emb}, 4 subjects with generally low accuracies (15-30\%) had higher accuracies than \texttt{linear subject} (even though the mean across subjects is lower). This suggests that group models could be successfully used for some subjects if those subjects could be identified. Compared to a subject-level model that might overfit to the noisy data of a bad subject, the group model might possess inherent regularisation due to having been trained on multiple subjects, and thus improve performance on bad subjects. As shown this is achieved at the expense of decreasing the accuracy of good subjects.

Strong negative correlations of -0.88 and -0.54 are obtained between \texttt{linear subject} subject-level accuracies and the change in accuracy achieved by the \texttt{non-linear group-emb} and \texttt{non-linear group-emb finetuned} models, respectively. Comparing finetuning to from-scratch subject-level models (\texttt{linear subject}), only 2 high-accuracy subjects are slightly worse, and generally low/mid-accuracy subjects show more improvement than high-accuracy subjects (Figure~\ref{fig:accuracy_changes}).

We analysed our main findings on another publicly available visual MEG dataset with 92 different images (15 subjects, and 30 trials per image) \citep{cichy2014resolving}. This dataset is the same as the 92-image dataset used in the previous chapter. Linear subject-level models achieved 35\% accuracy, whereas a linear group model without embeddings had 12\%, and a nonlinear group model with embeddings had 28\%. Thus we can see that our approach behaves similarly on this dataset, improving a lot over the naive group baseline, but not quite achieving the performance of the linear subject-level models. Finetuning the group model separately on individual subjects achieved 38\% accuracy surpassing from-scratch subject-level models.

In summary, these results suggest the following recommendations for decoding MEG task data. \begin{enumerate}
    \item Subject embeddings and non-linearity should be used for achieving good group models.
    \item Group-level models can be used to improve over subject-level models on low-performance subjects.
    \item For the best subject-level performance, the finetuning approach should be used, benefitting low-performance subjects the most.
\end{enumerate}

\subsection{Insights into the embedding-aided group model}
\label{ssec:embeddings}
For the embedding-aided group-level setup (\texttt{non-linear group-emb}), 4 further models were trained for 5-fold cross-validation. The training and validation sets still contained 80\% and 20\% of trials respectively, and the splitting was done so that all of the data appears exactly once in the validation set across the 5 folds, for each subject and class. Average accuracy was 37.4\% (as opposed to the 38\% reported in Figure~\ref{fig:group_acc}), with a 95\% confidence interval of 0.8\%. Thus, the proposed group-level model is robust to different random seeds and dataset partitions. More extensive robustness analysis is omitted due to computational constraints.

In non-linear subject-level models (\texttt{non-linear subject}), accuracy improves as we use fewer convolutional layers, whereas for non-linear group-level models (\texttt{non-linear group-emb}) using more layers improved accuracy (see Table~\ref{table:layer_acc}). Thus, subject-level models seem to rely more on the fully-connected block as they are unable to extract good features, and group-level models rely more on the convolutional block to learn shared features across subjects. To have a fair comparison between the two approaches we selected the best number of layers for both individual and group-level models. To be clear, because of how we perform the temporal downsampling after the convolutional layers (described in Section~\ref{ssec:group_models}), using fewer convolutional layers increases the overall parameter count because the fully-connected block has to be enlarged. Thus, the group model (with 6 conv layers), is about 2.5x smaller than the subject-level models (with 3 conv layers). However, \texttt{non-linear group-emb finetuned} models achieve higher accuracy than from-scratch subject-level models \texttt{linear subject}. This shows that, when they are initialised well (with a group model trained on multiple subjects), even subject-level models can benefit from non-linearity and more convolutional layers.

\begin{table}[t!]
        \begin{center}
            \begin{tabular}{lrrrr}
            %\toprule
                & linear subject  & nonlinear subject &  nonlinear group-emb  \\ \midrule
                
                \bf 3 conv layers & 0.45 & 0.39 & 0.22  \\ 
                \bf 6 conv layers & 0.41 & 0.25 & 0.38

                %\bottomrule
            \end{tabular}
        \end{center}
        \caption{\label{table:layer_acc} Effect of number of convolutional layers on the validation accuracy of two subject-level and one group-level model.}
\end{table}

\begin{figure}[!t]
  \centering
  \includegraphics[width=0.8\linewidth]{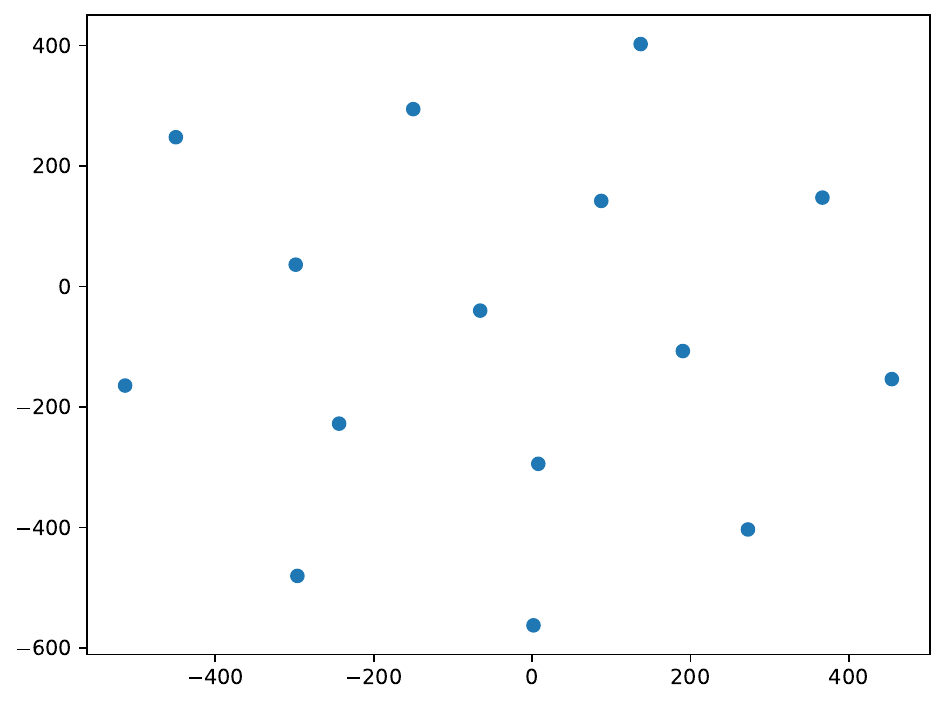}
  \caption{2D t-SNE projection of the subject embeddings in the trained \texttt{non-linear group-emb} model.}
  \label{fig:tsne}
\end{figure}

We tried different approaches to understand how subject embeddings help the \texttt{nonlinear group-emb} model. A clustering or 2D projection of the embedding space such as PCA or t-SNE \citep{van2008visualizing} did not show any clusters (see Figure~\ref{fig:tsne}). This is likely to be a consequence of only having 15 subjects since cases where such visualisations work well \citep{liu2017visual} typically have thousands of dimensions (e.g. words in word-embeddings). To assess whether the embeddings simply encode which subjects are good, we transformed the embeddings with PCA and correlated all components with the accuracies across subjects. We found no significant correlations; therefore, embeddings do not appear to encode information about subject-level accuracy. To assess how much embeddings contribute to a trained model, we tried both setting the embeddings to zero and shuffling them. The validation accuracy of \texttt{nonlinear group-emb} decreased to 10\% for both approaches. This is a 28\% reduction from the original accuracy. Thus, embeddings encode crucial information to aid decoding, but the \texttt{nonlinear group-emb} is still better than chance without them.

\begin{figure}[!t]
  \centering
  \includegraphics[width=0.8\linewidth,trim={2cm 8cm 2cm 2cm}]{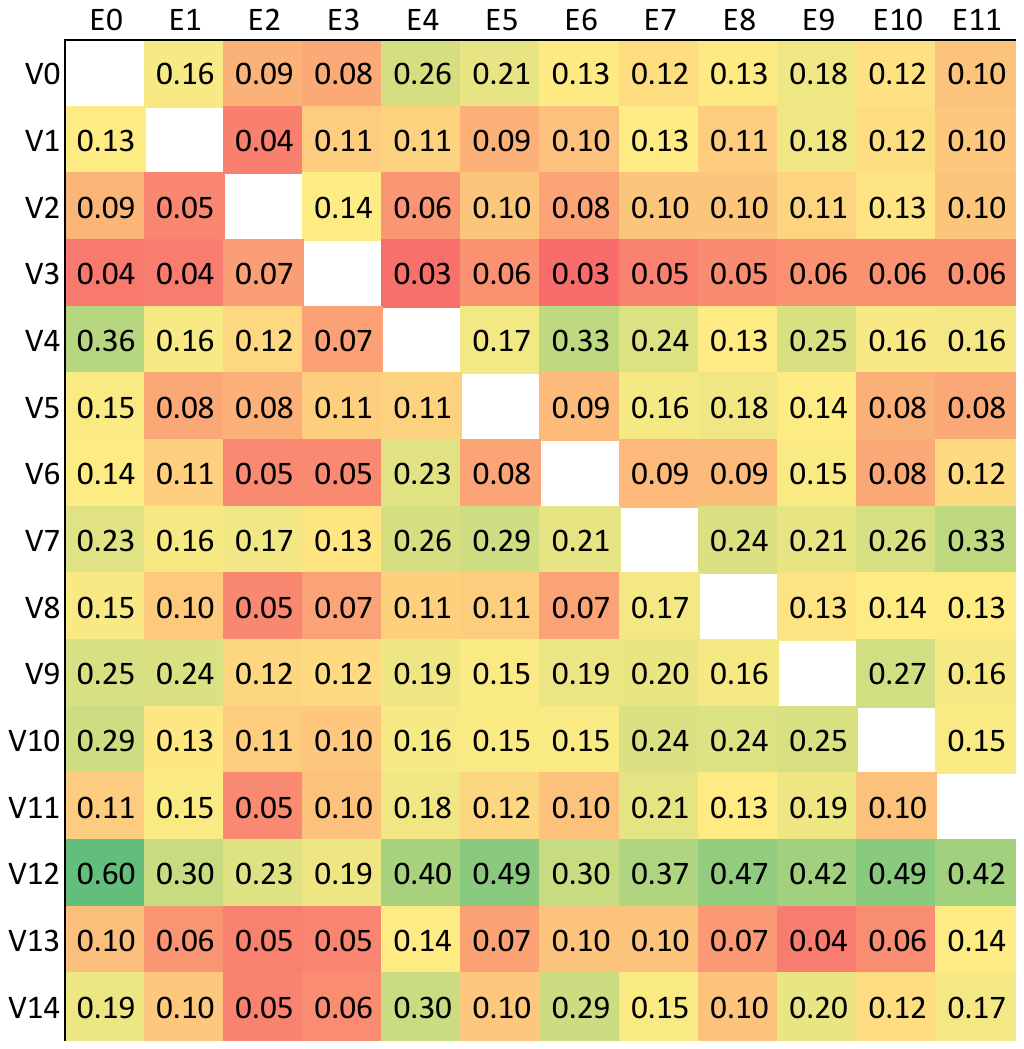}
  \caption{Subject embedding confusion matrix from the trained \texttt{non-linear group-emb} model. Columns (E0-E14) refer to subject embedding indices and rows (V0-V14) refer to subject validation sets. Greener shading (higher values) shows subjects with higher retained accuracy when their embeddings are swapped.}
  \label{fig:embeddings}
\end{figure}

To gain further insight into the learned subject embeddings we computed accuracy on each subject’s validation data using other subjects’ embeddings. In the resulting subject-by-subject confusion matrix the value in the \textit{i}-th row and \textit{j}-th column shows how well the embedding of subject \textit{i} can be replaced with the embedding of subject \textit{j}  (Figure~\ref{fig:embeddings}). After division with the original accuracies, the metric shows how much accuracy can be retained when swapping subject embeddings. Some subjects’ embedding cannot be replaced by others (e.g. subject 3), and some subjects’ embedding can be more easily replaced (e.g. subject 12). Conversely, some subjects’ embeddings are more general as they can replace many others (e.g. subject 14), and some are less general (e.g. subject 2). We tried clustering this matrix, and looked at correlation with both embedding distance and subject accuracy, however no meaningful results were found.

Training with an embedding dimensionality of 3 and 14, resulted in 20\% and 38\% accuracy, respectively. We tried these two settings to see how embedding size in the lower and upper limits influences performance. As an embedding dimensionality of 14 performs the same as 10, we could draw the conclusion that 10 is not a limiting factor. From the much worse result with an embedding dimensionality of 3, we could draw the conclusion that compressing the embedding representations too much is not possible. As with the clustering analysis, this is likely to be due to having few subjects.

\subsection{Leave-one-subject-out evaluation}
\label{ssec:generalization}
To this point, we have reported results for across-subject decoding, in which we use a single group decoding model that generalises across (and within) subjects; an approach that is, for example, relevant when one wants to gain neuroscientific insights that generalise to the group level. In this section, we report leave-one-subject-out (LOSO) cross-validation results; which is relevant, for example, when one wants to develop BCI methods that work on previously unseen subjects. Movement classification is one such application where it would be beneficial to be able to use a decoder trained on other subjects in a zero-shot setting. We also analyse how performance improves when we allow models to use increasing amounts of data (finetuning) from the left-out subject. We compare the LOSO and finetuning performance of \texttt{nonlinear group}, \texttt{nonlinear group-emb}, and \texttt{linear subject}. The \texttt{linear subject} approach serves as a baseline and it is only trained on the left-out subject. Thus, in the LOSO (zero-shot) setting, this model has chance level, since no training is performed on the left-out subject.

\begin{figure}[!t]
  \centering
  \includegraphics[width=0.8\linewidth]{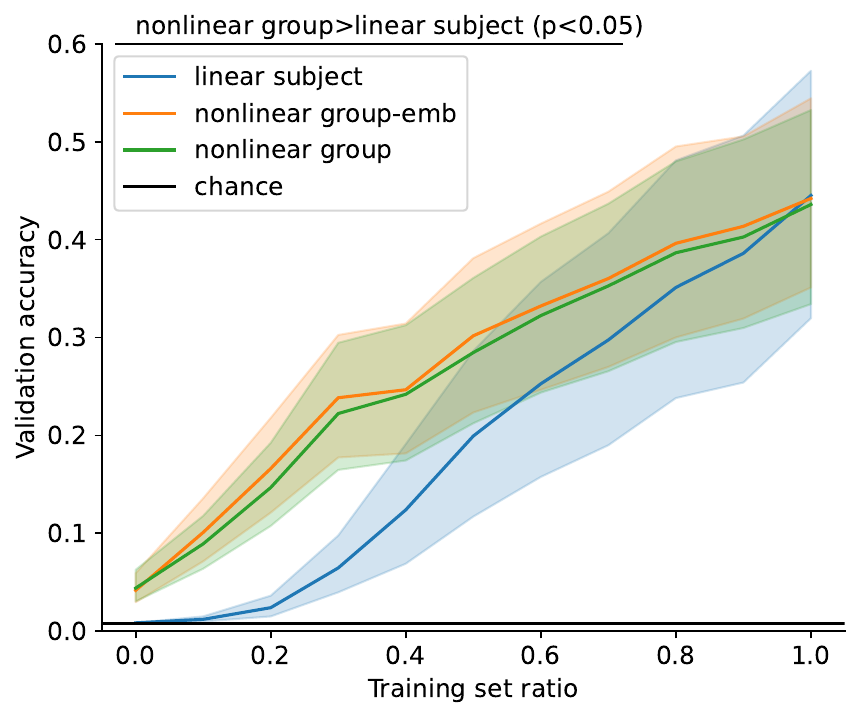}
  \caption{Generalisation and finetuning on left-out subjects. The horizontal axis shows the amount of training data used from the left-out subject; a training set ratio of 0 corresponds to a zero-shot approach. \texttt{Linear subject} is trained from scratch, while \texttt{nonlinear group-emb} and \texttt{nonlinear group} are initialised with the trained non-linear group-level model with and without embeddings, respectively. The 95\% confidence interval of the accuracy across left-out subjects is shown with shading.}
  \label{fig:generalization}
\end{figure}

When training \texttt{nonlinear group-emb} the left-out subject’s embedding was initialised randomly. In the LOSO (zero-shot) evaluation, both group models achieve 5\% accuracy (Figure~\ref{fig:generalization}). Up to the case when 70\% of the training data of the left-out subject is used, both group models are much better than \texttt{linear subject} (\(p<0.05\), corrected for multiple comparisons). This is expected and the benefit of group-level models in LOSO analysis has been previously established \citep{elango2018sequence}. Thus, to achieve the same level of performance as linear subject much less data is needed when finetuning a group model. Unsurprisingly, the \texttt{nonlinear group-emb} model does not improve over the naive model (\texttt{nonlinear group}), but is, importantly, not worse. As opposed to the finetuning setup in Figure~\ref{fig:group_acc}, when adapting to new subjects, better group performance does not translate to better finetuning performance. We think this is because when adapting to a new subject, that subject’s embedding was randomly initialised, and thus it has to be learned during the finetuning. This is a limitation of our approach.

\begin{figure}[!t]
\centering
\begin{subfigure}[t]{0.48\textwidth}
  \centering
  \includegraphics[width=1.0\linewidth]{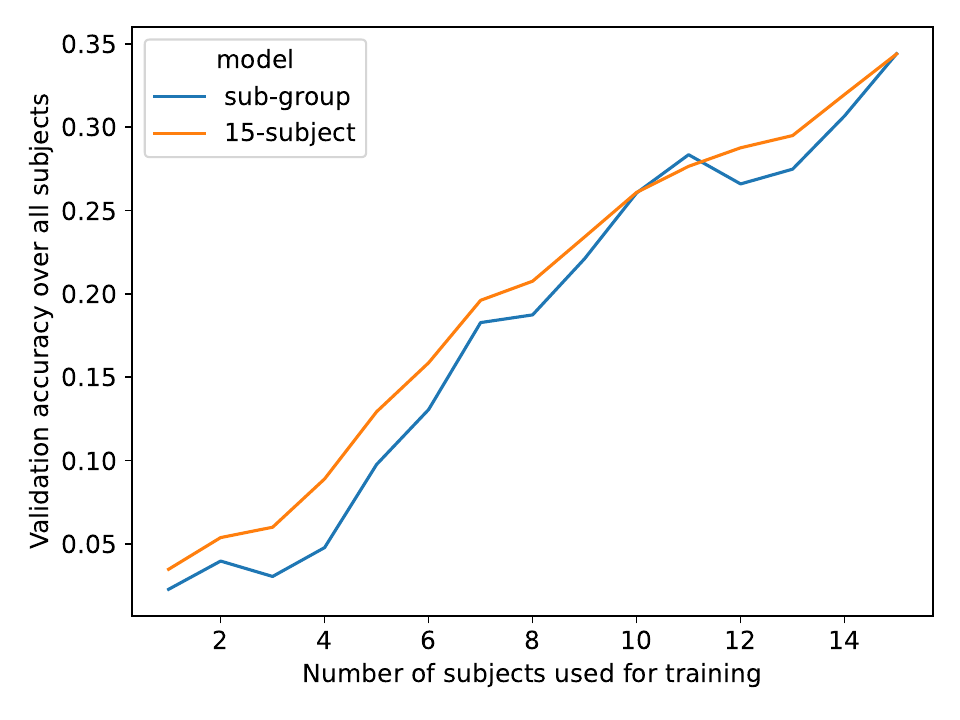}
  \caption{Evaluation over all subjects}
  \label{fig:sub_group}
\end{subfigure}\hfill
\begin{subfigure}[t]{0.52\textwidth}
  \centering
  \includegraphics[width=0.95\linewidth]{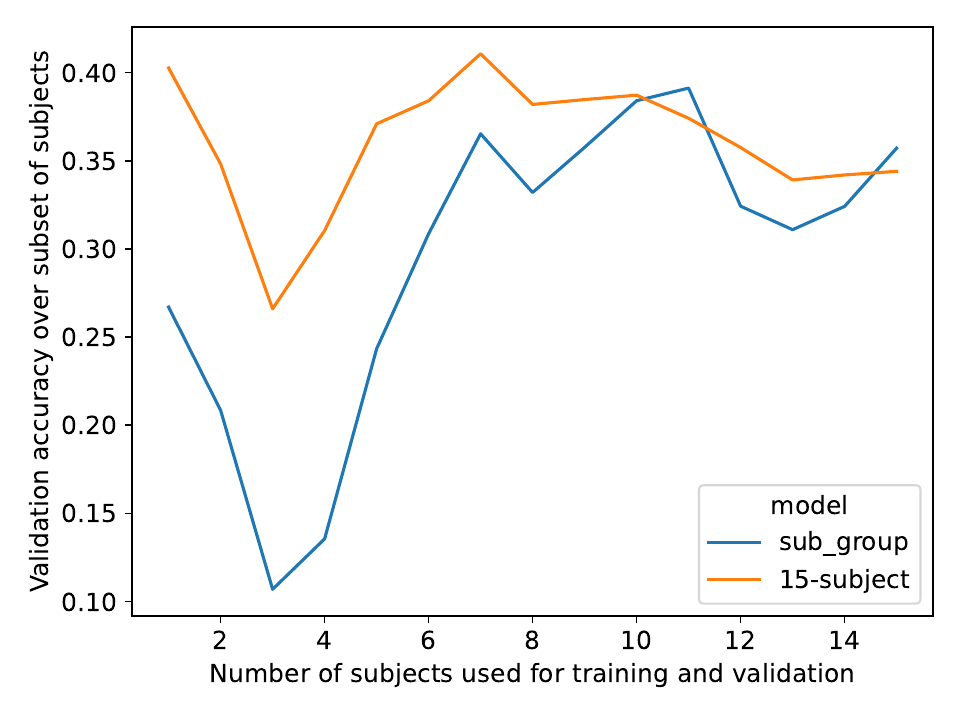}
  \caption{Evaluation over subsets of subjects}
  \label{fig:sub_group_persub}
\end{subfigure}
\caption{(a) Validation accuracy over all subjects with respect to increasing the subset of subjects used for training the sub-group model (blue line) on the horizontal axis. The 15-subject model (orange line) is our standard \texttt{non-linear group-emb} model trained on all subjects. (b) Validation accuracy over the subset of subjects used for training the sub-group model (blue line). The 15-subject model (orange line) is our standard \texttt{non-linear group-emb} model trained on all subjects. The 15-subject model is evaluated on the same increasing sets of subjects as used for the sub-group models.}
\end{figure}

To see what effect increasing the number of subjects has on group model performance we trained 15 embedding-aided sub-group models with increasing number of subjects, i.e. 1 subject, 2 subjects, …, 15 subjects. We used the same hyperparameters as for our original \texttt{non-linear group-emb}. We then evaluated each sub-group model on the validation set of the subjects it was trained on (Figure~\ref{fig:sub_group_persub}). The resulting validation accuracy is shown in the plot as a function of the number of subjects in the sub-group. One downside of this approach is that the order in which we add the subjects to the sub-group models matters a lot because of the high between-subject variability. However, to test multiple orderings we would have to run hundreds of trainings which is not possible under our computational constraints. Nevertheless, we compared our increasing subject-number sub-group models with the theoretical best performance achieved by the group model trained on all subjects (15-subject). We can see that the gap between the full group model and the restricted sub-group models generally tightens as we increase the number of subjects used for training. It is difficult to draw strong conclusions without repeating this analysis with different permutations of subjects.

An alternative visualisation for the previous analysis is to keep the validation set fixed, i.e. always compute validation performance on the validation set of all subjects (Figure~\ref{fig:sub_group}). To provide the theoretical maximum from the 15-subject group model we took its performance on the respective subjects (e.g. in the case of 2 subjects, the first 2 subjects), and replaced the other subjects with a 1/118 (chance) accuracy value. This again shows a slight tightening between the full group model and the restricted sub-group models as we increase the number of subjects. Notably, there is a dip in performance when we add subject 12 to the group model as this subject had a particularly bad performance.

\subsection{Model-level PFI}
\label{ssec:globalPFI}

\begin{figure}[!t]

  \centering
  \includegraphics[width=0.8\linewidth]{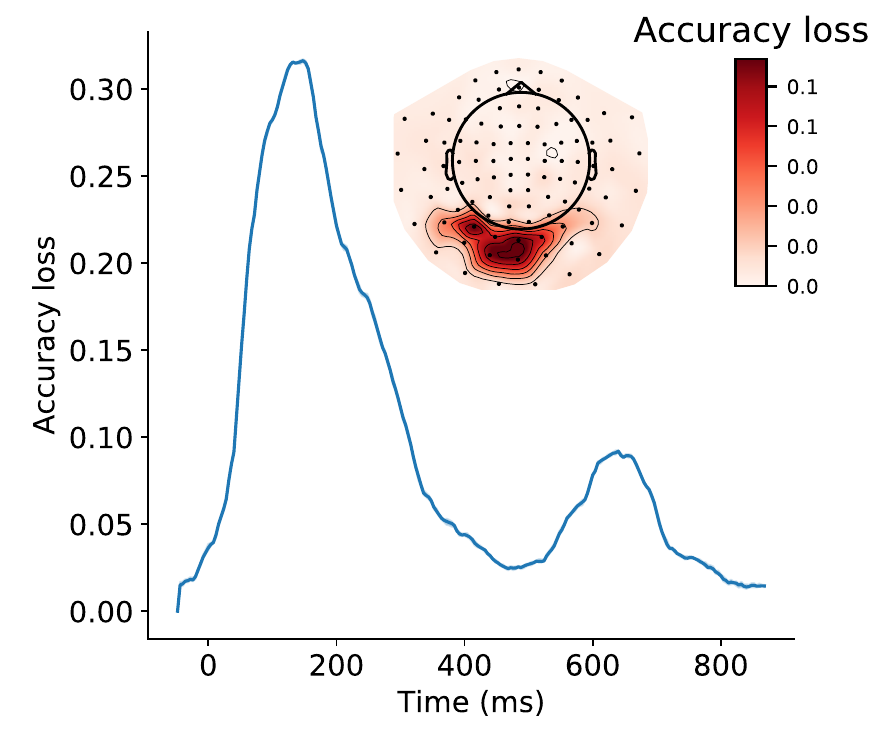}
  \caption{Temporal (line) and spatial (sensor space map) PFI for the trained \texttt{non-linear group-emb} model. For temporal PFI accuracy loss (vertical axis) is plotted with respect to time since visual image presentation (horizontal axis). Shading shows the 95\% confidence interval which is not visible due to low variability. For spatial PFI, darker red shading is equivalent to higher accuracy loss.}
  \label{fig:global_PFI}

\end{figure}

An established critique of deep learning models applied to neuroimaging data is the lack of interpretable, neuroscientific insight they provide about the underlying neural processes that drive the decoding \citep{murdoch2019definitions}. To gain such insights, it is useful to assess the time- and space-resolved information/discriminability within trials. Figure~\ref{fig:global_PFI} shows the temporal and spatial PFI of the trained \texttt{nonlinear group-emb} model. To make the results robust and smooth, the shuffling for temporal PFI was applied to 100 ms windows, and magnetometers and gradiometers in the same location were shuffled together for spatial PFI. Time windows or channels with higher accuracy loss than others are interpreted as containing more information about the neural discriminability of the visual images. This indicates when and where information processing related to the presented images is happening in the brain.

Temporal PFI shows a large peak around 150 ms which is in line with previous subject-level PFI results on this dataset (see Chapter~\ref{Chap3}. After this, the information content rapidly decreases, with a second, smaller peak around 650 ms, which could correspond to a brain response following the end of image presentation at 500 ms. Spatial PFI shows that the most important channels are in the back of the head in the sensors in visual areas, as expected for a visual task.

\begin{figure}[!t]
\centering
\begin{subfigure}{0.5\textwidth}
  \centering
  \includegraphics[width=0.95\linewidth]{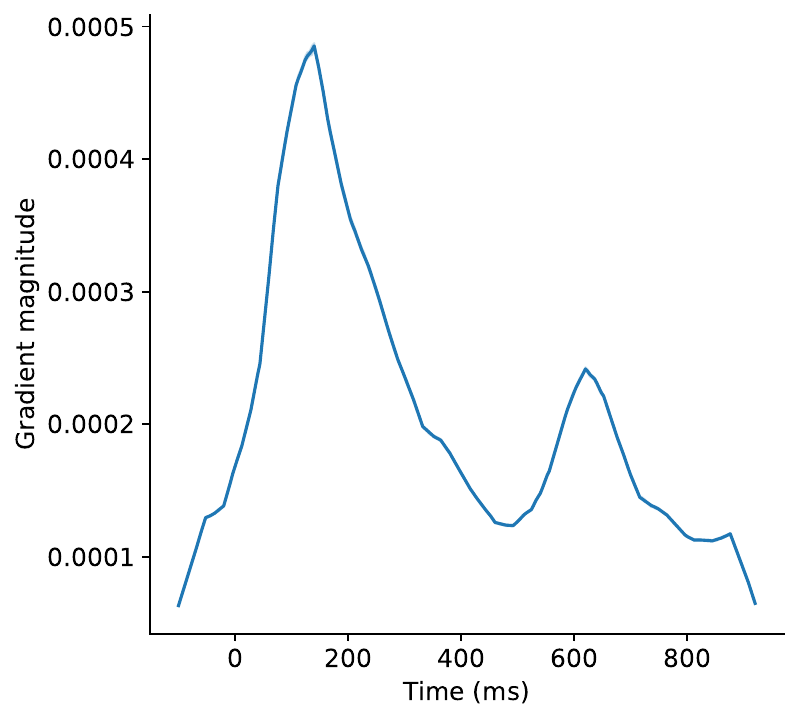}
  \caption{Temporal gradient analysis}
  \label{fig:temporal_gradient}
\end{subfigure}%
\begin{subfigure}{0.5\textwidth}
  \centering
  \includegraphics[width=0.95\linewidth]{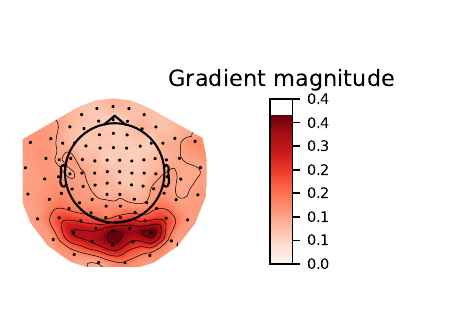}
  \caption{Spatial gradient analysis}
  \label{fig:spatial_gradient}
\end{subfigure}
\caption{Using gradient analysis by backpropagating the loss to randomly initialised inputs with the trained \texttt{non-linear group-emb} model. In (a) we can see the temporal profile of the gradients averaged over channels. In (b) we can see the spatial profile of the gradients averaged over time.}
\label{fig:gradient_analysis}
\end{figure}

We found good agreement between the PFI analysis and the alternative approach of a gradient-based analysis often used in deep learning models (Figure~\ref{fig:gradient_analysis}). In this analysis a salience map is obtained by backpropagating to randomly initialised inputs. We smoothed the temporal profile with the same window size as for the PFI analysis. Temporally we can see that the agreement between the two methods is high, with peaks aligning very well (less than 10ms difference). Spatially the two methods do show some differences, but overall gradient analysis still points to the most important information being in the visual cortex. For a full explanation of this method please see Section~\ref{ssec:interpret_deep}.

\subsection{Kernel analysis}
\label{ssec:kernelPFI}

To provide further insight into our trained \texttt{non-linear group-emb} model, we next show that interpretable spatial, temporal, and spectral information can be obtained by analysing the learnt weights. This analysis becomes possible because we use a multi-layered neural network, and there is no equivalent analysis that we could do in a classical linear model. When using deep learning it is important to ask how the trained model arrives at the information presented in Section~\ref{ssec:globalPFI}. We can leverage the structure of the model, i.e. the successive layers, and the filters in the convolutional layers can be regarded as individual computational units. The aim here is to understand the model itself and how it represents and processes the data internally. This is in line with previous efforts showing how successive layers in a deep convolutional model align with the visual system of the brain \citep{kriegeskorte2015deep}.

\begin{figure}[!t]
\centering
\begin{subfigure}{0.5\textwidth}
  \centering
  \includegraphics[width=1.0\linewidth]{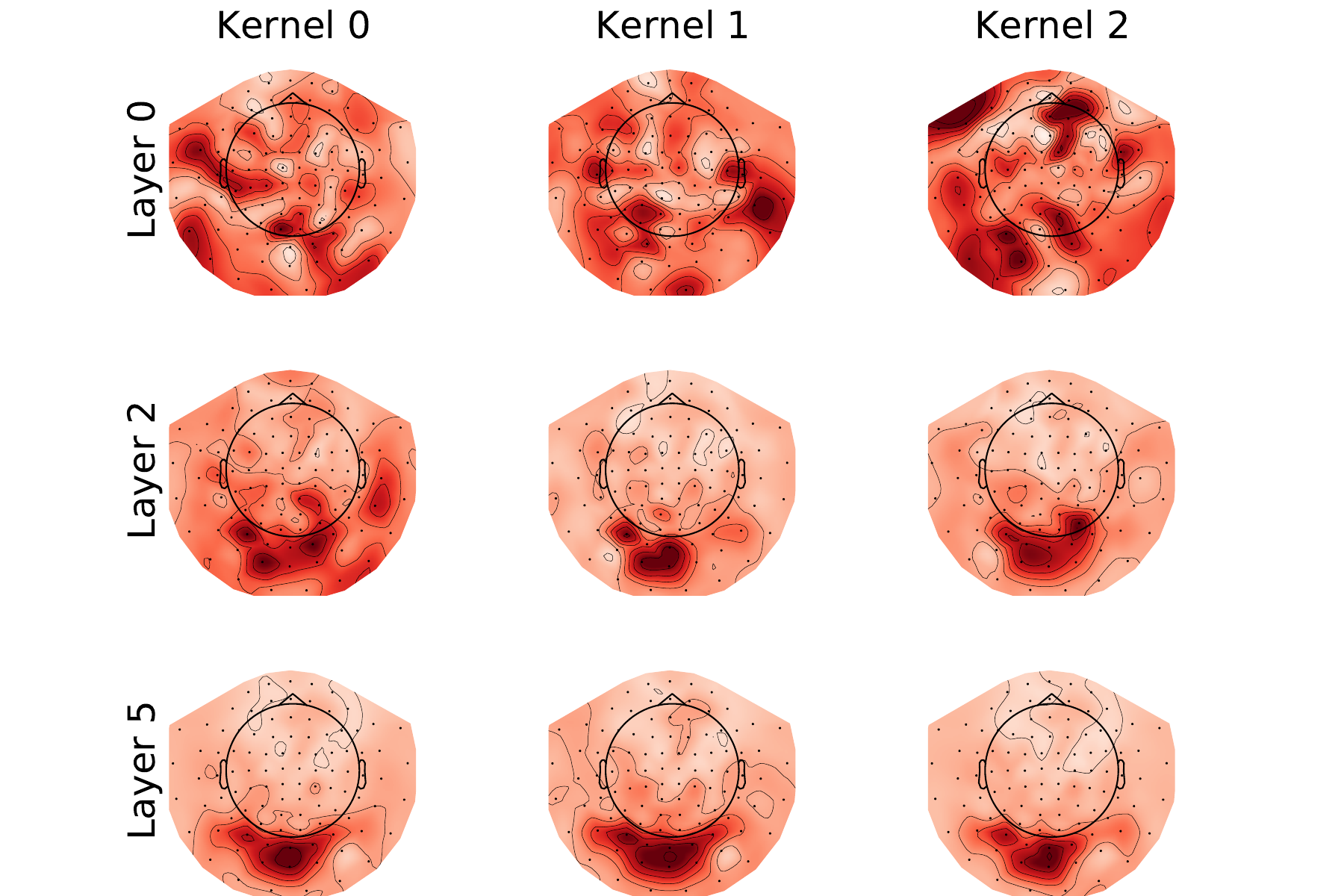}
  \caption{Spatial PFI}
  \label{fig:spatial_pfi}
\end{subfigure}%
\begin{subfigure}{0.25\textwidth}
  \centering
  \includegraphics[width=1.0\linewidth]{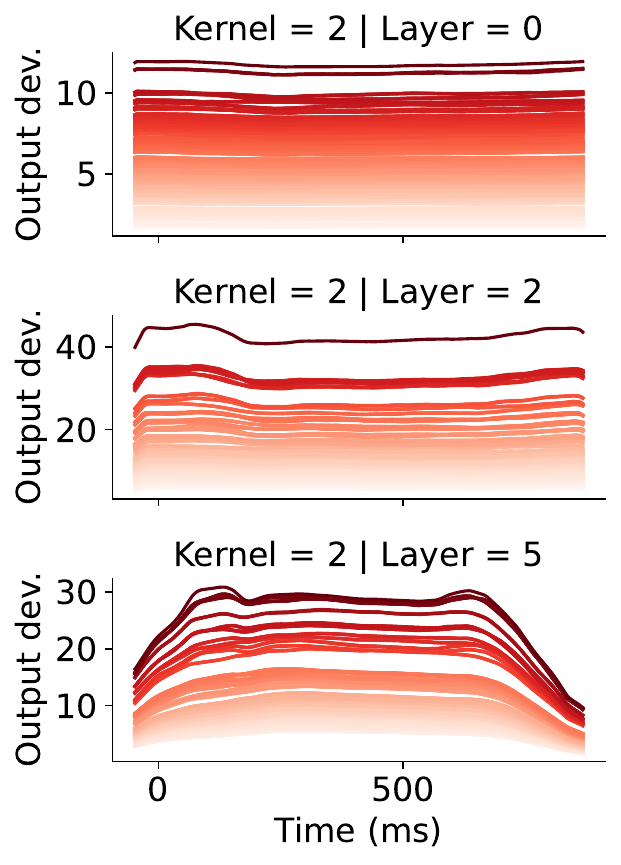}
  \caption{Spatiotemporal PFI}
  \label{fig:spatiotemporal_pfi}
\end{subfigure}%
\begin{subfigure}{0.24\textwidth}
  \centering
  \includegraphics[width=1.0\linewidth]{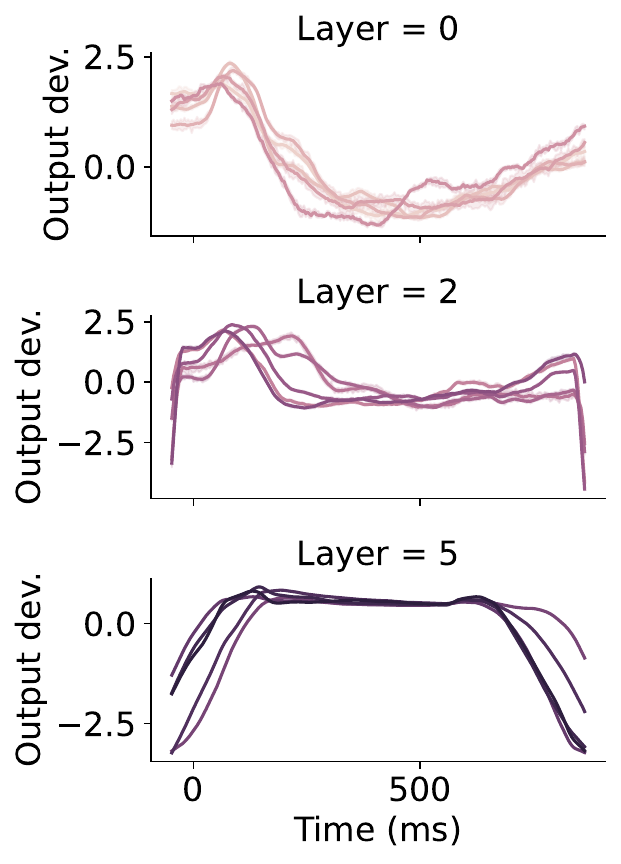}
  \caption{Temporal PFI}
  \label{fig:temporal_PFI}
\end{subfigure}
\caption{Spatio-temporal insights can be obtained using PFI. Spatial (a), channel-wise temporal (b), and temporal (c) PFI across \texttt{non-linear group-emb} kernels within 3 layers (rows). For spatial PFI, kernels are plotted separately; whereas for temporal PFI, 5 kernels (lines) are plotted together. Channel-wise temporal PFI shows the temporal PFI of each channel for Kernel 2. Channel colouring is matched to the corresponding spatial PFI map, and darker reds mean higher output deviation. For temporal PFI, output deviation is normalised. The horizontal axis shows the time elapsed since the image presentation, for both temporal PFI types. 95\% confidence intervals are shown with shading. }
\label{fig:spatiotemporal}
\end{figure}

Figure~\ref{fig:spatiotemporal} shows results for just 3 of the 6 convolutional layers, with all 6 layers shown in Figure~\ref{fig:kernel_spatial_pfi_30} and Figure~\ref{fig:spatiotemporal_ap}. Kernels within a layer tend to have similar temporal sensitivity, and hence we only show 5 out of over 1e5 total kernels (Figure~\ref{fig:temporal_PFI}). Output deviations are standardised to compare temporal PFI across kernels with different output magnitudes. In the early layers, sensitivity peaks around 100 ms (as in Figure~\ref{fig:global_PFI}), then rapidly decreases, eventually climbing again slowly. Kernels in early layers have somewhat random spatial sensitivity (Figure~\ref{fig:spatial_pfi}), but this gets narrowed down to channels over the visual cortex in deeper layers, with some differences between individual kernels. This sensitivity is similar to the spatial features that were shown to be most informative for classification performance (see Figure~\ref{fig:global_PFI}).

Figure~\ref{fig:spatiotemporal_pfi} shows the temporal profile of the spatial PFI. This is achieved by limiting the shuffling to 100 ms time windows and 4-channel neighbourhoods (3 closest channels for each channel) at a time, which is then repeated across all time points and channels. This shows spatial sensitivity does not seem to change with time; i.e. the most important channels are always the same, also observed in previous spatiotemporal PFI analyses of this dataset presented in Chapter~\ref{Chap3}.

\begin{figure}[!t]
\centering
\begin{subfigure}{0.45\textwidth}
  \centering
  \includegraphics[width=0.8\linewidth]{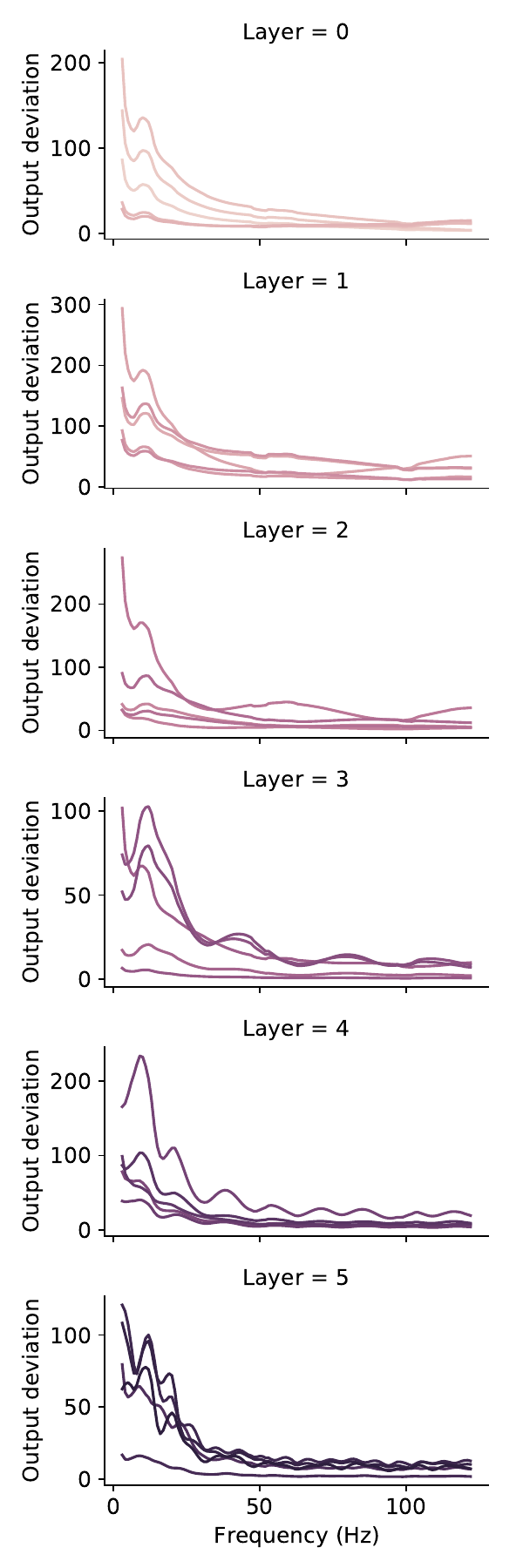}
  \caption{Spectral PFI}
  \label{fig:spectral_pfi_ap}
\end{subfigure}%
\begin{subfigure}{0.45\textwidth}
  \centering
  \includegraphics[width=0.88\linewidth]{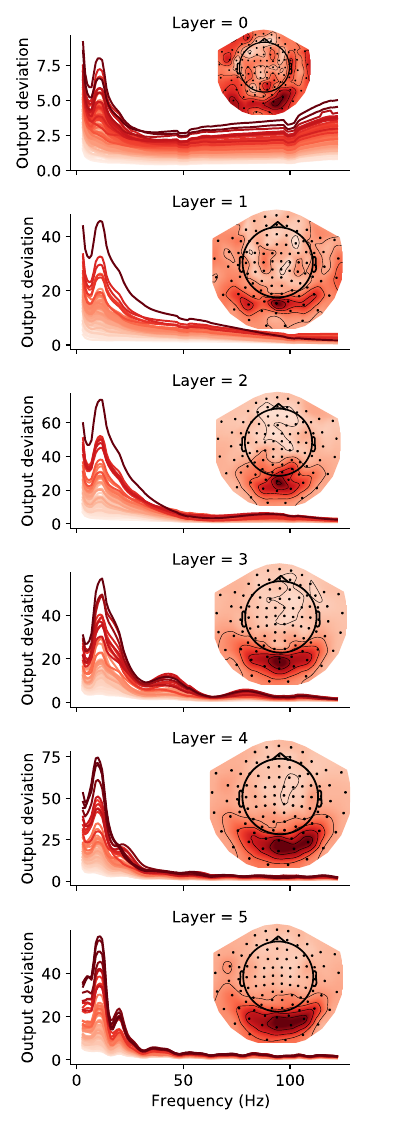}
  \caption{Channel-wise Spectral PFI}
  \label{fig:spatiospectral_chn_ap}
\end{subfigure}
\caption{Frequency sensitivity of kernels via spectral PFI (a), channel-wise spectral PFI (b) of the trained \texttt{non-linear group-emb} model in 6 layers (rows). Kernels are plotted together (lines) for spectral PFI. Each channel-wise spectral PFI plot is for 1 kernel, where lines show the spectral PFI of corresponding channels in the topomap. 95\% confidence interval is shown with shading for spectral PFI. Due to small variability across permutations, this is barely visible.}
\label{fig:spatiospectral_ap}
\end{figure}

In neurophysiology, we are often interested in the oscillatory content of the signal, and what/how specific frequencies are associated with certain tasks, here, decoding of visual stimuli. To this end, we use PFI in the spectral domain, where it is used to measure the change in kernel output to perturbations in specific frequency bands (Figure~\ref{fig:spectral_pfi_ap}). Across all layers and kernels, the profile has a $1/f$ (frequency) shape with a clear peak at 10 Hz. These are common features of the MEG signal \citep{demanuele2007distinguishing, drewes2022individual}, indicating that the spectral sensitivity of the kernels coincides with the power spectra of the data. Our previous model-level spectral PFI analysis in Chapter~\ref{Chap3} did not show the 10Hz peak so clearly because of lower sampling rates. In Figure~\ref{fig:spatiospectral_chn_ap}, we also looked at the spectral PFI of 4-channel neighbourhoods and found that kernels are sensitive to the same channels (in the visual area) across all frequencies, with these channels having larger 10 Hz peaks. Further spectral investigations are presented in Appendix~\ref{appendix_group}.

In summary, the analyses presented in this section show that the kernels are sensitive to interpretable temporal, spatial, and spectral features of the MEG data. Specifically, we have shown that kernels are sensitive to channels in the visual area, with this sensitivity getting more focused in deeper layers. Kernels are also sensitive to the 10Hz peak of the MEG data, and the temporal sensitivity shows a peak at 100-150ms in early layers.

\section{Discussion}
\label{sec:discussion}

In this chapter, we focused on across-subject decoding, motivated by the fact that group-level models that perform well in this manner can be useful for gaining neuroscientific insights that are relevant at the group level. In this setting, our proposed deep learning-based group-level model outperforms naïve group models and achieves similar performance to subject-level models, but with three key benefits. First, it provides potentially richer insights at the group level. As nonlinear models are required for the group-level decoding to work, we had to use PFI, and showed that it is effective in the case of nonlinear group-level models. Second, there is potential for the group-level model to outperform subject-level models when larger population datasets are available. Third, a group-level model can be used to initialise subject-level models surpassing the performance of subject-level models initialised randomly. We have shown how subject embeddings and non-linearity are crucial for this. These are important insights towards the goal of using group models in decoding neuroimaging data, which would allow for better use of this inherently limited resource.

Interestingly, we found that at the subject level, linear models perform better than their nonlinear counterparts. Although some studies have found deep learning to improve over simpler linear models, this improvement is often marginal \citep{cooney2019classification, schirrmeister2017deep}. Such results are difficult to generalise across different MEG datasets, due to variability in tasks, the number of subjects, and the amount and quality of data \citep{schirrmeister2017deep}.

Other than being useful for fine-tuning, our embedding-aided group model can be useful in the case of much larger datasets, where we cannot afford to have a separate model for each subject. As we have shown, even in this limited dataset with 15 subjects, the group model can provide improvement in a few subjects. Our results suggest follow-up studies to understand why some subjects performed better or worse. A current limitation of our approach is that it is still worse than subject-level models (on average).

We have demonstrated the use of PFI on group models to obtain insight into which time points and channels contributed to the decoding and to obtain meaningful information encoded in convolutional kernels. PFI can also provide group-level temporal, spatial and spectral information by averaging over linear subject-level models. Here our aim was to show that PFI works similarly well in the case of nonlinear group-level models. Using this and other methods, such as representational similarity analysis, neuroscientific investigations can be performed at the group level using a single model, instead of averaging over individual subject models. We note that one downside of PFI is that the absence of influence on the output does not necessarily mean that a specific channel or time window does not carry information about the target variable. When applying PFI to kernel outputs it is unclear how to summarise and visualise this information across millions of kernels. This is one downside of using deep learning models with such a large parameter space. As future work it may be useful to assess the usage of subject embeddings with PFI.

While the across-subject decoding we focus on in this work is most relevant to situations where we want to obtain insights at the group-level, other applications, such as BCIs that need to work well on previously unseen subjects, may be more appropriately evaluated using leave-one-subject-out (LOSO) evaluation. In this context, we found that using subject embeddings did not improve performance. Exploiting subject embeddings in a pure LOSO framework is not trivial, as some additional approach is needed to initialise/learn the embedding of the left-out subject in an unbiased manner. We have not tried to only optimise the subject embedding while freezing the rest of the model. While computationally less expensive, this is not expected to be as good as optimising the whole model (and subject embedding) on the new subject, which we have presented in Section~\ref{ssec:generalization}. In larger datasets with more subjects, between-subject similarities in the embeddings could be exploited and different heuristics explored, e.g. initialising the embedding with the average of all learned subject embeddings. However, research aimed at improving performance in new subjects often leverages transfer learning in some way, where a limited amount of data from the new subject can be used \citep{zubarev2019adaptive}. In this scenario, we think our across-subject group model could be helpful, by, for example, using the limited data from the new subject or by learning a useful embedding for the new subject in an unsupervised manner. As we have shown in Section~\ref{ssec:group_models}  this could be especially useful for subjects with low performance.

As opposed to a naive continuation of the trends in Figure~\ref{fig:generalization}, we expect that with more trials, the gap between group initialisation and training from scratch would continue, up to some limit. We believe that the reason why the gap closes at 100\% training data is due to the ratio of training and validation sets and the low number of examples. The small validation set (6 examples per class) is probably not representative of the full data distribution.

We expect the subject embedding and group modelling to generalise to different task and recording modalities (EEG, fMRI, etc.) because they face similar decoding challenges. The specific Wavenet-based model is readily generalisable to other electrophysiological data such as EEG and electrocorticography (ECoG), because of the same temporal dynamics they capture. Further research is needed into deep learning models capable of implicitly learning inter-subject variability. An important question is whether scaling up models on large datasets would achieve this goal.

\chapter{Forecasting MEG signals}
\label{Chap5}

In previous chapters, we have presented methods for dealing with within-subject and between-subject variability in MEG decoding. When addressing such variability, we assumed that all data came from the same experiment and scanner, which is a serious limitation of these approaches. Single experiments usually investigate only a few research questions with limited dataset sizes. By utilising multiple datasets collected by different researchers, we could potentially achieve the scale required for deep learning to be truly applicable. Training a single model on multiple datasets allows us to apply it to various encoding and decoding paradigms, which could be especially useful for experiments with small datasets.

However, there are two major challenges to overcome. First, the decoding models used so far are not well suited for generalisation across datasets, as not all data is recorded with decoding in mind, such as rest data. It would also be difficult to include the vastly different experimental stimuli from every dataset into a single decoding model. Second, variability in electrophysiology becomes even more problematic, as we now must address variability not only between subjects but also between different experimental setups and scanners.

For the first issue, the title of this chapter (forecasting) provides a natural solution. As discussed in Chapter~\ref{Chap2}, forecasting of multivariate time series is a general task that can be formulated for any M/EEG dataset. Thus, we need not be concerned with specific experimental paradigms or recording modalities. Moreover, by using an unsupervised modelling framework, we can potentially learn shared representations across datasets, which could then prove useful for specific tasks such as encoding or decoding. For the second issue, a solution that often works well in deep learning is scale. That is, by using more and more data and sufficiently expressive models, variability can be implicitly learned and modelled. Unsupervised modelling (e.g., forecasting) also enables larger scale, by utilising all timepoints for training, rather than using only a subset as in encoding or decoding

In this chapter, we explore how to design deep learning forecasting models that can reproduce spatiotemporal dynamics and various other properties of MEG data. We present both Wavenet-based \citep{oord2016wavenet} and Transformer-based \citep{Vaswani:2017} models, comparing them with standard linear autoregressive (AR) modelling on MEG data.

We show that Transformer-based models provide better modelling capabilities than Wavenet and linear AR models by reproducing the HMM statistics of real data and evoked activity in task data. Through a series of ablations, we demonstrate which aspects of the Transformer-based models enable these improvements. In the case of the Transformer, our design includes a novel application of tokenisation methods, allowing such a model developed for the discrete domain of language to be applied to continuous multichannel time series data.

We also extend the forecasting framework to work with condition labels as inputs, enabling better modelling (encoding) of task data. Finally, we present a method for transforming  a forecasting model into a generative decoder through the use of Bayes’ theorem. 

To be clear, we do not apply our methods to multiple datasets, as our aim was to develop proof-of-concept models and analyse them on a reasonably sized dataset. Testing on multiple datasets at scale is left for future work.

\section{Introduction}

Unsupervised learning provides a dataset-agnostic method for learning shared representations. Within unsupervised learning, we can further differentiate between methods aiming to learn interpretable representations and purely data-driven approaches. The goal of interpretable models is to provide neuroscientific insights into electrophysiology data in an unsupervised manner. This is especially useful for rest data, where there is no external stimulus or behaviour linked to the brain activity. Models designed without focusing on interpretability can still be analysed using the techniques mentioned in Section~\ref{ssec:interpret_deep}. However, such models are primarily used to generalise over multiple heterogeneous datasets and provide a pre-trained foundation model. By leveraging large amounts of data, the hope is that the model will be capable of generalising to new data types and provide improvement over a model trained on a single, small dataset. This is especially useful for BCI settings.

The concept of using vast amounts of data to boost performance in downstream tasks originates from deep learning. Perhaps the most successful recent example is that of large language models, trained on diverse data sources and demonstrating enhanced capabilities over task-specific models in a multitude of language-related tasks (e.g., translation, summarisation) \citep{brown2020language}. This can also be viewed as a form of transfer learning. Zero-shot performance is obtained when no fine-tuning is done for the downstream task.

Several factors enabled the success of large language models, including data scale, model size, fast GPUs, and effective neural network architectures \citep{kaplan2020scaling, fedus2022switch, sutton2019bitter}. To adopt this paradigm for electrophysiology data, the primary obstacles are the model architecture and data size. In this chapter, we focus on the former. Unfortunately, the number of brain-recording datasets is limited due to the high cost of data collection. Recordings are often not publicly released because of privacy concerns. To achieve data scale comparable to language modelling, lowering the financial barrier to collecting brain data and advocating for public release is needed. While language data is freely available online, brain data is far more difficult to find via automated scraping, has much higher dimensionality requiring more storage and download bandwidth, and is far more heterogeneous due to differences in scanners and experiments.

Our focus in this chapter is designing general models well-suited to multichannel timeseries that can scale effectively. We also focus on using forecasting models, which are causal and can generate data recursively, as they achieve a good balance between interpretability and scalability. While some unsupervised models aimed at neuroscientific investigations have been proposed \citep{gohil2022mixtures} here we focus on reviewing more data-driven self-supervised approaches.

\subsection{Self-supervised learning}

Self-supervised learning (SSL) has emerged as a promising approach for learning useful representations from unlabelled electrophysiological data. SSL reformulates an unsupervised learning problem as a supervised one by exploiting inherent structure in the data to generate "pseudo-labels". In the context of electrophysiology, recent works have proposed SSL tasks tailored to the temporal and multivariate nature of neural time series data \citep{banville2021uncovering,kostas2021bendr,wang2023brainbert}.

\citet{banville2021uncovering} investigate three SSL tasks for learning from unlabelled EEG recordings. Each task is trained via a contrastive loss function, where the model learns to pull positive pair examples closer in a representation space while pushing negative pairs apart. They demonstrate that the representations learned via SSL on unlabelled EEG data transfer well to supervised downstream tasks, consistently improving over limited label training and matching full supervision performance.

Building on this \citet{kostas2021bendr} propose combining self-supervised contrastive learning with Transformer networks to enable pre-training on large amounts of unlabelled EEG data. Their approach, BErt-inspired Neural Data Representations (BENDR), uses a Transformer encoder architecture applied to learned representations of raw EEG segments. A technical description of Transformer models is provided in Section~\ref{ssec:transformers}. First, a temporal convolutional network extracts initial representations of the EEG time series, referred to as BENDR features. Next, a Transformer encoder module takes the BENDR features as input. Contiguous segments of the BENDR representations are randomly masked, and the model is trained via a contrastive loss to predict the original features. Fine-tuning the pretrained model significantly improves performance on supervised EEG analysis tasks compared to training just on the downstream datasets.

\section{Methods}

In our quest for designing expressive forecasting models of MEG data, we can look to artificial intelligence domains with similar characteristics, such as audio or natural language processing. These domains share some similarities with MEG data, like the sequential nature of the modality. However, while audio data is also a continuous timeseries, it only contains a single channel and comes at a much higher sampling rate compared to M/EEG data. Language data is perhaps even more different as its timeseries are comprised of distinct units (words) from a finite vocabulary set. As such, starting with models developed for these domains and adapting them to handle the nuances of M/EEG data is a promising approach. Indeed, in this chapter we adapt Wavenet, originally developed for forecasting audio data \citep{oord2016wavenet}, and GPT-2, originally developed for forecasting language \citep{radford2019language}.

\subsection{Wavenet}
\label{ssec:full_wavenet}

\begin{figure}[!t]
    \centering
    \includegraphics[width=0.95\textwidth]{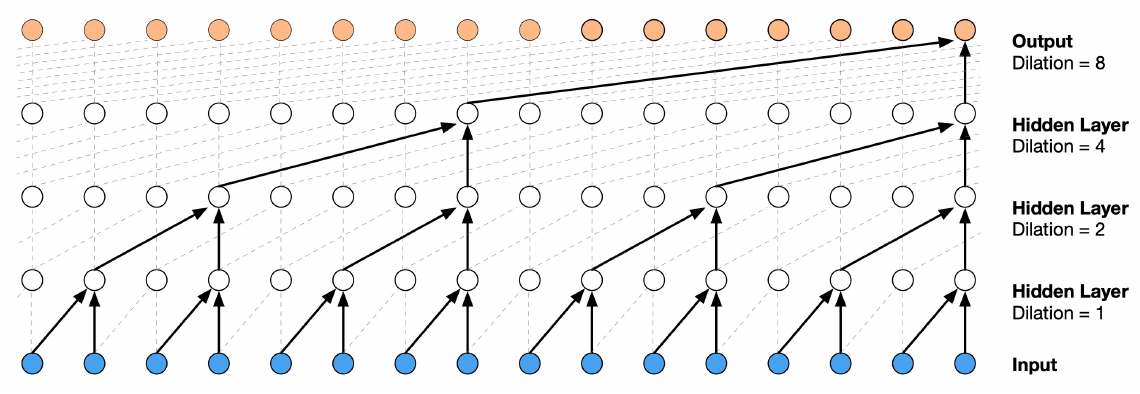}
    \caption{A stack of dilated convolutions, the core architecture of Wavenet. The dilation factor is doubled in successive layers. Figure from \citet{oord2016wavenet}.}
    \label{fig:wavenet}
\end{figure}

Here we describe the Wavenet architecture (Figure~\ref{fig:wavenet}) used in the original paper \citep{oord2016wavenet}, and how we adapted it for electrophysiological data. Wavenet models the conditional probability of each time sample given all preceding samples autoregressively:

\begin{equation}
p(\mathbf{X}) = \prod_{t=1}^{T} p(\mathbf{x}_t | \mathbf{x}_1, ..., \mathbf{x}_{t-1})
\end{equation}

where $\mathbf{x}_t$ is the sample at time $t$ and $T$ is the total sequence length. Unlike our simplified Wavenet used in Chapter~\ref{Chap4} which outputs point estimates of the continuous value of the data at the next timepoint, the full network predicts a categorical distribution over tokenised samples using a softmax output layer. Throughout this chapter we use tokenisation and quantisation interchangeably. Both have the aim of discretising a continuous quantity into a finite set of distinct bins/levels/tokens.

In the original paper, the audio waveform is tokenised using a quantisation to 8 bits following a $\mu$-law companding transform \citep{lewis1997law}:

\begin{equation}
    f(\mathbf{x}_t) = \mathrm{sign}(\mathbf{x}_t)\frac{\ln(1 + \mu |\mathbf{x}_t|)}{\ln(1 + \mu)}
\end{equation}

where $\mu$ controls the number of quantisation levels, set to 255 as in the original Wavenet. $f(.)$ is applied to each value of $\mathbf{x}_t$ independently. This nonlinear transformation improves reconstruction versus uniform quantisation of the raw input, as it skews the distribution such that more levels are allocated to smaller magnitudes. For MEG data, we observe similar benefits when applying this transform prior to quantisation. Note that the input must be scaled to $(-1,1)$ first, and clipping outliers above some threshold helps ensure a more uniform mapping.

Critically, tokenisation, in this case through quantisation, enables modelling of probability distributions over data and sampling, instead of just point estimates from MSE-based training. Cross-entropy loss also avoids the mean-prediction bias induced by MSE \citep{banville2021uncovering}.

\begin{figure}[!t]
    \centering
    \includegraphics[width=0.95\textwidth]{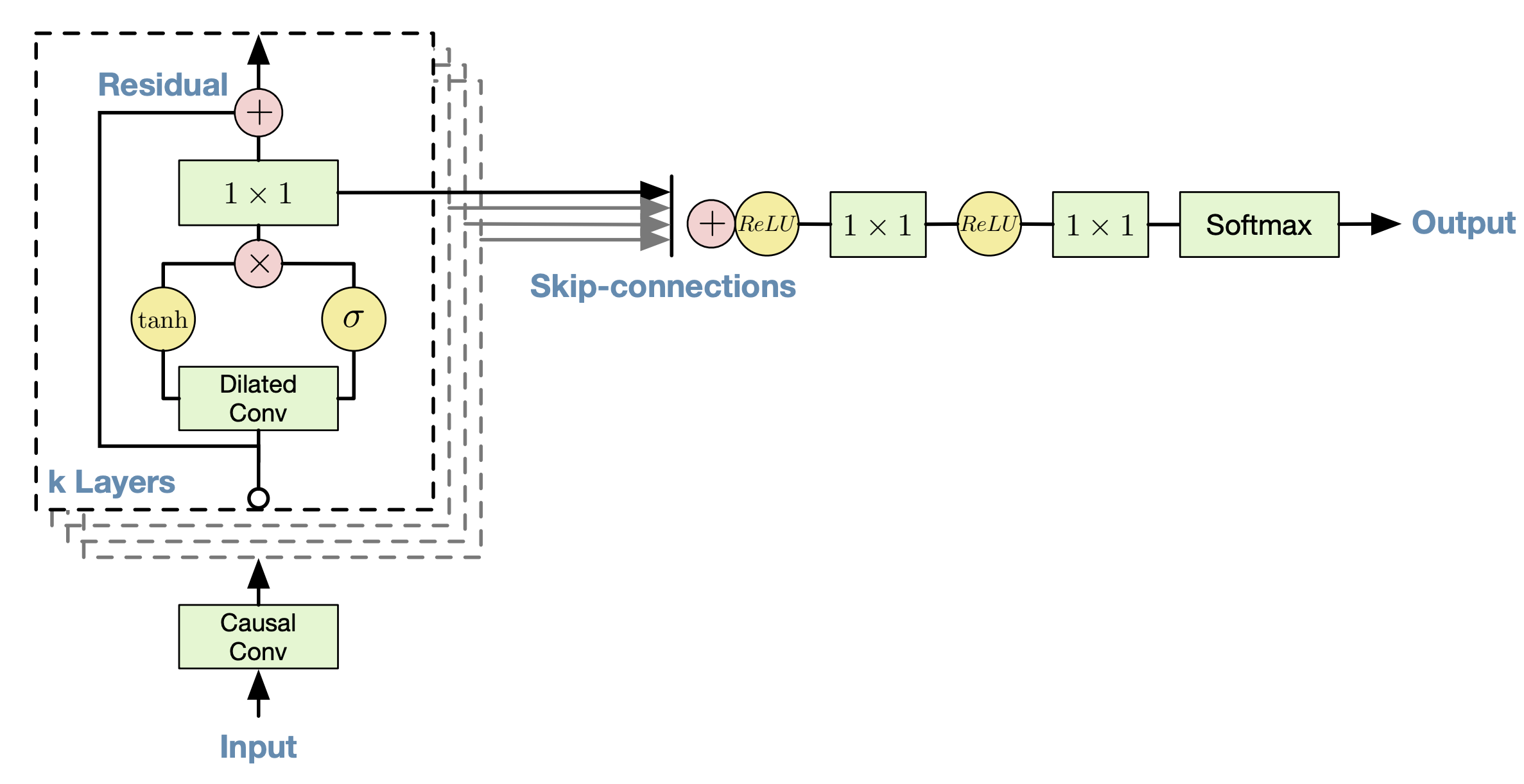}
    \caption{Overview of the full Wavenet architecture with gated dilated convolutions and residual connections. Figure from \citet{oord2016wavenet}.}
    \label{fig:wavenet_layers}
\end{figure}

Wavenet uses nonlinear activation functions and skip connections between layers (Figure \ref{fig:wavenet_layers}). The computations for layer $l$ are:

\begin{align}
\label{eq:wavenet_layer}
\mathbf{Z}^{(l)} &= \tanh\left(\mathbf{W}^{(l)}_{f}*\mathbf{H}^{(l)}\right) \odot \sigma\left(\mathbf{W}^{(l)}_{g}*\mathbf{H}^{(l)}\right) \\
\mathbf{S}^{(l)} &= \mathbf{W}^{(l)}_{s} * \mathbf{Z}^{(l)} \\
\mathbf{H}^{(l+1)} &= \mathbf{W}^{(l)}_{r} * \mathbf{Z}^{(l)} + \mathbf{H}^{(l)}
\end{align}

where $*$ is convolution, $\odot$ is element-wise multiplication, $\sigma$ is the sigmoid, $\tanh$ is the hyperbolic tangent function. $\mathbf{W}^{(l)}_{f}$ and $\mathbf{W}^{(l)}_{g}$ are the filter and gate convolutions. $\mathbf{S}^{(l)}$ is the skip connection output, and $\mathbf{H}^{(l+1)}$ is the residual layer output. Importantly, the residual and skip convolutions ($\mathbf{W}^{(l)}_{r}$ and $\mathbf{W}^{(l)}_{s}$)  use 1x1 kernels, while the initial two are dilated.

The skip outputs are summed across layers and passed through further 1x1 convolutions:

\begin{align}
    \mathbf{S} &= \sum_{l=1}^L\mathbf{S}^{(l)}\\
    \mathbf{Y} &= \mathrm{Conv1x1}\left(\mathrm{ReLU}\left(\mathrm{Conv1x1}\left(\mathrm{ReLU}\left(\mathbf{S}\right)\right)\right)\right) \\
    \hat{\mathbf{X}}_{T+1} &= \mathrm{Softmax}(\mathbf{Y})
\end{align}

By setting the channels of the last convolution to the number of tokens, i.e. the vocabulary size which in this case is the number of quantisation bins  ($Q=256$), $\mathbf{Y}$ represents logits over tokens and $\hat{\mathbf{X}}_{T+1} \in \mathbb{R}^{C \times Q}$ gives the predicted distribution at $T+1$. Cross-entropy loss can then train the model to accurately forecast future timesteps.

\subsection{Multi-channel Wavenet}

When adapting Wavenet to M/EEG, a key challenge is the multi-channel nature of the data. We devise two versions: \texttt{WavenetFullChannel} as univariate, and \texttt{WavenetFullChannelMix} as multivariate. In both, each channel is transformed and tokenised independently to form the input to the models.

In \texttt{WavenetFullChannel}, we first apply an embedding layer to the tokenised data, learned separately per channel. The embedding layer represents each discrete bin as a high-dimensional continuous vector, enabling powerful representations in the convolutional layers whose input channels match the embedding size. To be clear in this univariate approach the same model is applied to each channel. However, a different embedding layer is learned for each channel, meaning that for example the quantised value of 0.42 in channel x will have a different vector representation than in channel y. This helps the model differentiate between channels.

The embedding operation is given below:

\begin{align}
\forall c \in {1, 2, \dots, C}: \mathbf{X}_{e}^{(c)} &= \mathbf{W}^{(c)} \mathbf{X}^{(c)}  \\
    \mathbf{H}_0 &= \mathrm{Concatenate}(\mathbf{X}_{e}^{(1)}, \mathbf{X}_{e}^{(2)}, \dots, \mathbf{X}_{e}^{(C)})
\end{align}

Here, $\mathbf{X}^{(c)} \in \mathbb{R}^{Q \times T}$ is the tokenised one-hot input and $\mathbf{W}^{(c)} \in \mathbb{R}^{E \times Q}$ is the embedding layer of channel $c$ mapping tokens $Q$ to embeddings of size $E$. $\mathrm{Concatenate}$ concatenates along the channel dimension.

$\mathbf{H}_0 \in \mathbb{R}^{C \times E \times T}$ is the resulting input to Wavenet with $C$ as the batch dimension. Thus, the same model is applied independently to each channel in parallel. At output, a distribution is predicted simultaneously for each channel at $T+1$. The model is optimised to accurately predict all channels.

\texttt{WavenetFullChannelMix} includes an extra linear layer after summing the skip representations to mix information across the channel dimension:

\begin{align}
    \mathbf{S} &= \sum_{l=1}^L\mathbf{S}^{(l)} \\
    \mathbf{S} &= \mathbf{S}\mathrm{.permute}(1, 2, 0) \\
    \mathbf{S}_{out} &= \mathbf{S}\mathbf{W}_m
\end{align}

where $\mathbf{W}_m \in \mathbb{R}^{C \times C}$ is the mixing weight matrix. The permutation is needed to apply the projection to the appropriate channel dimension. After this $\mathbf{S}_{out}$ is permuted back to the original dimension order and the rest proceeds identically to \texttt{WavenetFullChannel}.

In the original Wavenet, audio generation can be conditioned on additional inputs through embedding-based global conditioning or time-aligned local conditioning. For some experiments, we augment the model with local features of task stimuli or subject labels, first embedded into continuous vectors:

\begin{align}
    \mathbf{H}_y &= \mathbf{Y}\mathbf{W}_y \\
    \mathbf{H}_o &= \mathbf{O}\mathbf{W}_o \\
    \mathbf{H}_{c} &= \mathrm{Concatenate}(\mathbf{H}_y, \mathbf{H}_o)
\end{align}

where $\mathbf{Y} \in \mathbb{R}^{T \times N}$ contains the condition index $n \in (1, \dots, N)$ at each time point, and $\mathbf{O} \in \mathbb{R}^{T \times S}$ contains the subject index $s \in (1, \dots, S)$ at each time point $t \in (1, \dots, T)$. $\mathbf{W}_y \in \mathbb{R}^{N \times E_n}$ and $\mathbf{W}_o \in \mathbb{R}^{S \times E_s}$ are embedding matrices mapping the labels to learned continuous vectors of size $E_n$ and $E_s$, respectively. The subject index is the same across time points of the recording from the same subject. The condition index is set to the (visual) stimuli presented (e.g., one of the 118 images in \citet{cichy2016comparison}), for exactly those time points when the stimulus is on. At any other time, the task condition embedding $\mathbf{H}_y$ is set to 0.

$\mathbf{H}_{c}$ is the conditioning vector fed into Wavenet at each layer, modifying Equation~\ref{eq:wavenet_layer} to:

\begin{align}
\mathbf{Z}^{(l)} &= \tanh\left(\mathbf{W}^{(l)}_{f}*\mathbf{H}^{(l)} + \mathbf{W}^{(l)}_{c}*\mathbf{H}_{c}\right) \odot \sigma\left(\mathbf{W}^{(l)}_{g}*\mathbf{H}^{(l)} + \mathbf{W}^{(l)}_{c}*\mathbf{H}_{c}\right) 
\end{align}

where $\mathbf{W}^{(l)}_{c}$ (1x1 convolution) projects $\mathbf{H}_{c}$ before adding it to the input representation. This conditions the prediction on both past brain activity and stimuli:

\begin{align}
    p(\mathbf{X}|\mathbf{Y}, \mathbf{O}) = \prod_{t=1}^{T} p(\mathbf{x}_t | \mathbf{x}_1, ..., \mathbf{x}_{t-1}, \mathbf{y}_{1}, ..., \mathbf{y}_{t-1}, \mathbf{o}_{1}, ..., \mathbf{o}_{t-1})
\end{align}

In single-subject models we only use the task labels $\mathbf{Y}$.

The full Wavenet architecture can either be interpreted as forecasting with extra conditioning or as a generative encoder augmented with past brain activity. In addition, the probabilistic formulation allows converting the model into a decoder using Bayes' rule, enabling both forecasting and decoding within the same framework:

\begin{align}
     p(Y|X) &= \frac{p(X=x|Y) p(Y)}{p(X=x)} 
\end{align}

where $X$ is the random variable representing the data, $Y$ is the random variable representing the task labels, and $x$ is a particular sample of $X$. $p(Y)$ is the task label prior distribution which in the 118-image dataset is uniform. $p(X=x|Y)$ is the likelihood of the data given the label which we get from the above formulation of Wavenet. The only tricky part is $p(X=x)$ as this requires marginalisation over $Y$. In the case of the 118-image dataset this means that we have to run the trained model with all of the possible task labels to obtain $p(X=x)$:

\begin{equation}
    p(X=x) = \sum_{i=1}^{N}{p\left(X=x|Y=i\right)p\left(Y=i\right)}
\end{equation}

Thus, in a single self-supervised deep learning model we have flexibly encapsulated forecasting, encoding, and decoding, all three of the main modelling methods of M/EEG data. This unification of modelling approaches was inspired by a GitHub repository applying similar ideas to images\footnote{\url{https://github.com/cheind/autoregressive}}. The inverted decoder formulation also allows for iterative estimation of $p(Y|X)$ at each timestep. The author of the GitHub repository has applied this method to estimating the probability of image labels (digits 0 to 9) from pixel images, as more and more of the image was fed into the model.

\subsection{GPT2}
\label{ssec:transformers}

While Wavenet is an effective model for forecasting time series, it may be that other types of architectures are better suited for multichannel data. The dilated convolutional architecture, while fast and parameter-efficient, might limit the model's expressivity, particularly when scaling up on multiple datasets. Indeed, in recent years there has been a second deep learning revolution driven by the Transformer architecture \citep{Vaswani:2017}. Unlike the weight sharing and autocorrelation inductive biases (priors) of convolutional models, Transformers have two key architectural priors. First, they are sequential models operating on a discrete set of input tokens (e.g. words), mapped to continuous embeddings. Second, their primary mode of processing these representations is the attention mechanism. This allows Transformers to model complex dependencies across long sequences without regard to their distance in the input or output. This provides a more flexible inductive bias well-suited to language modelling and other tasks involving highly structured sequential data \citep{Devlin:2019,Radford:2019}.

Since their introduction, Transformers have become the dominant model architecture for natural language processing (NLP). Models like BERT \citep{Devlin:2019}, GPT-2 \citep{radford2019language}, and GPT-3 \citep{brown2020language} have achieved state-of-the-art results on a wide range of NLP benchmarks. The success of Transformers for language has led researchers to apply them to other sequential modelling tasks. For time series forecasting, Transformer-based models offer several potential advantages over RNNs and temporal convolutional networks. The self-attention mechanism provides direct connectivity between any two time steps, capturing long-range dependencies. Pre-trained representations like BERT can inject useful inductive biases from language modelling. The parallelisable architecture allows more efficient computation compared to recurrent models.

Early explorations of Transformers for time series have shown promising results. \cite{zhou2021informer} adapted the self-attention mechanism for long-range forecasting and demonstrated state-of-the-art performance on multiple public datasets. As with NLP, we expect Transformer models to become a leading approach for time series modelling \citep{wen2022transformers}.

Thus, we set out to design a Transformer model suited for M/EEG data, while keeping the key elements that made it successful in language modelling. Specifically, we use GPT-2, a popular autoregressive Transformer variant. When adapting GPT-2 to continuous multivariate time series, the main challenges are at the input and output layers interfacing the model with the data. We describe GPT-2 for language modelling first, then present our modifications. A particularly detailed visual description of GPT2 is given in \citet{alammar2019illustrated}.

\begin{figure}[!t]
\begin{subfigure}{0.25\textwidth}
  \centering
  \includegraphics[width=1.0\linewidth]{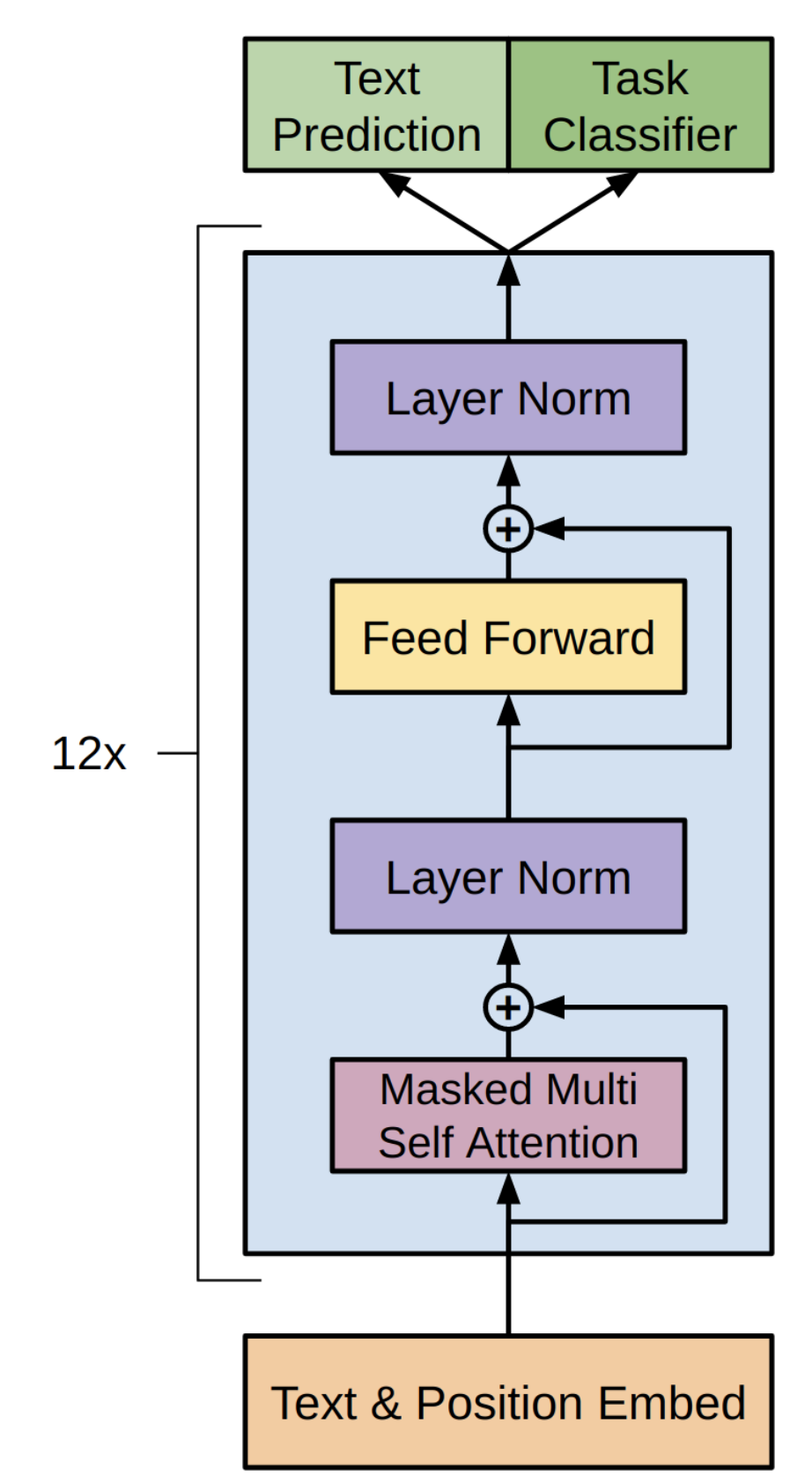}
  \caption{GPT2 architecture}
\end{subfigure}%
\begin{subfigure}{0.75\textwidth}
  \centering
  \includegraphics[width=0.9\linewidth]{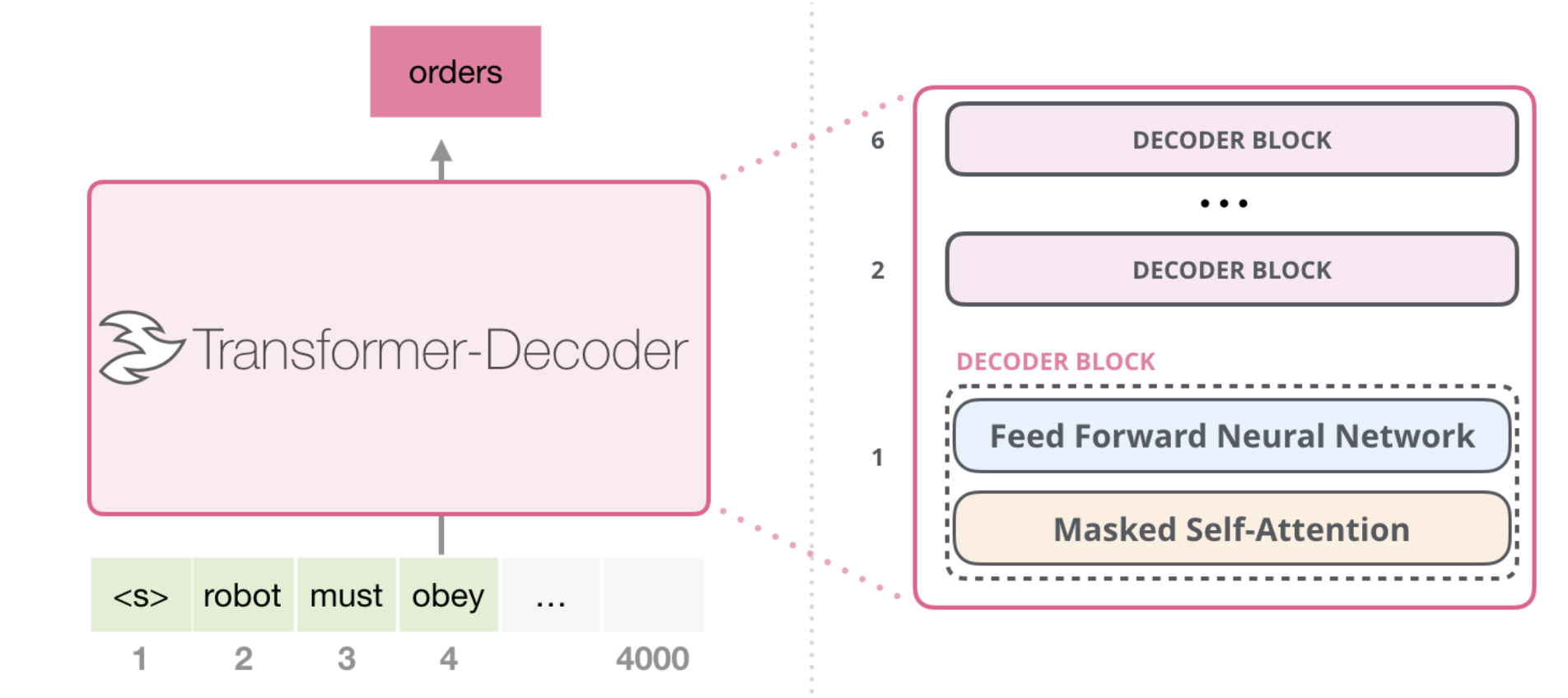}
  \caption{GPT2 architecture (alternative visualisation)}
\end{subfigure}
\caption{Two visualisations of the core GPT2 architecture for language modelling. Figures from \citet{Radford:2018} (left) and \citet{alammar2019illustrated} (right).}
\label{fig:gpt2_visu}
\end{figure}

GPT2 utilises a multi-layer Transformer decoder\footnote{Here the term decoder is used as an architecture type in deep learning, i.e., autoregressive forecasting, as opposed to a decoder of brain data.} architecture (Figure~\ref{fig:gpt2_visu}). Each layer contains two sublayers: a multi-head self-attention mechanism and a position-wise feedforward network. Residual connections and layer normalisation are employed around each sublayer. The self-attention mechanism allows the model to attend over previous positions when generating the next token. It operates on query $\mathbf{Q}$, key $\mathbf{K}$, and value $\mathbf{V}$ projections of the layer input $\mathbf{H}^{(l)} \in \mathbb{R}^{T \times E}$:

\begin{equation}
\text{Attention}(\mathbf{Q},\mathbf{K},\mathbf{V}) = \mathrm{softmax}{\frac{\mathbf{Q}\mathbf{K}^T}{\sqrt{E}}}\mathbf{V}
\end{equation}

where $E$ is the dimension of the feature space inside the model, also called hidden dimension, often set to the embedding size. $\mathbf{Q}$, $\mathbf{K}$, and $\mathbf{V}$ are computed by:

\begin{align}
\mathbf{Q}^{(l)} &= \mathbf{H}^{(l)}\mathbf{W}^{(l)}_Q \\ 
\mathbf{K}^{(l)} &= \mathbf{H}^{(l)}\mathbf{W}^{(l)}_K\\
\mathbf{V}^{(l)} &= \mathbf{H}^{(l)}\mathbf{W}^{(l)}_V
\end{align}

where $\mathbf{W}^{(l)}_Q, \mathbf{W}^{(l)}_K, \mathbf{W}^{(l)}_V \in \mathbb{R}^{E \times E}$ are learned projections. A crucial function of the attention mechanism is comparing different vector representations of elements in the input sequence through the dot-product similarity measure. Most Transformer variants use multi-head attention computing attention separately over multiple distinct feature partitions, concatenating the outputs:

\begin{align}
\forall i \in N: \mathbf{H}^{(l)}_i &= \mathbf{H}^{(l)}[:, i:i+d] \\ 
\mathbf{Z}^{(l)}_i &= \mathrm{Attention}(\mathbf{H}^{(l)}_i\mathbf{W}_{Q_i}, \mathbf{H}^{(l)}_i\mathbf{W}_{K_i}, \mathbf{H}^{(l)}_i\mathbf{W}_{V_i}) \\
\mathrm{MultiHeadAttention}(\mathbf{H}^{(l)}) &= \mathrm{Concatenate}(\mathbf{Z}^{(l)}_1,\dots,\mathbf{Z}^{(l)}_N)\mathbf{W}^{(l)}_O
\end{align}

where $N$ is the number of attention heads, and $d=\frac{E}{N}$ is the feature dimension of a single head. Note that the feature dimension depends on the number of heads, as the dimensionality of all heads has to sum up to $E$. $\mathbf{W}_{Q_i}, \mathbf{W}_{K_i}, \mathbf{W}_{V_i}$ are the projections for head $i$, and $\mathbf{W}^{(l)}_O$ is an output projection.

The feedforward layer applies two affine transforms with ReLU activation:

\begin{equation}
\text{FFN}(\mathbf{Z}) = \mathrm{ReLU}(\mathbf{ZW}_1 + \mathbf{b}_1)\mathbf{W}_2 + \mathbf{b}_2
\end{equation}

This allows learning nonlinear representations of the input at each position. Altogether a GPT2 layer is the combination of the self-attention and feedforward layers:

\begin{align}
\mathbf{H}^{(0)} &= \mathbf{X}\mathbf{W}_e + \mathbf{W}_{p} \\
\mathbf{Z}^{(l)} &= \mathrm{LN}(\mathbf{H}^{(l)} + \mathrm{MultiHeadAttention}(\mathbf{H}^{(l)})) \\
\mathbf{H}^{(l+1)} &= \mathrm{Dropout}(\mathrm{LN}(\mathbf{Z}^{(l)} + \mathrm{FFN}(\mathbf{Z}^{(l)}))) \\
\hat{\mathbf{Y}} &= \mathrm{softmax}(\mathbf{H}^{(L)}\mathbf{W}^T_e)
\end{align}

where $\mathbf{W}_e \in \mathbb{R}^{Q \times E}$ embeds the discrete tokens $\mathbf{X} \in \mathbb{R}^{T \times Q}$ into $E$ dimensions.
$\mathrm{LN}$ is Layer Normalisation, a regularisation technique which normalises all activations within a layer to zero mean and unit variance. $\mathbf{W}_{p} \in \mathbb{R}^{T \times E}$ contains positional encodings, providing the model with sequential order information. This is needed as GPT2 lacks recurrent or convolutional elements. Each vector in $\mathbf{W}_p$ indexed by $t \in (1, \dots, T)$ contains a distinct $E$-dimensional representation of position $t$. The output $\mathbf{H}^{(L)}$ is projected back to the vocabulary via the transpose embedding matrix (weight tying). Alternatively, a separate output projection can be learned. The softmax output gives a token probability distribution.

GPT-2 is trained via supervised learning to predict the next token given previous context, minimising cross-entropy loss between model outputs $\hat{\mathbf{Y}}$ and ground truth targets $\mathbf{Y}$. To enable autoregressive training, $\mathbf{Y}$ is set to $\mathbf{X}$ shifted one timestep ahead. Crucially, to prevent information leakage from future timesteps ${t+1, \dots, T}$, causal masking is applied in each self-attention layer, setting outputs that would reveal future information at position $t$ to zero.

The mask $\mathbf{M} \in \mathbb{R}^{T \times T}$ is a lower triangular matrix:

\begin{equation}
\mathbf{M}_{ij} = \begin{cases}
0 & \text{if } i < j\\
-\infty & \text{if } i \geq j
\end{cases}
\end{equation}

\subsection{Channel-independent GPT2}

To apply GPT2 to our continuous multichannel time series data, we take a similar approach as with Wavenet by tokenising each channel independently using the same method as before. This serves as our equivalent of the discrete set of tokens in language modelling. The same GPT2 model is applied to each channel in parallel by setting the channel dimension as the batch dimension. We call this \texttt{ChannelGPT2}.

The input to the model includes the position embeding as well as subject and task-stimulus embeddings. We also add a label/embedding telling GPT2 which channel the current time series is coming from:

\begin{align}
    \mathbf{H}^{(0)} &= \mathbf{X}\mathbf{W}_e + \mathbf{W}_{p} + \mathbf{Y}\mathbf{W}_y + \mathbf{O}\mathbf{W}_o + \mathbf{W}_{c}
\end{align}

where $+$ denotes element-wise addition, $\mathbf{X} \in \mathbb{R}^{C \times T \times Q}$ is the tokenisd input, $\mathbf{W}_{c} \in \mathbb{R}^{C \times T \times E}$ are the learned channel embeddings of size $E$, which are distinct for each channel $c \in {1, \dots, C}$ but constant across time $t$. $\mathbf{Y}$ and $\mathbf{O}$ are the task and subject index matrices, mapped to their respective embeddings. As with the positional encoding $\mathbf{W}_p$, we simply add all embeddings (task, subject, channel) into a single representation. Note that instead of having channel-specific embeddings of the tokenised input $\mathbf{X}$ we learn the same mapping $\mathbf{W}_e \in \mathbb{R}^{Q \times E}$ across channels. Channel information is provided to the model through the channel embeddings.

A serious limitation of this channel-independent GPT2 model is that when predicting a single channel, it does not receive information from other channels. This is analogous to a univariate autoregressive model and ignores crucial cross-channel dependencies in the data. To be clear we often use the term univariate AR modelling in the sense that a separate AR model is trained on each channel. In the case of channel-independent Wavenet and GPT2 models, we train one and the same model on all channels.

\subsection{Flat GPT2}

In the image domain, tokenisation is often abandoned, and a linear projection directly maps image patches to continuous vector representations \citep{dosovitskiy2020image}. Similarly, \citet{nie2022time} have designed a channel-independent Transformer architecture applied to overlapping patches of continuous time series for forecasting. While this facilitates the input, without tokens categorical outputs cannot be generated. As discussed, maintaining operations over tokens and categorical outputs are desirable GPT2 features for M/EEG data. This is because we would like to output probability distributions and train using the cross-entropy loss.

The tokenisation can happen either before or after mixing information across channels. The latter matches GPT2's original design. One example of this is vector quantisation, which is used to tokenise multiple channels in Jukebox, a successful autoregressive Transformer model used on audio data \citep{dhariwal2020jukebox}. Before training the Transformer, a hierarchical VQ-VAE (vector quantized variational autoencoder \citep{van2017neural}) learns discrete codes (tokens) from raw audio. Once trained, VQ-VAE can map a continuous time series to a discrete token sequence $\mathbf{z}$. In the second step of Jukebox, the VQ-VAE is kept fixed, and the discrete tokens are used to learn an autoregressive Transformer. 

Importantly, VQ-VAE is applied to single-channel audio to compress the temporal dimension into discrete codes. For our application we would primarily want to apply vector quantisation to the channel dimension, to have a discrete token at each timestep, or perhaps across a few timesteps. While an adaptation of this could work on MEG data, we opted for a simpler non-deep learning method.

In \texttt{FlatGPT2} we apply vector tokenisation on small groups of channels using the Residual Quantiser algorithm \citep{babenko2014additive} from the faiss library\footnote{\url{https://github.com/facebookresearch/faiss/wiki/Additive-quantizers}}. By using 30 channel groups (buckets) we obtain 30 tokens per timestep, which is already a 10-fold reduction of the original dimension space. However, to have a single token per timestep (as in language modelling) we flatten the feature dimension (buckets) when feeding tokens to GPT2, hence the name \texttt{FlatGPT2}. Our total sequence length then becomes $B \cdot T$, where $B$ is the number of buckets and $T$ is the number of timesteps. This approach is also motivated by the observation that language models include extra information such as context within the sequence, instead of the feature space. Thus, when predicting the token of bucket $b$, we treat the previous timesteps of the other buckets as contextual information. For brevity and because \texttt{FlatGPT2} showed mostly negative results we omit the full mathematical description which can be found in Appendix~\ref{ssec:flatgpt}. Our main results and discussion focuses on the channel-independent GPT2 approach.

\subsection{Model interpretation}

To evaluate whether Wavenet and GPT2 models accurately capture brain dynamics beyond just predictive performance, we develop several analysis techniques to interrogate what these models learn.

\paragraph{Data generation} As mentioned in Section \ref{ssec:ar_generation}, generating new data from a trained model can reveal its capabilities. Different models have distinct generation procedures. Linear AR models take Gaussian noise as input and generate one timestep at a time. Gaussian noise is added to the output, which is appended to the input sequence. This recursive process is described by:

\begin{equation}
\label{eq:iir}
    \mathbf{x}_t = \boldsymbol{\epsilon}_t + f({\mathbf{X}_{t-K:t-1}})
\end{equation}

This intuitively treats the model $f$ as a black-box infinite impulse response (IIR) filter, where $\boldsymbol{\epsilon}_t \sim \mathcal{N}(0, 1)$, and $K$ is the receptive field size. These models can also be analysed as finite impulse response (FIR) filters by removing recursion and using only noise inputs at each timestep: $\mathbf{x}_t = f(\boldsymbol{\epsilon}_{t-K}, \dots, \boldsymbol{\epsilon}_{t-1})$.

For tokenised models (Wavenet and GPT2), we generate data by sampling from the predicted probability distribution and recursively feeding the sample back as input. Sampling can be done via argmax, top-p, top-k, or full distribution sampling. Argmax selects the bin/token with the highest probability, while top-k orders outputs by probability and samples from the top $k$ \citep{Holtzman:2020}. Top-p samples from the ordered outputs whose cumulative probability mass exceeds $p\%$ \citep{Holtzman:2020}. Full distribution sampling treats the distribution as categorical and samples directly. While this makes sense intuitively, top-p and top-k sampling can often work better in practice by avoiding generation of low-probability tokens, and thus reducing noise.

We compare generated timeseries to real/simulated data using power spectral density (PSD), covariance, and Hidden Markov Model (HMM) statistics. For task-conditioned models, we assess reconstruction of task-dependent dynamics by feeding in task labels during generation and examining evoked responses. To evaluate how well models capture task activity, we apply standard decoding models (e.g., linear classification) to generated trials and compare performance to real data. We also evaluate the generalisability of decoders trained on generated data to real data. Strong similarity in these metrics would indicate accurate modelling of task responses.

By removing certain model components and evaluating performance, ablation studies assess the contribution of different architectural factors. We perform ablations on linearity, conditioning embeddings, input length, univariate/multivariate modelling, and sampling strategies.

\section{Results}

As our dataset of choice, we used the continuous 118-image data from \citet{cichy2016comparison}. For each subject, the data was bandpass filtered between 1 and 50 Hz, and a notch filter was applied to remove line noise. Subsequently, independent component analysis (ICA) artifact rejection was performed with a dimensionality of 64. Components were visually inspected for each subject, and those that exhibited clear artefactual features (e.g. eye or cardiac signals) were removed. The data was then downsampled to 100 Hz. The continuous data was split into non-overlapping validation, test, and training sets. The validation and test sets included 4 trials of each of the 118 conditions, while the training set contained the remaining 22 trials. This non-overlapping uniform splitting of the continuous data was possible due to the experimental setup during data recording.

For each model other than \texttt{FlatGPT2}, the data from each channel was tokenised independently to 256 bins using a quantisation via the mu-law algorithm discussed in Section~\ref{ssec:full_wavenet}. To achieve uniform quantisation, we first standardised each continuous-data channel, clipped values higher than 4 or lower than -4, applied per-channel maximum absolute scaling to map the data to the range (-1, 1), and finally applied the mu-law transform and 8-bit quantisation.

Our aim was to evaluate several models and methods on this dataset. Due to computational constraints and limited iteration speed over experiments and methods, all experiments in the following sections were performed on a single representative subject, except in Section \ref{ssec:group_forecasting} where we explore our models on all 15 subjects.

\begin{sloppypar}
We compared the performance of linear AR models, Wavenet-based models, and GPT2-based models. We trained univariate AR(255) models on each channel. Note that we did also assess multivariate AR models (results not shown), but this did not improve performance compared to the univariate AR. We trained \texttt{WavenetFullChannel} with a matched receptive field of 255, two stacks of dilation blocks (7 layers per block, doubling dilation factors), 256 hidden channels, 1024 skip channels, no dropout, and a 20-dimensional task embedding. \texttt{WavenetFullChannelMix} had the same architecture but 128 hidden channels and 512 skip channels. We used early stopping on the validation set. This means that we ran training until overfitting was observed, and then analysed the model version with the lowest validation loss. All our analyses were performed on the distinct test set.
\end{sloppypar}

Our Channel-independent GPT2 (\texttt{ChannelGPT2}) had a variable receptive field between 128 and 256. This means that during training the model encountered examples that had a sequence length between 128 and 256, rather than all examples having the same length. GPT2 is normally trained to output all timesteps in a sequence of length $T$, given previous timesteps. However, this means that for the second timestep, the receptive field is only 1. Ideally, we wanted to match the training setup of our Wavenet models, where the receptive field is always 256. However, this would significantly slow down training as the whole forward and backward pass must be recomputed at each timestep. We opted for a trade-off, where we set the minimum receptive field to 128, ensuring efficient training and that the model is not trained to predict shorter sequence lengths. Hyperparameters for \texttt{FlatGPT2} are given in Appendix~\ref{ssec:flatgpt_results}.

The embedding size of all inputs (token vocabulary, position, task, channel) was set to 96, and we used 12 GPT2 layers, with 12 attention heads. We used Huggingface's implementation\footnote{\url{https://github.com/huggingface}}, so the rest of the parameters were the same as in their configuration. Dropout was set to 0 and we used early stopping on the validation set.

On average both the mu-law, and the residual tokenisation achieved low reconstruction error. We tested the reconstructed data by performing evoked analysis, and classification of the task responses, and achieved comparable performance to the raw data (results not shown). Thus, both types of tokenisation add negligible quality loss to the data.

Forecasting performance in terms of token accuracy and mean-squared error is given in Appendix~\ref{ssec:flatgpt_results}. We found that these metrics are not very useful for comparing different models. What we are really interested in is how well they can generate the underlying spatiotemporal dynamics, which we investigate next.

\subsection{Generating MEG data}

For deep learning models we used top-p sampling with $p=80\%$ (unless otherwise noted in the figure caption) to recursively generate data. We generated 3600 seconds with all models. For models that have task-conditioning (all except AR(255)) we use the task label timeseries from the training set.

\begin{figure}[!t]
\centering
\begin{subfigure}{0.49\textwidth}
  \centering
  \includegraphics[width=1.0\linewidth]{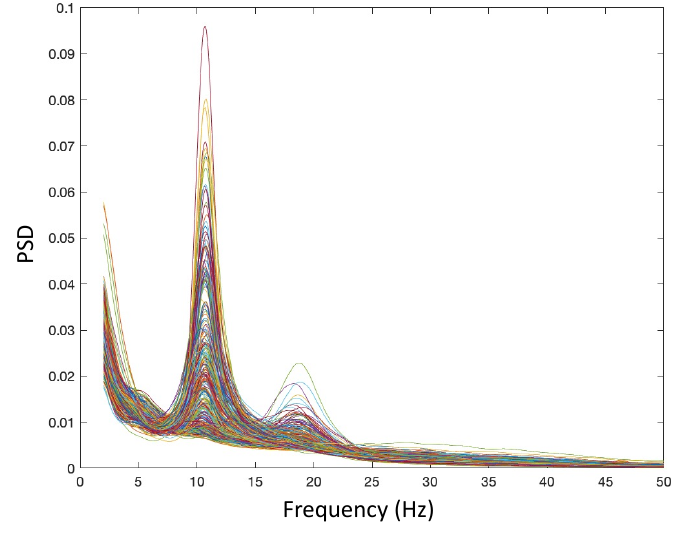}
  \caption{Data}
  \label{fig:data_psd}
\end{subfigure}%
\begin{subfigure}{0.49\textwidth}
  \centering
  \includegraphics[width=1.0\linewidth]{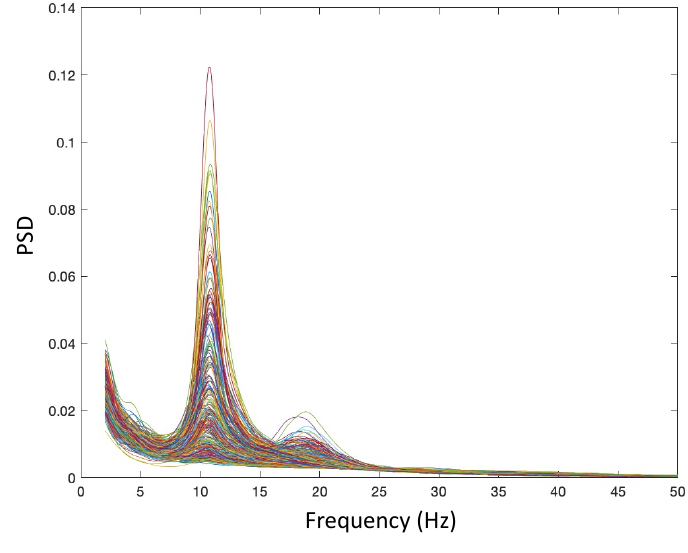}
  \caption{AR(255)}
  \label{fig:ar_psd}
\end{subfigure}
\begin{subfigure}{0.49\textwidth}
  \centering
  \includegraphics[width=1.0\linewidth]{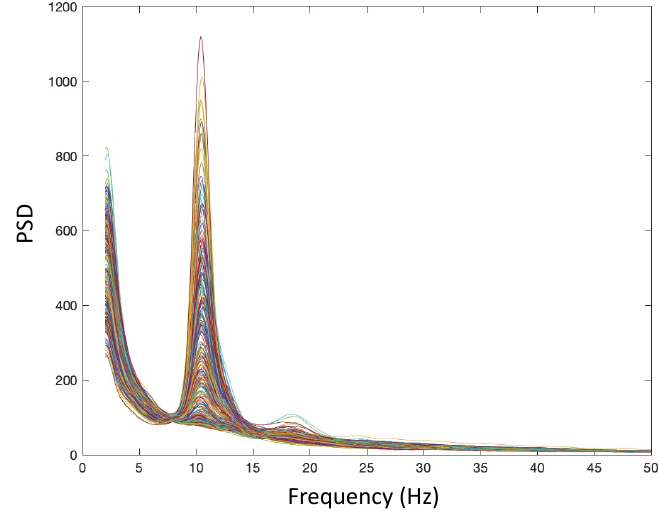}
  \caption{\texttt{WavenetFullChannel}}
  \label{fig:wavenefulltchannel_psd}
\end{subfigure}%
\begin{subfigure}{0.49\textwidth}
  \centering
  \includegraphics[width=1.0\linewidth]{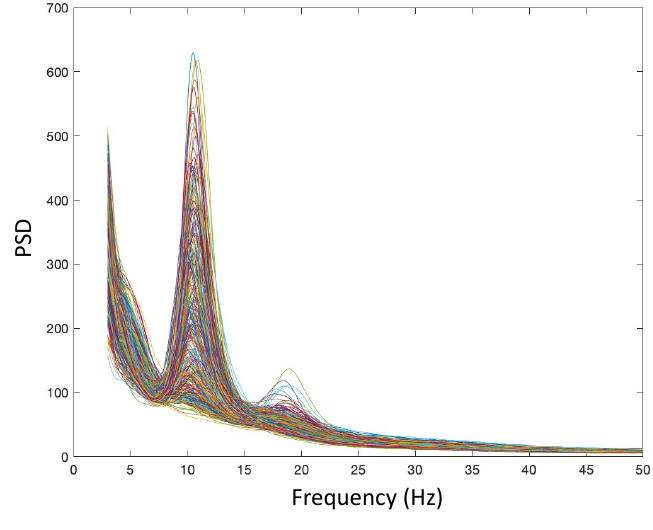}
  \caption{\texttt{ChannelGPT2}}
  \label{fig:gpt2_psd}
\end{subfigure}%
\caption{Comparison of generated data power spectral density (PSD) across channel-independent models. Each line represents a different MEG channel.}
\label{fig:psd_compare1}
\end{figure}

Generated token sequences are first de-tokenised and then the power spectral density (PSD) is computed on the continuous data. Figure~\ref{fig:psd_compare1} compares the PSD of the generated data across our channel-independent models. AR(255) clearly reproduces the MEG data PSD, and \texttt{WavenetFullChannel} and \texttt{ChannelGPT2} also do a good job with slight differences. All models capture the characteristic $1/f$ shape, and peaks at 10 and 19 Hz, likely related to alpha and beta band activity. Notably, \texttt{WavenetFullChannel} has reduced power at the 19 Hz peak which could indicate issues in capturing higher frequency dynamics.

\begin{figure}[!t]
\centering
\begin{subfigure}{0.49\textwidth}
  \centering
  \includegraphics[width=1.0\linewidth]{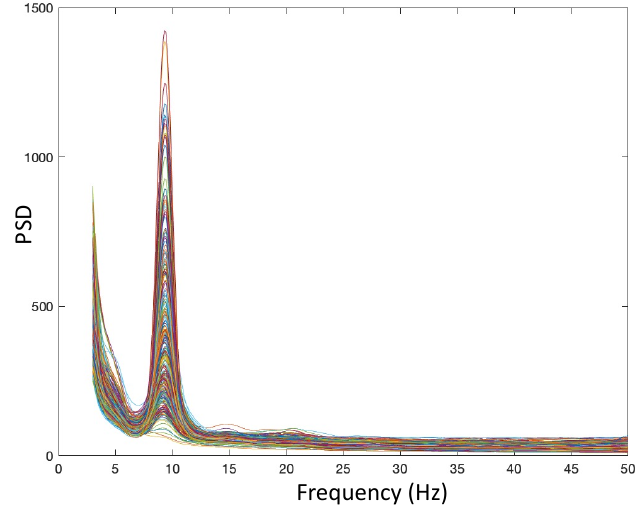}
  \caption{\texttt{WavenetFullChannelMix} p=0.72}
  \label{fig:wavenetfullchannelmix_p72_psd}
\end{subfigure}%
\begin{subfigure}{0.49\textwidth}
  \centering
  \includegraphics[width=1.0\linewidth]{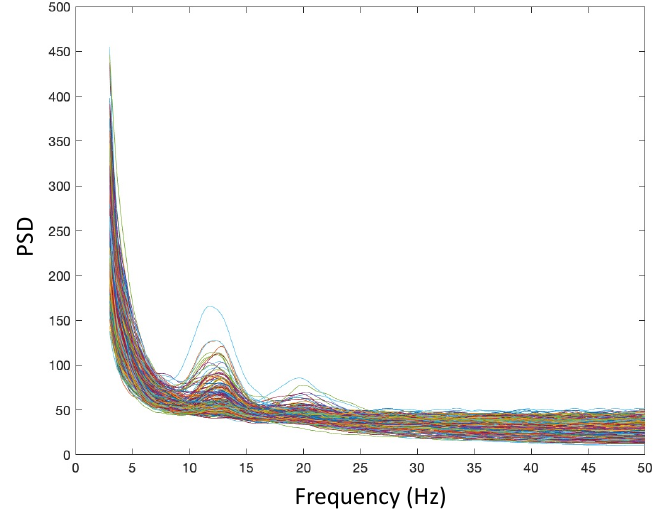}
  \caption{\texttt{WavenetFullChannelMix} p=0.80}
  \label{fig:wavenetfullchannelmix_p80_psd}
\end{subfigure}
\begin{subfigure}{0.49\textwidth}
  \centering
  \includegraphics[width=1.0\linewidth]{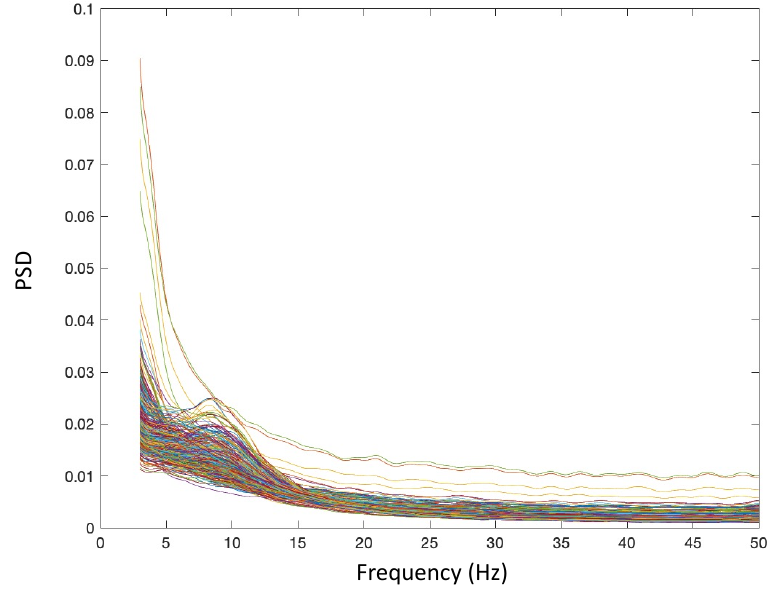}
  \caption{\texttt{FlatGPT2} p=0.8}
  \label{fig:gptflat_p90_psd}
\end{subfigure}%
\begin{subfigure}{0.49\textwidth}
  \centering
  \includegraphics[width=1.0\linewidth]{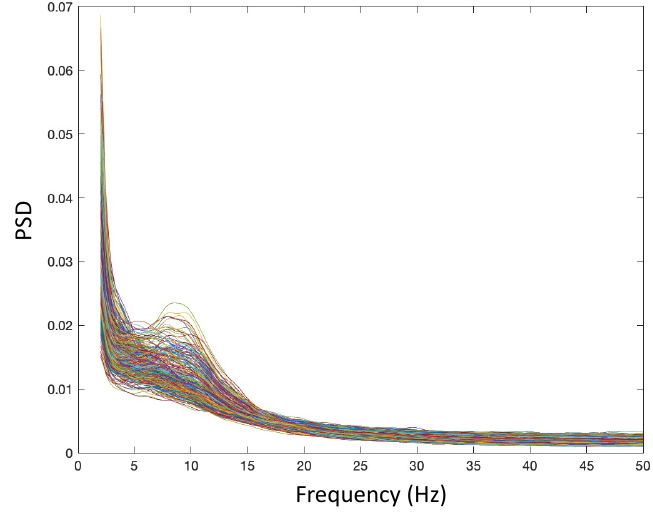}
  \caption{\texttt{FlatGPT2} p=0.95}
  \label{fig:gptflat_p95_psd}
\end{subfigure}
\caption{Comparison of generated data PSD across two channel-mixing models with varying top-p values. Each line represents a different MEG channel.}
\label{fig:psd_compare2}
\end{figure}

Looking at channel-mixing models in Figure~\ref{fig:psd_compare2}, the results are more mixed. We explored two settings of the top-p parameter, and this has a large effect on the quality of the PSD of the generated data. Even slight modifications (e.g., 0.72 vs. 0.8 for \texttt{WavenetFullChannelMix}) result in large differences in the frequency of the two peaks, and also the width of the peaks. This highlights the sensitivity of these models to sampling hyperparameters. Ultimately both top-p values provide subpar PSD's compared to channel-independent models, likely due to overfitting as channel-mixing models lack the implicit regularisation of modelling each channel separately. For \texttt{FlatGPT2} the situation is even worse as the PSD looks much noisier and the frequency of the peaks does not match the true data. As shown previously, the PSD can already be well captured by a linear (univariate) AR model. The added complexity of channel-mixing may introduce suboptimal loss minima where the PSD is not captured as well. Thus it is important to compare models on alternative measures where AR(255) might perform worse given its simplicity.

\begin{figure}[!t]
\centering
\begin{subfigure}{0.49\textwidth}
  \centering
  \includegraphics[width=0.9\linewidth]{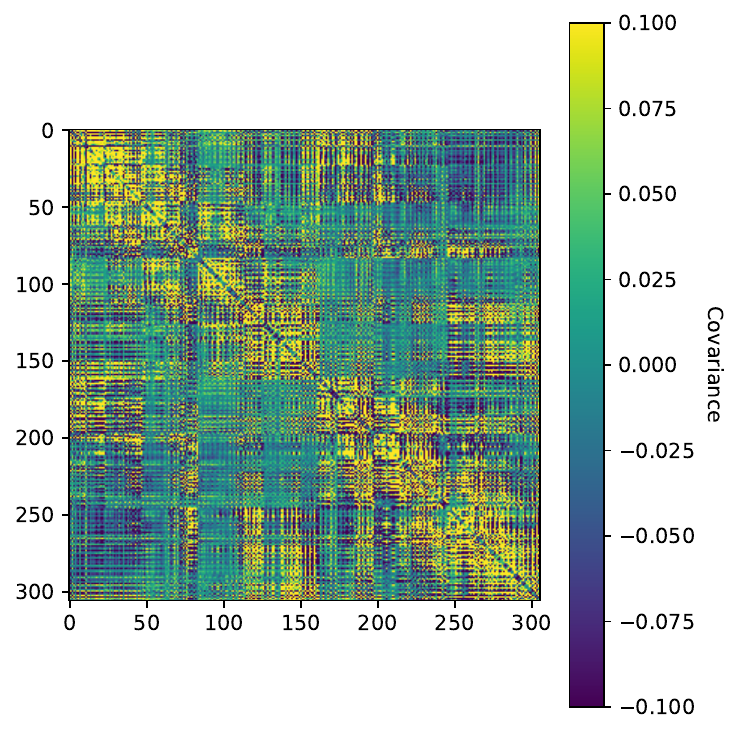}
  \caption{Data}
  \label{fig:data_cov}
\end{subfigure}%
\begin{subfigure}{0.49\textwidth}
  \centering
  \includegraphics[width=0.7\linewidth]{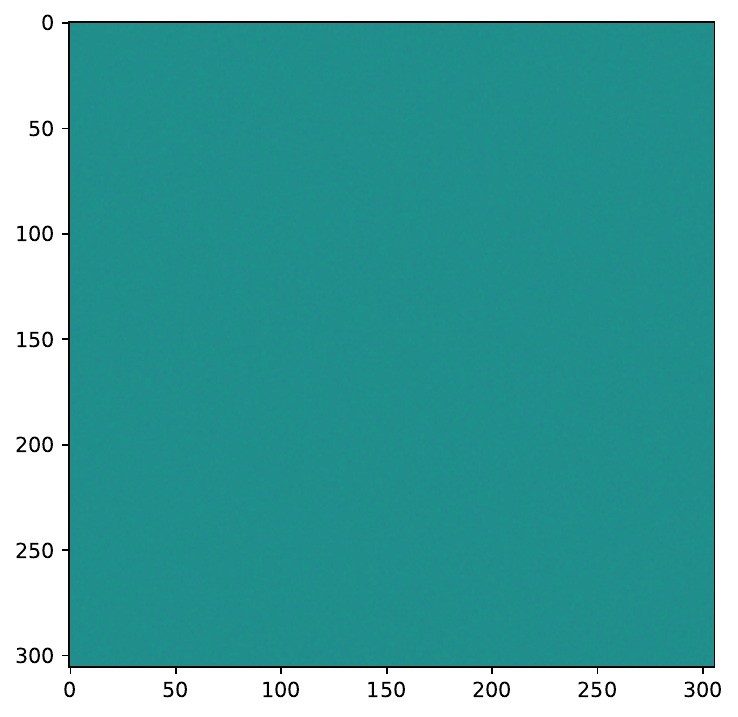}
  \caption{AR(255)}
  \label{fig:ar_cov}
\end{subfigure}
\begin{subfigure}{0.49\textwidth}
  \centering
  \includegraphics[width=0.7\linewidth]{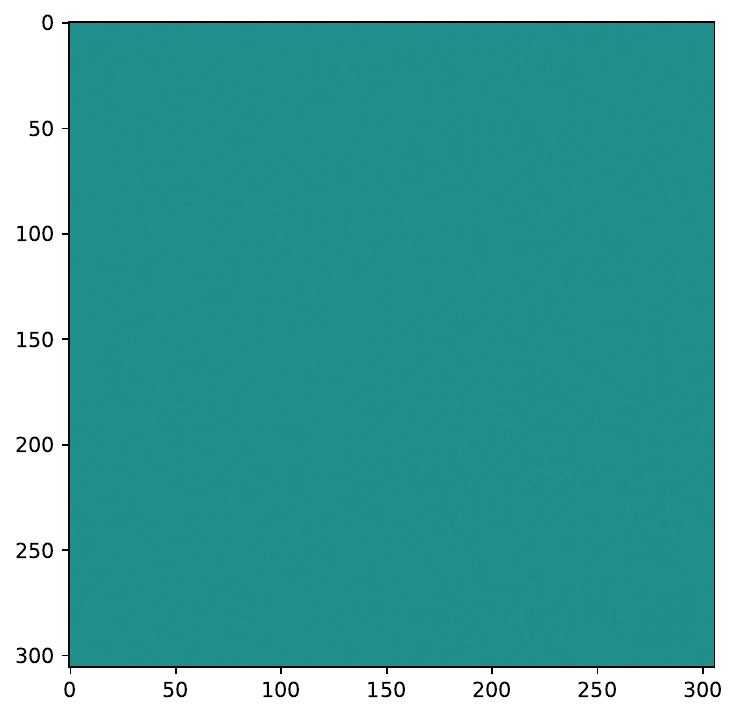}
  \caption{\texttt{WavenetFullChannel}}
  \label{fig:wavenetfullchannel_cov}
\end{subfigure}%
\begin{subfigure}{0.49\textwidth}
  \centering
  \includegraphics[width=0.7\linewidth]{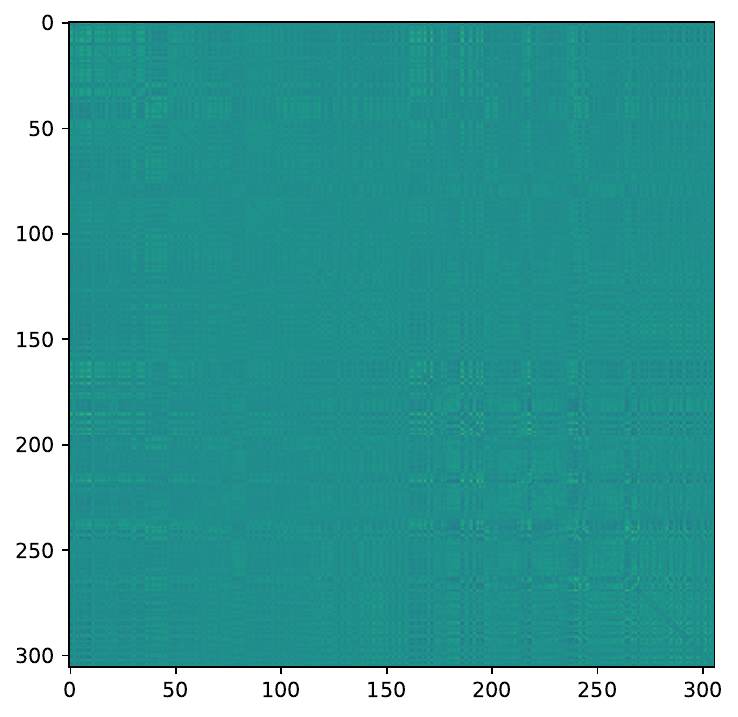}
  \caption{\texttt{ChannelGPT2}}
  \label{fig:gpt2_cov}
\end{subfigure}
\begin{subfigure}{0.49\textwidth}
  \centering
  \includegraphics[width=0.7\linewidth]{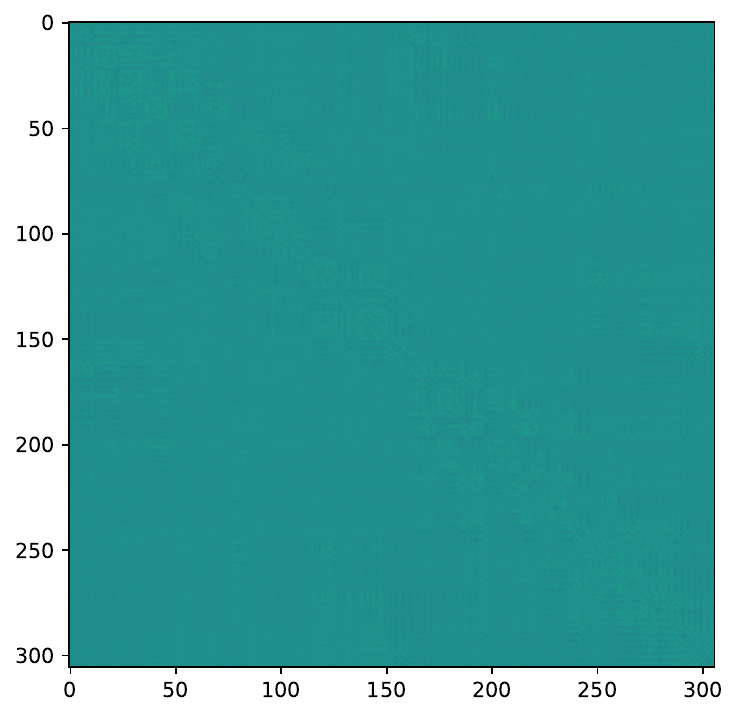}
  \caption{\texttt{WavenetFullChannelMix}}
  \label{fig:wavenetfullchannelmix_cov}
\end{subfigure}%
\begin{subfigure}{0.49\textwidth}
  \centering
  \includegraphics[width=0.7\linewidth]{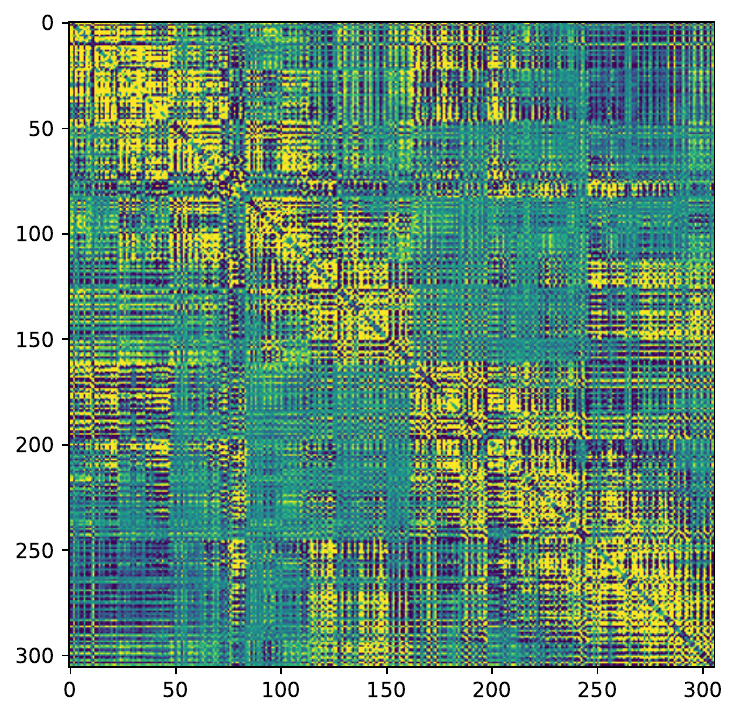}
  \caption{\texttt{FlatGPT2}}
  \label{fig:gpt2flat_cov}
\end{subfigure}
\caption{Covariance of generated data between channels (vertical and horizontal axes). All plots have the same scaling as (a).}
\label{fig:generated_cov}
\end{figure}

As the PSD is a channel-independent measure, we next looked at generated data covariance which captures the interactions between different channels (Figure~\ref{fig:generated_cov}). This reveals that the only model capable of closely matching the data covariance is \texttt{FlatGPT2}. All other models produce data with covariances much closer to 0. This is perhaps expected for channel-independent models which generate data independently for each channel, but somewhat surprising for \texttt{WavenetFullChannelMix}. Even though \texttt{FlatGPT2} may not produce accurate spectral data, by having information about other channels in the input it does an excellent job at capturing covariance. This highlights the trade-offs between different model architectures.

\subsection{HMM statistics of generated data}

\begin{figure}[!ht]
    \centering
    \includegraphics[width=0.85\textwidth]{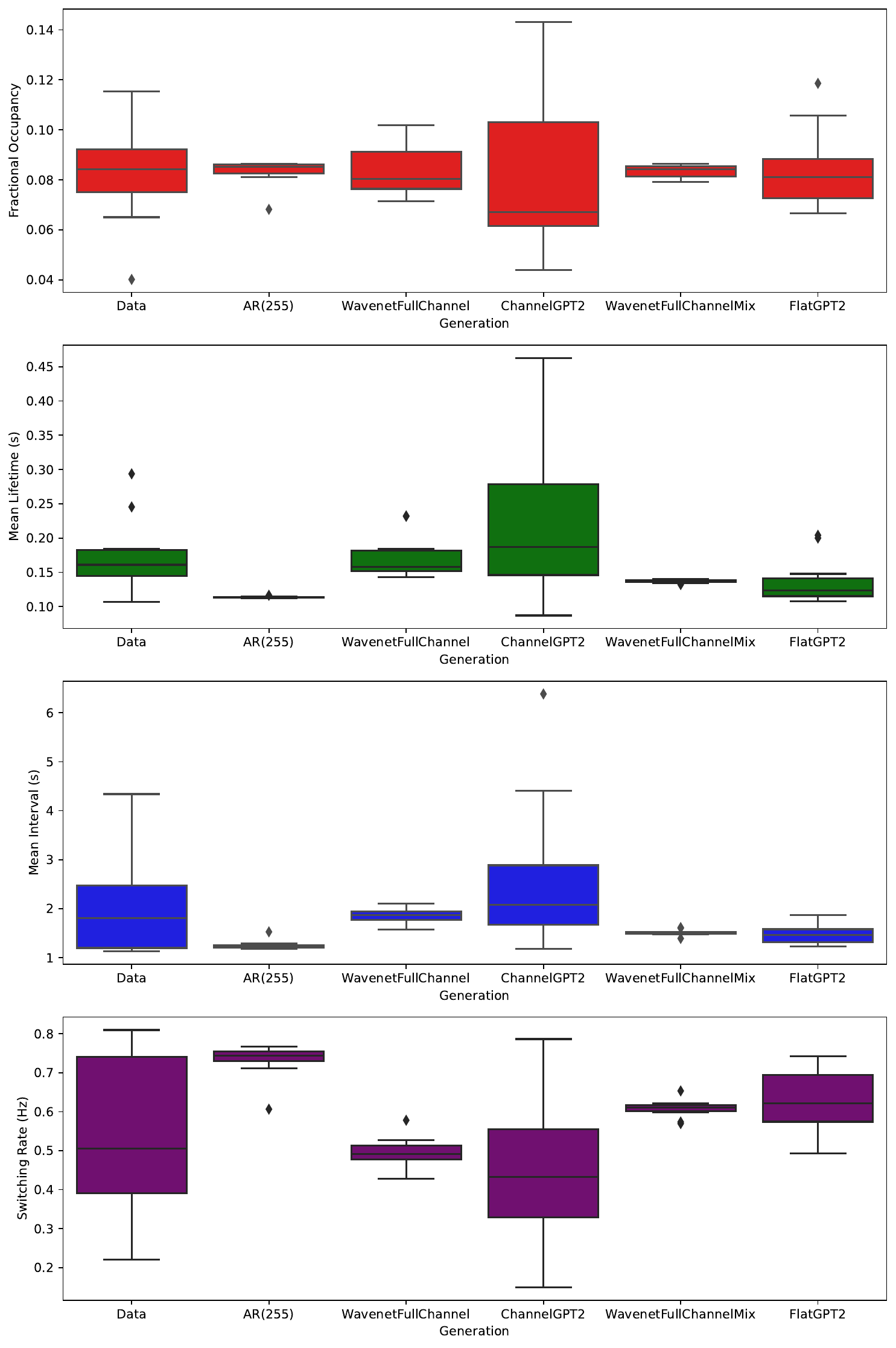}
    \caption{Distribution across states of 4 HMM statistics (rows) for each model and data (columns). The distribution mean and variance is computed across the individual values of the 12 states. Lower variance means states with more homogeneous statistics.}
    \label{fig:hmm_stats_all}
\end{figure}

Next, we looked at how well the generated data matches real data in terms of HMM statistics. HMMs are useful for unsupervised discovery of discrete states underlying timeseries data \citep{rabiner1989tutorial, vidaurre2018discovering}. We fit a separate 12-state time-domain embedding HMM (TDE-HMM) to each multivariate generated timeseries \citep{vidaurre2018spontaneous}. We used the osl-dynamics package \citep{gohil2023osl}, and set the number of embeddings to 15, the PCA projection dimensionality of the channels to 80 and the sequence length to 2000. We trained the HMMs for 20 epochs with an initial learning rate of 0.02, and extract four different summary statistics from the inferred state timecourse. The distributions of these summary statistics over the 12 states across models are shown in Figure~\ref{fig:hmm_stats_all}. Note that since a separate HMM is trained for each model, the states are not matched between models. Thus, we look at the distribution over states, rather than individual states.

\begin{figure}[!ht]
    \centering
    \includegraphics[width=0.88\textwidth]{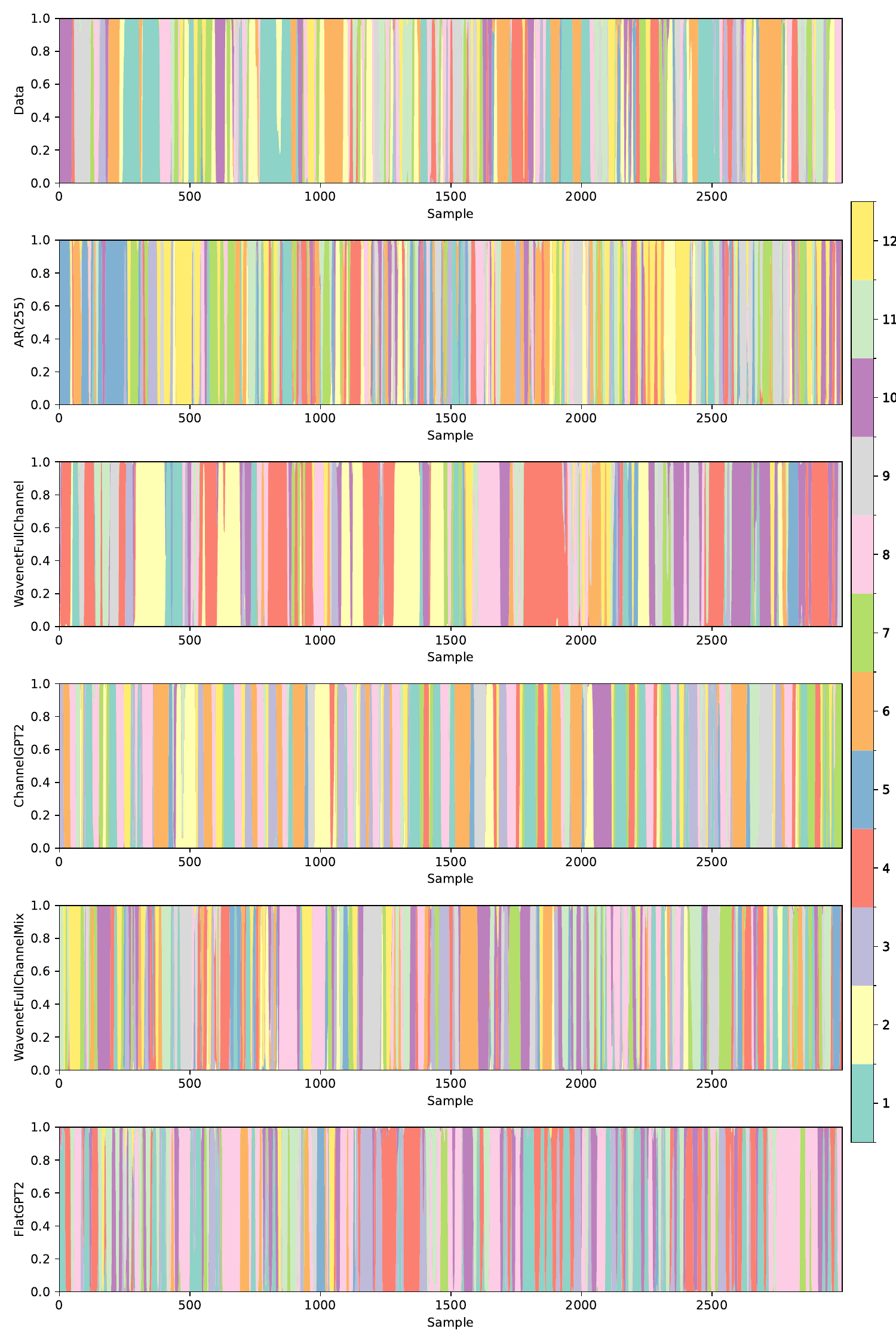}
    \caption{Example state timecourses from the HMMs trained on each model's generated data (rows). Each state is represented by a different colour. Note that state indices and timecourses are not matched across models.}
    \label{fig:alps}
\end{figure}

Across the four summary statistics we can see that the real data has high variance in the distribution over states. AR(255) and \texttt{WavenetFullChannelMix} fail to produce data with variable state statistics, and even the mean over states is not captured well. \texttt{WavenetFullChannel} does a great job at capturing the mean of the state distributions, but still produces data with relatively invariant states. \texttt{ChannelGPT2} seems to best capture the distributions across all four statistics, especially for the mean interval and switching rate. This shows that Transformer-based models can generate data that better matches the HMM-inferred dynamics of real MEG data. Example state timecourses generated from all models are plotted in Figure~\ref{fig:alps} to qualitatively illustrate the differences in the generated dynamics.

\begin{figure}[!t]
  \centering
  \begin{subfigure}{0.33\textwidth}
    \centering
    \includegraphics[width=1.0\linewidth]{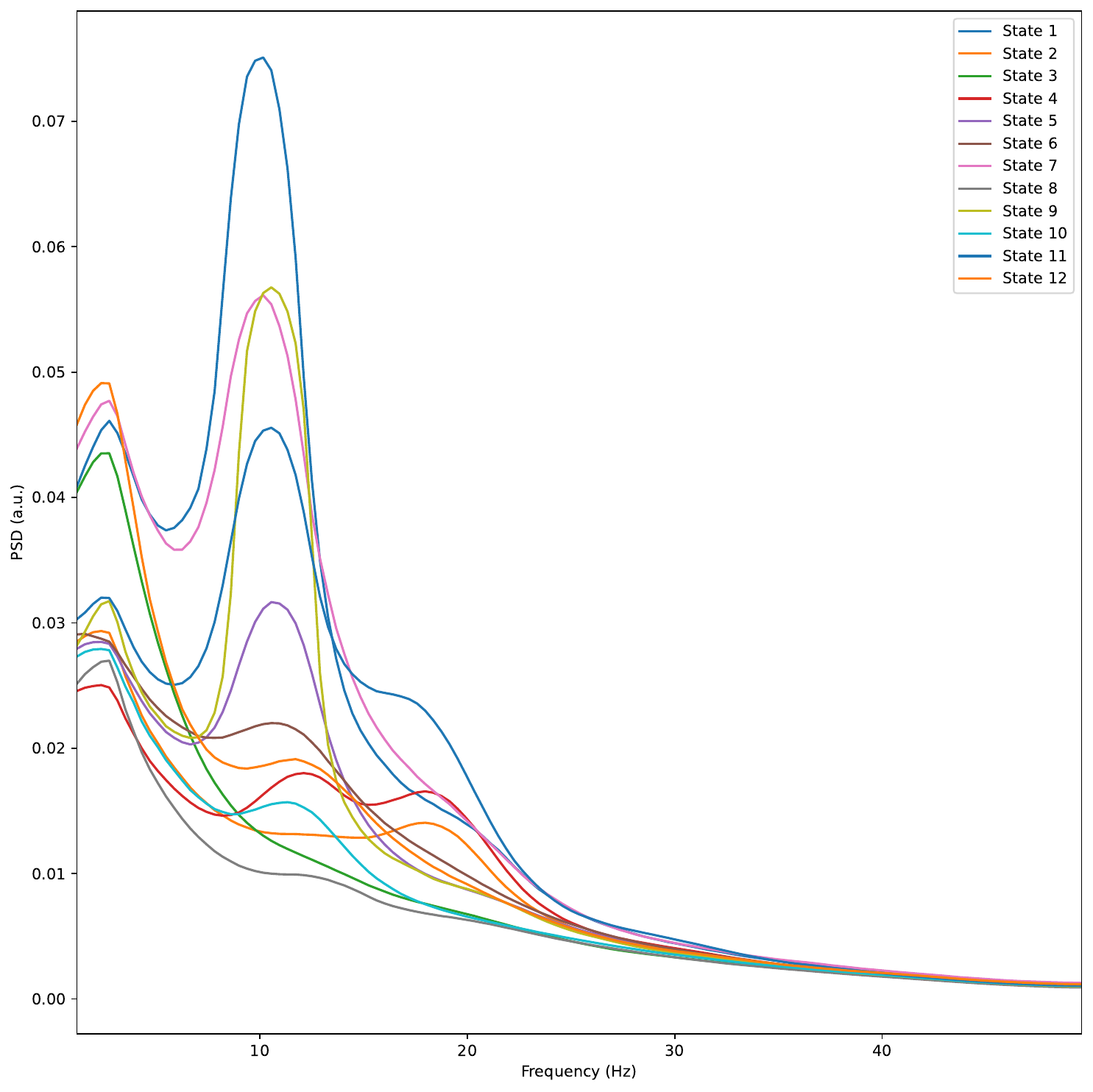}
    \caption{Data}
    \label{fig:data_hmm_psd}
  \end{subfigure}%
  \begin{subfigure}{0.33\textwidth}
    \centering
    \includegraphics[width=1.0\linewidth]{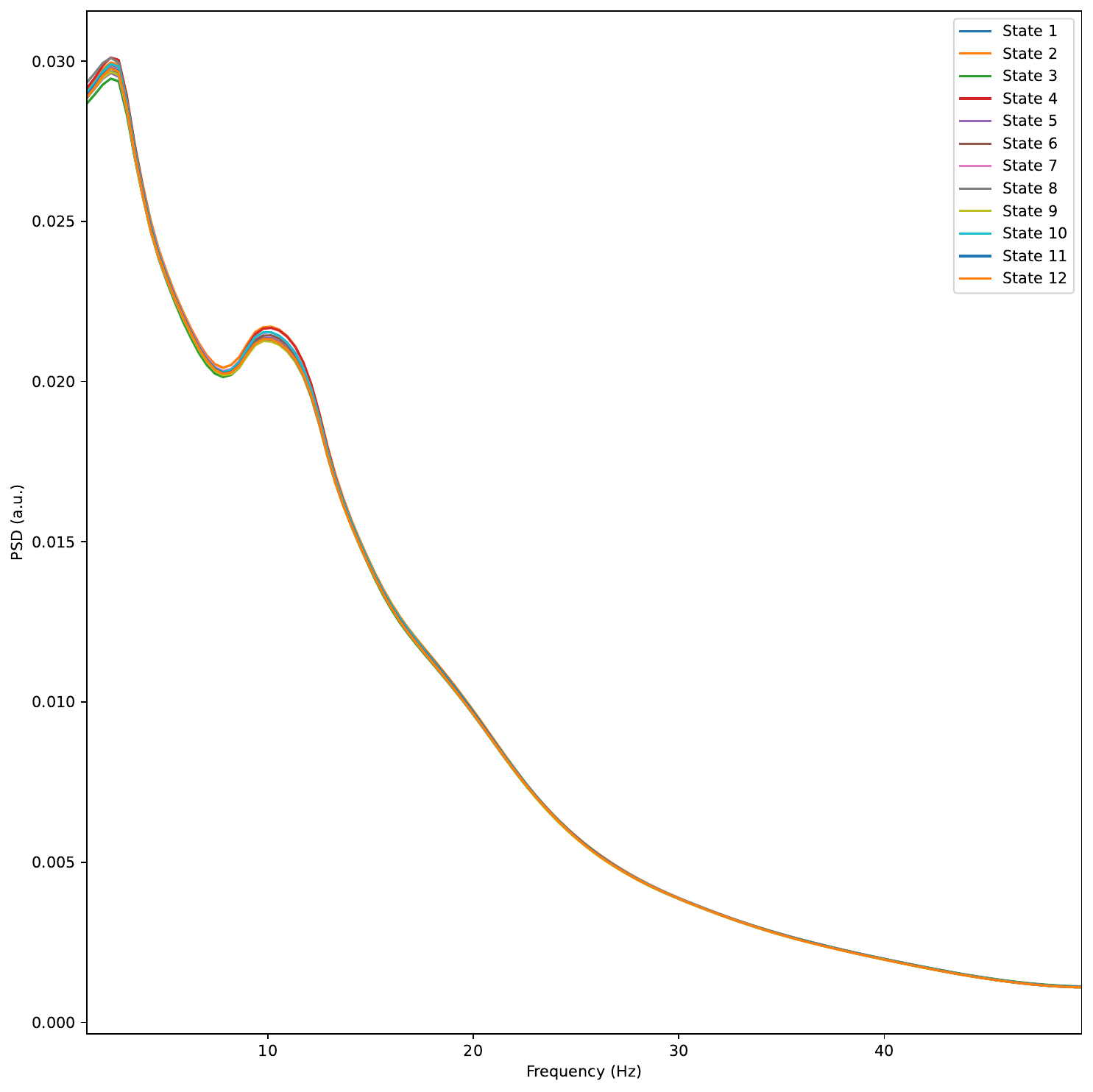}
    \caption{AR(255)}
    \label{fig:ar_hmm_psd}
  \end{subfigure}%
  \begin{subfigure}{0.33\textwidth}
    \centering
    \includegraphics[width=1.0\linewidth]{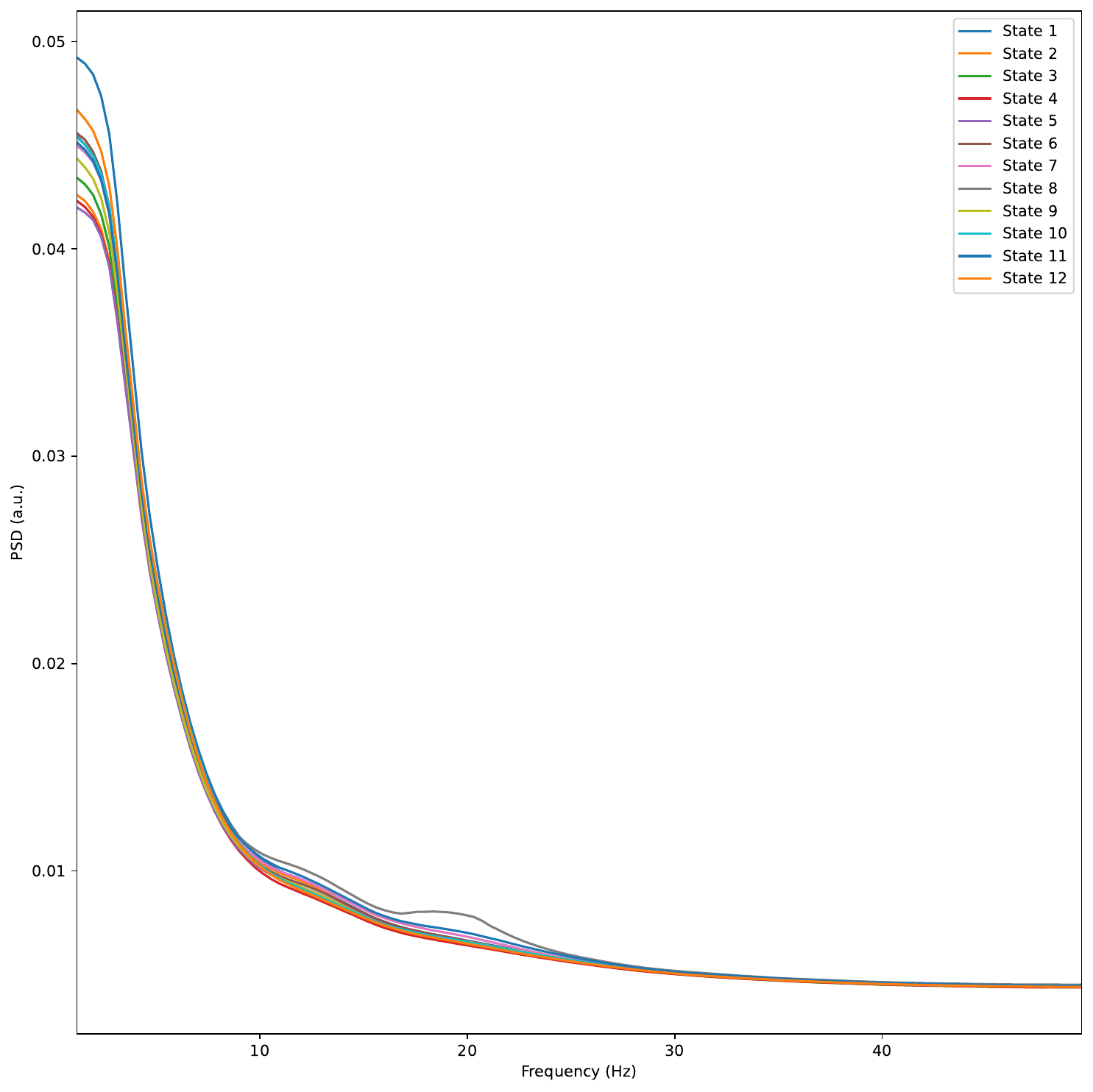}
    \caption{\texttt{WFCM}}
    \label{fig:wavenet_hmm_psd}
  \end{subfigure}
  \begin{subfigure}{0.5\textwidth}
    \centering
    \includegraphics[width=0.66\linewidth]{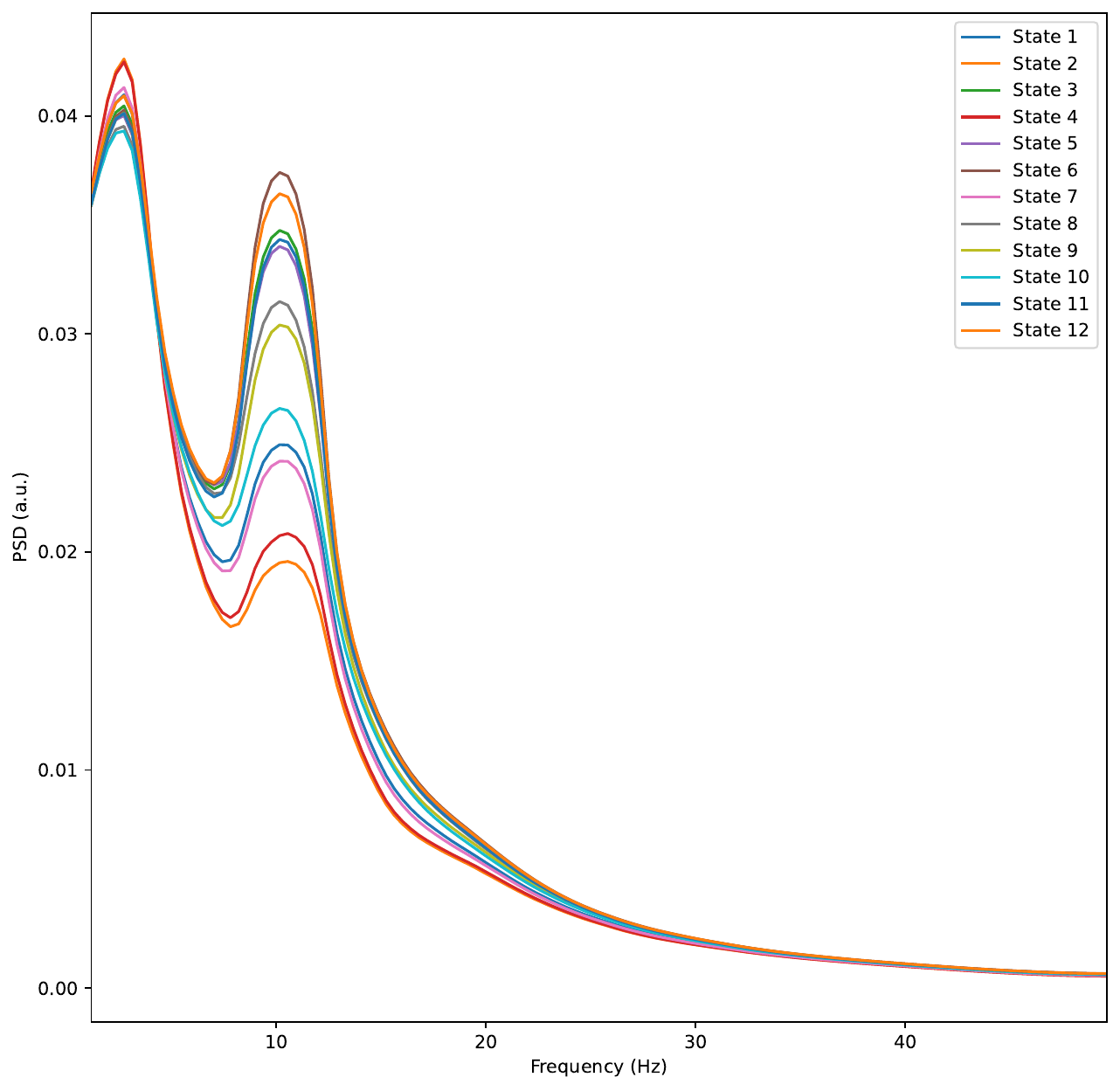}
    \caption{\texttt{WavenetFullChannel}}
    \label{fig:wavenetfullchannel_hmm_psd}
  \end{subfigure}%
   \begin{subfigure}{0.5\textwidth}
    \centering
    \includegraphics[width=0.66\linewidth]{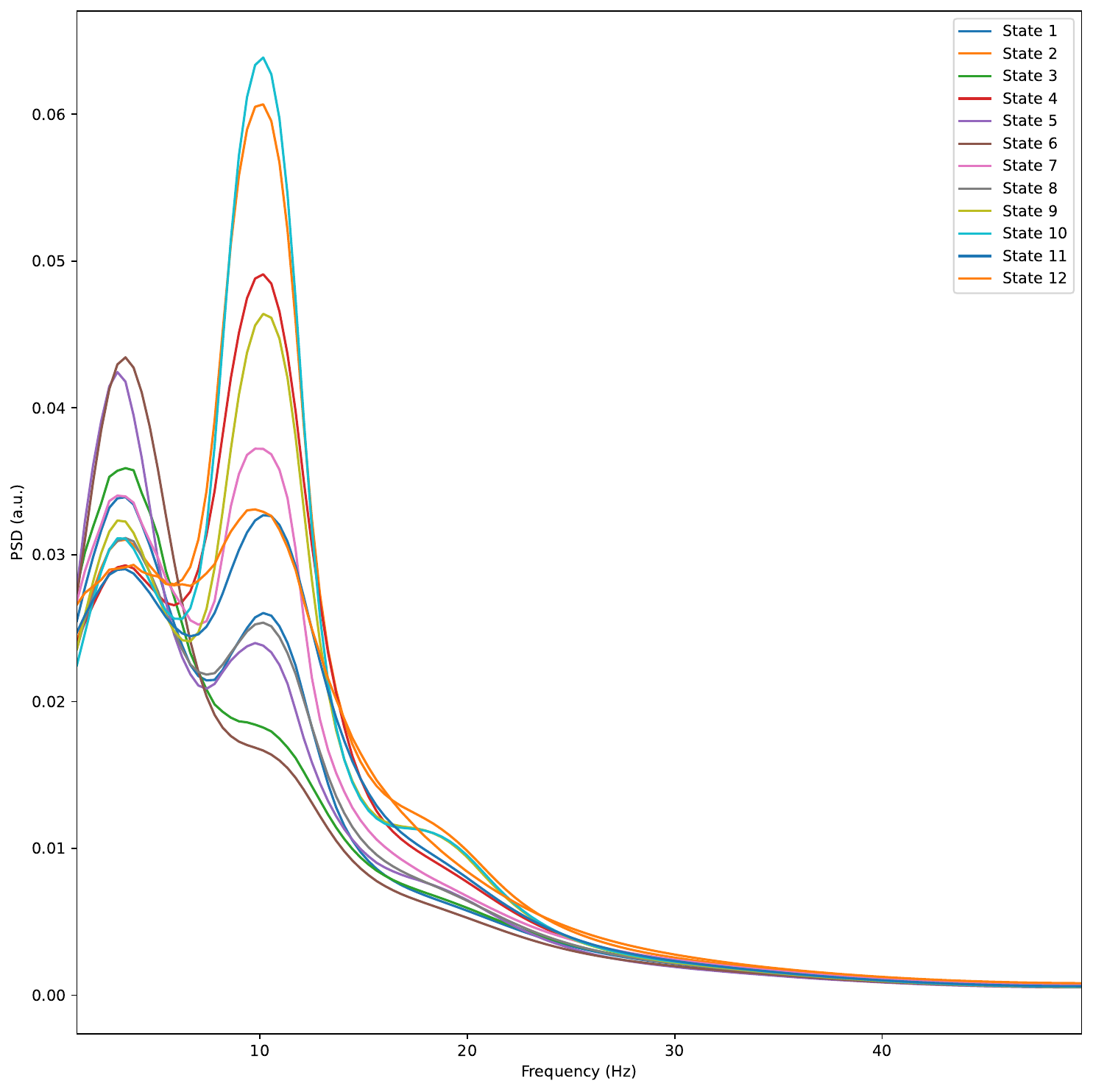}
    \caption{\texttt{ChannelGPT2}}
    \label{fig:gpt_hmm_psd}
  \end{subfigure}
  \caption{Power spectral density of HMM states inferred on the generated data of each model. \texttt{WFCM} refers to \texttt{WavenetFullChannelMix}. Each line is the PSD of a different state. Note that states are not matched across models. Horizontal axis represents frequency in Hz. \texttt{FlatGPT2} is omitted due to failing to generate data with PSD matching real data.}
  \label{fig:hmm_psd}
  \end{figure}

In addition to state statistics, we can also compute the power spectra of each state across the timeseries. In MEG data different states might capture oscillatory activity with specific frequencies. We plot the extracted power spectra from the inferred state time courses in Figure~\ref{fig:hmm_psd}. We can see that the HMM trained on the MEG data contains many states that capture the 10 Hz peak, with fewer states having a 20 Hz peak. It is also clear that the states of the HMM fitted to the \texttt{WavenetFullChannelMix} generated timeseries do not contain these spectral peaks. While the AR(255) does contain states with a 10 Hz peak, the shape does not match the data well, and also states do not show the same variability as in real data.

In contrast \texttt{ChannelGPT2}, matches the state PSDs of the real data very well, further demonstrating the superiority of Transformer models in capturing complex neural dynamics. While \texttt{WavenetFullChannel} also improves substantially over the AR(255) power spectra, it falls short in capturing the 20 Hz peak and the heterogeneity between states observed in the real data and the generated data of \texttt{ChannelGPT2}. This and previous analyses show that the combination of channel-independence and a Transformer-based architecture are critical for matching the dynamics of real data.

  %Not required: The power of each state can also be averaged over frequencies and plotted as a spatial map over the MEG channels. This reveals what the kind of distinct spatial patterns that HMMs states are capturing. We plot the power maps for each state for the previously discussed models in Figure~\ref{fig:hmm_powermap}. %these should be plotted separately for magnetometers and gradiometers

\subsection{Evoked analysis of generated data}
% HMM and evoked from data
% evoked response correlation topomap

The analyses in the previous section considered metrics for assessing the quality of an arbitrary generated timeseries, applicable to any M/EEG dataset. We can also leverage the experimental aspect of the \citet{cichy2016comparison} data and provide further focused insights on the task-related brain activity. As mentioned before, we used the task label timeseries from the training data when generating data with our models. If the models properly incorporate this conditioning, the generated data should reflect aligned task-related activity similar to real data.

By simple epoching of the generated timeseries based on the known task labels, we can compute evoked responses generated by our models. We do this for all models except AR(255) as it did not include task labels in its model. To compare the shape of average evoked responses, we average over all epochs in both real data and the generated timeseries. This results in data of shape $\bar{\mathbf{X}} \in \mathbb{R}^{C \times T}$ where $C=306$ is the number of channels and $T=1000$ ms is the trial/epoch length.

\begin{figure}[!t]
  \centering
  \begin{subfigure}{0.95\textwidth}
    \centering
    \includegraphics[width=1.0\linewidth]{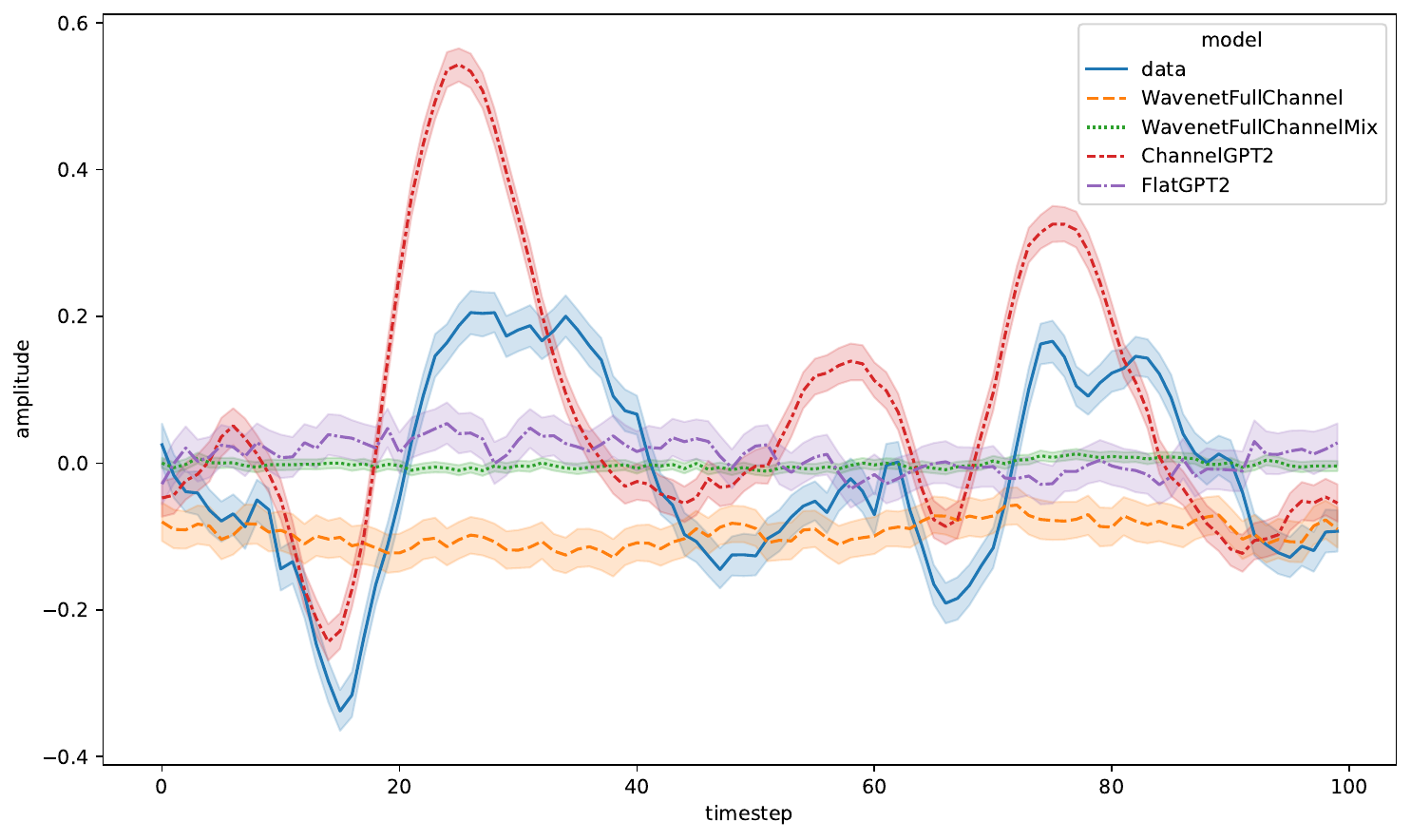}
    \caption{Frontal channel}
    \label{fig:evoked_frontal_comparison}
  \end{subfigure}
  \begin{subfigure}{0.95\textwidth}
    \centering
    \includegraphics[width=1.0\linewidth]{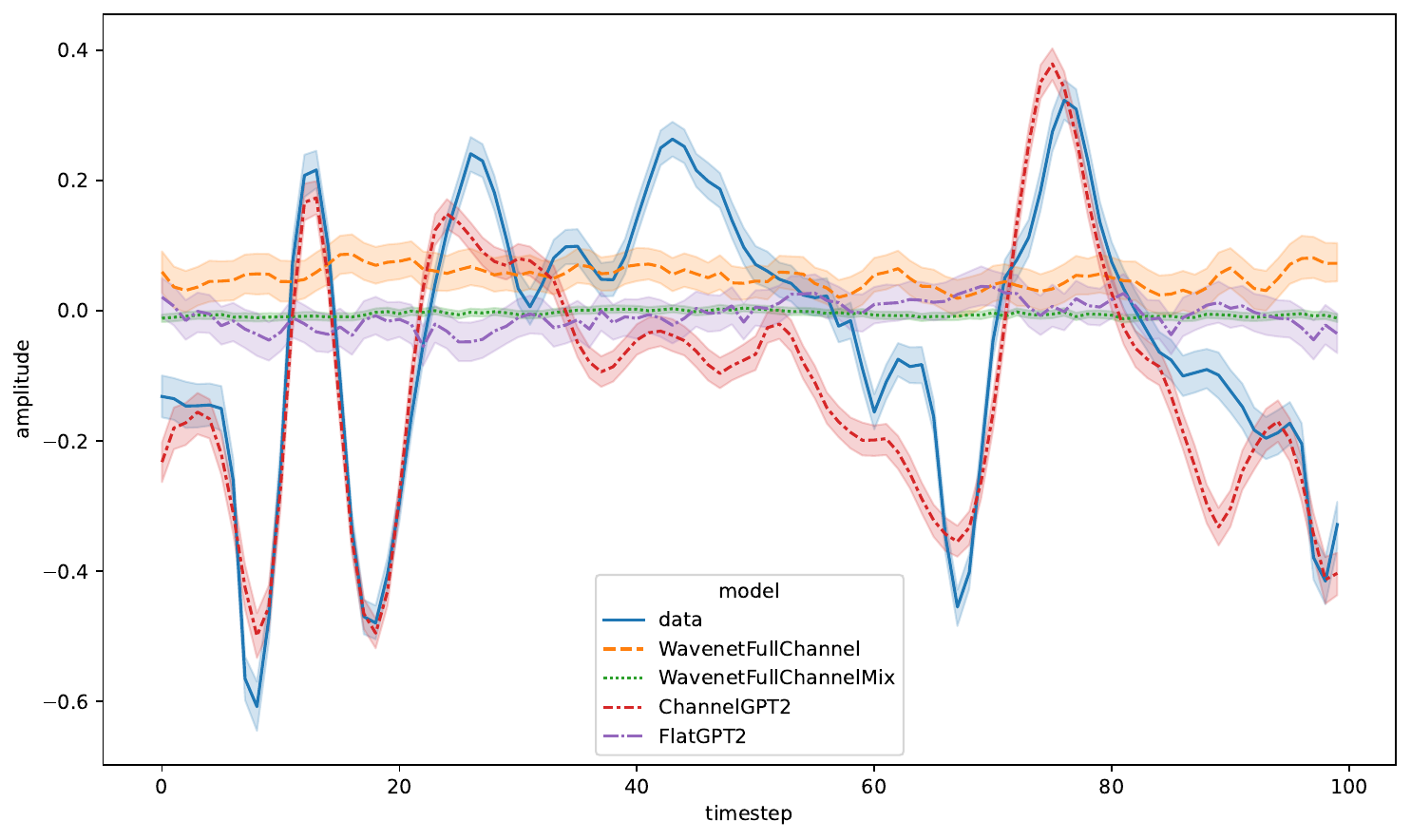}
    \caption{Visual channel}
    \label{fig:evoked_visual_comparison}
  \end{subfigure}
  \caption{Comparison of evoked timecourses of 2 channels across our task-conditioned models. The whole x-axis encompasses 1 second. Timestep 0 is when stimulus presentation starts, and timestep 50 (500 ms) is when it stops. The peak occurring after 50 timesteps indicates a visual response to the stopping of the stimulus (removal of the image). Shading indicates variability across trials.}
  \label{fig:evoked_comparison}
  \end{figure}

We visualise evoked responses across our models and the real data in a frontal and a visual channel in Figure~\ref{fig:evoked_comparison}. While both Wavenet models and \texttt{FlatGPT2} completely fail to capture the evoked timecourse, \texttt{ChannelGPT2} does a remarkably good job, especially in the visual channel. This is not surprising as the dataset is collected from a visual experiment, so most activity is visual. \texttt{ChannelGPT2} closely matches both the amplitude and the timing of the evoked response peaks across the whole 1-second epoch. Variability across trials is also well matched.

To quantify the similarity between real and generated evoked activity, we compute the correlation of the mean and variance (across individual epochs) of the evoked response for each channel separately. We plot the correlation values between the data and each model as a sensor space map, allowing insights into the spatial pattern of similarity. For the plots we average over magnetometers and gradiometers at the same location. 

\begin{figure}[!t]
  \centering
  \begin{subfigure}{0.24\textwidth}
    \centering
    \includegraphics[width=1.0\linewidth]{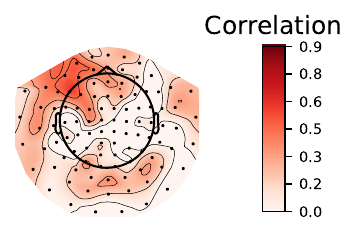}
    \caption{\texttt{WFC}}
    \label{fig:WavenetFullChannel_meancorr}
  \end{subfigure}%
  \begin{subfigure}{0.24\textwidth}
    \centering
    \includegraphics[width=1.0\linewidth]{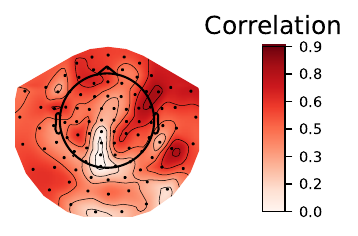}
    \caption{\texttt{WFCM}}
    \label{fig:WavenetFullChannelMix_meancorr}
  \end{subfigure}%
  \begin{subfigure}{0.24\textwidth}
    \centering
    \includegraphics[width=1.0\linewidth]{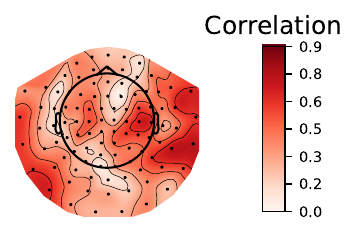}
    \caption{\texttt{FlatGPT2}}
    \label{fig:FlatGPT2_meancorr}
  \end{subfigure}%
  \begin{subfigure}{0.24\textwidth}
    \centering
    \includegraphics[width=1.0\linewidth]{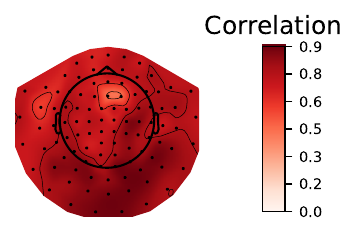}
    \caption{\texttt{ChannelGPT2}}
    \label{fig:ChannelGPT2_meancorr}
  \end{subfigure}%
  \caption{Correlation between the timecourses of the mean (over individual epochs) evoked responses from the real data and mean evoked responses generated by each model. The correlation values are visualised across sensors. \texttt{WFC} refers to \texttt{WavenetFullChannel} and \texttt{WFCM} refers to \texttt{WavenetFullChannelMix}. Darker reds indicate higher correlation.}
  \label{fig:mean_evoked_corrs}
  \end{figure}

Figure~\ref{fig:mean_evoked_corrs} shows the correlation between the timecourses of the mean (across trials) evoked responses obtained from the actual data and the mean evoked responses obtained from data generated by each model. By computing the correlation between the timecourses of each channel we can plot these correlation values as a sensor space map. As expected, \texttt{ChannelGPT2} generates data with evoked responses that have much higher correlation with evoked responses from real data, and slightly higher correlation in visual areas compared to other channels, matching the known topography of visual evoked responses. In other models the correlation is low, and spatially better in frontal areas, likely because the evoked responses here are noisier providing an easier fit.

\begin{figure}[!t]
  \centering
  \begin{subfigure}{0.24\textwidth}
    \centering
    \includegraphics[width=1.0\linewidth]{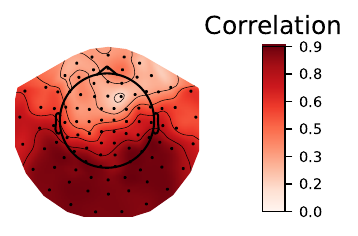}
    \caption{\texttt{WFC}}
    \label{fig:WavenetFullChannel_varcorr}
  \end{subfigure}%
  \begin{subfigure}{0.24\textwidth}
    \centering
    \includegraphics[width=1.0\linewidth]{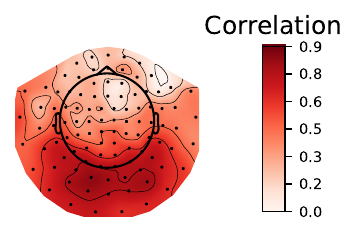}
    \caption{\texttt{WFCM}}
    \label{fig:WavenetFullChannelMix_varcorr}
  \end{subfigure}%
  \begin{subfigure}{0.24\textwidth}
    \centering
    \includegraphics[width=1.0\linewidth]{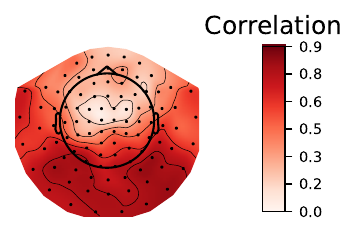}
    \caption{\texttt{FlatGPT2}}
    \label{fig:FlatGPT2_varcorr}
  \end{subfigure}%
  \begin{subfigure}{0.24\textwidth}
    \centering
    \includegraphics[width=1.0\linewidth]{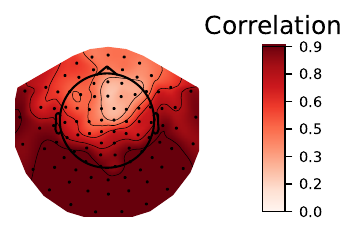}
    \caption{\texttt{ChannelGPT2}}
    \label{fig:ChannelGPT2_varcorr}
  \end{subfigure}%
  \caption{Correlation between the timecourses of the variance (over individual epochs) of the mean evoked responses from the real data and the variance of the mean evoked responses generated by each model. The correlation values are visualised across sensors. \texttt{WFC} refers to \texttt{WavenetFullChannel} and \texttt{WFCM} refers to \texttt{WavenetFullChannelMix}. Darker reds indicate higher correlation.}
  \label{fig:var_evoked_corrs}
  \end{figure}

\begin{sloppypar}
Figure~\ref{fig:var_evoked_corrs} shows the correlation between the variance (over individual epochs) timecourses of the mean evoked response obtained from the actual data and the evoked responses obtained from data generated by each model. Again, \texttt{ChannelGPT2} generates data that has the highest correlations with the real data, with higher values in channels in the back of the head, appropriately capturing the topography of response variability. Other models have similar spatial distribution, and notably \texttt{WavenetFullChannel} also produces evoked responses with variance partially matching the real data.
\end{sloppypar}

\begin{figure}[!t]
    \centering
    \includegraphics[width=0.85\textwidth]{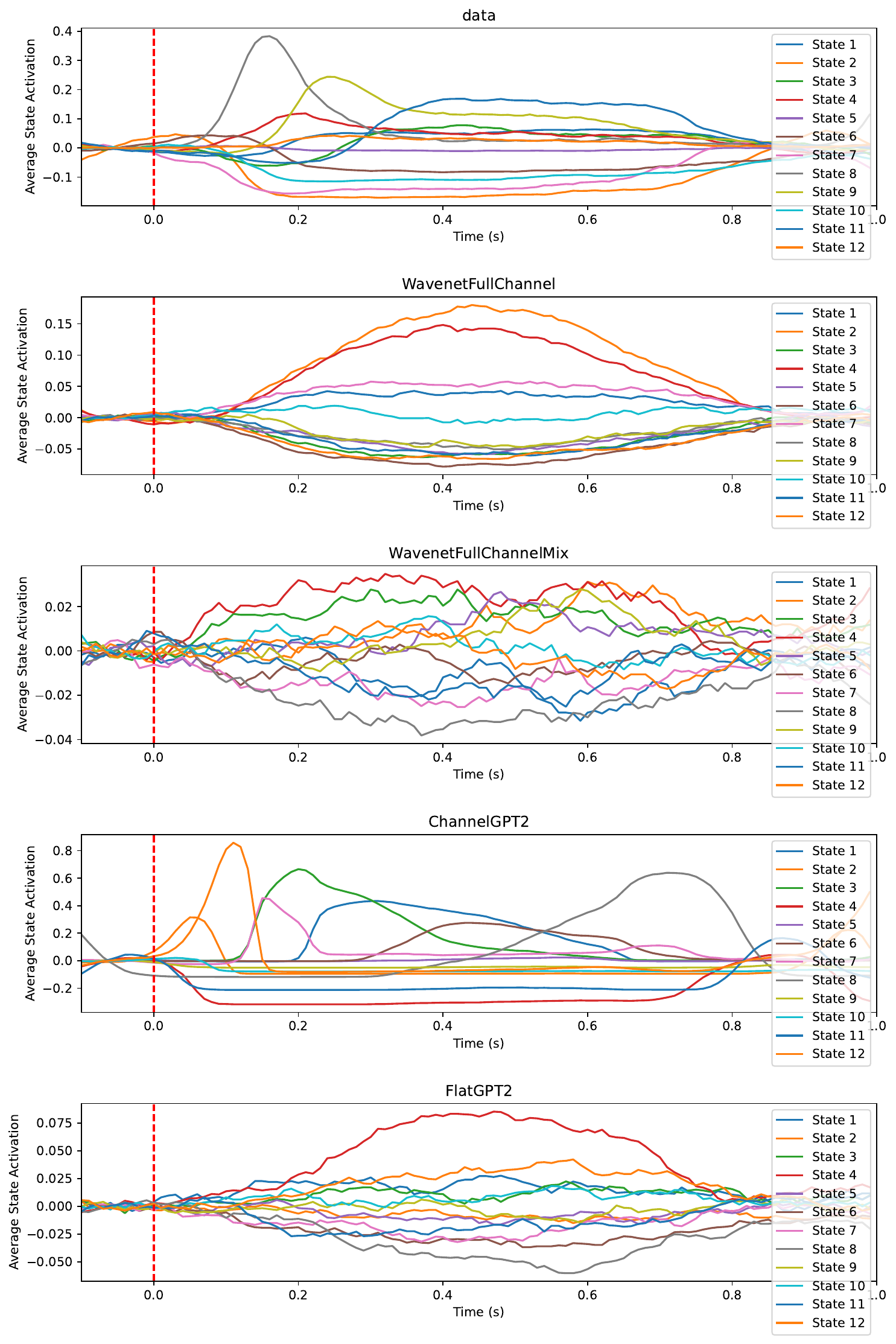}
    \caption{Evoked response state timecourses of HMMs trained on the MEG data and generated data from our task-conditioned models. Note that states are not matched between models. Image presentation starts at 0 seconds and ends at 0.5 seconds.}
    \label{fig:hmm_evokeds}
\end{figure}

Finally, a different way to assess task-related activity is to examine the evoked state timecourses from the HMMs fitted on the real and generated timeseries. Rather than looking at individual channels, this provides an overall view of which state gets activated when during individual trials. This is computed by simply epoching the state timecourse, and averaging over all trials. We plot these for the real data and each generated timeseries in Figure~\ref{fig:hmm_evokeds}. As expected, the HMM trained on models other than \texttt{ChannelGPT2} shows poor evoked state timecourses. \texttt{ChannelGPT2} generated data produces states with similar evoked dynamics and variability as the real data.

In summary, by leveraging the experimental nature of the MEG dataset, we evaluated how well different models generated task-evoked responses and dynamics. Across standard evoked analysis and discovered brain states, the Transformer-based \texttt{ChannelGPT2} model produced accurate task-related activity closely matching real MEG recordings. This further demonstrates its ability to generate physiologically grounded and experimentally relevant MEG timeseries. While we have not tested it directly for encoding by comparing individual trials with real data, our generation results show promise for encoding applications.

\subsection{Group modelling}
\label{ssec:group_forecasting}
% evoked responses from generated data comparison with gpt-subject
% HMM transfer of evoked state timecourses and comparison with gpt2-subject + trial variability
% put channel embedding results here (briefly)
% very briefly mention that flatgpt2 was scaling well to more data and psd improved, but evoked stuff didn't

Up to this point, all trainings and analyses were done on MEG data from a single subject. We now look at whether adding more data improves modelling and generation. This is in line with the overall goal of training such foundational forecasting models on multiple large datasets. Here we take a first step in exploring this by scaling \texttt{ChanelGPT2} and \texttt{FlatGPT2} to the 15 subjects in the \citet{cichy2016comparison} data, and calling these  \texttt{ChanelGPT2-group} and \texttt{FlatGPT2-group}, respectively. For adapting to multiple subjects and to capture variability over subjects, we used subject embeddings as described in the Methods.

We used the same hyperparameters as for the single-subject trainings, except for the following modifications. For \texttt{ChanelGPT2-group} we increased the embedding size to 240. For \texttt{FlatGPT2-group} we increased it to 480 and increased the number of layers and attention heads to 12. Dropout within the GPT2 model for \texttt{FlatGPT2-group} was set to 0.1. Both \texttt{ChanelGPT2-group} and \texttt{FlatGPT2-group} proved difficult to overfit, meaning that using more data acted as a regulariser, and we stopped training when validation losses did not improve for 5 consecutive epochs.

\begin{figure}[!t]
    \centering
    \includegraphics[width=0.95\textwidth]{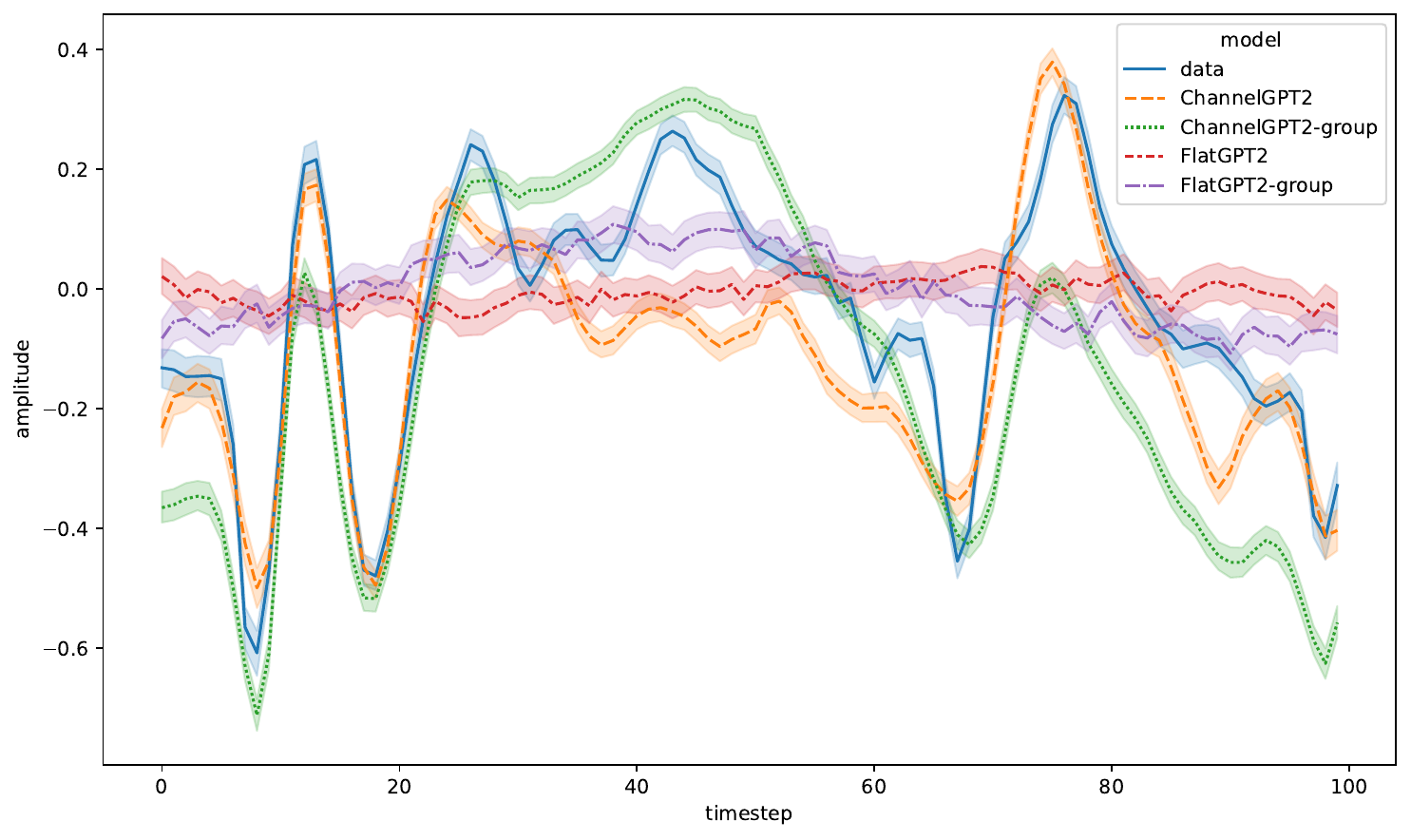}
    \caption{Comparison of evoked responses in a visual channel across single-subject and group models. The horizontal axis encompasses 1 second, where timestep 0 is the stimulus onset and timestep 50 is stimulus offset. Shading indicates 95\% confidence interval across trials.}
    \label{fig:gpt_group_evokeds_visual}
\end{figure}

We were interested in whether evoked responses improve even further when using more data. To compare with the single-subject training we generated data using the subject embedding of that subject assuming that the model learned to condition its predictions on the subject labels. We compare the evoked response of single-subject and group models for one 1 visual channel in Figure~\ref{fig:gpt_group_evokeds_visual}. \texttt{FlatGPT2-group} failed to produce sensible evoked responses similar to the single-subject \texttt{FlatGPT2}. We found that generally \texttt{ChannelGPT2-group} produces evoked responses that are more smoothed than the single-subject model. We hypothesise this is partly because the model learns to generate data that is closer to the average statistics over subjects, and while it can adapt its generation based on the subject label, it is not perfect.

\begin{figure}[!t]
  \centering
  \begin{subfigure}{0.49\textwidth}
    \centering
    \includegraphics[width=1.0\linewidth]{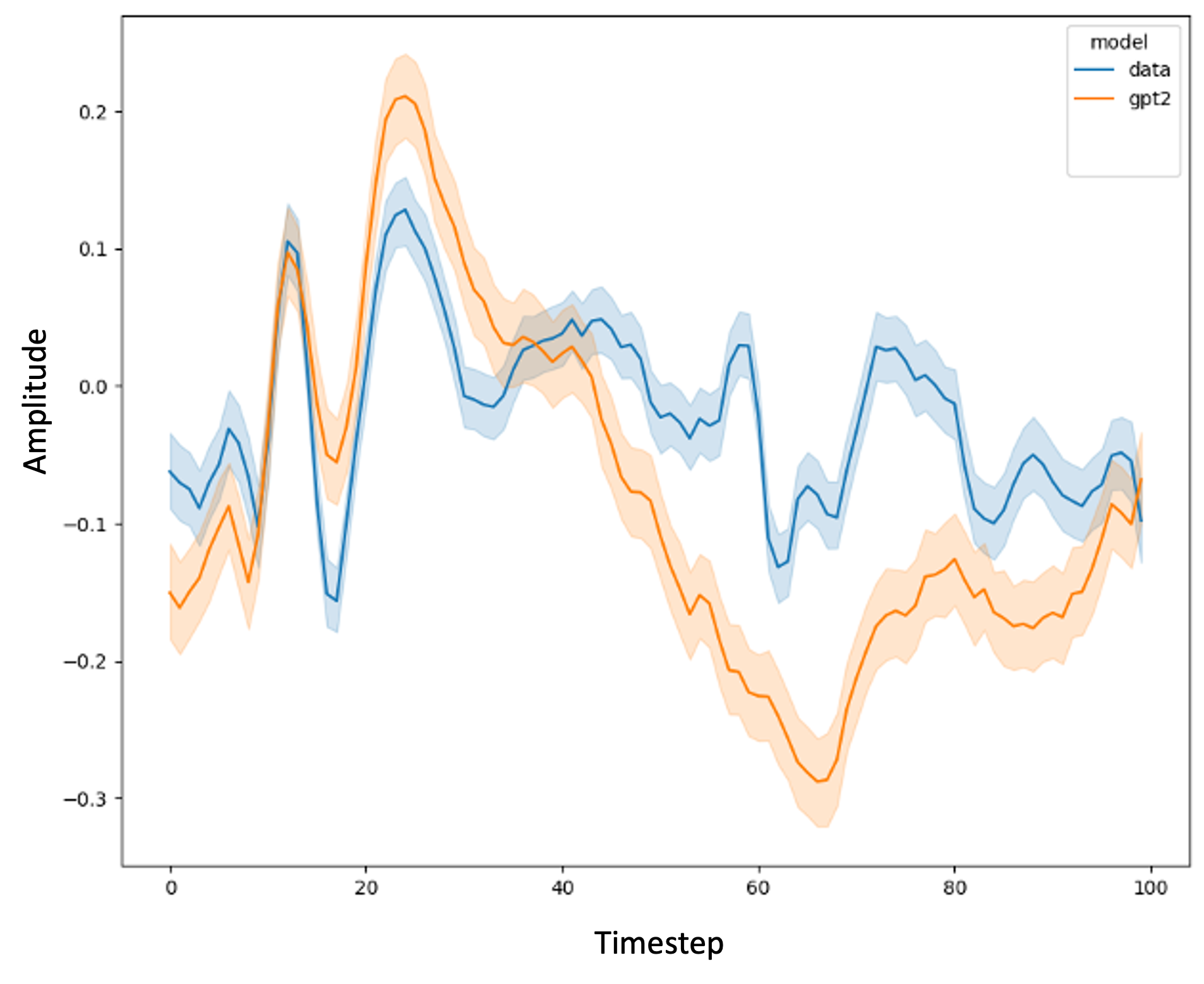}
    \caption{Visual channel}
    \label{fig:group_data_evoked1}
  \end{subfigure}
  \begin{subfigure}{0.49\textwidth}
    \centering
    \includegraphics[width=1.0\linewidth]{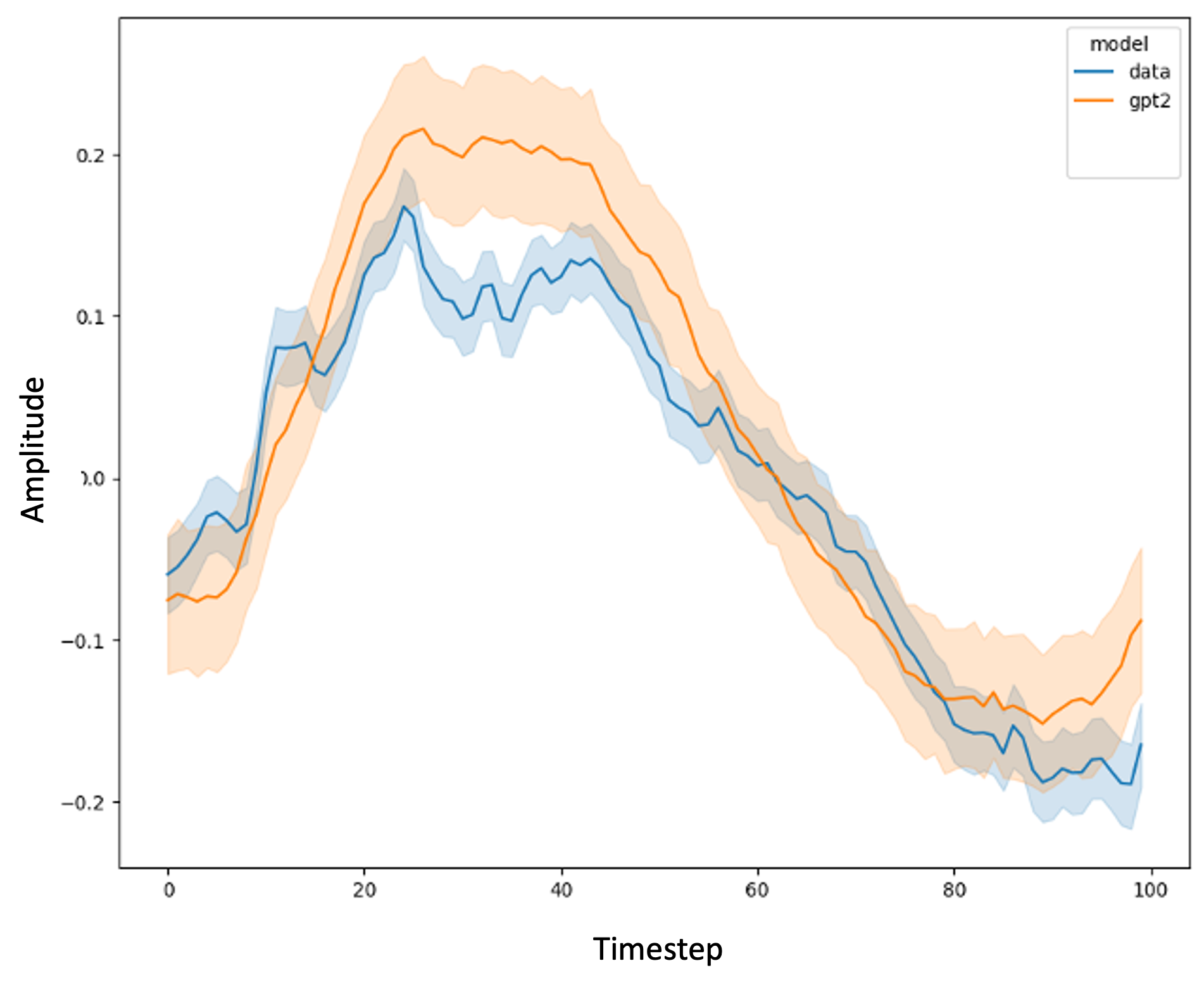}
    \caption{Frontal channel}
    \label{fig:group_data_evoked2}
  \end{subfigure}
  \caption{Comparison of evoked responses averaged across all subjects in the data (blue line) and the generated data from \texttt{ChannelGPT2-group} (orange line). The horizontal axis encompasses 1 second, where timestep 0 is the stimulus onset and timestep 50 is the stimulus offset. Shading indicates 95\% confidence interval across trials.}
  \label{fig:group_evoked_2channels}
  \end{figure}

To test our hypothesis regarding \texttt{ChannelGPT2-group} generating more of an average across subjects, we generated data for all subjects (using appropriate subject embeddings) and compared the grand average evoked responses with those extracted from the MEG data of all subjects. Two channels are plotted in Figure~\ref{fig:group_evoked_2channels}. The evoked response averaged over all subjects is much noisier because of the high between-subject variability. However, we can see that indeed \texttt{ChannelGPT2-group} can generate this well, perhaps slightly smoother than the real data. Comparing these plots with Figure~\ref{fig:gpt_group_evokeds_visual}, it is also clear that it adapts its generation well to a specific subject compared to the group average.

\begin{figure}[!t]
  \centering
  \begin{subfigure}{0.8\textwidth}
    \centering
    \includegraphics[width=1.0\linewidth]{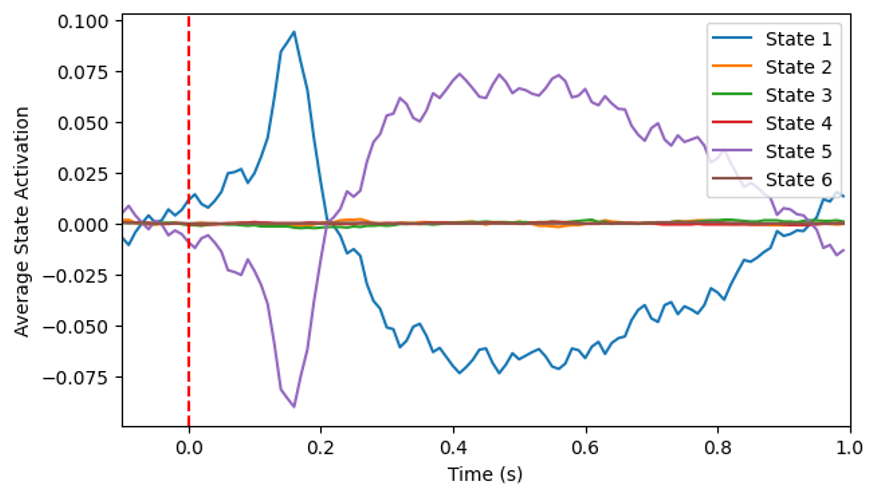}
    \caption{Data}
    \label{fig:data_datahmm}
  \end{subfigure}
  \begin{subfigure}{0.8\textwidth}
    \centering
    \includegraphics[width=1.0\linewidth]{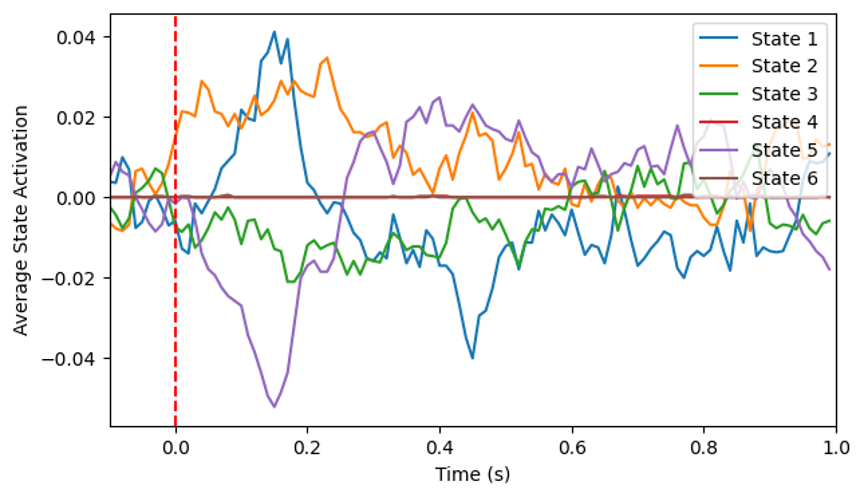}
    \caption{\texttt{ChannelGPT2-group}}
    \label{fig:gpt2_datahmm}
  \end{subfigure}
  \caption{Comparison of evoked state timecourses inferred from the data of all subjects and from the generated data of \texttt{ChannelGPT2-group} for all subjects. State indices are matched between the two plots, as the same fitted HMM model was used.}
  \label{fig:transfer_hmm_evoked}
  \end{figure}

A further way to test alignment between group-level evoked responses is to fit an HMM on the data of all subjects, and then infer state timecourses with this model on the generated data of all subjects from \texttt{ChannelGPT2-group}. By taking this approach we can directly match the evoked state timecourses between the real and generated timeseries. We trained an amplitude-envelope HMM (AE-HMM) with 6 states \citep{quinn2019unpacking} and show results in Figure~\ref{fig:transfer_hmm_evoked}. Two states that show strong activation during real task data show similar temporal signatures and amplitude changes in the generated data, albeit slightly noisier. In the generated data there are two additional states which seem to get activated during the trial. This indicates that while \texttt{ChannelGPT2-group} can capture some of the state-level dynamics, there is room for improvement.

\begin{figure}[!t]
  \centering
  \begin{subfigure}{0.33\textwidth}
    \centering
    \includegraphics[width=1.0\linewidth]{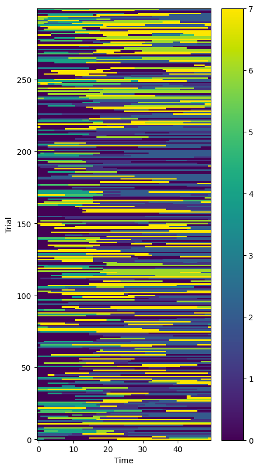}
    \caption{Data}
    \label{fig:data_trialvar}
  \end{subfigure}%
  \begin{subfigure}{0.33\textwidth}
    \centering
    \includegraphics[width=1.0\linewidth]{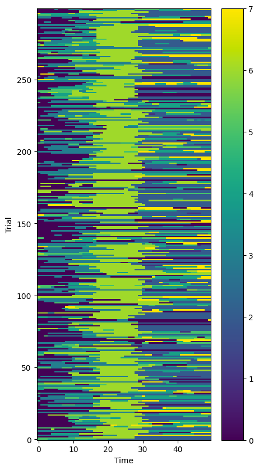}
    \caption{\texttt{ChannelGPT2}}
    \label{fig:gpt2_trialvar}
  \end{subfigure}%
  \begin{subfigure}{0.33\textwidth}
    \centering
    \includegraphics[width=1.0\linewidth]{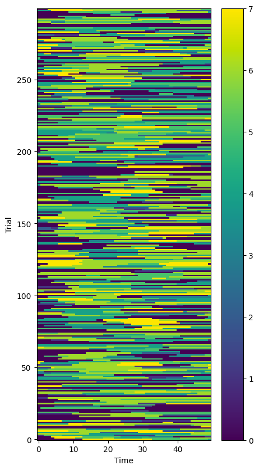}
    \caption{\texttt{ChannelGPT2-group}}
    \label{fig:gpt2group_trialvar}
  \end{subfigure}
  \caption{Comparison of trial-level variability in the evoked state timecourses of an HMM trained on real data and applied to the generated data of \texttt{ChannelGPT2} and \texttt{ChannelGPT2-group}. Different colours represent different states (matched across models). Individual trials however are not matched and we cannot compare the plots at the trial-level, only as an aggregate visualisation of variability across trials.}
  \label{fig:hmm_trialvar}
  \end{figure}

Finally, we examine the variability in state time courses over individual trials. For this we trained an 8-state HMM on the real data of a single subject, and inferred the state timecourses on both the single-subject \texttt{ChannelGPT2} and \texttt{ChannelGPT2-group} generated data, obtaining matched states. We hypothesised that even if the average evoked responses are similar to the real data, GPT2 may not able to generate trials with variability in the temporal activation of states. Figure~\ref{fig:hmm_trialvar} shows that this is indeed true for the single-subject \texttt{ChannelGPT2} generated data. \texttt{ChannelGPT2-group} responses seem to include much higher temporal variability in state activations, though still falling short of the real data. This indicates that the model can capture some trial-to-trial variability through its exposure to multiple subjects, but has difficulty fully matching the complexity of real neural data. More data may be needed to improve this aspect of generation.

In summary, in the last few sections we showed that deep learning models, and in particular a channel-independent Transformer-based model can reproduce spatial, temporal, and spectral signatures of real data both at single-subject and group-levels. We were next interested whether such a model can aid in specific tasks, for example decoding of experimental conditions in the \citet{cichy2016comparison} data.

\subsection{Classification of generated data}
% classification accuracy between data, gpt-subject, and gpt-group
% transferring classifier from generated to real data with increasing number of trials
% very briefly generative classification performance

While there are multiple ways a forecasting model could be used to aid decoding of task labels, here we opted for two approaches, leaving more complicated methods to future work. We first investigated whether the task responses produced by \texttt{ChannelGPT2} can be classified with performance comparable to trials of real data. This also further tests how well the model captures spatiotemporal task-related activity and information. The benefit of this approach is that if similar performance is obtained, then \texttt{ChannelGPT2} could simulate an arbitrarily large number of trials to potentially improve decoding of real data through pretraining on this simulated data. This is a form of transfer learning where the decoding model, not the forecasting model, is transferred.

\begin{sloppypar}
We generated 20 trials for all 118 conditions for 1 subject with both \texttt{ChannelGPT2} and \texttt{ChannelGPT2-group}. We trained separate linear neural network models described in Chapter~\ref{Chap3} on the real data (20 trials/condition) and the generated datasets, with an appropriate 4:1 train and validation set ratio. This achieved 17.6\%, 1.9\%, and 7.2\% validation accuracy for the real data, \texttt{ChannelGPT2}, and \texttt{ChannelGPT2-group}, respectively. Thus the group model generates more classifiable subject-specific task-responses, but still does not reach the classifiability of real data. This and previous analyses indicate the group model successfully leverages larger datasets to produce more accurate task-related activity.
\end{sloppypar}

\subsection{Transfer learning}

A key advantage of generated data is the ability to generate virtually infinite amounts. We generated additional datasets with 40 and 60 trials/condition using \texttt{ChannelGPT2-group}. Training a decoder on these achieved 21.7\% and 44.2\% accuracy, respectively, exhibiting linear scaling of classification performance with simulated data amount. Critically, we assessed whether this simulated data can pretrain classifiers for transfer learning. We first pre-trained the neural network decoder on the 20-, 40-, and 60-trial generated datasets, then finetuned it (trained it further) on the MEG data (20 trials/condition). As the simulated data used for pre-training increased, accuracy of the finetuned model improved rapidly. Zeroshot (no finetuning) performance on real data was above chance with 2\%, 3\%, and 4\% accuracy, for increasing pretraining data quantities. Final accuracies after finetuning were 19.5\%, 21.5\%, and 23\%, respectively. Thus, each additional 20 simulated trials/condition improved final decoding by \textasciitilde2\%. These results are summarised in Table~\ref{table:transfer_results}.

\begin{table}[!t]
\centering
        \begin{tabular}{l|cm{4cm}}
        %\toprule
           \bf Trained on (no. trials)  &\bf Tested on MEG (20) & \bf Tested on GPT2 (same no. trial data)   \\ \midrule
            
            MEG (20) & 17.6 & -   \\ 
            GPT2 (20) & 2 & 7.2   \\
            GPT2 (40)  & 3 & 21.7 \\
            GPT2 (60)  & 4 & 44.2  \\
            GPT2 (20) + MEG (20) & 19.5 & - \\ 
            GPT2 (40) + MEG (20) & 21.5 & - \\ 
            GPT2 (60) + MEG (20) & 23 & - \\ 
            
            %\bottomrule
        \end{tabular}
    \caption{\label{table:transfer_results} Summary of transfer learning results. The first column shows the data used for training the decoder, with the number of trials per condition shown inside the parenthesis. GPT2 refers to the \texttt{ChannelGPT2-group} generated data, while GPT (.) + MEG (20) is the fine-tuned decoder on the MEG data. The other two columns represent the validation data on which the decoder performance is shown. Accuracy values are provided in percentages. Chance level is $100/118$.}
\end{table}

Finally, we also tried obtaining a decoding model directly from the \texttt{ChannelGPT2-group} forecasting model using Bayes' theorem, as described in Section~\ref{ssec:full_wavenet}. We found limited 5\% accuracy over 1 subject's validation set (versus 40-50\% with a discriminative decoder). This generative decoding approach may require larger datasets or more sophisticated architectures.

In summary, while generated data did not match real data in decodability when the number of trials was matched, the group model produced classifiable responses capturing key features, improving substantially over a single-subject model. Further, its simulated responses could improve decoding of real data through pretraining, demonstrating the utility of forecasting models for transfer learning. There is clear promise in scaling up simulated datasets to improve MEG decoding.

\subsection{Ablation experiments}
% 2 paragraphs max about experiments and results
% can show relevant figures as it does not contribute to word count

% random conditions / timings, etc.
% no channel / no cond embedding
% channel embedding plots

Ablation studies are a common approach in machine learning to understand model behaviour by selectively removing or altering components of the model \citep{meyes2019ablation}. We performed ablation experiments with \texttt{ChannelGPT2} to investigate how well it can generate task-related brain activity under varied conditions without further training.

\begin{figure}[!t]
    \centering
    \includegraphics[width=0.95\textwidth]{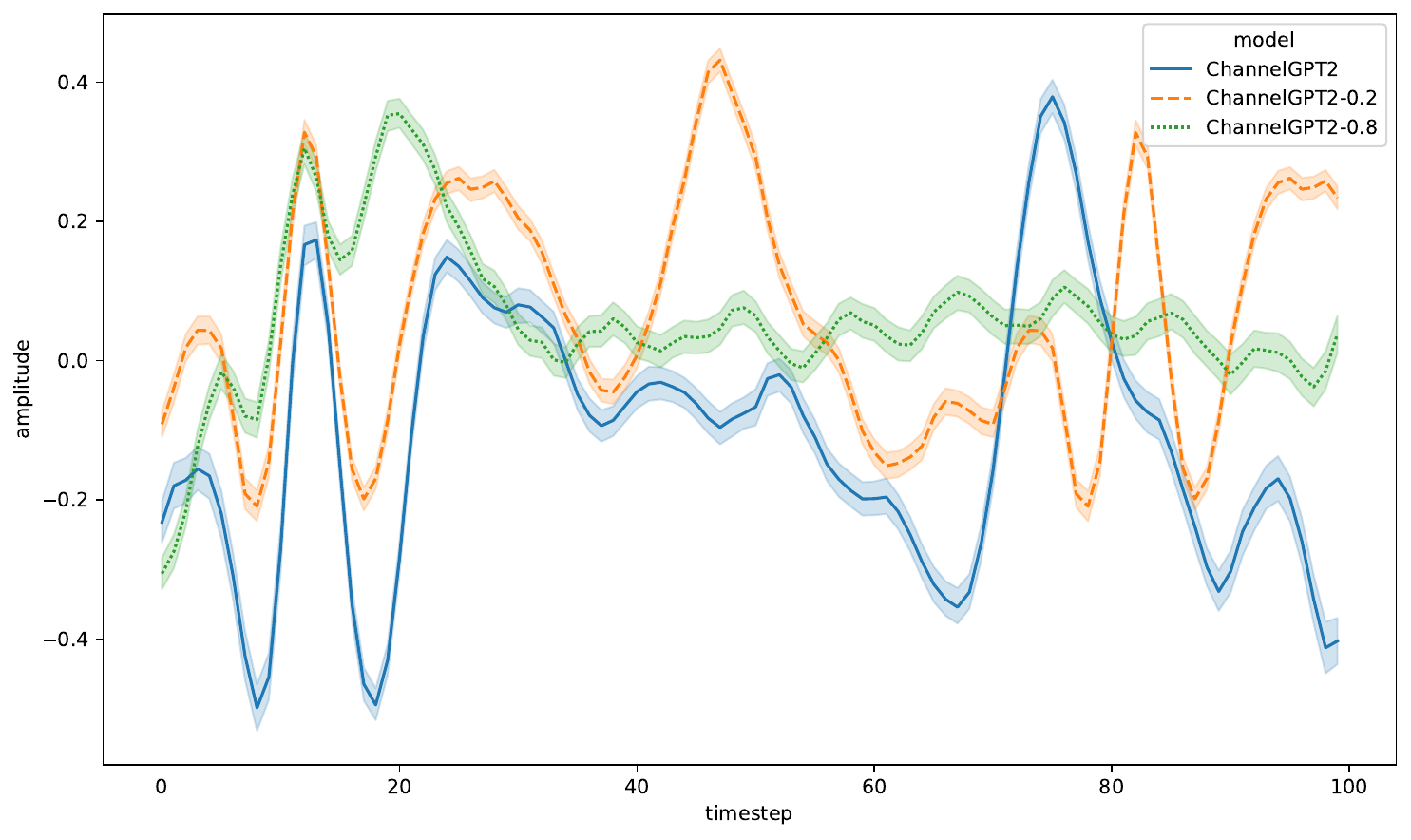}
    \caption{Evoked responses generated by \texttt{ChannelGPT2} for trials of 0.2 s (orange), 0.5 s (blue), and 0.8 s (green). The model was trained only on data containing trials of 0.5 s but adapts appropriately to the different durations.}
    \label{fig:evoked_random_timing}
\end{figure}

First, we evaluated the model's ability to adapt to different trial durations. The original \texttt{ChannelGPT2} was trained on trials lasting 0.5 seconds. We generated data using the same model but with trial durations of 0.2 s and 0.8 s. As shown in Figure \ref{fig:evoked_random_timing}, \texttt{ChannelGPT2} accurately adapted to the shorter and longer trials. The evoked responses matched the expected timecourses, with appropriate truncation or lack of second peaks due to stimulus offset. This demonstrates the model's ability to generalise to varied trial durations despite being trained on a fixed duration.

\begin{figure}[!t]
  \centering
  \begin{subfigure}{0.63\textwidth}
    \centering
    \includegraphics[width=1.0\linewidth]{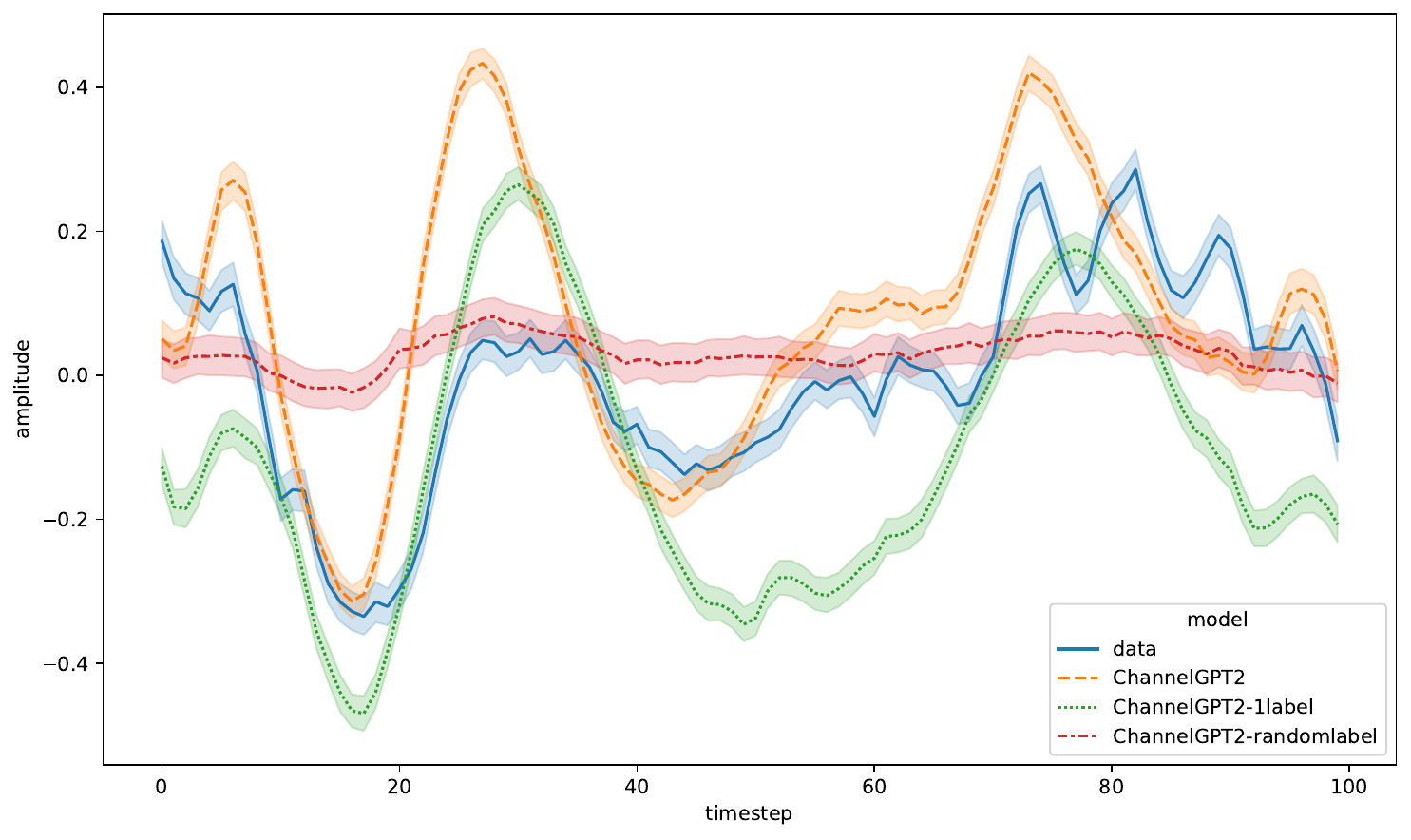}
    \caption{Channel 1}
    \label{fig:evoked_random_label_0}
  \end{subfigure}
  \begin{subfigure}{0.63\textwidth}
    \centering
    \includegraphics[width=1.0\linewidth]{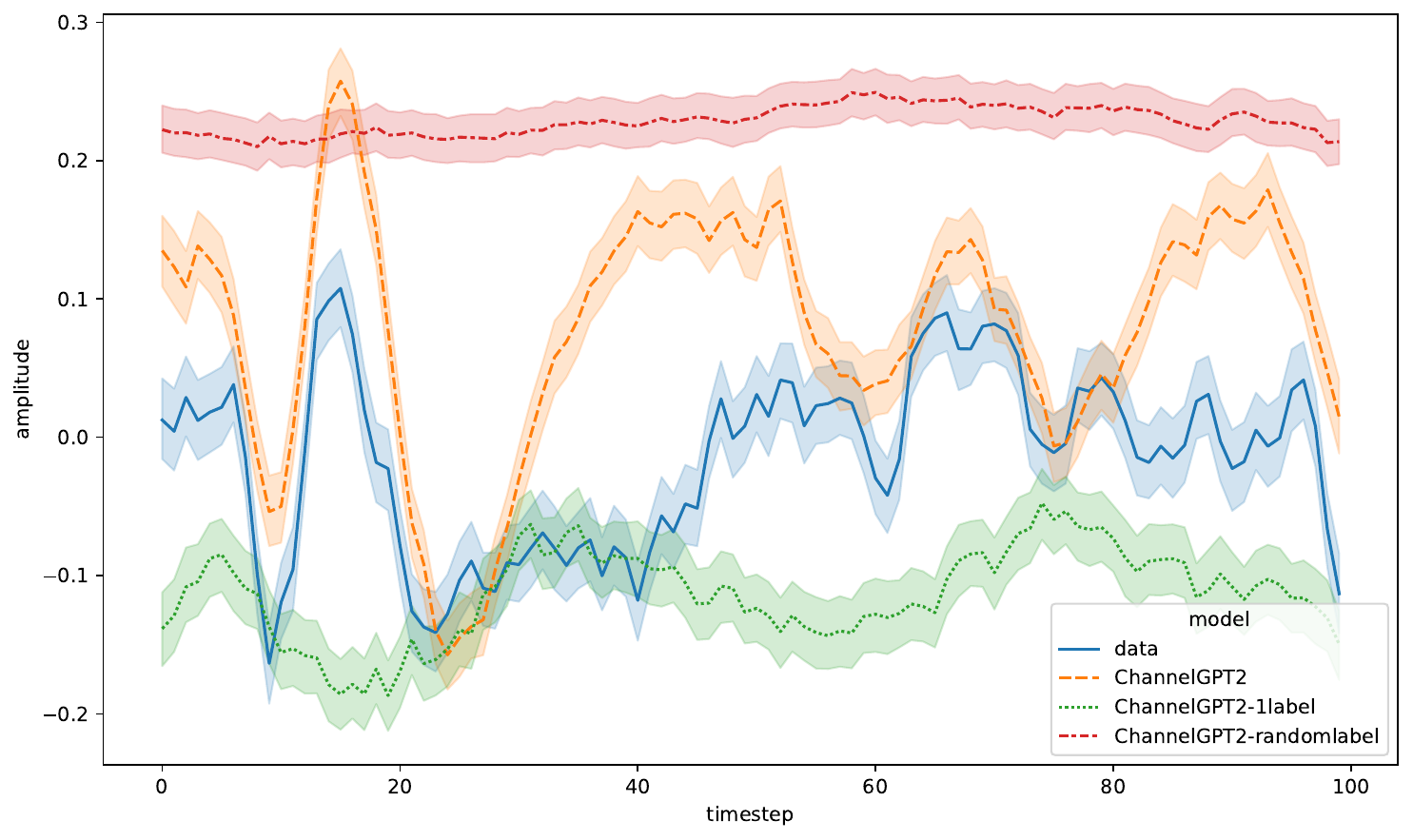}
    \caption{Channel 2}
    \label{fig:evoked_random_label_102}
  \end{subfigure}
  \begin{subfigure}{0.63\textwidth}
    \centering
    \includegraphics[width=1.0\linewidth]{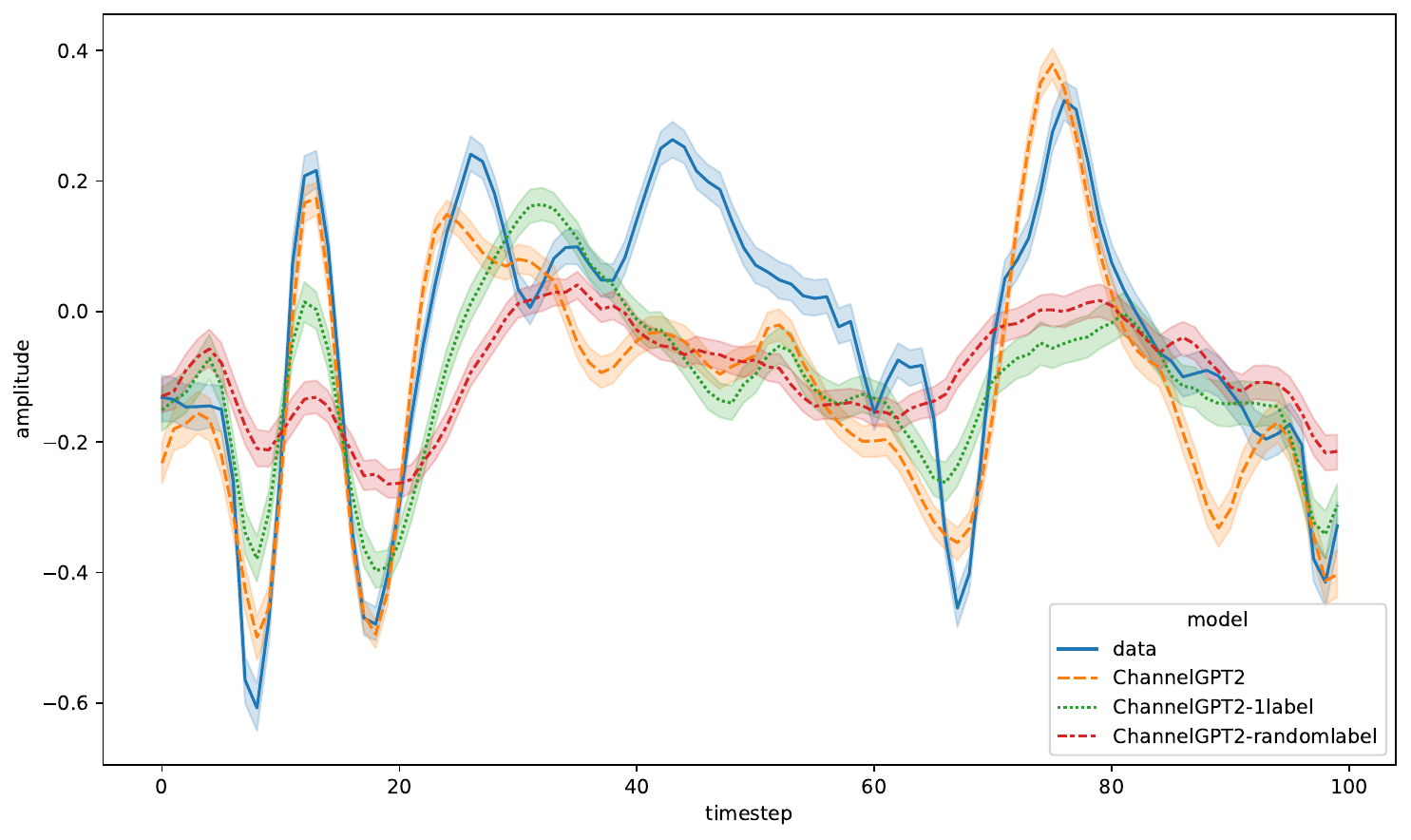}
    \caption{Channel 3}
    \label{fig:evoked_random_label_265}
  \end{subfigure}
  \caption{Evoked responses for models trained with shuffled or single condition labels, indicating reliance on semantic content. Three representative channels are presented. See main text for an explanation of model types. Timestep 0 is the stimulus onset and timestep 50 is the stimulus offset.}
  \label{fig:evoked_random_label}
  \end{figure}

Next, we performed two experiments to determine whether \texttt{ChannelGPT2} relies solely on timing information or also utilises the semantic content of the condition labels. First, we trained a model (\texttt{ChannelGPT2-randomlabel}) where the condition labels were shuffled randomly during training, breaking the semantic alignment between labels and evoked responses. Second, we trained a model (\texttt{ChannelGPT2-1label}) using a single condition label for all trials. This tests whether the model cheats by learning an average evoked response instead of adapting to each condition.

\begin{sloppypar}
As evident in Figure \ref{fig:evoked_random_label}, both models failed to generate distinct evoked responses for different semantic conditions. This demonstrates that \texttt{ChannelGPT2} leverages both timing and semantic information in the conditioning labels, rather than simply learning a stereotyped temporal template. Quantitatively, evoked response correlation with real data dropped to 44\% and 56\% for \texttt{ChannelGPT2-randomlabel} and \texttt{ChannelGPT2-1label}, respectively, compared to 74\% for the full \texttt{ChannelGPT2}. Both the visual analysis and the correlation numbers indicate that \texttt{ChannelGPT2-1label} was somewhat closer to matching \texttt{ChannelGPT2}.
\end{sloppypar}

\begin{figure}[!t]
  \centering
  \begin{subfigure}{0.33\textwidth}
    \centering
    \includegraphics[width=1.0\linewidth]{forecast_figures/gpt_p80_psd.pdf}
    \caption{\texttt{ChannelGPT2}}
    \label{fig:gpt_p80_psd}
  \end{subfigure}%
  \begin{subfigure}{0.33\textwidth}
    \centering
    \includegraphics[width=1.0\linewidth]{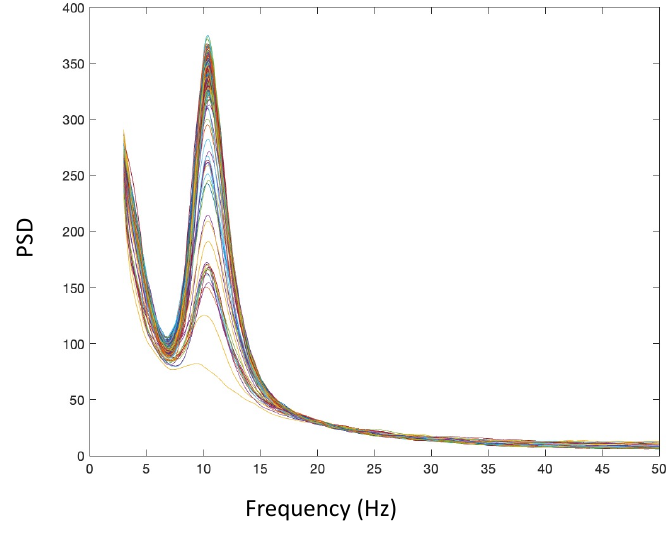}
    \caption{\texttt{ChannelGPT2} no $\mathbf{W}_c$}
    \label{fig:no_channelemb}
  \end{subfigure}%
  \begin{subfigure}{0.33\textwidth}
    \centering
    \includegraphics[width=1.0\linewidth]{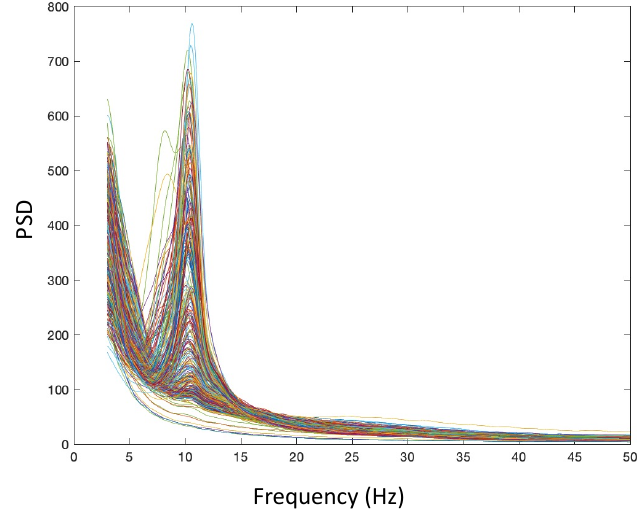}
    \caption{\texttt{ChannelGPT2} no $\mathbf{W}_y$}
    \label{fig:no_condemb}
  \end{subfigure}
  \caption{Generated power spectra for full model (left) versus ablations. Both channel (middle) and condition embeddings (right) are critical for accurate spectral content.}
  \label{fig:embedding_ablation}
  \end{figure}

\begin{figure}[!t]
  \centering
  \begin{subfigure}{0.33\textwidth}
    \centering
    \includegraphics[width=1.0\linewidth]{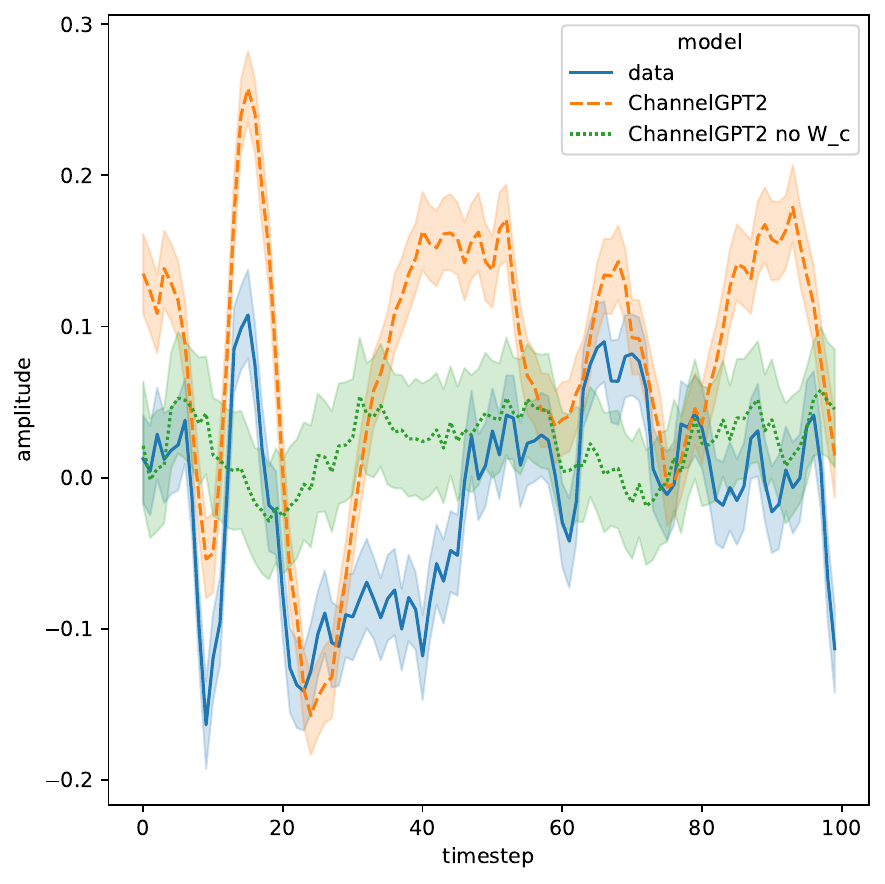}
    \caption{Channel 1}
    \label{fig:gpt_ablation_102}
  \end{subfigure}%
  \begin{subfigure}{0.33\textwidth}
    \centering
    \includegraphics[width=1.0\linewidth]{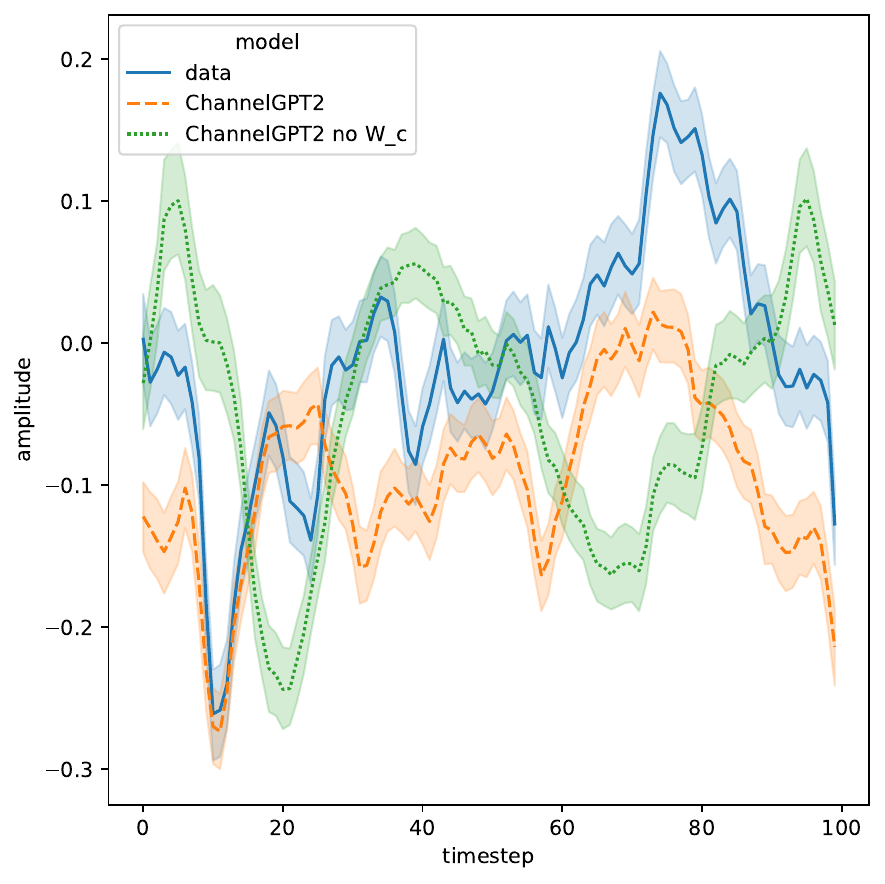}
    \caption{Channel 2}
    \label{fig:gpt_ablation_200}
  \end{subfigure}%
  \begin{subfigure}{0.33\textwidth}
    \centering
    \includegraphics[width=1.0\linewidth]{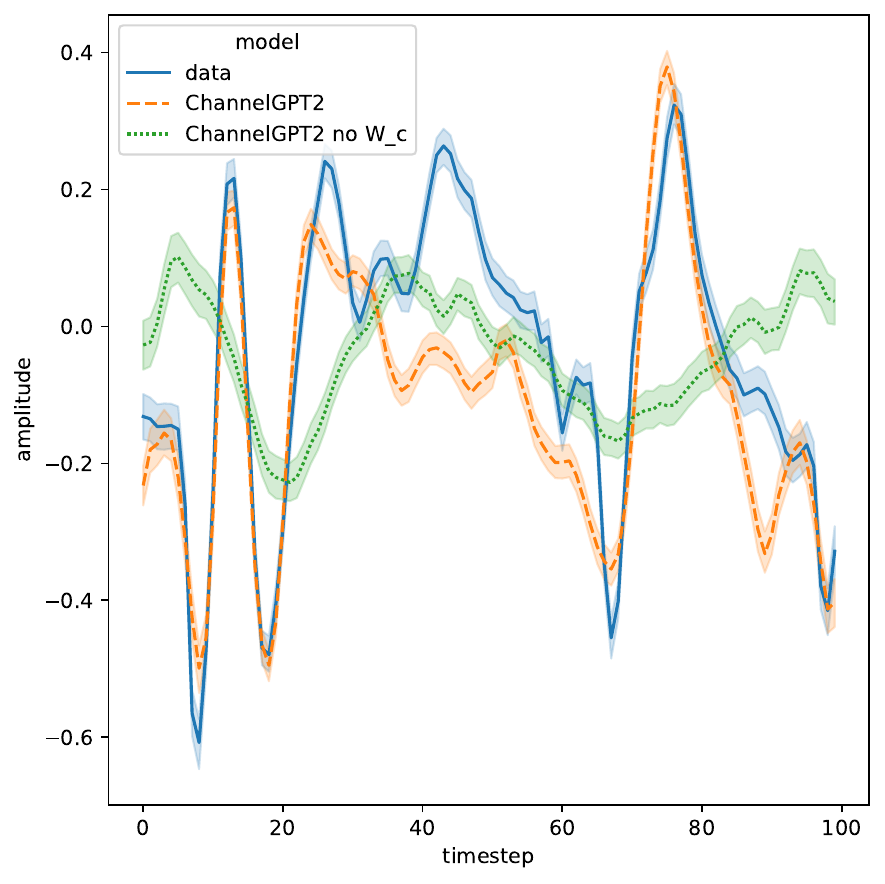}
    \caption{Channel 3}
    \label{fig:gpt_ablation_265}
  \end{subfigure}
  \caption{Comparison of generated evoked responses from \texttt{ChannelGPT2} and the model with ablated channel embeddings (\texttt{ChannelGPT2} no $W_c$) across 3 representative channels. Without channel embeddings the model fails to adapt evoked responses to different channels. Timestep 0 is the stimulus onset and timestep 50 is the stimulus offset.}
  \label{fig:embedding_ablation_evoked}
  \end{figure}

We also investigated the contributions of the channel and condition embeddings, by training two separate ablated models. As shown in Figure \ref{fig:embedding_ablation}, removing the channel embeddings resulted in very similar PSD across channels in the generated data, indicating the model relies heavily on these embeddings to adapt generation per channel. The evoked responses in Figure \ref{fig:embedding_ablation_evoked} confirm that without channel embeddings, variability between channels is reduced. Removing the condition embeddings resulted in noisier power spectra of the generated data and no 20 Hz peak.

  \begin{figure}[!t]
  \centering
  \begin{subfigure}{0.49\textwidth}
    \centering
    \includegraphics[width=1.0\linewidth]{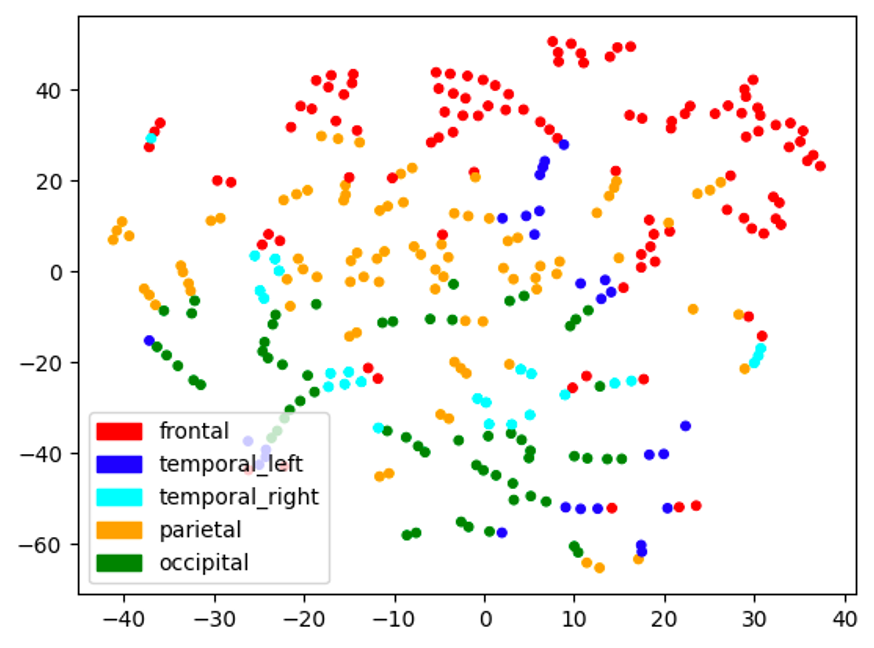}
    \caption{t-SNE}
    \label{fig:tsne_embs}
  \end{subfigure}%
  \begin{subfigure}{0.49\textwidth}
    \centering
    \includegraphics[width=1.0\linewidth]{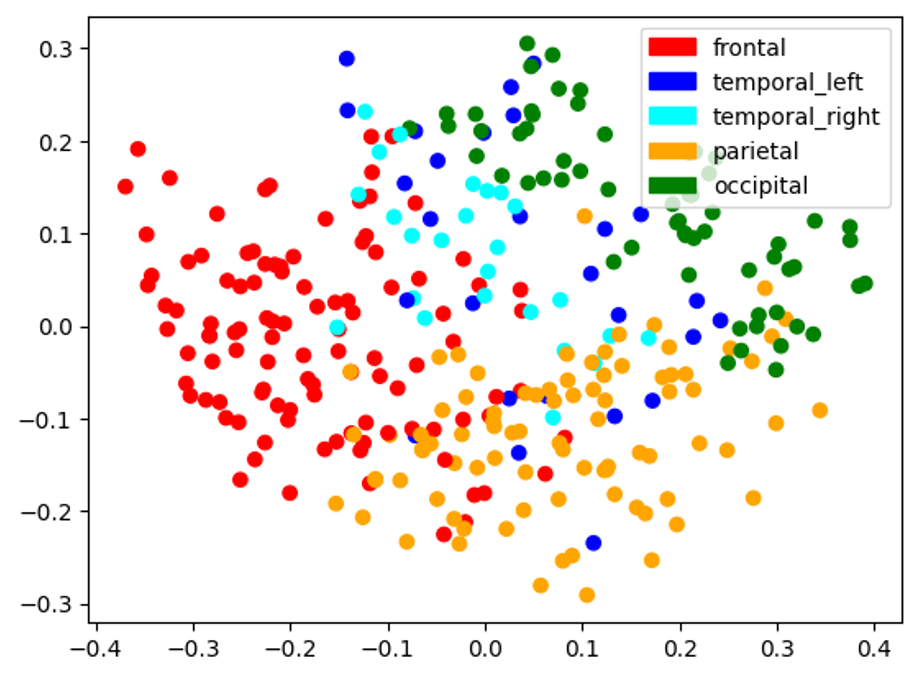}
    \caption{PCA}
    \label{fig:tsne_pca}
  \end{subfigure}%
  \caption{2D projection of the channel embeddings from \texttt{ChannelGPT2-group} with t-SNE (left) and PCA (right). Channels are coloured by their location on the scalp grouped into 5 major brain areas.}
  \label{fig:tsne_channelemb}
  \end{figure}

Finally, we found that the channel embeddings encode spatial relationships, as sensors that are near to each other in the real sensor montage tend to have more similar embeddings. This is shown through a t-SNE and PCA projection of the embedding space in Figure~\ref{fig:tsne_channelemb}. Correlation between pairwise Euclidean distances of channels in physical space and embedding space was 0.45 (see Figure~\ref{fig:emb_distances} in the Appendix).

\section{Discussion}
% 1 short paragraph of extra things we tried, e.g. different models, longer receptive field
% brief summary and limitations

In this chapter, we have presented our initial efforts at developing a general forecasting model for M/EEG data. After carefully evaluating the trade-offs between various modelling approaches, we settled on two main architectures: one based on Wavenet \citep{oord2016wavenet}, and one based on GPT-2 \citep{radford2019language}. These models have proven successful in the audio and natural language domains, which share similarities with the time series nature of brain signals. We systematically compared different variants of our proposed models on both simulated and real M/EEG datasets.

We found that on real MEG data the forecasting performance was comparable between Wavenet and AR models according to next-timestep prediction metrics (results in Appendix). This suggests such metrics may be limited in their ability to effectively evaluate model dynamics beyond one-step prediction. Generated data analysis provided more discerning model comparisons. While the channel-independent AR and Wavenet models accurately reproduced the overall power spectral density, only the Transformer-based models captured more abstract multivariate statistics like inter-channel covariance and HMM state dynamics.

Critically, the \texttt{ChannelGPT2} model-generated data closely matched real MEG recordings across both temporal and spectral domains. Analysis of the discovered latent brain states showed \texttt{ChannelGPT2} reproduced variable oscillatory states similar to those inferred from human recordings \citep{vidaurre2018discovering}. Each state captured distinct spectral content, while the linear and Wavenet-based models failed to achieve this degree of heterogeneity in their dynamics. It is possible that this does not indicate a failing of the Wavenet architecture, but rather that different conditioning methods may be needed. One such approach that we have not tested is using the same type of channel embeddings as for \texttt{ChannelGPT2}.

The Deep Recurrent Encoder (DRE) proposed by \citet{chehab2021deep} is a highly relevant architecture to our approaches, as it demonstrates the advantages of modelling spatiotemporal dynamics for encoding neural data. DRE aims to predict MEG brain responses to visual word stimuli. Standard linear encoding models like temporal receptive fields (TRFs) face limitations in capturing the rich nonlinear dynamics, variability, and interactions inherent in MEG signals. DRE seeks to address these challenges by leveraging a convolutional LSTM architecture to model the intricate spatiotemporal neural dynamics across subjects.

While motivated as an encoding model, DRE can also be viewed through the lens of forecasting, with the addition of auxiliary task features. Forecasting holds inherent advantages over pure encoding, as it enables reconstructing real data and modelling complex spatiotemporal relationships, beyond just learning abstract representations.

Multiple analyses consistently demonstrated \texttt{ChannelGPT2}'s strengths in realistic conditional timeseries generation. \texttt{ChannelGPT2}-generated evoked responses had high correlation to real MEG data. However, modelling single-trial variability and between-subject differences remain challenging areas needing further work. Scaling to multiple subjects showed promise. The model was able to adapt its generated data based on the input subject label and generate task trials with variability more similar to real recordings than a single-subject model.

One consideration in our modelling approach is the use of sensor-space data as opposed to source space. Forecasting in the source space could more accurately capture the statistics of the data and inherently deal with between-subject differences. However, the measured raw data is in the sensor-space, and we wanted our models to receive this as input without any additional transformation (to source space). This delegates all modelling to the model itself. Similarly we let the model learn between-subject structure in a data-driven manner through the subject embeddings. We leave source space forecasting for future work.

Ablation studies quantified the importance of channel embeddings and task conditioning for accurate MEG modelling. Removing channel embeddings resulted in near identical generation across sensors, failing to capture spatial heterogeneity. Analysis of \texttt{ChannelGPT2}'s channel embeddings revealed spatial relationships between sensors were learned, with proximal channels having more similar embeddings. With incorrect or with no task labels, \texttt{ChannelGPT2} produced noisy evoked responses, indicating the model leverages both timing and label semantics. Furthermore, the model trained on 0.5s trials only, was able to produce reasonable
responses to longer or shorter trials, showcasing generalisation. These results demonstrates the value of multi-faceted conditioning for realistic brain data modelling.

A key investigation involved analysing the classification of generated data according to the condition labels. The trials generated by the group-level model were classified with much higher accuracy (closer to real data) than those of the single-subject model. We also demonstrated that generated data can improve decoding of real trials via transfer learning \citep{torrey2010transfer}, with benefits scaling with generated data quantity. The classifiability of generated trials and transfer learning results highlight the utility of forecasting models like \texttt{ChannelGPT2} for decoding real MEG data. Further analysis could involve permutation feature importance of the decoding model trained on generated data to gain insights into learned representations. Transfer learning also requires more thorough evaluation across diverse decoding tasks. It would be important to also investigate other more direct finetuning or transfer learning approaches of the forecasting model akin to vision or language domains. These could involve additional output layers and losses for finetuning on downstream tasks.

Overall, the proposed analyses enable thorough interrogation of forecasting model dynamics beyond standard predictive metrics. However, experiments were limited to a single dataset, lacking evaluation across heterogeneous datasets and tasks. Testing on more diverse and larger-scale datasets with multiple recording systems and experimental paradigms is needed to fully validate transfer learning capabilities for forecasting, encoding and decoding. Applying the models to different modalities like EEG would also be informative of generalisation.

The full potential of self-supervised learning is only realised with large-scale data. This remains challenging for brain imaging compared to vision and language. Lowering barriers to data access and promoting data sharing is critical to realise the promise of foundation models in neuroimaging \citep{poldrack2014making}.

Investigation of the proposed models on simulated data could  shed light on which model features are necessary for good modelling. In Appendix~\ref{ssec:simulated_results} we provide some insights into how Wavenet models are better able to capture distinct oscillatory activity in simulated data compared to  linear AR models.

A core limitation of the channel-independent GPT2 model is no direct leveraging of cross-channel information for each sensor prediction. Our \texttt{FlatGPT2} approach incorporating this performed worse. Different architectures or more data may enable proper utilisation of cross-channel dependencies. We tried various other approaches to mixing channel information beyond those reported, without success. For the Wavenet model, we incorporated all channels in the input by concatenating embeddings, and for the GPT2 models, we tried mixing channels with convolutions. We tried concatenating the output of each channel and then predicting from this shared output using a different projection for each channel. We also attempted to increase receptive field, dropout, and model size. One limitation in our approaches is the use of a next-timestep prediction loss. Future work should continue exploring architectures and different self-supervised or multi-timestep losses to leverage cross-channel information and improve modelling capabilities. In the context of BCIs, additional losses could be incorporated to predict mental states or decoding targets instead of future timesteps.

We did not analyse the inner representations of \texttt{ChannelGPT2} to explain its predictive abilities. Attention weight and activation visualisations could provide insights into important input features \citep{vig2019bertviz}. PFI analysis may also illuminate influential temporal, spatial, and spectral input features for forecasting.

In conclusion, this work demonstrates that deep forecasting models can accurately reproduce complex neural dynamics of both ongoing and task-related activity and provides an extensive analysis methodology. Key limitations are small-scale experiments, the lack of working channel-mixing methods and multi-dataset testing. Future work should explore more flexible conditioning, study different self-supervised and transfer learning frameworks, and critically, apply similar analyses when scaling up across diverse, large electrophysiology datasets. This has the potential to enable powerful transfer learning and advance foundational brain modelling and decoding.

\chapter{Decoding thoughts}
\label{Chap6}
% map out each section before writing so that i don't write too much
% only things that are actually interesting, e.g. evoked timecourses are not that interesting, or figuring out the reasons behind the negative results

% in the main part of the chapter put only positive results

In the previous chapters, we have presented methodological advancements for dealing with various types of variability in M/EEG data. The development of these methods was inspired by the central thesis of improving the modelling and decoding of brain activity. Specifically, decoding in the context of brain-computer interfaces (BCI) aimed at communication is of particular interest. However, methodological advancements can only go so far, and we were also interested in tackling this challenge experimentally. A BCI is simply the real-time (online) application of a decoding algorithm to brain data that is being streamed continuously to the computer. Most often BCIs are aimed at decoding brain data into intents, actions, text, or speech. While in this work we have not built a real-time BCI, our data and offline methods are aimed at enabling such BCIs in the future.

For improving communication speed with BCIs, we posited two distinct solutions, applying decoding to inner speech, and improving decoding by collecting a large number of trials. We hope that our investigations will contribute to the field of neural speech prosthetics \citep{metzger2022generalizable}, which one day may be capable of restoring communication to people with locked-in syndrome, a condition where people are unable to move or speak. To be clear this chapter is in the spirit of proof of concept with the aim to gather preliminary evidence as to the extent to which it is possible to decode inner speech noninvasively using electrophysiology, ideally self-generated inner speech, but otherwise elicited.

Despite the prevalence of inner speech in everyday life, research on this has been limited, particularly when it comes to non-invasive methods \citep{panachakel2021decoding}. This chapter aims to fill this gap by using EEG and MEG to collect data from three different inner speech paradigms, and by conducting an initial decoding analysis. Specifically, we tested silent reading, repetitive inner speech, and generative inner speech tasks. We hypothesised that silent reading would yield the most decodable signals due to the visual presentation of words. While inner (covert) speech refers to the internal voice/monologue that most people possess and is a purely cognitive process, silent reading involves visual processing of the presented text and thus has additional sensory activity.

We collect a high number of inner speech trials from a few participants. Besides comparing across recording modalities we also compare across inner speech types. Our aim is to analyse the decodability of inner speech within each task and between tasks by the use of transfer learning. We find that in both EEG and MEG, silent reading can be decoded relatively well with 30-40\% accuracy across 5 words. However, the decoding performance of both types of inner speech is mostly at chance level. This prohibited further transfer learning investigations between tasks. While the inner speech results are primarily negative, we believe our exploration of data size and various decoding methods is valuable. The dataset itself is useful for the research community as it contains a much larger number of trials within one participant than any other inner speech dataset. Having multiple sessions also allows for testing across-session performance.

Finally, we systematically compare silent reading decoding performance within 3 participants across four non-invasive modalities. These are EEG, 2 types of MEG machines, Elekta and CTF, and optically-pumped magnetometers (OPMs). We also compare the spatiotemporal dynamics of silent reading between these modalities. This is especially aimed at validating OPMs as a new kind of non-invasive brain recording technology. We find comparable performance to EEG, but OPM performance did not reach traditional MEG.

\section{Introduction}

Inner speech, also known as verbal thinking or covert self-talk, refers to the internal monologue that occurs within one's mind. This phenomenon has been extensively studied in psychology and cognitive science \citep{alderson2015inner}. More recently, neuroimaging techniques have allowed researchers to examine the neural correlates of inner speech directly. As discussed by \citet{geva2018inner}, early 20th century studies used measurements of tiny muscle movements during imagined speech production to infer inner speech. However, the advent of modern neuroimaging techniques like MEG and fMRI has enabled more direct investigation of the brain regions involved in inner speech compared to overt speech.

As mentioned in Chapter~\ref{Chap1}, traditional brain-computer interface (BCI) systems are relatively slow and do not leverage inner speech, which has the potential to enable communication at natural speech rates. Some progress has been made in decoding visual stimuli during reading limited sets of words or sentences \citep{mugler2014direct, hulten2018cracking, moses2019real}, as well as decoding speech perception and overt speech production where muscle movements are present \citep{dash2020decoding, defossez2022decoding}. However, detecting brain signals associated specifically with inner speech remains challenging given the lack of external stimuli or produced behaviour to provide timing information.

While we focus here on the hard problem of non-invasive inner speech decoding, more tractable approaches exist. These include applying invasive methods to inner speech, as well as applying non-invasive methods to related tasks that share features with inner speech. Related tasks either provide external timing information through stimuli like silent reading/listening, or leverage produced behaviour like overt loud or silent speech. However, it is unclear how insights from these tasks might transfer to decoding inner speech itself. Disentangling task-related activity (e.g. visual, auditory, or motor) from pure inner speech is difficult. Similarly, models tuned to detect muscle activation during silent speech may not transfer well to pure inner speech decoding \citep{dash2020decoding}.

There are also experimental challenges inherent to studying inner speech. While people often experience spontaneous inner speech during mind wandering, there is no way to align such endogenous thought with recorded brain activity. This leaves two options: a purely unsupervised learning approach, or using more constrained, artificial experiments. The latter often employs visual or auditory cues to elicit inner speech in a time-locked manner. Some papers refer to cued silent reading tasks as a form of inner speech. However, we argue this confounds inner speech with concurrent visual processing of the words.

Pure inner speech paradigms can be achieved by using identical cues across different conditions. Brain responses to the cues themselves will be consistent, while activity varying between inner speech conditions can be disentangled. A limitation is that using identical cues provides no overt record of the specific condition of each trial. Participants can either be instructed on the inner speech to generate for each cue beforehand (\textit{repetitive} inner speech), or report the contents after each trial (\textit{generative} inner speech) \citep{jones2021note}. Both approaches enable studying inner speech independently from external stimuli or behaviour.

The aims here are multifaceted, but guided by advancing non-invasive BCI communication. We approach this challenge through comparing: \begin{enumerate}
\item Non-invasive recording modalities: EEG, MEG, OPM.
\item Types of inner speech: silent reading, repetitive, generative.
\item Data quantities: number of trials and sessions.
\item Decoding methods.
\end{enumerate}

Our focus is on enabling BCI applications rather than basic cognitive neuroscience of inner speech per se. We review relevant research on decoding inner speech using invasive and non-invasive neural recording methods in Section~\ref{sec:discussion_innerspeech}. Next, we describe our experimental paradigms and analysis methods for investigating non-invasive inner speech decoding.

\section{Methods}

\begin{figure}[!t]
  \centering
  \includegraphics[width=0.8\linewidth]{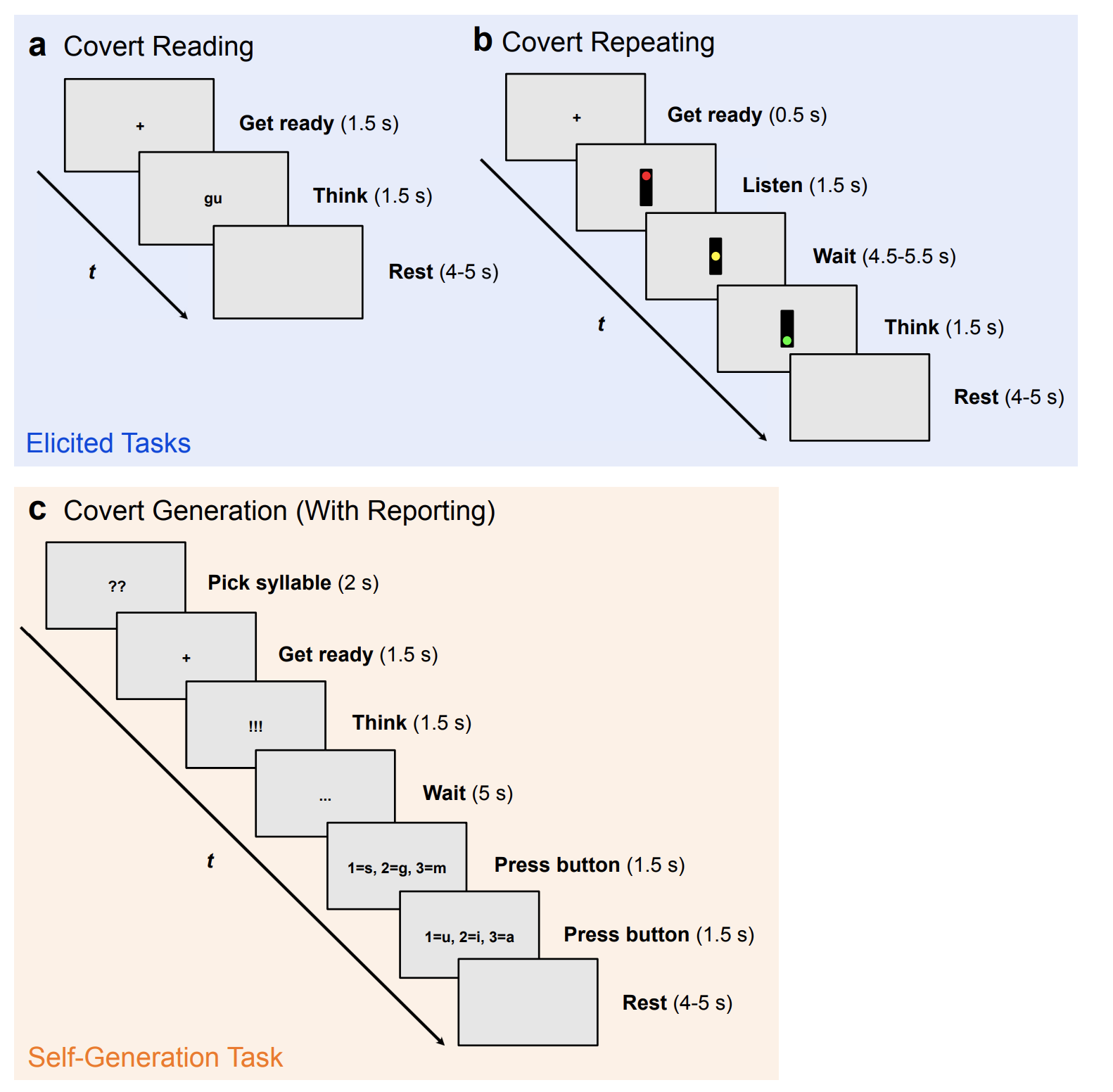}
    \caption{Visualisation of silent reading (a), repetitive (b) and generative inner speech (c) paradigms used in \citet{jones2021note}. Figure from \citet{jones2021note}.}
    \label{fig:oiwi_tasks}
\end{figure}

Our experimental paradigm follows previous efforts to delineate repetitive and generative inner speech. In a similar line of work \citet{jones2021note} investigate whether neural decoders trained on elicited inner speech data can be successfully transferred to decode self-generated inner speech (see Figure \ref{fig:oiwi_tasks} for task paradigms). Using fMRI data from one subject, the authors trained deep neural networks on a large dataset of elicited inner speech collected during covert reading and repeating tasks. They then tested these models on new self-generated inner speech data collected while the subject freely imagined syllables. The transferred decoders predicted unseen phonemes with high accuracy. The successful zero-shot task transfer demonstrates the viability of leveraging elicited speech to train models that can decode self-generated inner speech. This has practical significance for developing inner speech brain-computer interfaces, since elicited tasks allow collection of labelled training data even from locked-in patients.

While previous studies have investigated various language units such as characters, phonemes, words, and even whole phrases, we focused our experiments on the word level. A limited set of words already has direct benefits for potential patients. Working with words also allows scaling up to cover the entire language lexicon in future research. Since we are interested in potential clinical applications, we chose a set of words that could be most useful for patients: help, hungry, tired, pain, thirsty. Decoding at the individual rather than group level is critical for practical applications. Thus we conducted our experiments with a small number of participants. Selecting only 5 words also allows us to test how collecting a large number of trials per word can improve decoding performance.

We collected data across two versions of the experimental paradigm, adapting it as we gained results to better align with our objectives. In version 1, we collected 1-second silent reading trials followed by 4 consecutive 1-second repetitive and generative inner speech trials. We collected a small MEG dataset, concentrating efforts on obtaining a high number of EEG sessions from the same participant to assess between-session transferability. In version 2 we omitted inner speech entirely and collected only silent reading data across 4 modalities - two MEG systems (Elekta and CTF\footnote{\url{https://www.nottingham.ac.uk/research/groups/spmic/facilities/ctf-meg-scanner.aspx}}), EEG, and optically pumped magnetometers (OPMs). In version 1, we used an Elekta Neuromag Triux 306-channel system for MEG scans. We used a Neuroscan 64-channel cap for standalone EEG (standard 10-20 layout), with MEG-compatible Easycap EEG used for combined MEG and EEG.

For combined M/EEG, EEG ground and reference were on the left cheek and nose, respectively. For standalone EEG, the Cz and POz locations served as reference and ground, and we placed extra electrodes on the two mastoids. These could also serve as reference in offline analysis. Voltage and thus signal is always measured relative to a reference electrode in EEG. This means that the signal is the difference in voltage between the reference and other electrodes. Thus, the choice of reference greatly influences the characteristics of the recorded signal. It is best practice to place the reference electrode on the head but away from locations which might contain signals of interest. In our case both reference placements satisfy these criteria. While the signal shape and evoked responses will look different with different reference choices, this does not matter for decoding applications, as long as the signal of interest is not accidentally removed.

For most scans, we simultaneously collected electrooculogram (EOG) and electrocardiogram (ECG) data for easier artifact removal. ECG electrodes were placed on the wrists, with horizontal EOG on the outer side of the eyes and vertical EOG above and below the left eye. Electromyography (EMG) electrodes on the jaw monitored subtle mouth movements. Structural MRI scans were obtained for all participants in versions 1. During Elekta scans, we video recorded the mouth of participants to ensure no task-related motion. Eye tracking was performed for all MEG and EEG scans. We collected 5 minutes of resting state data before and after each scan. Stimuli were delivered via PsychoPy. The experiment was reviewed and approved by the Medical Sciences interdivisional research ethics committee at the University of Oxford (reference number R75957/RE001).

\subsection{Experimental paradigm}

\begin{figure}[!t]
    \centering
    \includegraphics[width=0.98\textwidth]{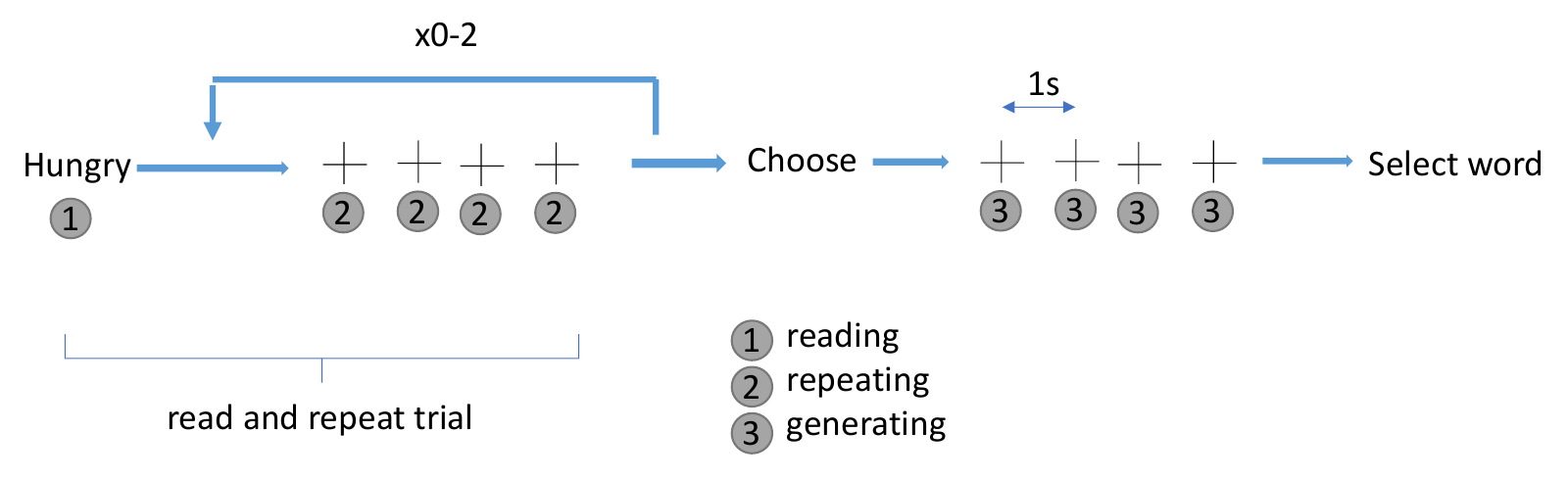}
    \caption{Paradigm for version 1 of our experiments. The participant silently {\it reads} `Hungry', then {\it repeats} it four times at 1-second intervals cued by crosses. This can repeat 0-2 times before {\it generating} a new word from the 5-word set at four 1-second cross cues, avoiding the previous read/repeat word(s).}
    \label{fig:version2_paradigm}
\end{figure}

\paragraph{Version 1} In the first version of the experiment, participants silently read words displayed individually on a screen, followed by 4 consecutive visual fixation-cross cues to covertly repeat the word (Figure \ref{fig:version2_paradigm}). This phase lasted approximately 5 minutes. In the next phase, participants continued reading and repeating words, but were now prompted after 0-2 read/repeat trials to imagine speaking a different word from the 5-word set (the generative inner speech task). 0-2 means that we randomly sampled either 0, 1, or 2 consecutive read and repeat trials. Similarly to the repetitive task we prompted the inner speech with 4 consecutive visual fixation-cross cues. Participants then indicated their imagined word with a button press. The random 0-2 read/repeat trials before each generative prompt ensured that participants did not pre-select words, better resembling unconstrained inner speech. We noticed that pure generative blocks let participants pre-plan words upon indicating the previous selection. Introducing random read/repeat trials limited this behaviour.

Each cross was displayed for 0.3 seconds followed by a 0.7 second-long blank screen. Word stimuli were displayed for 0.8-1.0 seconds, followed by a 0.8-1.0 second-long blank screen. Word order was randomised. The total second phase duration was approximately 50 minutes in 4 blocks with breaks between blocks. We collected simultaneous MEG and EEG data at the Oxford Centre for Human Brain Activity (OHBA). While we collected some combined MEG, we focused on obtaining multiple EEG-only sessions from one participant to assess between-session transferability.

\paragraph{Version 2} As the inner speech tasks yielded poor results, the second version focused solely on silent reading trials to collect more data within the 1-hour sessions. We expanded our aims and collected combined M/EEG at OHBA along with CTF-MEG and OPM data (same participants) in collaboration with the OPM Lab led by Matthew Brookes at the University of Nottingham. The simple paradigm displayed the 5 words randomly for 0.8-1.0 seconds with 0.8-1.0 second breaks. After every 10 trials, participants indicated the last word read to monitor attention.

\subsection{Analysis}
% include preprocessing
% simple methods for analysing/visualising the data

\begin{figure}[!t]
\begin{subfigure}{0.33\textwidth}
  \centering
  \includegraphics[width=1.0\linewidth]{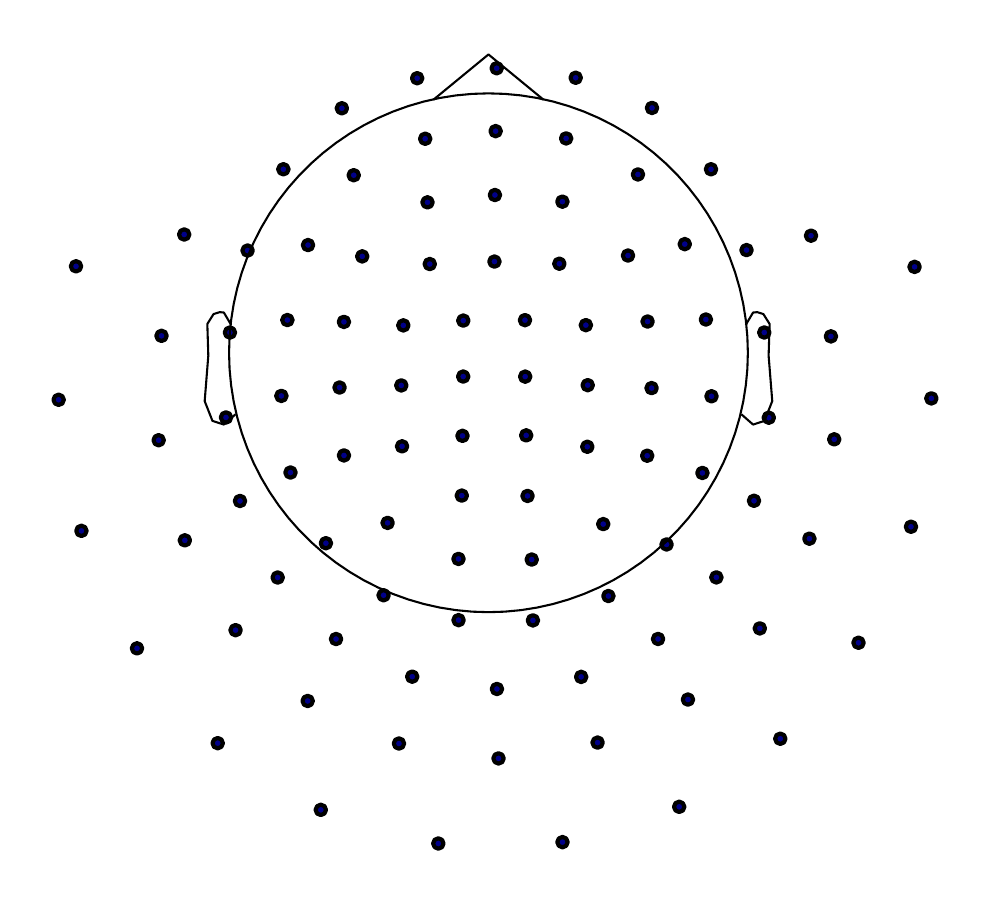}
  \caption{Elekta}
  \label{fig:elekta_sensors}
\end{subfigure}%
\begin{subfigure}{0.33\textwidth}
  \centering
  \includegraphics[width=1.0\linewidth]{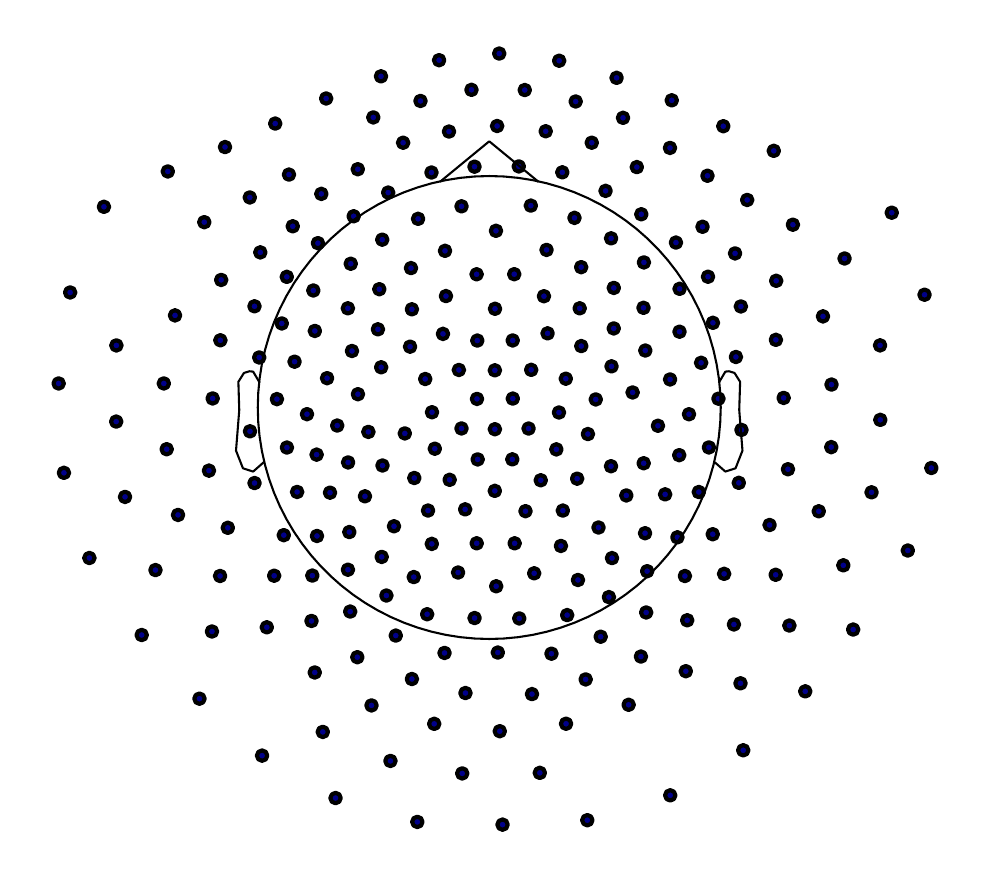}
  \caption{CTF}
  \label{fig:ctf_sensors}
\end{subfigure}%
\begin{subfigure}{0.33\textwidth}
  \centering
  \includegraphics[width=1.0\linewidth]{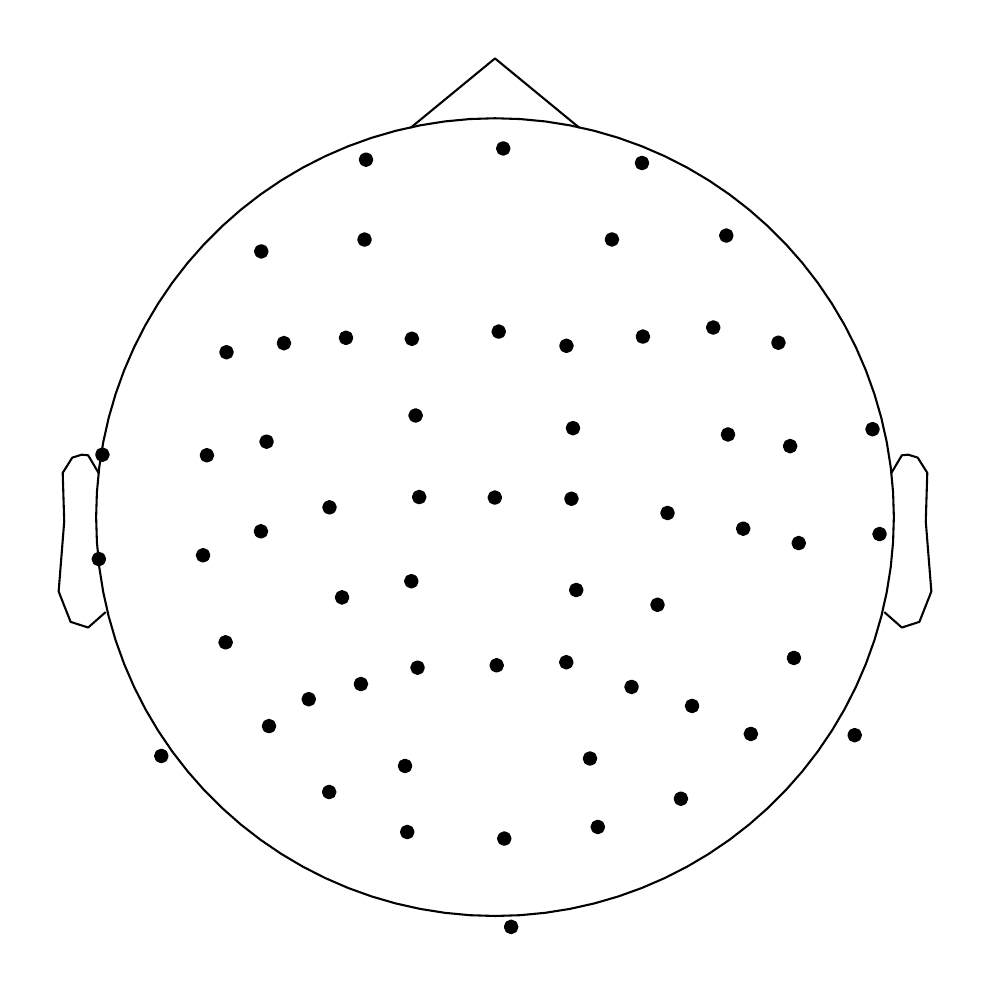}
  \caption{EEG}
  \label{fig:eeg_sensors}
\end{subfigure}
\caption{Sensor locations across scanning systems. CTF contained 1 gradiometer per location, while Elekta had 2 gradiometers and 1 magnetometer. Please note that OPM sensor layouts are reported in Section~\ref{ssec:data_stats}.}
\label{fig:sensor_plots}
\end{figure}

\paragraph{Data acquisition and preprocessing} % added paragraph to separate acquisition/preprocessing
Elekta and EEG data were acquired at a sampling rate of 1000 Hz with a built-in bandpass filter between 0.03 and 330 Hz, while CTF and OPM scans were acquired at a sampling rate of 1200 Hz with a built-in lowpass filter of 600 Hz. The Elekta system contained 102 magnetometers and 204 planar gradiometers (102 sensors x 2 orientations), totalling 306 channels. CTF contained 265 axial gradiometers. Sensor configurations for three modalities are illustrated in Figure~\ref{fig:sensor_plots}. OPM data were recorded using a variable number of triaxial magnetometers (typically 150-180) in 60 fixed scalp locations, measuring magnetic fields along orthogonal axes. Channel configurations for individual participants are reported in Section~\ref{ssec:data_stats}.

Elekta data were preprocessed using Maxwell filtering for movement compensation and signal space separation using the MaxFilter algorithm \citep{taulu2006spatiotemporal}. Noisy OPM channels identified during recording were removed prior to analysis. For all systems, data were bandpass filtered (1-25 Hz typically, with higher lowpass for some experiments), notch filtered at 50 and 100 Hz, and subjected to automated bad channel detection (except OPM) using the oslpy package\footnote{\url{https://github.com/OHBA-analysis/oslpy}} \citep{oslpy}. Bad segments were identified via a multi-pass procedure across progressively wider temporal windows (200 to 800 ms) with a significance threshold of 0.1. Independent component analysis was applied for dimensionality reduction (64 components for MEG/OPM, 32 for EEG). Components reflecting ocular or cardiac artifacts were removed before downsampling to 100 Hz and epoching.

An additional mean field correction was applied to OPM data after preprocessing:

\begin{align}
\mathbf{O} &= \begin{pmatrix} \mathbf{O}_x\ \mathbf{O}_y\ \mathbf{O}_z \end{pmatrix} \\
\mathbf{M} &= \mathbf{I} - \mathbf{O} \mathbf{O}^\dagger\\
\mathbf{X}_m &= \mathbf{M} \mathbf{X}
\end{align}

Where $\mathbf{O}_x$, $\mathbf{O}_y$, $\mathbf{O}_z$ are sensor orientation vectors, $\mathbf{O}$ stacks them vertically, $\mathbf{I}$ is the identity matrix, and $\mathbf{M}$ is the final mixing matrix applied to the data $\mathbf{X}$. The aim of this transformation is to remove any spatially homogeneous field not coming from the brain and is described in detail in \citet{tierney2021modelling}. Elekta data already includes such corrections in the MaxFilter algorithm, and CTFs do not need it as they measure the field gradient with gradiometers only.

Basic analyses involved comparison of evoked responses across channels, conditions, sessions, and modalities. Single-trial covariance matrices were also computed and visualised using t-SNE \citep{Maaten:2008}.

\paragraph{Decoding}
% main decoding methods tried

We employed several decoding methods for both the silent reading and inner speech tasks. A standard pipeline for silent reading involved our full-epoch LDA and neural network models from Chapter~\ref{Chap3}. For inner speech we tried several methods, but mainly used the covariance matrix over each trial as features for an LDA model. Channels were standardised prior to decoding. We tried concatenating across the 4 consecutive trials to form 4-second epochs, as well as averaging over trials, before decoding. Specific methods and hyperparameters are detailed in the Results.

\section{Results}
\subsection{Data statistics}
\label{ssec:data_stats}
% summarise for all versions

A total of 4 male participants (P2, P4, P5, P6) between the ages of 20 and 40 participated across the two versions of our study. The participant pool included both native and non-native English speakers, though all had C2 level proficiency in English. Participant 4 (P4) is the author of this thesis and a non-native English speaker.

\begin{figure}[!t]
\begin{subfigure}{0.33\textwidth}
  \centering
  \includegraphics[width=1.0\linewidth]{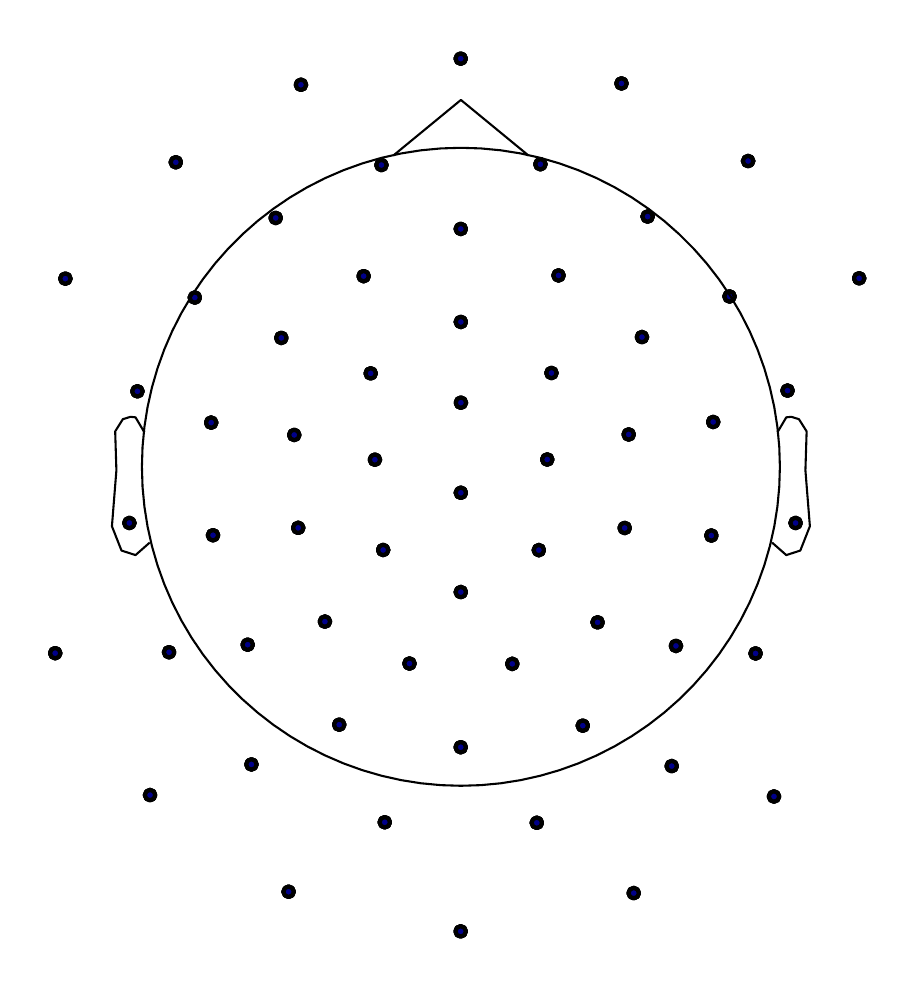}
  \caption{P4}
  \label{fig:opm_sensors_rich}
\end{subfigure}%
\begin{subfigure}{0.33\textwidth}
  \centering
  \includegraphics[width=1.0\linewidth]{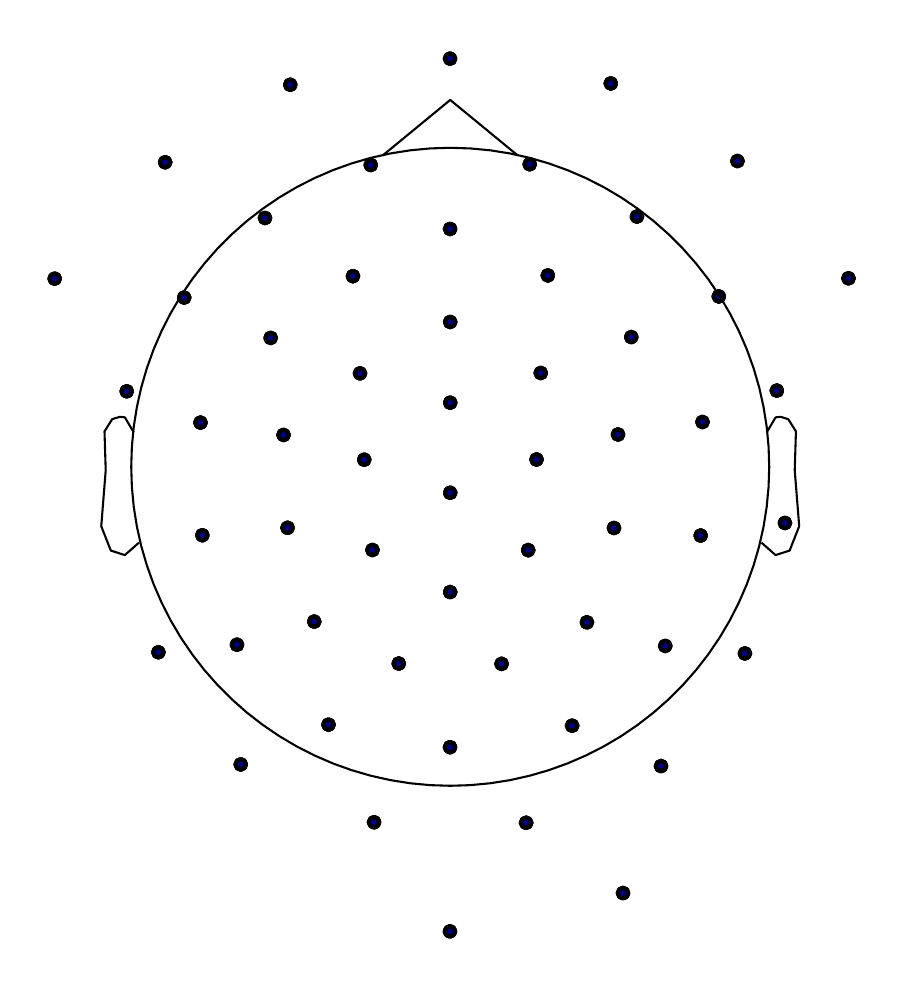}
  \caption{P5}
  \label{fig:opm_sensors_lukas}
\end{subfigure}
\begin{subfigure}{0.33\textwidth}
  \centering
  \includegraphics[width=1.0\linewidth]{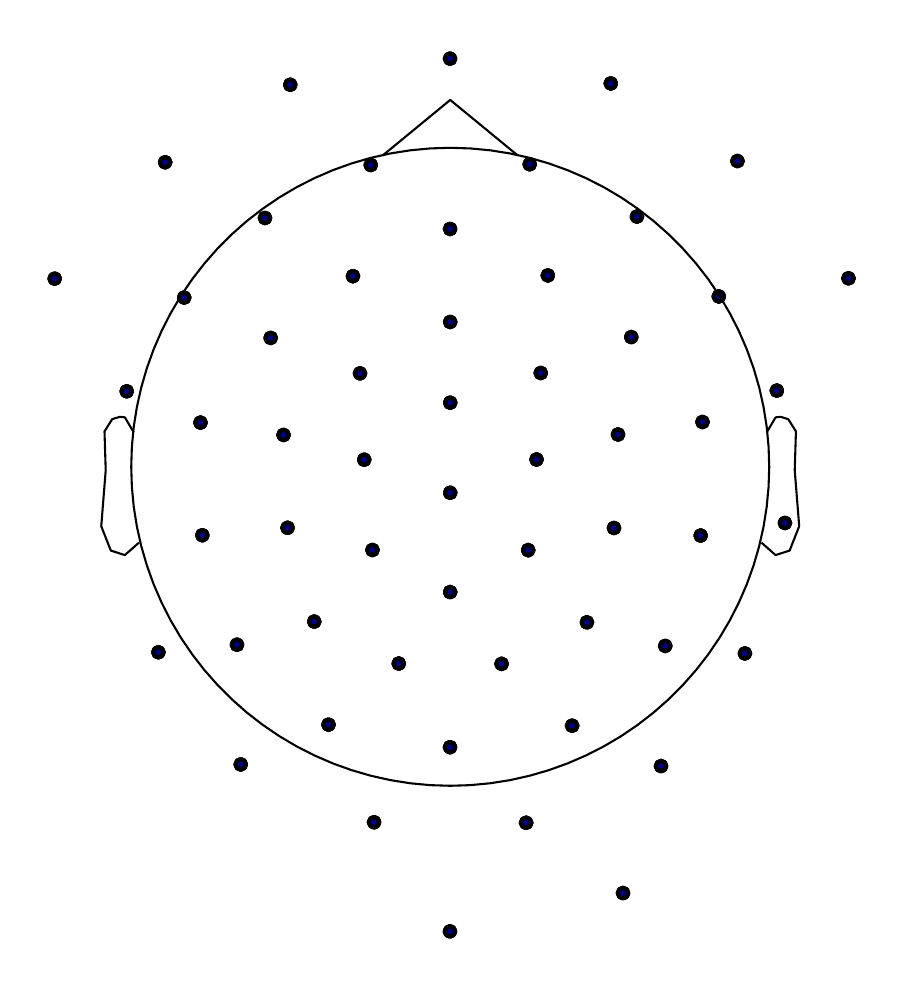}
  \caption{P6}
  \label{fig:opm_sensors_ryan}
\end{subfigure}
\caption{OPM sensor configurations across the three participants in version 2 of the experiment. Each location contained an OPM sensor measuring the magnetic field in three orthogonal directions. Sensor layouts and number of sensors are different due to technical difficulties with operating all sensors without overheating, excessive noise, or other issues.}
\label{fig:opm_sensors}
\end{figure}

The two versions of the experiment were conducted with different goals. Version 1 involved evaluating the feasibility of decoding inner speech in 3 participants. Since EEG and MEG provided comparable decoding accuracy, 10 EEG sessions were collected from P4 in version 1 to examine improvements from increased data size and test decoder adaptability across sessions. In version 2, the inner speech task was removed and silent reading trials were collected using combined M/EEG, CTF, and OPMs across 3 participants. P4 and P5 also participated in version 1. The high number of silent reading trials (1250 per session) enabled thorough investigation of this paradigm. The OPM sensor layouts are shown in Figure \ref{fig:opm_sensors}.

Tables \ref{table:data_stats2} and \ref{table:data_stats3} summarise the number of sessions and trials for the participants in each version. While target numbers of trials are reported, minor variations occurred due to randomisation. The extensive datasets collected enabled thorough investigation of silent reading and inner speech paradigms. The multiple sessions also allow examination of between-session and between-modality variability. Overall, the dataset provides a unique resource to advance decoding of covert speech from non-invasive electrophysiological signals.

\begin{table}[t!]
\centering
        \begin{tabular}{l|cc}
        %\toprule
             &\bf MEG & \bf EEG  \\ \midrule
            
            Participant 2 (P2) & \multicolumn{2}{c}{1}  \\ 
            Participant 4 (P4) & 1 & 10  \\
            Participant 5 (P5) & 1 & 1 \\ \midrule
            total silent reading trials & 519 & 2076  \\
            total repetitive inner speech trials & 2076 & 8304 \\
            total generative inner speech trials & 1920 & 7680 
            
            %\bottomrule
        \end{tabular}
    \caption{\label{table:data_stats2} Number of sessions for each participant in version 1 of the study (top 3 rows). Total number of trials is given across all sessions and participants. Number of trials may be slightly lower or higher than shown due to randomness. Note that for P2 we conducted a combined M/EEG session while for the other participants MEG and EEG scans were separate.}
\end{table}

\begin{table}[t!]
\centering
        \begin{tabular}{l|ccc}
        %\toprule
             &\bf combined M/EEG &\bf CTF &\bf OPM   \\ \midrule
            
            Participant 4 (P4) & 1 & 1 & 1  \\ 
            Participant 5 (P5) & 1 & 1 & 1  \\
            Participant 6 (P6) & 1 & 1 & 1 \\ \midrule
            total silent reading trials & 3750 & 3750 & 3750  \\

            %\bottomrule
        \end{tabular}
    \caption{\label{table:data_stats3} Number of sessions for each participant in version 2 of the study (top 3 rows). Total number of trials is given across all sessions and participants.}
\end{table}

\subsection{Data analysis}
%non-decoding stuff

\begin{figure}[!t]
    \centering
    \includegraphics[width=0.6\textwidth]{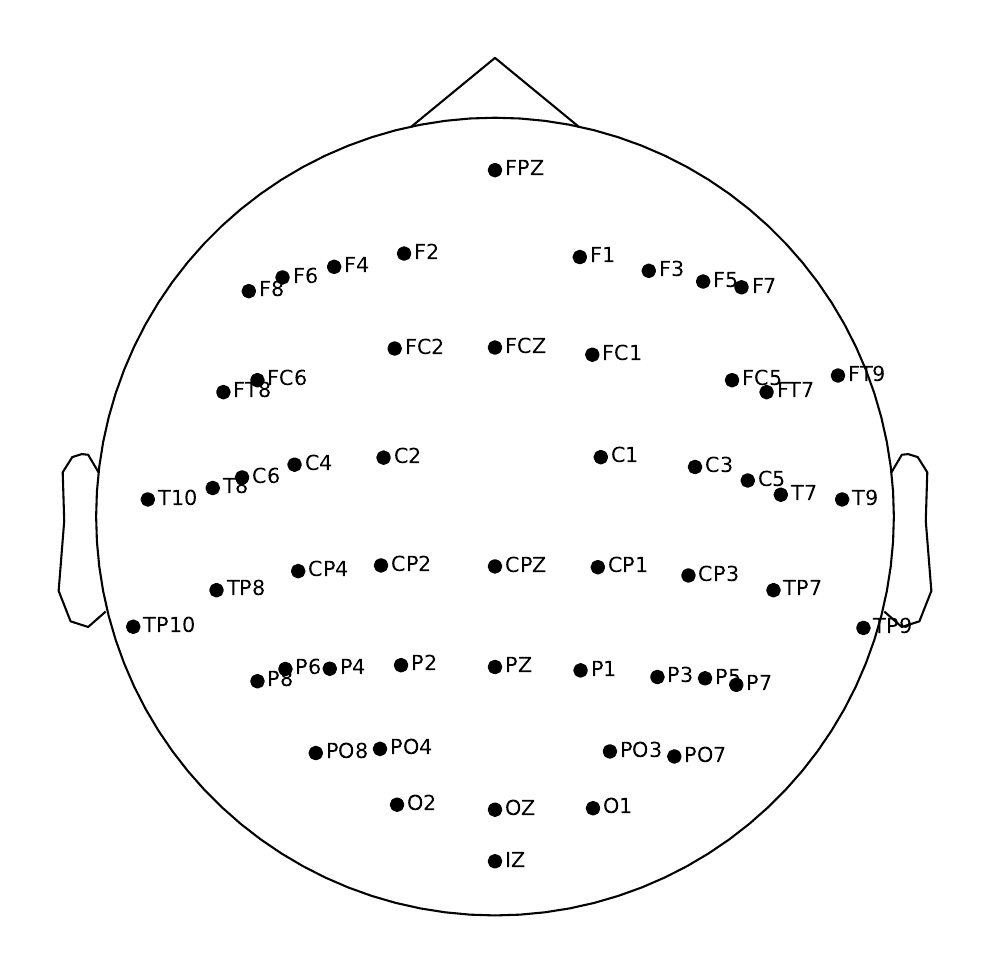}
    \caption{EEG electrode locations for P4 in version 1 of the experiment.}
    \label{fig:eeg_sensors_names}
\end{figure}

In this section we present our non-decoding analyses of the collected data. The aim of this analysis is to validate data quality and uncover any insights into differences between tasks. These visualisations were primarily conducted on the 10 EEG sessions of P4, as this participant had the most inner speech trials (version 1 of the study) and sessions, allowing investigations into between-session variability. For the visualisations in this section, no independent component analysis (ICA) artifact removal was performed on the data. We plot the electrode positions and their names over the scalp in Figure~\ref{fig:eeg_sensors_names}.

\begin{figure}[!t]
    \centering
    \includegraphics[width=0.75\textwidth]{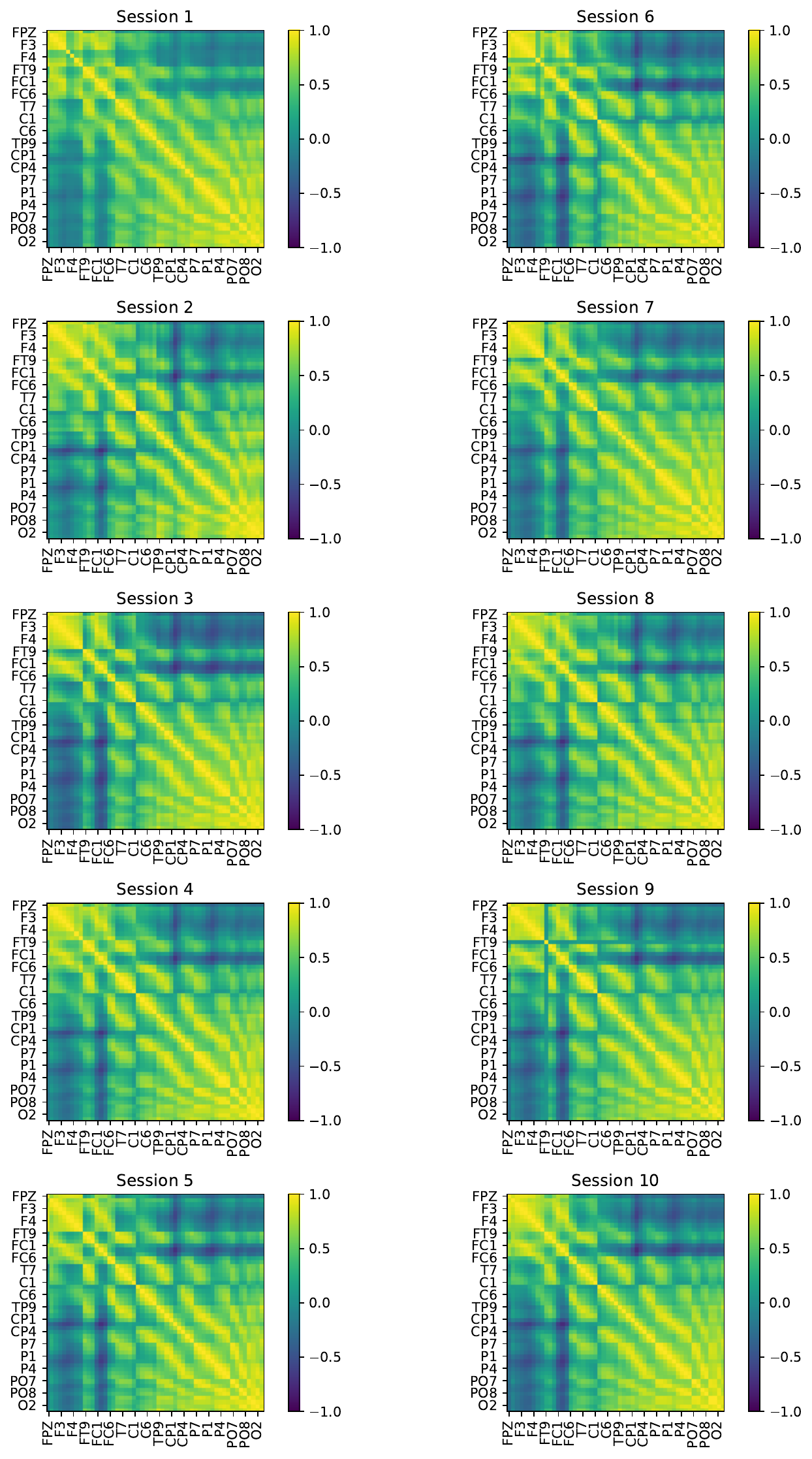}
    \caption{Averaged trial-covariances across the 10 EEG sessions of P4 in version 1 of the experiment. Each matrix represents a different session.}
    \label{fig:eeg_covariances}
\end{figure}

\begin{figure}[!t]
    \centering
    \includegraphics[width=0.5\textwidth]{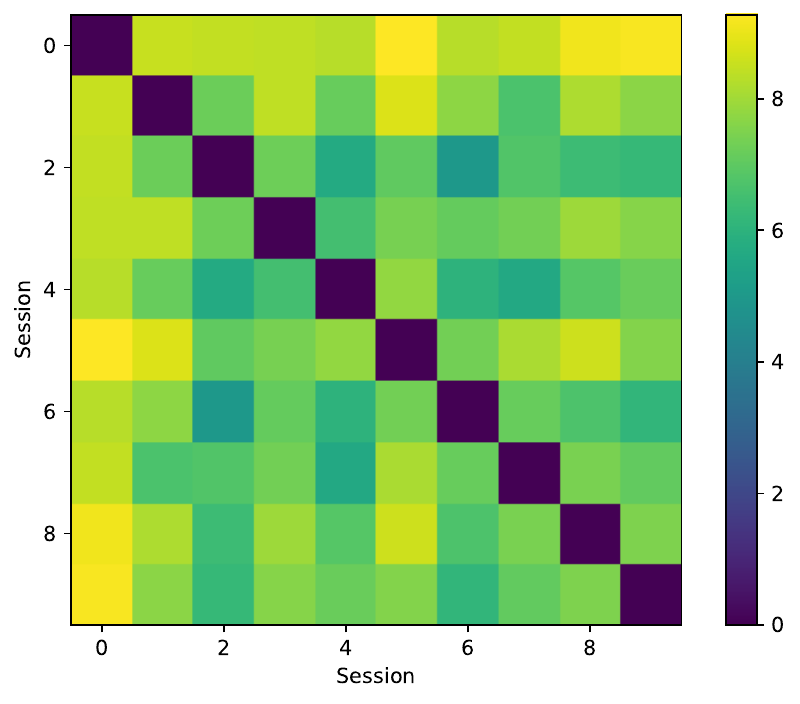}
    \caption{Riemann distance matrix between the average session-covariances across the 10 EEG sessions of P4 in version 1.}
    \label{fig:eeg_covariances_riemann}
\end{figure}

To visualise between-session variability, we compute the covariance across each 1-second inner speech trial, and average these across individual sessions. Figure~\ref{fig:eeg_covariances} exhibits the averaged covariance of each session. Channels demonstrate high covariance within brain regions, such as the frontal and visual areas. Across sessions, average covariances appear similar. To quantify similarity between sessions, we computed the Riemannian distances of the covariances between pairs of sessions for all possible pairs. This produces a session-by-session distance matrix (Figure~\ref{fig:eeg_covariances_riemann}). This can provide insight into between-session differences. For instance, the first session seems quite distant from the other sessions. This can mean that a decoder trained on other sessions may not perform very well on this session.

\begin{figure}[!t]
    \centering
    \includegraphics[width=0.99\textwidth]{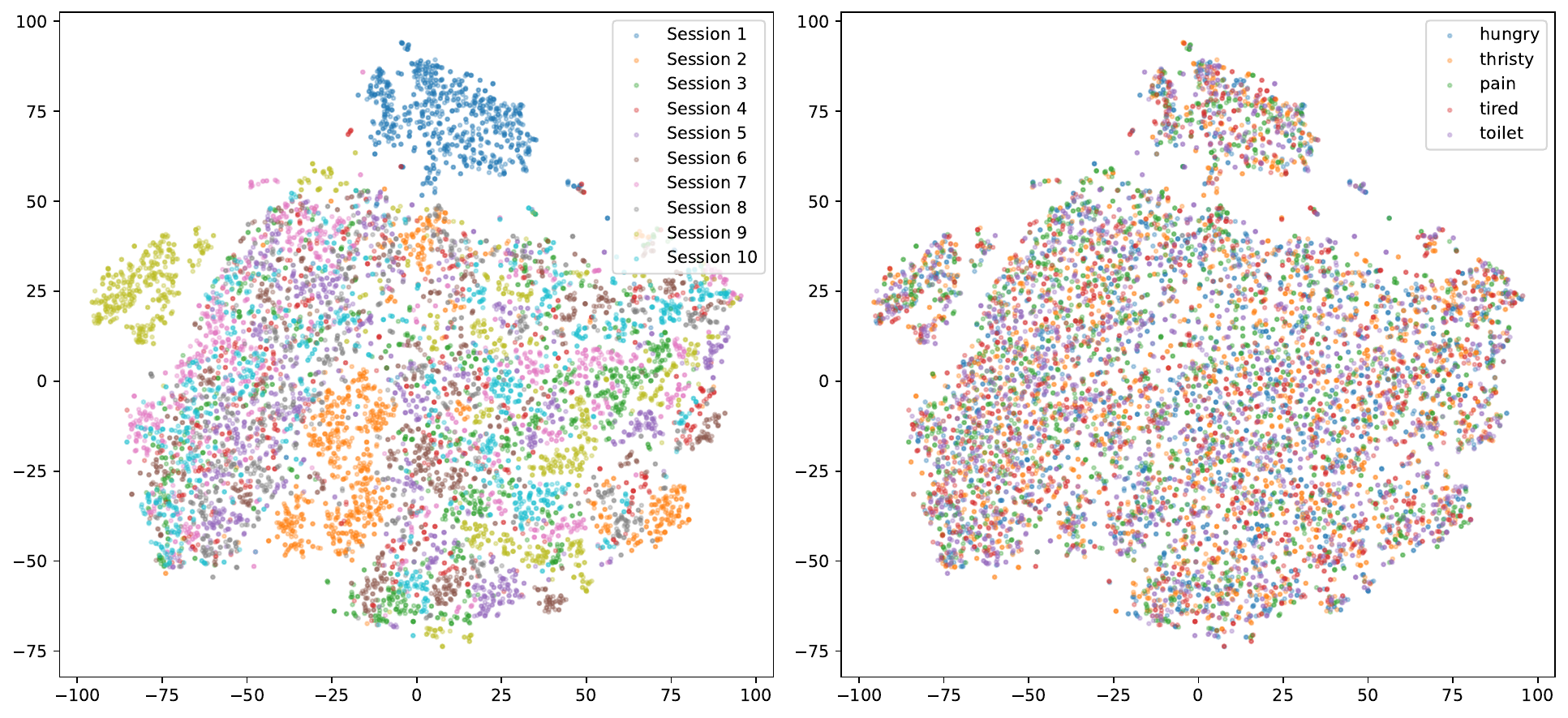}
    \caption{t-SNE projection of the per-trial covariances across the 10 EEG sessions of P4 in version 1. These are coloured according to the session label on the left, and according to the condition (word) on the right.}
    \label{fig:eeg_tsne}
\end{figure}

Finally, we investigated whether the covariance representations demonstrate interesting structure when visualised in 2D. To this end, we simply applied t-SNE to the individual trial covariances to project them into 2 dimensions and visualised the result. Figure~\ref{fig:eeg_tsne} portrays this projection with two types of labelling. When trials are labelled by their corresponding condition (word), no apparent clustering emerges. This further bolsters that differentiating between words in inner speech is challenging. Structure can be discerned in the projection when labelled by session. This is anticipated since trials within one session are more similar to each other than to other sessions. These findings imply that decoding inner speech may be an equally challenging endeavour. An investigation of evoked responses is provided in Appendix~\ref{ssec:inner_speech_evoked}.

\subsection{Decoding inner speech in experiment version 1}
While we collected reading data in all experiments, we will only present results from version 2, since that version was specifically aimed at analysing the silent reading task. In this section we will first analyse the MEG data, followed by results from the 10 EEG sessions of P4.

On the MEG data we attempted the methods from Chapter~\ref{Chap3}, such as sliding window LDA, full-epoch LDA, and the linear neural network. Running full-epoch models on the 1-second inner speech trials yielded chance-level results when decoding which word is being used in inner speech. We also attempted sliding-window decoding on a subset of MEG channels overlying the language area. However, the decoding accuracy timecourse exhibited substantial fluctuations and never exceeded 24\%, where 20\% is chance level. Thus, this is also a negative result.

\begin{figure}[!t]
    \centering
    \includegraphics[width=0.95\textwidth]{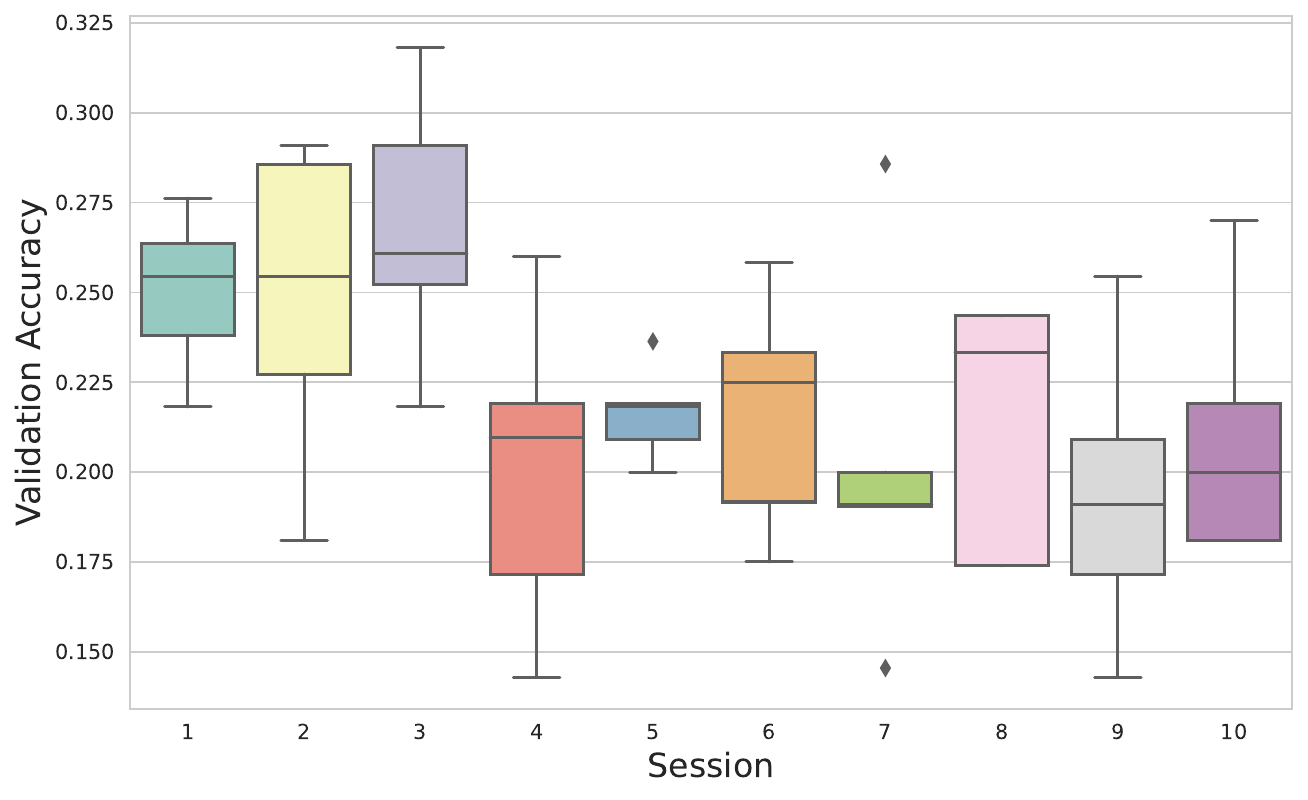}
    \caption{Validation accuracy distributions across the 5 folds of the 10 EEG sessions of P4 in experiment version 1. Separate LDA models are trained and evaluated on each fold and session to decode which of the 5 words is being used in the 1-second inner speech trials. Chance level is 0.2.}
    \label{fig:eeg_accs}
\end{figure}

On the EEG data of the generative inner speech trials of P4, LDA models (see Section~\ref{ssec:background_decoding}) were trained on each session utilising the channel-covariance over the 1-second epoch as features with 5-fold cross-validation. For this analysis preprocessing involved a 1-40 Hz bandpass filter and no ICA artefact removal was employed. Before computing covariance, trials were normalised to unit variance and zero mean. We found above 25\% validation accuracy in only 3 sessions (Figure~\ref{fig:eeg_accs}), with chance level being 20\%. However, when correcting for multiple comparisons none of the sessions had significantly better performance than chance. It may be that by running more cross-validation folds the performance in some sessions reaches significance. Decoding the repetitive inner speech trials, or both types together did not produce better results.

Nevertheless, we trained a single LDA model across the 3 best sessions, achieving 33\% cross-validation accuracy. The same per-session folds were employed as in the previous analysis. To train a single LDA model across sessions, we made some modifications to the decoding pipeline. Rather than using the 1-second trial, we utilised the entire 4-second epoch with the four consecutive cues to compute covariance. To account for between-session differences, the mean session-level evoked response was subtracted from each trial before covariance computation, and the mean session-level covariance was also subtracted from each trial's covariance. Since we evaluated many different methods on this data, and we selected these 3 sessions based on the previous analysis, there is a risk this result is inflated. Running the same decoding approach on all 10 sessions reduced cross-validated accuracy to 23.2\%.

While the EEG data provides promising results, with some sessions exhibiting above-chance decoding accuracy, we must be cautious about drawing robust conclusions, due to the limited performance and risk of overfitting. The limited inner speech performance precluded assessment of transfer between silent reading and inner speech tasks. Evaluating transferability between sessions was also not feasible as most sessions displayed chance-level performance.

\subsection{Decoding silent reading in experiment version 2}

\begin{figure}[!t]
    \centering
    \includegraphics[width=0.9\textwidth]{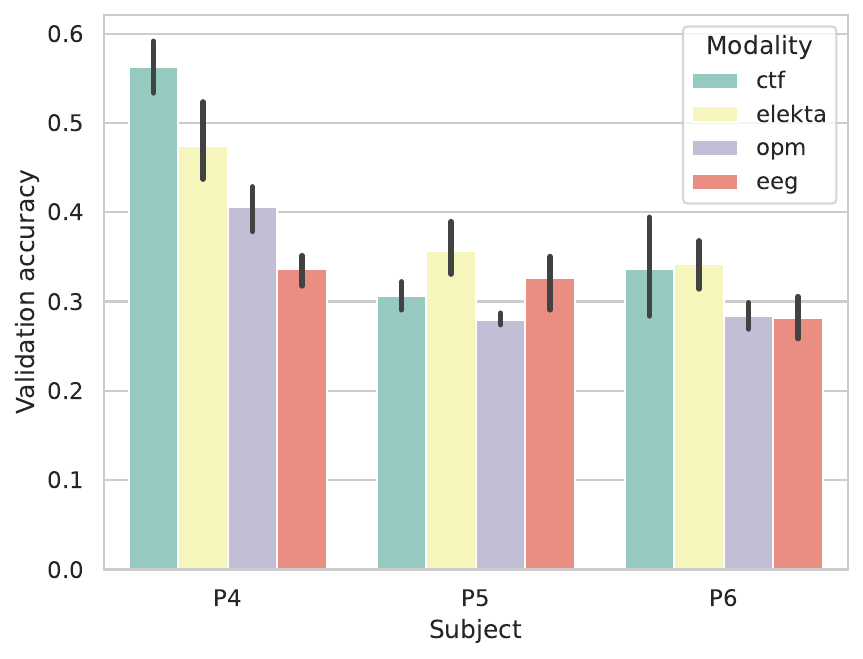}
    \caption{Validation accuracy (across 5 folds) for each session in experiment version 2. Separate LDA-NN (see Chapter~\ref{Chap3}) models are trained and evaluated on each fold and session to decode which word is presented during the 1-second trials. Black bars indicate 95\% confidence interval. Chance level is 0.2 due to having 5 words with equal trial counts.}
    \label{fig:reading_only_accs}
\end{figure}

In this version, we collected a substantial number of silent reading trials only from 3 participants, across 4 modalities. For each session, we implemented the LDA-NN approach from Chapter~\ref{Chap3} with 5-fold cross-validation (Figure~\ref{fig:reading_only_accs}). We utilised the 500 ms following word onset as our examples for decoding. For CTF, Elekta, and OPM data, the dimensionality of the LDA-NN reduction was set to 50, and for EEG to 20.

Accuracies are generally low, providing evidence that even with the visual component and numerous trials, silent reading is a challenging decoding task. P4 exhibits higher performance than the other two participants across all modalities except EEG. While CTF data achieved the best performance for P4, followed by Elekta, OPM and EEG, this is not the case for the other participants. Across modalities, P5 and P6 display more comparable performance. CTF and Elekta appear higher, while OPM and EEG are slightly lower but similar (especially for P6). The discrepancy between the CTF results for P4 and P6/P5 is particularly surprising. As depicted in Figure~\ref{fig:opm_sensors}, unfortunately, due to experimental difficulties, the OPM sensor coverage of the visual area was inferior in these participants compared to P4. This could potentially explain the lower OPM performance.

It is difficult to derive conclusions from these results, and the experiment should be replicated across more subjects to enable a more robust comparison between modalities. It seems that traditional MEG scanners exhibit the best performance, while EEG and OPM lag behind but are comparable. This provides evidence that challenging decoding tasks such as silent reading are feasible with OPMs. Further innovation and better spatial coverage should enhance OPM decoding performance to approach traditional MEG.

\begin{figure}[!t]
\begin{subfigure}{0.33\textwidth}
  \centering
  \includegraphics[width=0.98\linewidth]{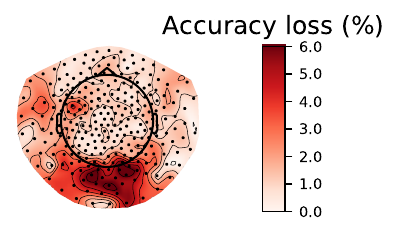}
  \caption{CTF}
  \label{fig:reading_only_ctf_pfich}
\end{subfigure}%
\begin{subfigure}{0.33\textwidth}
  \centering
  \includegraphics[width=0.98\linewidth]{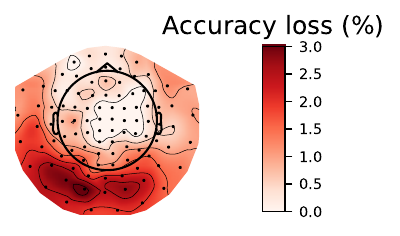}
  \caption{Elekta}
  \label{fig:reading_only_elekta_pfich}
\end{subfigure}%
\begin{subfigure}{0.33\textwidth}
  \centering
  \includegraphics[width=0.98\linewidth]{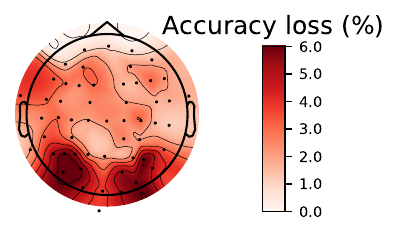}
  \caption{EEG}
  \label{fig:reading_only_eeg_pfich}
\end{subfigure}
\caption{Sensor importance maps averaged across subjects for 3 modalities in experiment version 2. The importance maps are obtained by running spatial PFI on the trained LDA-NN decoding models (see Chapter~\ref{Chap3} for methods). Darker red shading indicates higher accuracy loss and thus higher stimulus-related information content.}
\label{fig:reading_only_pfich}
\end{figure}

\begin{figure}[!t]
\begin{subfigure}{0.33\textwidth}
  \centering
  \includegraphics[width=0.98\linewidth]{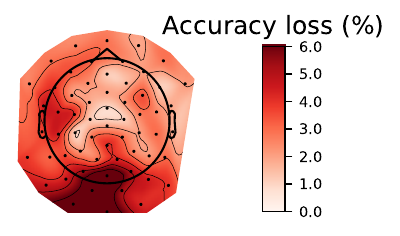}
  \caption{P4}
  \label{fig:reading_only_opm_pfich_rich}
\end{subfigure}%
\begin{subfigure}{0.33\textwidth}
  \centering
  \includegraphics[width=0.98\linewidth]{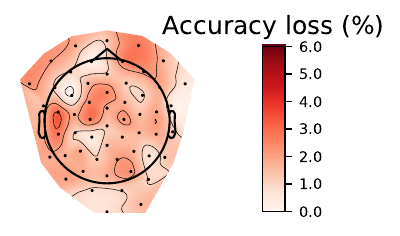}
  \caption{P5}
  \label{fig:reading_only_opm_pfich_lukas}
\end{subfigure}%
\begin{subfigure}{0.33\textwidth}
  \centering
  \includegraphics[width=0.98\linewidth]{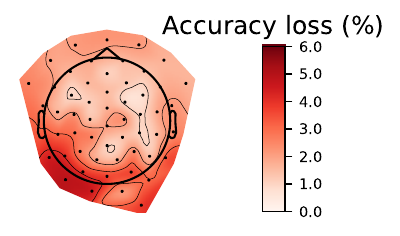}
  \caption{P6}
  \label{fig:reading_only_opm_pfich_ryan}
\end{subfigure}
\caption{Sensor importance maps across subjects on the OPM recordings of experiment version 2. The importance maps are obtained by running spatial PFI on the trained LDA-NN decoding models (see Chapter~\ref{Chap3} for methods). Darker red shading indicates higher accuracy loss and thus higher stimulus-related information content. Note that P5 and P6 had less channels available, hence the smaller topographic map.}
\label{fig:reading_only_pfich_opm}
\end{figure}

Next, we investigated the temporal and spatial PFI of the decoding models. We followed the methodology presented in Chapter~\ref{Chap3}.  We expect that PFI should appear similar across modalities. When plotting the spatial PFI for each modality we average across subjects and cross-validation folds. We utilised 20 permutations and set the number of nearby sensor locations for the spatial window to 4 in all modalities. Note this equates to 12 channels for Elekta, 8 channels for CTF, and 12 channels for OPMs, since these modalities contained multiple sensors at the same site. We depict the spatial PFI of Elekta, CTF, and EEG in Figure~\ref{fig:reading_only_pfich}. It is clear that the visual area drives decoding across all modalities. We plot the OPM accuracy loss maps separately for each participant due to variability in available sensors (Figure~\ref{fig:reading_only_pfich_opm}). This demonstrates similar visual importance in P4 and P6. PFI was ineffective in P5, possibly due to limited decoding performance.

\begin{figure}[!t]
\begin{subfigure}{0.33\textwidth}
  \centering
  \includegraphics[width=0.98\linewidth]{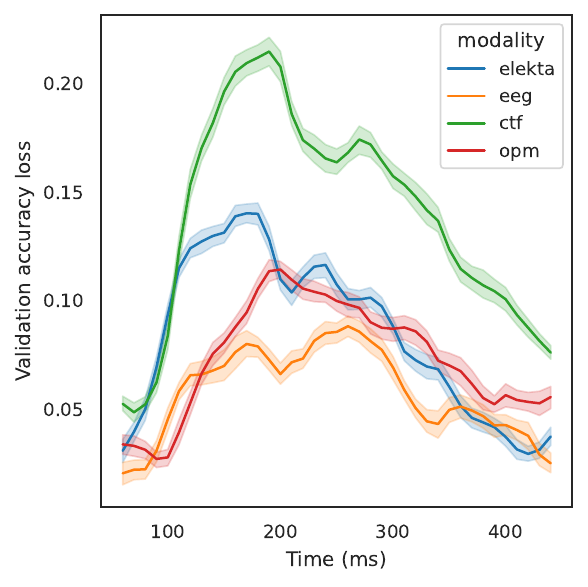}
  \caption{P4}
  \label{fig:reading_only_pfits_rich}
\end{subfigure}%
\begin{subfigure}{0.33\textwidth}
  \centering
  \includegraphics[width=0.98\linewidth]{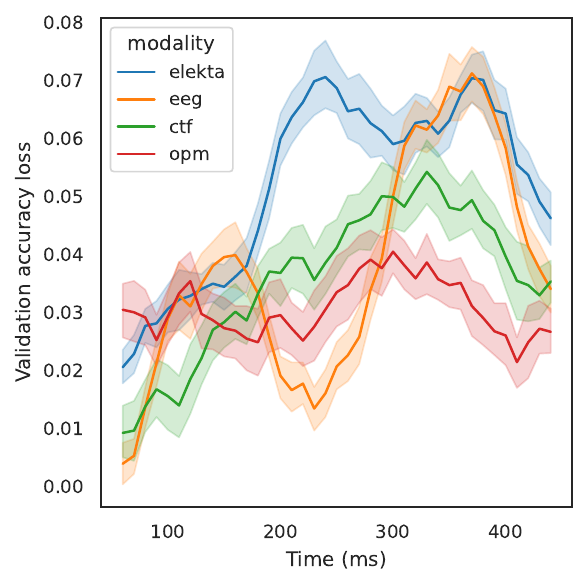}
  \caption{P5}
  \label{fig:reading_only_pfits_lukas}
\end{subfigure}%
\begin{subfigure}{0.33\textwidth}
  \centering
  \includegraphics[width=0.98\linewidth]{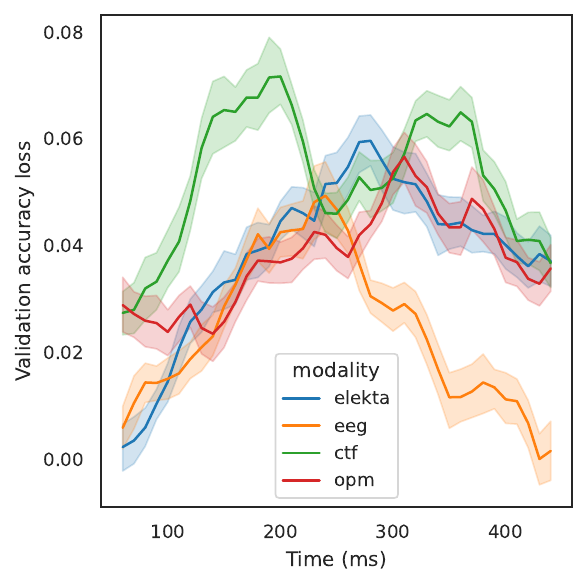}
  \caption{P6}
  \label{fig:reading_only_pfits_ryan}
\end{subfigure}
\caption{Temporal PFI across the 3 subjects (P4, P5, P6) and 4 modalities (lines with different colours) in experiment version 2. The timecourses are obtained by running temporal PFI on the trained LDA-NN decoding models (see Chapter~\ref{Chap3} for methods). Shading indicates 95\% confidence interval across PFI permutations. The horizontal axis indicates time since stimulus onset.}
\label{fig:reading_only_pfits}
\end{figure}

Finally, we illustrate temporal PFI across subjects and modalities in Figure~\ref{fig:reading_only_pfits}. We utilised 20 permutations and a temporal window of 100 ms. Since decoding is driven by the visual region, we would anticipate peak accuracy loss around 150ms, which is indeed evident for P4. This subject also exhibits much less noisy timecourses. Across modalities, timecourses appear similar. Notably, all modalities except OPM display a second, smaller peak around 250ms. In other subjects, accuracy loss peaks later in the trial. This could reflect slower reaction times when silently reading. Interestingly, the CTF data for P6 and the EEG data for P5 exhibit two distinct peaks. This could indicate decoding is driven by both visual word processing and language processing due to silent reading. These plots highlight substantial between-subject variability in task-related brain-activity.

\section{Discussion}
\label{sec:discussion_innerspeech}
% very briefly negative results and methods tried

In this discussion we will first present and review related works for decoding inner speech from invasive and non-invasive modalities. Then in Section~\ref{ssec:inner_speech_conclusion} we summarise our findings and conclude this chapter.

Typical experimental paradigms used in neuroimaging studies of inner speech include silent word repetition and silent reading, similar to ours. As reviewed by \citet{geva2018inner}, these studies consistently demonstrate that compared to overt speech conditions, inner speech leads to reduced activation in motor regions like primary motor cortex and auditory sensory areas like primary auditory cortex. Inner speech also shows less engagement of sensory feedback regions like the superior temporal sulcus. However, inner speech robustly activates left hemisphere language regions. The authors suggest this enhanced activation in phonological and semantic regions may serve as an automatic compensatory mechanism that augments inner speech performance given the reduced sensory and motor feedback.

However, accurately interpreting the results of neuroimaging studies on inner speech remains challenging. Methodological limitations include controlling for overt speech production during inner speech tasks and ensuring participants are actually using inner speech rather than alternative cognitive strategies. Research on visual and motor imagery highlights that imagery across modalities relies on networks similar to actual perception and action \citep{kosslyn1994image}, reflecting shared neural processing. However, identifying a consistent substrate for imagery across all sensory modalities has proven more difficult.

When comparing invasive and non-invasive methods, the only major differences are in the brain signals and noise characteristics of each recording modality. The cognitive process of interest - inner speech itself - remains constant across invasive and non-invasive recordings. Invasive methods can record neural activity with much higher spatial resolution and signal-to-noise ratio (SNR). Thus, the key question becomes determining how reduced resolution and SNR in non-invasive EEG or MEG impacts detectability of inner speech processes. We discuss relevant invasive research of inner speech next.

\subsection{Invasive methods}
\label{ssec:invasive}

Inner speech has been most successfully studied through invasive methods. Electrocorticography (ECoG), where electrode arrays are placed below the skull directly on the brain surface, is one such technique. More invasive methods involve implanting electrodes within cortical tissue to record single neuron activity.

\citet{martin2016word} demonstrate one of the first successful decodings of individual words during imagined speech from direct cortical recordings in humans. Their study design involved recording high gamma activity using ECoG during listening, overt speech, and imagined speech conditions for 6 different words. They developed a binary classification approach using support vector machines that incorporated dynamic time warping (DTW) to account for temporal variability in speech production. At the group level, classification accuracy was significantly above chance for imagined speech, with the best word pair reaching 88\% accuracy. However, across all word pairs accuracy was much lower, and performance was variable between subjects. Discriminative information was located primarily in the superior temporal gyrus, inferior frontal gyrus, and sensorimotor cortex, consistent with their role in speech processing \citep{hickok2007cortical}. This work can inform future noninvasive research on which brain areas to focus on and employ techniques like DTW to handle temporal variability.

Recent work by \citet{wandelt2022online} demonstrates the feasibility of decoding internal speech from single neuron activity in the supramarginal gyrus (SMG) and somatosensory cortex (S1) of a tetraplegic human participant. Their study design allowed comparison of neural activity between visual word reading, listening, vocalised, and inner speech across 8 words. The authors found individual SMG neurons showed selective tuning to specific words during the internal speech condition. Using these neural signals, they achieved up to 91\% decoding accuracy for internal speech words using a real-time setting.

Importantly, the authors found strong shared neural representations in SMG between internal speech, reading visually presented words, and vocalised speech production. This points to the involvement of common underlying cognitive processes between these tasks. Their decoder was also robust to different internal speech strategies, such as auditory vs. visual imagery, suggesting flexibility for individual mental strategies in future applications. Specifically, they tested the generalisation performance of the decoder between all tasks. Decoders trained on auditory cue trials were less generalisable to inner and vocalised speech than those trained on written cue trials. This shows silent reading brain activity may be closer to pure inner speech. While the cue modalities were separable during the cue-phase brain activity, they overlapped during subsequent phases. Thus, internal and vocalised speech representations may not be influenced by the cue modality, a promising result for repetitive inner speech paradigms.

In a different line of work, \citet{willett2021high} demonstrate the potential for real-time decoding of attempted handwriting movements from neural activity as a high-speed BCI. In this study, a participant with tetraplegia from spinal cord injury attempted to handwrite letters and words by imagining holding a pen and writing. Neural activity was recorded from intracortical electrode arrays implanted in the hand area of motor cortex. They found individual neurons showed selective patterns of activation for different handwritten letters, enabling reconstruction of pen trajectories. While not direct inner speech, it remains a purely imagined task with no external stimuli or produced behaviour. In online experiments, the participant achieved remarkable typing speeds of over 90 characters per minute, with over 94\% raw accuracy. While an invasive approach was used, the neural dynamics revealed in motor cortex potentially inform non-invasive BCI design. Attempted handwriting may be a promising paradigm for EEG and imaging BCIs if cortical patterns can be sufficiently resolved.

While limited in humans for ethical reasons, invasive recordings in neurological patients provide unparalleled detailed characterisation of the neurophysiology underlying inner speech phenomena. These studies unequivocally show that inner speech and movement imagery decoding are possible invasively, and demonstrate their potential in BCI applications. Next, we turn to non-invasive studies of inner speech.

\subsection{EEG}

While invasive studies are rare due to the nature of intracranial recordings, EEG studies of inner speech are also uncommon because of the difficulty in overcoming the low signal-to-noise ratio and spatial resolution inherent in scalp EEG recordings.

\citet{cooney2019classification} investigate using CNNs to classify imagined spoken word-pairs from EEG signals. Their dataset contained 6 Spanish words imagined by 15 subjects. All 15 possible word-pairs were extracted and EEG signals corresponding to an early imagined speech time-window were used. Results showed a deep CNN achieved the best average accuracy of 62.37\% across subjects and word-pairs. This performance however is still barely above chance level, indicating the ongoing difficulty of decoding imagined speech from noisy EEG recordings.

\citet{ling2019visual} investigate how visual words are represented in the brain using EEG-based decoding and image reconstruction techniques. Their study had 14 participants view 80 high-frequency nouns while recording EEG data. They then used multivariate pattern analysis on the EEG data to decode visual and orthographic properties of the words. Specifically, they were able to decode pairwise word discriminability well above chance across participants, with peak performance around 170ms after stimulus onset. They also applied representational similarity analysis to show the word decoding results correlated with visual and orthographic similarity but not semantic similarity. This is perhaps unsurprising as decoding visual activity is well-studied with EEG.

\subsection{MEG}

\citet{defossez2022decoding} present a novel method for decoding natural continuous speech from non-invasive MEG and EEG recordings. The authors leverage recent advances in self-supervised speech representation learning, specifically wav2vec 2.0 \citep{baevski2020wav2vec}, to obtain semantically meaningful speech embeddings from raw audio. These speech embeddings are aligned with M/EEG signals recorded while participants passively listened to audio samples. A joint CNN architecture with a contrastive loss is used to predict the speech embeddings from the neural signals. Without any individual calibration, their model can identify 3-second speech segments with up to 72.5\% top-10 accuracy across nearly 1,600 samples for MEG and 19.1\% across 2,600 samples for EEG.

For decoding inner speech, this study provides a promising framework to handle individual variability and extract meaningful speech features from limited data. The zero-shot decoding is particularly impressive, as it avoids constraints of classifiers trained on small stimulus sets. However, additional work is needed to apply this to inner speech due to the lack of audible signals for alignment. Methodologically, this work sits at the interface of our efforts in Chapters~\ref{Chap4} and \ref{Chap5}. They leverage both group modelling and training large models across multiple datasets. Their approach is well suited to incorporating forecasting or other self-supervised objectives. The contrastive loss allows for out-of-distribution decoding, as it is not limited by the categorical nature of standard classifiers. Furthermore by incorporating multiple feature extractors (e.g. CNNs trained on images), the same contrastive approach and brain model can be applied to various decoding tasks.

Direct MEG investigations of inner speech are limited. \citet{dash2020decoding-imagined} decode 5 imagined and overtly spoken phrases from MEG. Three decoding methods were tested: an artificial neural network (ANN) using statistical features, a CNN on time-frequency images, and a CNN with combined spatial, spectral and temporal features. The CNN approaches significantly outperformed the ANN, achieving up to 96\% accuracy for spoken phrases and 93\% for imagined phrases with the combined features. A key limitation is using phrases which likely contain more decodable information but are harder to scale up. A contrastive approach, or decoding at the phoneme/word level is more desirable.

\subsection{OPM-MEG}

While superconducting quantum interference devices (SQUIDs) are traditionally used for MEG, optically pumped magnetometers (OPMs) have recently emerged as a promising alternative for MEG measurements \citep{boto2018moving}. OPMs offer several advantages over SQUIDs including room-temperature operation, lower cost, higher sensitivity, and allow head motion \citep{boto2017new}. Their compact size also enables flexible sensor arrays that can be customized to target specific brain regions or conform to individual head shapes \citep{boto2018moving}.

A key application of MEG is non-invasive decoding of mental states and cognitive processes from neural activity patterns. Some initial studies have now explored using OPM-MEG for neural decoding. \citet{wittevrongel2021practical} demonstrate OPM-MEG enables robust single-trial analysis and real-time decoding for BCI applications. They compared OPM-MEG and EEG for decoding visual evoked responses, including event-related potentials/fields (ERPs/ERFs) to motion-onset and steady-state visual evoked potentials (SSVEPs) to flickering stimuli. For motion-onset, OPM-MEG and EEG showed similar ERP/ERF components (N/M200, P/M300) with comparable signal-to-noise ratios. For SSVEPs, OPMs had higher SNR in the high frequency range (25-29 Hz) while EEG was better at low frequencies (8-12 Hz). In a real-time SSVEP spelling task, OPM-MEG achieved 97.7\% average accuracy comparable to state-of-the-art EEG systems. These results validate OPM-MEG for robust single-trial decoding in BCI applications. The improved SNR and spatial resolution suggest OPMs could enable more advanced decoding capabilities. Their wearable and flexible nature makes them well-suited for practical BCI applications.

\subsection{Conclusion}
\label{ssec:inner_speech_conclusion}

This chapter presented an in-depth investigation into decoding inner speech and silent reading from non-invasive electrophysiological recordings. The key findings were the following. Silent reading of words could be decoded from EEG, MEG, and OPMs with 30-40\% accuracy across 5 words, driven by early visual processing. Comparing modalities showed traditional MEG had the best performance for decoding silent reading, while OPMs and EEG had lower but comparable accuracies. Inner speech decoding was mostly at chance levels in EEG and MEG across multiple decoding approaches. The highest accuracy reached for inner speech was 33\% across 3 EEG sessions using covariance features, but the validity of this result is debatable.

Our silent reading results demonstrate the feasibility of decoding visual representations of words from non-invasive recordings, consistent with prior EEG and MEG decoding studies \citep{chan2011decoding, ling2019visual}. The decoding appeared to be driven by early visual responses, with a later peak potentially reflecting higher-level language processing \citep{kutas1988event}. This late component merits further investigation as a marker of semantic processing. While more subjects are needed for a robust comparison, OPMs achieved decoding accuracy comparable to that of EEG. In one participant with good spatial coverage, OPM decoding performance was even better than EEG, highlighting their promise given advantages like wearability. Dense coverage of visual regions may be critical for the investigated decoding task. Our results also underscored the high between-subject variability, both in overall performance and in the timing of informative decoding features.

In contrast to silent reading, our extensive efforts to decode two types of inner speech were largely unsuccessful across EEG and MEG. While we explored various decoding algorithms and experimental designs, accuracy never substantially exceeded chance levels. This contrasts with more promising results from intracranial recordings in humans \citep{martin2016word, wandelt2022online}, and suggests non-invasive signals may not adequately capture the subtle dynamics of inner speech. There was also substantial between-session variability.

In addition to the analyses presented, numerous unsuccessful decoding approaches were pursued on the inner speech data. On the MEG recordings, these included logistic regression, CNNs, SVMs, concatenating or averaging consecutive trials, and per-session versus aggregated-session decoding. For EEG, besides the MEG-based methods, other unsuccessful attempts involved temporal alignment of trials, PCA denoising, Riemannian classifiers, baseline correction, and wavelet features. We tried several referencing approaches such as common average reference, mastoid references, and current source density estimation through the Laplacian method. We also tried wider bandwidth filters (up to 100Hz), however we noticed no improvement with the addition of gamma activity, even though it is reported in invasive inner speech studies. Noise and superposition in scalp measurements may overshadow weak gamma oscillations.

Several factors could underlie the difficulty of decoding inner speech non-invasively. Inner speech lacks the external stimuli and muscle activations present during overt tasks, reducing the signal-to-noise ratio. There is also high inter-individual variability in inner speech strategies. Here we focused on collecting large trial counts from a few participants rather than a small sample across many subjects. Our cross-cue paradigm may also induce visual confounds that overshadow inner speech signals. Having participants repeatedly imagine brief, single words likely differs from natural inner speech involving longer phrases. Further limitations of our work include the small number of participants and the small set of words.

Future investigations could explore alternative paradigms more representative of natural speech, such as imagining longer phrases or reading whole sentences silently. Transfer learning and self-supervision may help extract robust inner speech representations amidst noise \citep{defossez2022decoding}. Intracranial findings point to superior temporal, inferior frontal, and motor areas as promising decoding targets. For non-invasive BCIs, approaches beyond word-level decoding may be needed for inner speech-based communication, such as decoding phonemes, or imagined handwriting.

One downside of our task design is that the low-level characteristics of the visual appearance of the words might be a confounding factor in our silent reading results. Future work should consider using varied graphical representations of the same word across trials.

In summary, our results highlight the significant challenges in decoding inner speech correlates non-invasively compared to overt tasks. Substantial innovation in experiments and analyses will likely be essential to enhance the fidelity of decoded inner speech for BCIs. While current decoding performance was limited, our proof of concept work provides a useful platform with extensive trial counts for future efforts at modelling inner speech. Having multiple sessions allows for testing across-session generalisability. Neuroscientific understanding of inner speech may be deepened through comparing the different experimental paradigms.

\chapter{Discussion}
\label{Chap7}
% BRIEF
% joint discussion of all results in terms of variability issues and decoding for BCIs
% limitations and future directions

This thesis delves into the realm of brain modelling, targeting the enhancement of decoding performance and BCI communication speeds. The significance of this research becomes apparent when considering clinical populations, particularly individuals with locked-in syndrome, who heavily rely on such technology for communication. In the broader context, BCIs symbolise the culmination of the seamless integration of advanced technological tools into human lives.

Historically, human civilisation has witnessed an ongoing assimilation of tools to augment our capabilities. Although this has spanned millennia, the advent of computers and subsequently smartphones represented important leaps. These devices, coupled with the proliferation of wearables like smartwatches, underscore an evolving paradigm of human-machine symbiosis. Pre-computer age tools predominantly showcased mechanical prowess; however, with the dawn of the digital age, this shifted towards enhancing human cognitive capabilities. The field of artificial intelligence (AI) stands testament to this shift, wherein computers, with their unparalleled cognitive processing capabilities, are reshaping our understanding of intelligence \citep{letheren2020black, chollet2019measure}.

An evident lack remains in the realm of interfaces bridging human cognition with these technological advancements. Predominant tools still rely heavily on manual interaction, be it through keyboards or touch displays. Thus, there is a need for a more direct, and arguably more intuitive interface, directly with the human brain. The path to BCI improvement can be split into two primary avenues. The first involves the development of sophisticated hardware that facilitates more detailed brain recordings. The second encompasses the design of innovative methods capable of circumventing the constraints of current hardware. Our research aligns with the latter approach. Aiming for maximal societal impact, our focus was on leveraging non-invasive technologies, with a particular emphasis on electrophysiology, given its fast temporal dynamics.

Delving deeper into current non-invasive hardware, two frontiers emerge in the pursuit of BCI enhancement. The first encompasses purely software-centric solutions, extensively explored in Chapters \ref{Chap3}, \ref{Chap4}, and \ref{Chap5}. The second frontier pertains to experimental methodologies, explored in Chapter \ref{Chap6}. Our work addressed critical challenges in both domains, though the actual application to end-users is left for future exploration.

BCI technology hinges on machine learning methods. To harness the full potential of such methodologies, complex, nonlinear models are indispensable, as are large datasets. A notable challenge with non-invasive electrophysiology is the pronounced variability across time, participants, and tasks. Most task datasets possess limited data from individual participants, and often lack a diverse participant pool to encapsulate the full spectrum of brain variability. Thus, a pragmatic approach to BCI improvement involves crafting methods adept at navigating variability within constrained datasets.

\section{Variability within individuals}

Chapter \ref{Chap3} embarked on addressing variability intrinsic to individual brains. Our findings show the efficacy of linear decoding on full epochs of stimulus presentation. A crucial revelation was the performance improvement gained from the integration of a dimensionality reduction technique targeting the channel dimension during supervised training of the decoder. To understand these improvements better we should analyse the challenges in the application of machine learning to M/EEG data. At the outset, the data size is modest, encompassing 30 examples for each condition across a total of 118 conditions. The 306 MEG channels exhibit substantial covariance, thereby presenting a high-dimensional input with correlated features. It is for this reason that PCA frequently emerges as a method to reduce the dimensionality of the channel space.

While PCA effectively decouples the channels, it might inadvertently eliminate task-specific information due to its unsupervised nature. Consequently, it stands to reason that executing a similar dimensionality reduction but within the decoding objective yields superior results. Once such a projection is learned, it mirrors the utility of PCA in feature extraction. These supervised features are then amenable to integration with any conventional model, such as Linear Discriminant Analysis (LDA). Alternative methodologies for supervised dimensionality reduction, such as the Riemannian classification method, do exist \citep{barachant2014meg}, however, they are particularly effective when the number of classes is minimal, a scenario that diverged from the datasets explored in this thesis.

Our approach is contingent on the availability of ample data, as corroborated by the limited performance observed in the small replay dataset. When operating in data-scarce environments, the \textit{curse of dimensionality} becomes a considerable challenge, due to using features extracted from the entire epoch. In such cases methods for extracting higher-level features, such as power in different frequency bands may prove better \citep{higgins2022relationship, hu2011feature}. This is evidenced by the long list of various decoding features proposed in the BCI literature \citep{panachakel2021decoding}. A further significant limitation of our work is the lack of application across diverse tasks and modalities, such as EEG.

A parallel challenge with deploying machine learning models on high-dimensional inputs is the loss of neuroscientific interpretability. We demonstrated the versatility of Permutation Feature Importance (PFI) as a tool that can be employed across various input dimensions—temporal, spatial, or spectral—to glean insights into task-associated brain activity patterns. The adaptability of PFI is commendable, allowing for analyses on a per-participant or per-condition basis. Through the window size parameter, it offers the flexibility to strike a balance between noise mitigation and pattern resolution. However, PFI does not ensure that identified patterns truly encompass task-related information, due to how model optimisation works. Nonetheless, empirical analyses have affirmed its congruence with direct methods, such as sliding window analysis.

\section{Modelling variability between individuals}

Chapter \ref{Chap4} shifted to another form of variability. Beyond the confines of intra-participant variability, inter-participant variability manifests itself in terms of anatomical differences, functional localisation variations, and divergent neural dynamics \citep{saha2020intra}. Such heterogeneities often prohibit the creation of universally applicable decoding models that exhibit consistency across individuals. From a BCI perspective, the ability to amalgamate data from a multitude of participants would be an invaluable asset. It paves the way for deploying a pre-trained model on a new individual's data without any finetuning.

The subject embedding technique investigated in Chapter \ref{Chap4} emerged as a salient solution. It circumvents individual variability by learning a low-dimensional representation for each participant \citep{chehab2021deep}. By integrating this embedding as an input to a shared decoder across participants, we were able to approach the performance levels of subject-specific models. This showcases the potential for more potent applications of deep learning by capitalising on multi-subject datasets. Yet, this methodology warrants more exhaustive assessments, especially on larger datasets that contain a broader set of participants.

Our findings also underscored a crucial observation: while nonlinearity might not improve performance in single-subject scenarios, it becomes indispensable for incorporating information from subject embeddings into a group model. We believe that in single-subject scenarios, the limited dataset size might constrain the applicability of nonlinear models. Furthermore, our research illuminated the applicability of PFI to kernels within a convolutional network, thereby facilitating the extraction of interpretable importance maps, particularly in the spectral domain. This is in line with the efforts of the field of interpretable artificial intelligence \citep{linardatos2020explainable}.

Our approach has significant limitations. We did not venture into assessing the method across different task types. Nonetheless, the efficacy of subject embedding has received validation in recent works across different datasets \citep{defossez2022decoding, chehab2021deep}. An intriguing question that emerges is whether, beyond a critical threshold of participants, the inherent variability can be implicitly modelled by a sufficiently complex model.

Another point of contention is the observed decrease in performance when training models on data from one participant and subsequently deploying it on another. We did not observe any marked improvement in such scenarios. Although the subject embedding technique was not conceptualised for these specific applications, their utility in BCI contexts is undeniable. Recent advancements in the field have witnessed the exploration of alternative methodologies, such as the subject-specific layer, which projects input data into a standardised space \citep{defossez2022decoding}. However, this approach also requires training on each new participant. For achieving genuine zero-shot performance, the development of models that innately learn inter-subject variability may be needed.

\section{Towards foundational electrophysiology models}

Chapter~\ref{Chap5} embarks on addressing the inherent challenges of integrating deep learning methodologies with electrophysiological data. Several difficulties emerge, stemming from the use of different scanners, varied tasks, and diverse experimental setups. However, a common denominator across these scenarios is the presence of multichannel time series data with a high sampling rate. Drawing inspiration from the recent strides in large language models, where a single sequence model can handle a plethora of language-related tasks, one might envisage creating an analogous model for electrophysiological data. Pursuing this line of thought offers two distinct advantages. Firstly, such a model could harness data spanning multiple datasets, providing a robust framework to model variability. Secondly, drawing a parallel to language models, if we can build a deep learning model that can 'understand' brain data, it should inherently possess the ability to execute auxiliary tasks, be it encoding or decoding brain signals.

The versatility of such models could manifest in their ability to perform tasks either with or without dedicated fine-tuning for specific downstream applications. For example, our proposed Transformer-based model includes forecasting, encoding associated with task stimuli, and decoding based on Bayes' theorem. However, the decoding performance was somewhat limited, suggesting that transfer learning, facilitated through fine-tuning, might emerge as a better avenue. We did not experiment with this approach in our research, but our results showed that the performance of downstream decoding can be boosted by simulating training data. This is an alternative application of foundation models to downstream tasks.

A significant portion of our research efforts was channelled into analysing the input structure of M/EEG data. Our goal was to discern the types of architectures or inductive biases that might be optimally suited to model such data. Given the inherently sequential nature, coupled with the unparalleled performance of Transformer-based models across diverse sequential data modalities—ranging from language to audio and video—we gravitated towards the GPT2 model, the autoregressive forecasting variant of the Transformer. When compared with conventional CNN-based architectures, such as Wavenet, our findings revealed that the Transformer has the ability to generate data that exhibited a higher congruence with real data. This suggests a superior capability of the Transformer in mirroring the dynamics inherent to real-world data.

A crucial limitation of our approach is the inability to inherently accommodate information from different channels. In essence, our method could be characterised as univariate, although channel embeddings played a pivotal role in tailoring the model to individual channels. Our endeavours to include multiple channels into the input were met with limited success. We think that maintaining the innate inductive biases of Transformers, which emphasise 1D sequence modelling on embeddings of discrete tokens, is paramount. While our FlatGPT2 model did not achieve good performance, alternative strategies might hold promise. For instance, one might consider adapting the neural network-driven vector quantisation techniques exemplified by models like Jukebox \citep{dhariwal2020jukebox}. While this model used a vector-quantised variational autoencoder (VQ-VAE, \citet{van2017neural}) in the time domain, adaptation to the channel dimension should be possible.

Some of our findings substantiated that predicting the next timestep may not serve as a robust measure of model performance. Future research should contemplate adopting multi-timestep or contrastive loss frameworks. A plausible strategy could involve deploying the VQ-VAE model across both channel and temporal dimensions, aiming to distill a more coarse sequence of discrete tokens. Nevertheless, any quantisation-centric approach must carefully consider reconstruction error. We posit that a significant portion of the signal dynamics should be entrusted to the Transformer, given its adeptness in capturing complex dynamics.

A notable challenge with forecasting models tailored for electrophysiology is the absence of external data. Intrinsically, brain activity is influenced by a plethora of external stimuli and physiological processes, many of which elude the experimenter. Consequently, the task of forecasting suffers from uncertainties. However, large language models have demonstrated remarkable efficacy, despite the fact that the vast expanse of text on the internet is not typically conditioned on the underlying motivations or contexts of its human authors. This again motivates the need to scale large electrophysiology models.

A constraint in our modelling approach is its reliance on categorical task stimuli labels. Such an approach, while effective in our context, does not readily lend itself to scalability across diverse tasks and datasets. However, it is conceivable to construct robust representations tailored for various stimulus modalities—ranging from images to audio. These representations can then serve as conditioning embeddings. As shown by \citet{defossez2022decoding}, tools such as wav2vec \citep{baevski2020wav2vec} can be leveraged to derive informative representations of auditory stimuli.

Our results in Chapter~\ref{Chap5} were confined to a single dataset. A pivotal avenue for future research would be the rigorous evaluation of these models across an array of datasets and tasks. Scaling, spanning datasets, tasks, modalities, and broader research domains, via transfer learning, emerges as a promising strategy. This approach holds the potential to navigate inter-dataset variability, in pursuit of versatile, generalisable foundational and decoding models.

\section{Probing the limits of non-invasive BCIs}

Transitioning to the experimental frontier, we recognise that the efficacy of a model, regardless of its sophistication, can be substantially undermined by limitations in data volume and quality. This understanding motivated our experimental explorations in the concluding Chapter~\ref{Chap6}. While recent invasive studies have shown impressive communication rates, especially with naturalistic paradigms like imagined speech and handwriting, their non-invasive BCI counterparts appear to lag behind. 

Prevailing EEG-based BCI methodologies predominantly revolve around the relatively slower P300 and SSVEP paradigms \citep{guan2004high, icscan2018steady}. These artificial paradigms emerged due to the noisiness of the EEG signal. To build robust BCIs researchers needed to resort to the strong signals of the visual cortex. Motor imagery has also shown promise but is usually limited to a handful of decodable classes \citep{saha2020intra, halme2018across}. A more positive outlook on these BCI methods is that by being smart about experimental paradigms EEG-based BCIs can achieve previously unimaginable performance.

Our research endeavoured to analyse the feasibility of inner speech in non-invasive modalities and to discern the potential enhancement in performance with increasing data volume. Regrettably, even with hundreds of trials from a single word, the decoding performance for inner speech hovered around chance levels. It is also worth noting that these analyses were conducted at the session level. We did not venture into cross-session or cross-participant decoding. While the decoding of silent reading showed potential, our investigations revealed that this was driven predominantly by the visual processing associated with word presentation. Therefore it is crucial for future research to investigate the shared representations of silent reading, listening, vocalised and inner speech in non-invasive modalities.

The limited success in inner speech decoding experiments could be attributed to the fact that the experimental paradigm did not emulate naturalistic phenomena. This was primarily due to the repetitiveness of tasks and the focus on isolated words. A better approach would involve evaluating decoding efficacy during free dialogue or imagination. It may well be that by using a different set of words which are phonetically or semantically more dissimilar, decoding performance would increase. While counter-intuitive, collecting inner speech data across a much larger set of words in naturalistic settings, akin to the listening experiments in \citet{defossez2022decoding}, could also enable better performance. The categorical classification approach could be replaced by contrastive objectives with word embeddings, leveraging the representational space of language models.

An overarching limitation that permeated all our research was the confinement to small datasets with homogeneous tasks. Several other limitations warrant mention. Across all chapters, the cohort of human participants was relatively modest in size. This was particularly pronounced in the multimodal decoding experiments, which compared EEG, MEG, and OPM recordings. This requires expanded replication to enable robust conclusions. The restricted sensor coverage in our experiments limited information content in OPMs. In future OPM studies dense coverage of visual and language areas should be ensured. With advancements leading to more comprehensive sensor coverage, OPMs could potentially eclipse traditional MEG systems, especially if they facilitate the deployment of custom sensor arrays \citep{boto2018moving}.

\section{The future of brain modelling for BCIs}

Future research should focus on scaling across multiple facets: datasets, participant cohorts, and the transferable knowledge across studies. This will pave the way to more effectively capture and model the intrinsic variability embedded within cognitive neural dynamics. The progression of self-supervised and few-shot learning paradigms, particularly tailored for electrophysiology data, holds the promise of revolutionising the domain. Such advancements will be crucial in harnessing the full potential of limited labelled datasets \citep{banville2021uncovering}.

The establishment of expansive, open-access corpora comprising raw neural recordings is pivotal. These repositories, encompassing data from a multitude of diverse studies, could serve as a bedrock for training robust, generalisable models. Realising this vision necessitates fostering collaborative data-sharing initiatives that operate within the confines of privacy regulations \citep{poldrack2014making}. Additionally, the development and deployment of automated annotation and processing tools could further streamline this process. A salient research trajectory involves testing whether foundational models can enhance the sample efficiency and adaptability of BCIs, spanning tasks, sessions, and participants. Such advancements could have profound implications, significantly elevating the quality of life for clinical populations.

Another integral component revolves around rapid modalities of imagined communication, akin to those explored in invasive data studies. Imagined paradigms have received limited attention in non-invasive research, primarily due to the challenges posed by faint signals relative to ongoing brain activity and noise. Several prospective directions have been identified, such as delving into various facets of inner speech, e.g., imagining mouth movements or auditory imagination of speech, and even delving into different linguistic units, e.g., phonemes. Different modalities, like imagined handwriting, too, need deeper exploration through non-invasive methods. While in silent speech paradigms EMG is more practical for BCI applications, inner speech lacks any motor output and thus can only be detected in the brain. However, augmenting inner speech with motor imagery (e.g. imagining mouth movements), could provide stronger signals in the motor cortex compared to pure inner speech.

We are particularly inspired by the contrastive methodology delineated by \citet{defossez2022decoding}. While adapting to inner speech presents challenges, we posit that a closed-loop setup augmented with real-time feedback could show promise. In such a paradigm, participants would commence by reading a word aloud, followed by auditory exposure to the word's pronunciation at a designated cue. As the experiment progresses, the audio rendition could be mixed with noise or its volume incrementally reduced. The objective is to seamlessly transition the participant to rely predominantly on inner speech, especially as the audio cue becomes increasingly unintelligible.

Leveraging the CLIP approach \citep{radford2021learning} on this data could be instrumental. This involves distilling positive samples by identifying audio snippets that exhibit maximal similarity with the inner speech representations. While in the beginning of the experiment the methodology would be similar to the one used in \citet{defossez2022decoding}, as the participant switches to inner speech the contrastive learning through paired audio segments becomes challenging.

Our experimental endeavours highlight that the most significant constraints might be deeply rooted in the hardware. Non-invasive electrophysiological recordings may be too noisy, thereby challenging the extraction of salient signals associated with inner speech. We believe that significant strides in hardware innovation and recording modalities are imperative. Given the empirical evidence supporting the efficacy of inner speech decoding in invasive modalities, the key is overcoming challenges posed by factors like signal-to-noise ratio and spatial resolution.

More theoretical and experimental work should be conducted to understand the channel information capacity of non-invasive modalities. While here we think that noise or superposition of signals drowns out any inner speech-related activity, it would be important to be able to show this. A possible experiment could involve simultaneous ECoG and EEG recordings during an inner speech task to directly compare the synchronised signals from the two modalities. In addition, through source modelling, we could pin down inner speech activity from ECoG and subject it through a forward head model to see how signals manifest on the scalp. This would provide more concrete evidence whether the channel capacity of EEG is the limiting factor in detecting inner speech.

One promising innovation in recent times has been the advent of OPMs. These devices, in theory, possess a superior signal-to-noise ratio, and thanks to engineering advancements, a larger number of sensors can be densely packed on the human scalp. Yet, there remains a divide between theoretical potential and actual performance, necessitating further engineering innovations. As our silent reading experiments have shown, even the most advanced OPM systems, as of now, do not quite match traditional MEG scanners. It is conceivable that even future OPMs might fall short in discerning the subtle signals of inner speech. Hence, pushing the envelope in non-invasive techniques will invariably hinge on pioneering research spanning the complex physics and biology of the human brain.

\section{Conclusion}

In summary, this thesis presents a range of methodological innovations and empirical insights, each crafted to enable subsequent research in non-invasive brain decoding. At the crux of our research lies an integrated modelling paradigm, encapsulating the multifaceted variability inherent to neural dynamics. Our findings underscore the significance of both advanced hardware capabilities and powerful data-driven models in the pursuit of enhanced BCI communication speeds.

As the boundaries between technology and human cognition continue to blur, the onus is on the scientific community to harness these advancements, ensuring they are both accessible and adaptable. By carefully addressing the challenges and harnessing the opportunities that lie ahead, we believe that the horizon holds the promise of a world where BCIs are not just niche assistive tools but an integral extension of human expression and capability.

\renewcommand{\bibname}{Bibliography}
\bibliography{ml}
\bibliographystyle{apalike}

\begin{appendices}

\chapter{Interpretable full-epoch decoding}
\section{Results}

\subsection{Multiclass versus pairwise decoding}
\label{ssec:multiclass_vs_pairwise}

We have demonstrated that multiclass full-epoch models are better than sliding window models while maintaining the same level of spatiotemporal information. Here we wish to highlight an additional advantage of using multiclass full-epoch models. Researchers frequently use pairwise models to analyse the representational differences between individual conditions or groups of conditions, such as in representational similarity analysis (RSA). However, this approach can be computationally intensive, especially when dealing with a large number of classes.

Here, we show how we can utilise a single trained multiclass full-epoch LDA-NN model to predict pairwise accuracy scores. This is done by iteratively taking all pairs of conditions, computing the predicted probabilities across all classes for each trial, and selecting the condition with the higher probability (of the two conditions) as the predicted class. By comparing this to the ground-truth labels, we can obtain pairwise accuracy scores for each pair of conditions $i,j$:

\begin{equation}
\mathrm{accuracy}_{ij} = \frac{1}{N}\sum_{n=1}^{N}\left[\mathrm{argmax}_{k \in {i,j}} p(y_n=k \mid \mathbf{X}_n;\theta)\right] == y_n
\end{equation}

where $N$ is the number of trials across these two conditions, $\mathbf{X}_n$ is the feature vector for trial $n$, $y_n$ is the true class label for trial $n$, $p(y_n=k \mid \mathbf{X}_n;\theta)$ is the predicted probability of class $k$ for trial $n$. $\mathrm{argmax}$ selects the condition ($i$ or $j$) with the higher predicted probability for trial $n$.

\begin{figure}[!t]
  \centering
  \includegraphics[width=1.0\linewidth]{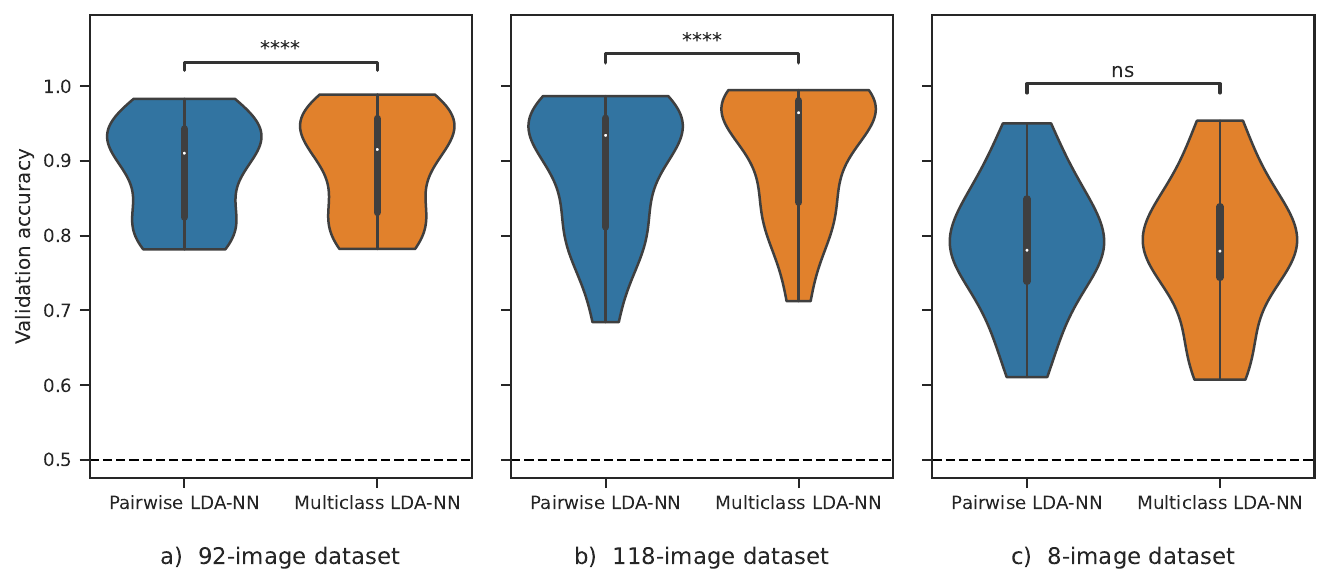}
    \caption{Comparison of pairwise full-epoch LDA-NN models (blue) with multiclass models evaluated for pairwise classification (orange) across the three datasets. In all datasets except the 8-image dataset, multiclass models evaluated in a pairwise fashion are significantly better (****, p<1e-4). The violin plot distributions are shown over the mean individual subject performance. The dashed line represents chance level.}
    \label{fig:pairwise_comparison}
\end{figure}

In Figure~\ref{fig:pairwise_comparison}, we compared the results of this method with those obtained by training individual pairwise (full epoch LDA-NN) models as is typical in the literature. For the 92 and 118-image datasets, the multiclass model achieved modest, but significant higher pairwise accuracy than the individual pairwise models. The difference was not significant for the 8-image datasets. Therefore, using a multiclass model can yield pairwise results that are similar to or even better than those obtained from individual pairwise models. This provides a much more efficient way of obtaining pairwise accuracies for the purposes of RSA.

This approach is useful for RSA and reduces computation time by approximately half the number of conditions in the data, as pairwise models reuse this data for training while a multiclass model uses it only once. Although the data must still be reused for evaluation, we can assume that evaluation is much faster than training. The slight increase in performance when using multiclass models could be because decoding many classes together helps to better constrain the relationship between features and class labels compared to doing 2 classes at a time.

\chapter{Group-level decoding}
\label{appendix_group}
\section{Methods}
\subsection{Model analysis}
\label{ssec:appendix_model_analysis}

In \textit{Kernel FIR Analysis}, we investigate the frequency characteristics of the convolutional kernels. Random noise is fed into a trained model, and the power spectral density of the output of specific kernels is computed to assess their finite impulse response (FIR) properties.

\begin{align}
\mathbf{X} &\sim \mathcal{N}(0, 1) \\
y &= f(\mathbf{X}; \theta)_{l, i, o}
\end{align}

where $y$ is the output of the kernel in layer $l$, applied to input channel $i$, contributing to output channel $o$. $f$ is the trained model with parameters $\theta$. Then we compute the PSD of $y$ to assess the kernel's spectral properties. For group models we add the learned subject embedding to $\mathbf{X}$ as usual. An issue with this method is that there are thousands of kernels in each layer, and for visualisation purposes we simply do a random sampling of these kernels. Alternatively, this method can be applied to whole feature channels as well, instead of individual kernels. Note that this method does not only assess the specific kernel's spectral properties as the input to a kernel within the model has been transformed by previous layers as well. However, we found this to produce more interesting visualisations than simply computing the PSD from the kernel weights directly. This is because of our architectural choice of having only kernels of size 2.

\section{Results}

\subsection{Kernel analysis}

Kernel FIR analysis shows the power spectra of kernels’ outputs when input examples are Gaussian noise (Figure~\ref{fig:kernel_fir_anal}). See Appendix~\ref{ssec:appendix_model_analysis} for method description. This analysis is answering a different question about the kernels in WaveNet compared to spectral PFI, which asks what frequency content of the input are kernels most sensitive to. In contrast, Kernel FIR analysis asks what are the input-output filtering characteristics of kernels. This provides more insight into how successive layers in Wavenet build up more and more complex filters. The subject embedding was set to a subject with average accuracy. The power spectra were normalised to make visual comparisons between kernels easier. Since the WaveNet architecture uses dilated filters with only 2 values per filter, early layers show broad filtering characteristics, but already in layer 2, more emphasis is put on lower frequencies. In deeper layers, filters (kernels) become more tuned to specific frequencies, generally below 20Hz. This is in line with the spectral properties of MEG data as discussed above. Both the spectral PFI and kernel FIR analysis show that there is significant variability between the spectral information encoded by various kernels. 

\begin{figure}[!t]
  \centering
  \includegraphics[width=1.0\linewidth]{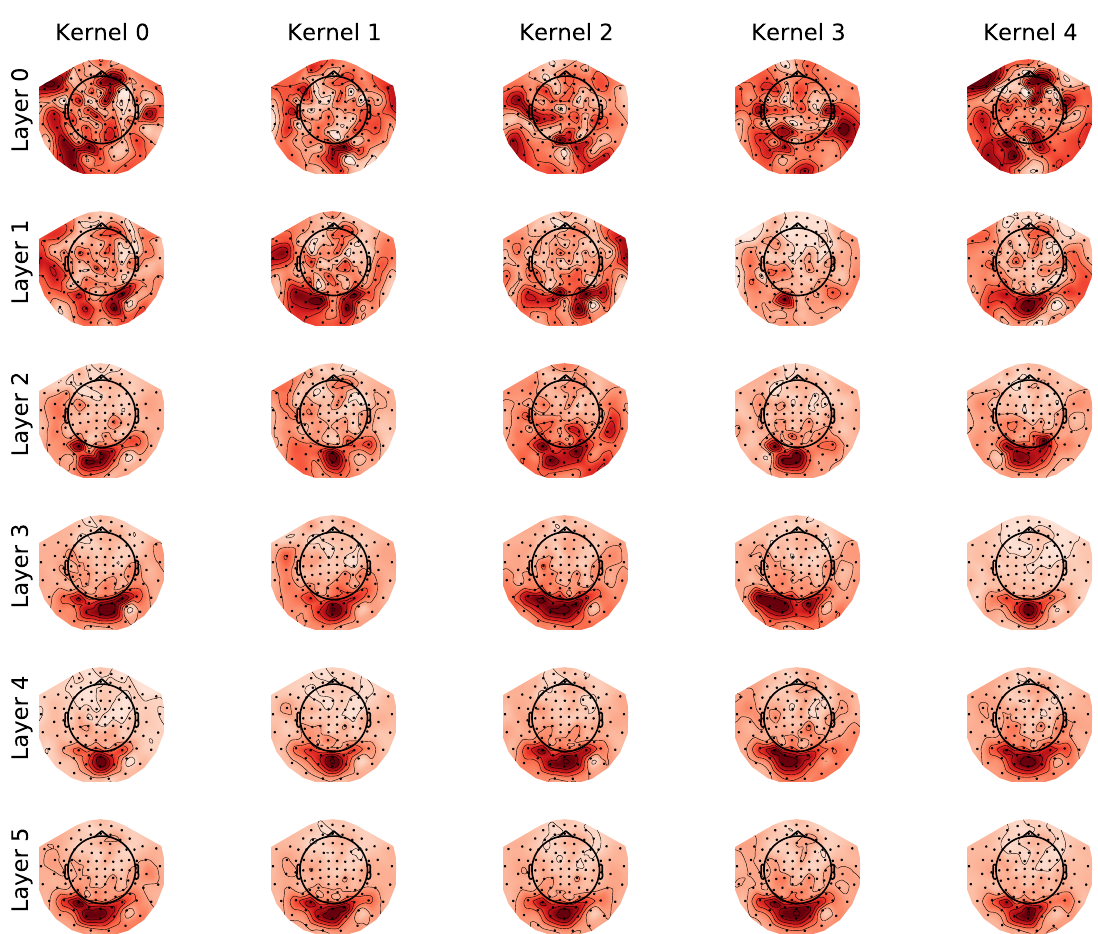}
    \caption{Spatial PFI across 6 layers (rows) in the trained \texttt{non-linear group-emb} model, with 5 kernels per row. Darker reds mean higher output deviation.}
    \label{fig:kernel_spatial_pfi_30}
\end{figure}

\begin{figure}[!t]
\centering
\begin{subfigure}{0.5\textwidth}
  \centering
  \includegraphics[width=0.8\linewidth]{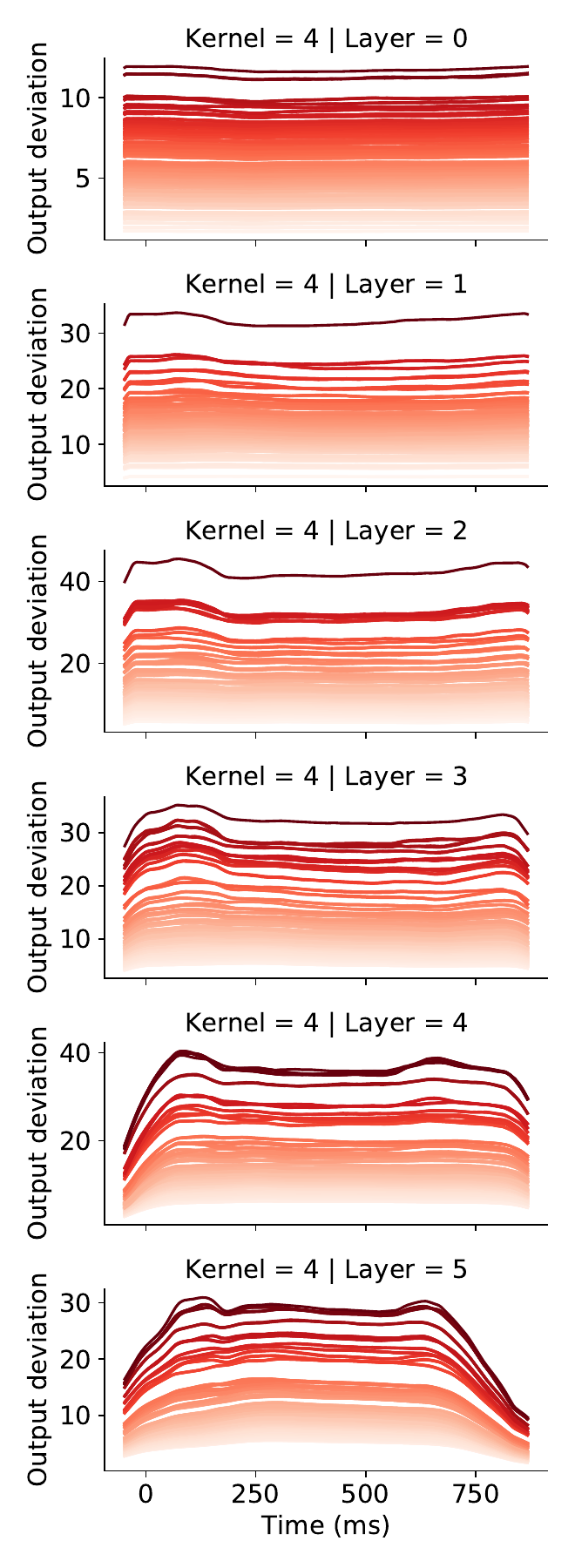}
  \caption{Channel-wise Temporal PFI}
  \label{fig:spatiotemporal_pfi_ap}
\end{subfigure}%
\begin{subfigure}{0.5\textwidth}
  \centering
  \includegraphics[width=0.8\linewidth]{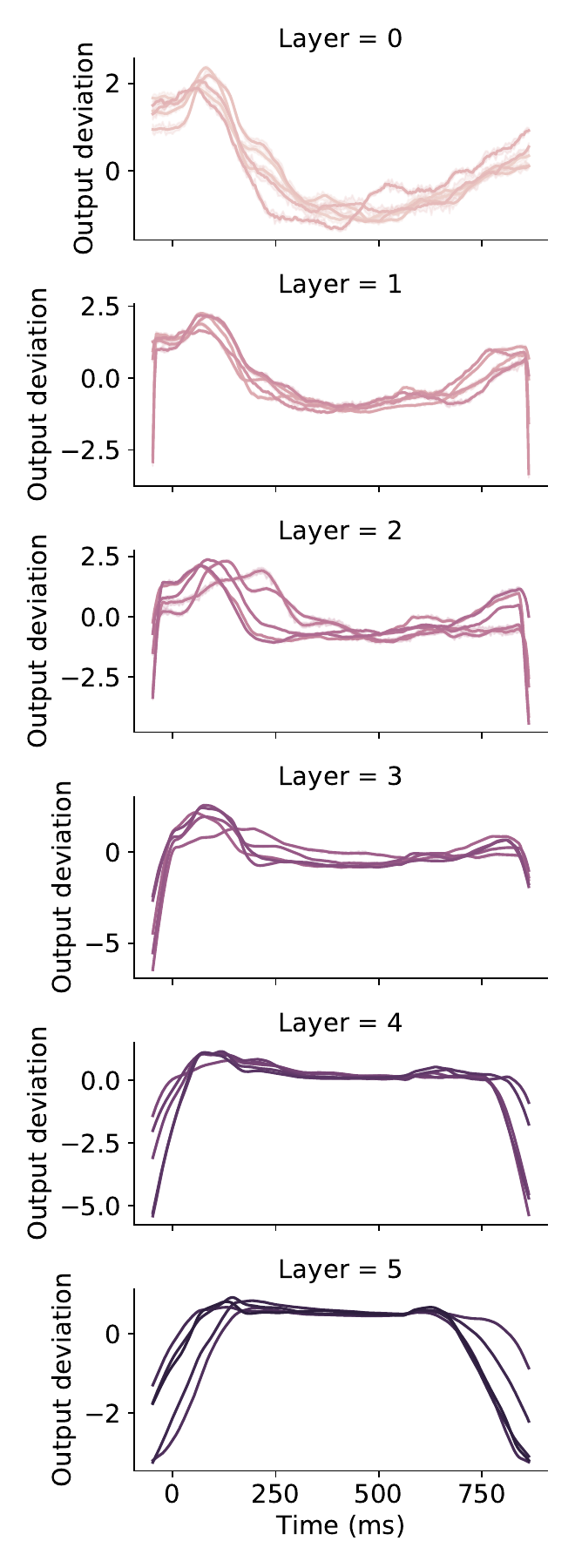}
  \caption{Temporal PFI}
  \label{fig:temporal_PFI_ap}
\end{subfigure}
\caption{Channel-wise temporal PFI (a), and temporal PFI (b) across kernels of the \texttt{non-linear group-emb} model in 6 layers (rows). For temporal PFI 5 kernels (lines) are plotted together. Channel-wise temporal PFI shows the temporal PFI of each channel for Kernel 5. Channel colouring is matched to the corresponding spatial PFI map, and darker reds mean higher output deviation. For temporal PFI output deviation is normalised. The horizontal axis shows the time elapsed since the image presentation for both temporal PFI types. 95\% confidence interval is shown with shading.}
\label{fig:spatiotemporal_ap}
\end{figure}

\begin{figure}[!t]
  \centering
  \includegraphics[width=0.9\linewidth]{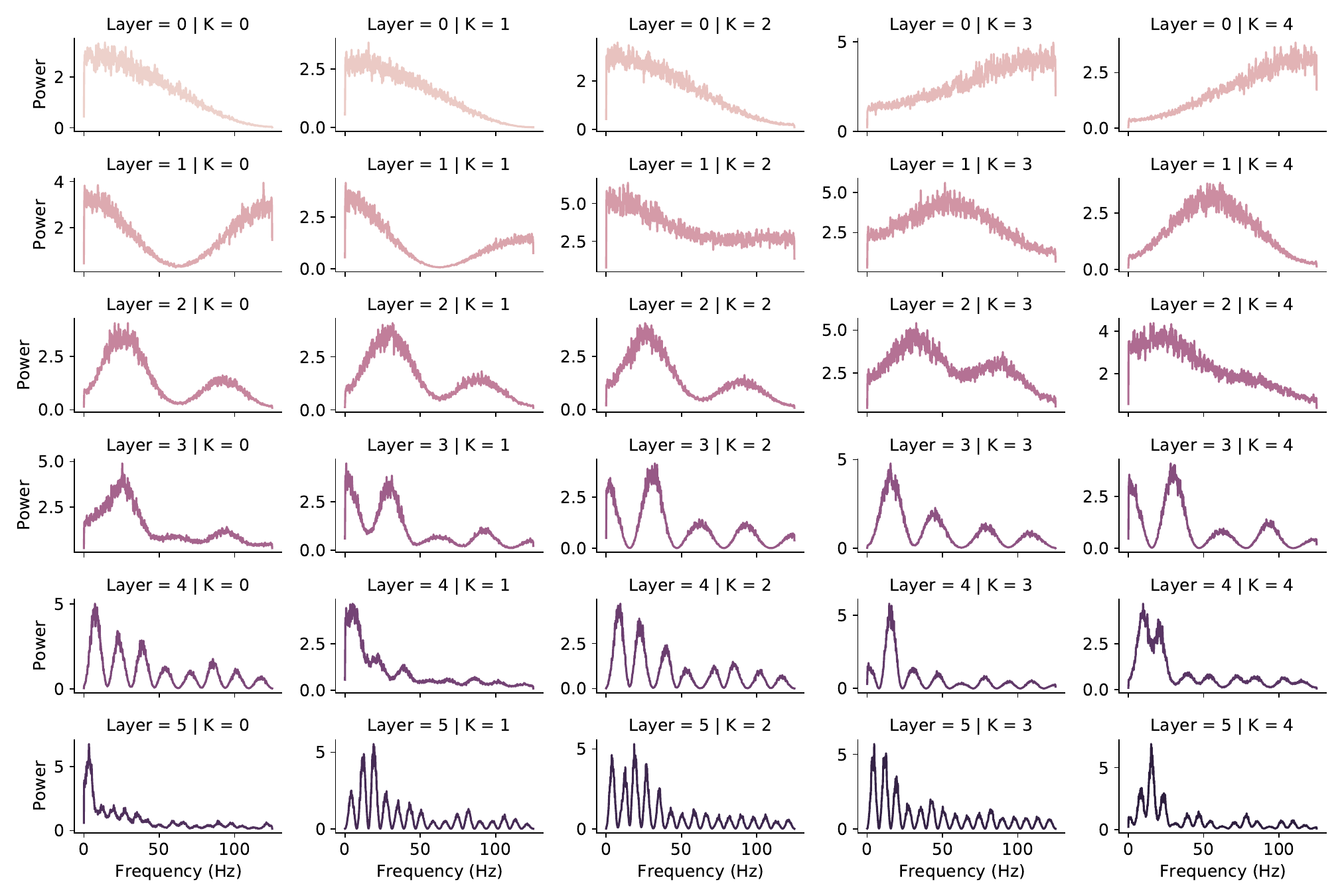}
    \caption{Frequency characteristics of 5 kernels across 6 layers (rows) via kernel FIR analysis in the trained \texttt{non-linear group-emb} model. The power spectra are normalised.}
    \label{fig:kernel_fir_anal}
\end{figure}

\chapter{Forecasting MEG signals}

\section{Methods}
\subsection{Simple Wavenet}
As described in Section~\ref{ssec:nn_ar}, a natural nonlinear extension of the linear multivariate autoregressive (AR) model is a convolutional neural network. Here we take inspiration from Wavenet \citep{oord2016wavenet} and design a simplified version for forecasting multichannel MEG data. This model is largely similar to the convolutional block of the Wavenet Classifier used in Chapter~\ref{Chap4} (Figure \ref{fig:wavenet}). For \texttt{SimpleWavenet}, we only use the dilated convolutional layers applied directly to the raw continuous data, with the inverse hyperbolic sine as the activation function.

The network receives 512 timesteps as input, roughly 2 seconds at a sampling rate of 250 Hz. The model starts with a 1x1 convolution (kernel size of 1) which applies the same linear transformation at each time point, serving to increase the channel dimension (projecting from 306 channels to 612). This is followed by 9 dilated convolutional layers (Figure \ref{fig:wavenet}), where the number of channels is kept the same, but the dilation rate increases by a factor of 2 in each successive layer. Finally, the channel dimension is reduced back to the original size with another 1x1 convolution. As discussed in Chapter~\ref{Chap4}, using 9 layers provides a receptive field of exactly 512 timesteps, so the output is a single $\mathbb{R}^C$ representing the prediction vector of the continuous values of the data at the timestep:

\begin{align}
    \hat{\mathbf{x}}_{T+1} &= \mathrm{SimpleWavenet}(\mathbf{X}) \\ 
    \mathcal{L}_{MSE} &= \frac{1}{C} \sum_{i=1}^C (x_{t+1,i} - \hat{x}_{t+1,i})^2
\end{align}

where $\mathbf{X} \in \mathbb{R}^{T \times C}$ is the MEG input segment of length $T$ and $C$ channels, $\hat{\mathbf{x}}_{T+1}$ is the predicted activity at time $T+1$, and $\mathbf{x}_{T+1}$ is the true brain activity. Compared to the Wavenet Classifier in Chapter~\ref{Chap4}, the main differences are the removal of the fully-connected block and the use of MSE loss $\mathcal{L}_{MSE}$ for forecasting instead of classification.

Normally we allow full mixing between channels since this is the default behaviour of standard convolutional layers, essentially implementing a fully-connected network across the channel dimension. We call this model type "multivariate." In some experiments we fit separate models to each channel, prohibiting cross-channel mixing. The ensemble of these channel-specific models is called a "univariate" model. This follows the nomenclature of univariate and multivariate AR models.

\subsection{FlatGPT2}
\label{ssec:flatgpt}

Directly vector quantising 300 channels to any vocabulary size would result in poor reconstruction. In \texttt{FlatGPT2} we perform the tokenisation on small groups of channels instead. First, we compute the covariance over channels in the training data. Then, we apply K-means clustering \citep{hartigan1979algorithm} on the covariance matrix to group channels into buckets. This ensures that each bucket contains channels with high covariance. This is important because tokenising a feature space (group of channels) with high covariance can be done with fewer tokens while maintaining low reconstruction error. We set the number of clusters ($B=30$) based on manual tuning on the training data. Each cluster/bucket can contain a variable number of channels, usually between 5 and 20.

After assigning channels to buckets we apply the Residual Quantiser algorithm \citep{babenko2014additive} from the faiss library\footnote{\url{https://github.com/facebookresearch/faiss/wiki/Additive-quantizers}} to each bucket $b$ separately. This is a powerful additive quantiser \citep{liu2015improved} that achieves good reconstruction error with a relatively small vocabulary size $V$. Note that the total number of tokens, i.e. the vocabulary size will be $BV$, since we have $B$ quantisers. Once fit to the training data the quantiser is fixed and can be applied to new data.

Mathematically, the covariance is obtained by:
\begin{align}
\forall i,j \in {1,\dots,C} \quad \mathbf{C}_{ij} = \frac{1}{T} \sum_{t=1}^{T}(x_{t,i} - \mu_i)(x_{t,j} - \mu_j)
\end{align}

Where $x_{t,i}$ is the $i^{th}$ channel at timestep $t$, $\mu_i$ is the mean of channel $i$ over all timesteps, and $C$ is the total number of channels. $\mathbf{C}$ is a symmetric matrix, and thus the feature and variable dimensions of K-means are the same. K-means computes buckets $\mathcal{C}_1,\dots,\mathcal{C}_B$ which partition channels $C$ into distinct sets with high within-bucket covariance.

The residual quantiser $Q_b$ learns a codebook $\mathbf{C}_b \in \mathbb{R}^{V \times |\mathcal{C}_b|}$ for each bucket $\mathcal{C}_b$:

\begin{align}
\forall t \in {1,\dots,T} \quad z_{t,b} = Q_b(\mathbf{x}_{t,b}; \mathbf{C}_b)
\end{align}

Where $z_{t,b}$ is the quantised representation (token/code) at timestep $t$ of the channels $\mathbf{x}_{t,b} \in \mathbb{R}^{|\mathcal{C}_b|}$ in $\mathcal{C}_b$. The encoding in the quantiser is sequential, thus at stage $m$ of the encoding of $\mathbf{x}_{t,b}$, the quantiser picks the entry $i_m$ that best reconstructs the residual of $\mathbf{x}_{t,b}$ w.r.t. the previous encoding steps:
\begin{equation}
i_m = \underset{j}{\mathrm{argmin}} || \mathbf{T}_m(j) - (\mathbf{x}_{t,b} - \mathbf{T}_1[i_1] + ... + \mathbf{T}_{m-1}[i_{m-1}]) ||^2
\end{equation}

where $\mathbf{T}_m$ is a table of size $K_m$ containing $|\mathcal{C}_b|$ dimensional vectors. For notational simplicity we omit the index $b$ from $i_m$ and $\mathbf{T}_m$ in the above. The quantisation provides a vector $[i_1, ..., i_M]$, where each element $i_m$ comes from a set of size $\lceil\log_2(K_m)\rceil$ bits. This bit vector representation can be easily transformed to token indices ranging from 1 to $V=\sum_{m=1}^M \lceil\log_2(K_m)\rceil$. Note that this table description of the discrete code is just a different representation of the overall codebook $\mathbf{C}_b$. A code $[i_1, ..., i_M]$ can be reconstructed to obtain $\hat{\mathbf{x}}_{t,b}$ by retrieving the corresponding vectors from $(\mathbf{T}_1, \dots, \mathbf{T}_m)$ and adding them up. Reconstruction error is computed by comparing $\hat{\mathbf{x}}_{t}$ and $\mathbf{x}_{t}$.

By using 30 buckets we obtain 30 tokens per timestep, which is already a 10-fold reduction of the original dimension space, but we have not reached our initial goal of having 1 token per timestep. To achieve this, we flatten the feature dimension (buckets) when feeding tokens to GPT2, hence the name \texttt{FlatGPT2}. Our total sequence length then becomes $B \cdot T$, where $B$ is the number of buckets and $T$ is the number of timesteps. This approach is also motivated by the observation that language models include extra information such as context within the sequence, instead of the feature space. Thus, when predicting the token of bucket $b$, we treat the previous timesteps of the other buckets as contextual information.

We also add an extra separator token $z_{sep}$ between sequences of buckets corresponding to the same timestep to facilitate distinction between the bucket and time dimensions. An input sequence to \texttt{FlatGPT2} consists of tokens $z_{t,b}$ following a fixed order:

\begin{align}
    \mathbf{z} = (&z_{sep}, z_{t=1,b=1}, z_{t=1,b=2}, \dots, z_{t=1,b=B},\\ &z_{sep}, z_{t=2,b=1}, z_{t=2,b=2}, \dots, z_{t=2,b=B},\\ &z_{sep}, \dots, \dots, z_{t=T,b=B})
\end{align}

For each codebook $\mathbf{C}_b$ a separate embedding $\mathbf{W}_{e,b} \in \mathbb{R}^{V \times E}$ is learned. As in \texttt{ChannelGPT2} we add the appropriate conditioning embeddings to the input embedding with appropriate flattening across the channel/bucket dimension:

\begin{align}
    \mathbf{H}^{(0)} &= \mathbf{Z}\mathbf{W}_e + \mathbf{W}_{p} + \mathbf{Y}\mathbf{W}_y + \mathbf{O}\mathbf{W}_o + \mathbf{W}_{c} + \mathbf{W}_{t}
\end{align}

where $+$ denotes element-wise addition and $\mathbf{Z} \in \mathbb{R}^{(B+1)T \times V}$ is the one-hot version of $\mathbf{z}$. The task labels $\mathbf{Y}$ can vary across time, but are the same across the buckets of one timepoint. $\mathbf{W}_{c}$ now contains distinct embeddings of buckets $b \in (1, \dots, B)$, which are the same across timesteps. We also augment the input with $\mathbf{W}_{t}$, containing distinct embeddings for timesteps $t \in (1, \dots, T)$, which are the same across buckets. This is the timestep version of $\mathbf{W}_{c}$.

As usual, the model is trained to autoregressively predict the next token in the sequence given all previous inputs. At timestep $t$ and bucket $b$ the model has access to the tokens $\mathbf{z}_{1:t-1}$ from all buckets (and thus information from all channels), and the tokens $\mathbf{z}_{t,1:b}$, and has to predict token $z_{t,b+1}$. The buckets of the same timestep are predicted sequentially, thus, bucket ordering could influence results. We use an arbitrary bucket ordering and do not experiment with different orderings of the input sequence.

Note that at the last bucket $B$ in each timestep the prediction should be token $z_{sep}$, however, we simply discard this prediction during loss computation, as we do not require the model to predict separator tokens. The structure of the sequence already constrains the predictions such that a new timestep begins after every $B$ tokens. Conversely, when computing the prediction at input token $z_{sep}$, the target is the token with bucket $b=1$ of the next timestep. This is useful as in theory we could start the recursive generation of data with a single $z_{sep}$ token.

At the output, the transpose of $\mathbf{W}_e$ can be used to predict probabilities over the vocabulary, or a separate linear projection can be learned. Note that because each codebook $\mathbf{C}_b$ has a separate vocabulary of size $V$ assigned to it, we can speed up the output softmax by only computing probabilities over codes/tokens in $\mathbf{C}_b$ instead of the total vocabulary of size $BV$.

\texttt{FlatGPT2} contains important hyperparameters that affect design choices and performance (Table~\ref{table:flatgpt_params}). Increasing the number of buckets $B$ improves reconstruction error, as the vector quantiser has to quantise less channels, but increases the length of the input sequence to \texttt{FlatGPT2}, and the total size of the vocabulary $BV$. The number of code tables $M$ and the number of bits per code table define the size of the vocabulary $V=\sum_{m=1}^M K_m$. These were manually tuned, but generally, we observed that using fewer code tables with a higher number of bits achieves lower reconstruction error. For example, a vocabulary size of 16 bits can be achieved with both two 8-bit code tables and four 4-bit code tables. The trade-off is that using fewer code tables (with more bits) significantly increases computation time. Increasing the vocabulary $V$ (through the number of code tables and bits per table) improves reconstruction error, as more codes are available for quantising a bucket of channels. However, this increases the total vocabulary $BV$ of the model, resulting in a larger model.

In summary, key modifications compared to \texttt{ChannelGPT2} include vector quantisation (tokenisation) of channel groups, and flattening the channel dimension into the sequence. While in theory we could have flattened the full channel dimension without bucketing, this would have resulted in a 10x longer sequence length. However, we are limited by memory constraints since a standard GPT2 model scales quadratically with the sequence length. Memory-efficient Transformer variants are an active research area \citep{kitaev2020reformer, beltagy2020longformer, wang2020linformer}, but they have other drawbacks, and we leave their application to M/EEG data to future work.

\begin{table}[t!]
\centering
        \begin{tabular}{l|cc}
        %\toprule
           \bf Description  &\bf Parameter & \bf Typical value   \\ \midrule
            
            Number of buckets & $B$ & 30  \\ 
            Number of code tables & $M$ & 2  \\
            Number of bits per code table & $\lceil\log_2(K_m)\rceil$ & 7 \\
            Vocabulary size per bucket & $V=\sum_{m=1}^M K_m$ & 16384

            %\bottomrule
        \end{tabular}
    \caption{\label{table:flatgpt_params} Hyperparameters of the vector quantisation part of \texttt{FlatGPT2}.}
\end{table}

\subsection{Simulation}

To understand the learning mechanisms of Wavenet and GPT2 models and compare them against the linear AR model, we generate simulated time series with controllable properties. Simulations allow us to precisely define key aspects of the data, such as frequency of underlying signals and signal-to-noise ratio (SNR). The simulated time series comprise a finite set of events that govern local dynamics (over hundreds of milliseconds) in the time course. Each event type manifests as a damped oscillation at a specific frequency. The sequence of events (i.e., which event follows the current event) is determined by a transition probability matrix containing conditional probabilities between all events. The lifetime of each event is randomly sampled from a Gamma distribution. For each event, a different 2nd-order autoregressive (AR(2)) model produces the damped oscillations using the following parameters:

\begin{align}
    \phi_1 &= 2\cos\left(\frac{2\pi f}{S}\right)\\
    \phi_2 &= -1
\end{align}
where $f$ is the desired oscillation frequency and $S$ is the sampling rate in Hz.

To simulate damping of the AR-generated oscillations over time $t$, we use:
\begin{equation}
    x_t = z_t e^{-\lambda t}
\end{equation}
where $\lambda$ controls the damping rate. After generating the time series with damped oscillations and event transitions, we apply the inverse hyperbolic sine function to each time step independently to introduce non-linearity. Finally, independent Gaussian noise $\eta_t \sim \mathcal{N}(0, \sigma_\eta^2)$ is added to each time step.

\begin{figure}[!t]
    \centering
    \includegraphics[width=1.0\textwidth]{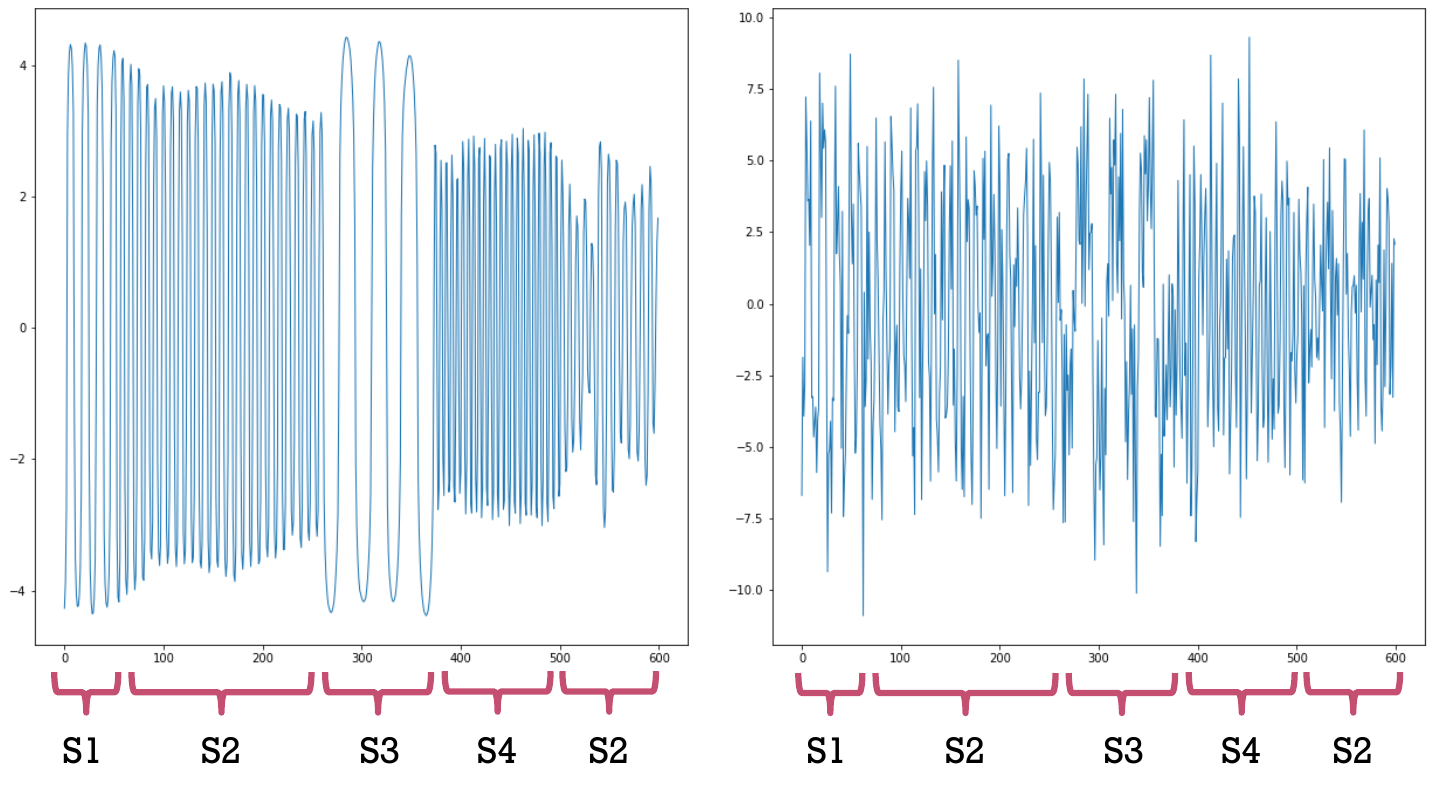}
    \caption{Sample simulated timeseries with four events ($S_1=8$ Hz, $S_2=17$ Hz, $S_3=30$ Hz, $S_4=45$ Hz) at 250 Hz sampling rate. Each event has a different lifetime and AR process noise. The timeseries is shown before and after adding Gaussian noise on the left and right, respectively. The horizontal axis denotes timesteps.}
    \label{fig:simulation}
\end{figure}

In summary, a simulated time series $\mathbf{x}$ is constructed as follows. Let $s \in {1,2,\dots,K}$ indicate an event, with $K$ total event types. Event transitions follow a transition probability matrix $P$, where $P_{ij} = p(s_t = j | s_{t-1} = i)$ gives the probability of transitioning from event $i$ to $j$.

The full generative model is:

\begin{align}
s &\sim \mathrm{Categorical}(P_{s_{t-1}}) \quad if \quad t-1 = T_{s_{prev}} \\
T_s &\sim \mathrm{Gamma}(\alpha, \beta) \\
z_t &= \phi_{1,s} z_{t-1} + \phi_{2,s} z_{t-2} + \epsilon_{s,t} \label{eq:event_ar} \\
x_t &= \sinh^{-1}\left(z_t e^{-\lambda t}\right) + \eta_t
\end{align}

where 

\begin{align}
\epsilon_{s,t} &\sim \mathcal{N}(0, \sigma_s^2)\\
\eta_t &\sim \mathcal{N}(0, \sigma_\eta^2)
\end{align}

The event-specific AR process in Equation~\ref{eq:event_ar} with parameters $\phi_{1,s}, \phi_{2,s}, \sigma_s^2$ generates values over timesteps $t = 1, \dots, T_s$. A new event is only sampled at timestep $T_s$. The Gamma distribution shape and rate parameters are $\alpha$ and $\beta$, respectively. Figure \ref{fig:simulation} shows a sample simulated timeseries before and after adding noise. We perform simulation experiments with univariate, single-channel data. Quantisation can be applied to simulated data similarly to real data.

\section{Results}

\subsection{\texttt{SimpleWavenet} on simulated data}
\label{ssec:simulated_results}

As a preliminary step, we evaluated our models on simulated data, where we could freely control the characteristics of the data and analyse how well the models could reproduce these features. In particular, we aimed to determine whether the continuous-data version of Wavenet (\texttt{SimpleWavenet}) could provide improved performance over a linear autoregressive (AR) model.

In order to obtain useful results with \texttt{SimpleWavenet}, it was crucial to set the signal-to-noise ratio (SNR) of the simulated data to approximately 1. This was observed empirically through extensive experimentation. While we do not have a good explanation for this, we believe that the point-estimate prediction of Wavenet make the model dynamics very sensitive, and it can easily diverge during generation. We generated single-channel simulations with 4, 8, and 12 distinct states, with state frequencies as follows:

\begin{enumerate}
\item 4 states: 10, 24, 36, 45 Hz
\item 8 states: 10, 14, 18, 22, 26, 33, 38, 45 Hz

\item 12 states: 8, 11, 14, 17, 20, 23, 26, 29, 35, 38, 41, 45 Hz
\end{enumerate}

The state timecourse is generated by a probability transition matrix with state lifetimes sampled from a Gamma distribution, and each state plays out one of the oscillations above. For the rest of this section we use the terms events and states interchangeably. These frequency bands cover the physiologically relevant ranges typically observed in magnetoencephalography (MEG) data \citep{baillet2017magnetoencephalography}. The total simulation duration was 3000 seconds at a 250 Hz sampling rate. The gamma distribution used for sampling event lifetimes had a shape and scale parameter of 10, yielding a probability density function with a peak around 90 ms, consistent with observed state lifetimes in empirical MEG data \citep{vidaurre2018spontaneous}. State transition probabilities were drawn from a uniform distribution.

The noise for the AR(2) models was sampled from a uniform distribution between 0.8 and 1.0. The damping exponent was 0.005 and the variance of the added Gaussian noise was 2.5. Simulated data was split into training and validation sets with a 4:1 ratio. The training set was z-transformed to have zero mean and unit variance. The validation set was standardised using the same parameters. We trained an AR(64) model and a \texttt{SimpleWavenet} with 8 hidden channels on the training data and evaluated on the validation set. \texttt{SimpleWavenet} was trained until validation loss stabilised.

\begin{figure}[!t]
    \centering
    \includegraphics[width=0.95\textwidth]{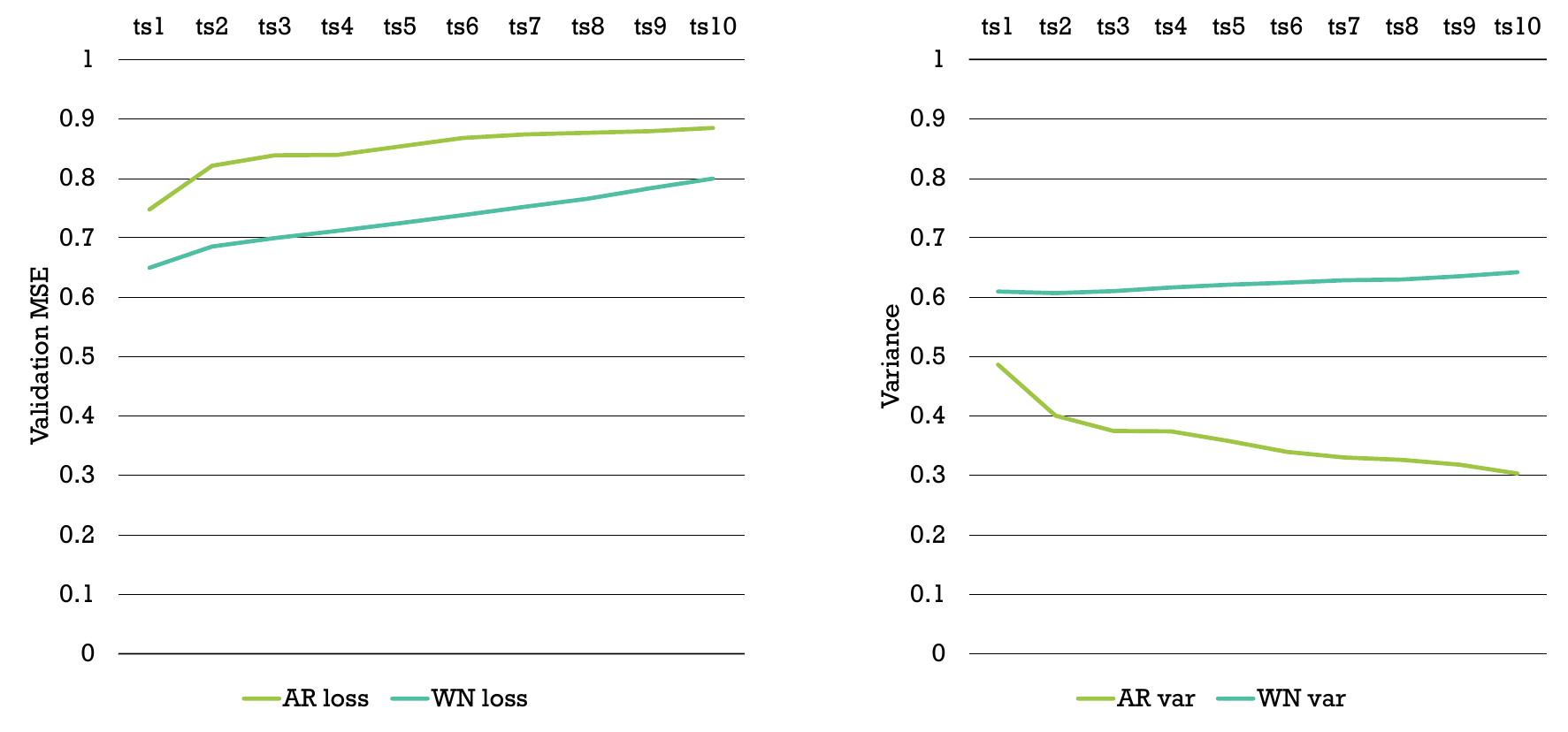}
    \caption{MSE loss (left) and variance of predictions (right) for AR and Wavenet (WN) models. Performance is shown for recursive generation across future timesteps (horizontal axis, \textit{ts}). Trainings were run on the simulated data with 8 states.}
    \label{fig:simulation_losses}
\end{figure}

Figure~\ref{fig:simulation_losses} compares the mean squared error (MSE) loss and variance of predictions for the AR and Wavenet models across multiple future timesteps, obtained by recursive generation without additional noise. \texttt{SimpleWavenet} achieved lower loss across all horizons, with variance close to the true data variance of 1. As expected, loss increased with longer prediction horizons. AR performance did not improve with higher model orders, suggesting \texttt{SimpleWavenet}'s superior performance stems from nonlinearity.

\begin{figure}[!t]
    \centering
    \includegraphics[width=1.0\textwidth]{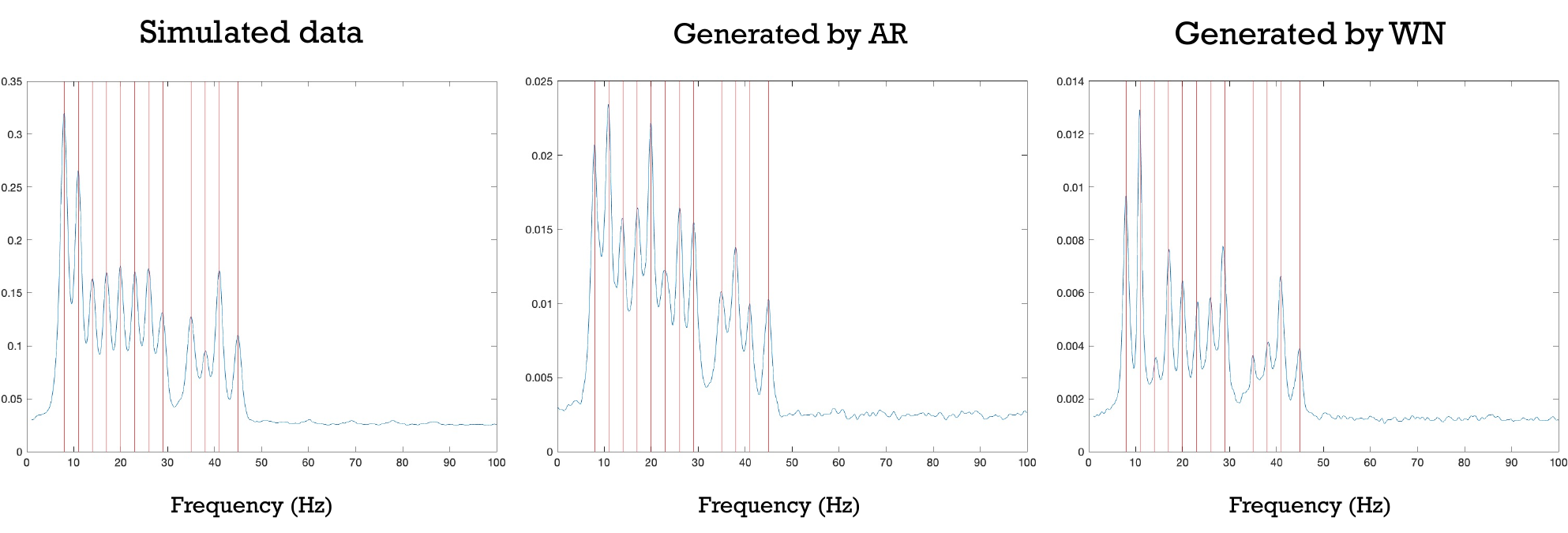}
    \caption{Power spectra for simulated data (left), AR-generated data (middle), and Wavenet-generated data (WN, right). Vertical lines indicate ground-truth state frequencies.}
    \label{fig:generation}
\end{figure}

We generated 1000 seconds of data from the models trained on the 12-state simulations and computed the power spectral density using Welch's method \citep{welch1967use}. Figure~\ref{fig:generation} shows the resulting power spectra. Both models accurately reproduced the frequency profile, with clear peaks at the true state frequencies.

\begin{figure}[!t]
    \centering
    \includegraphics[width=1.0\textwidth]{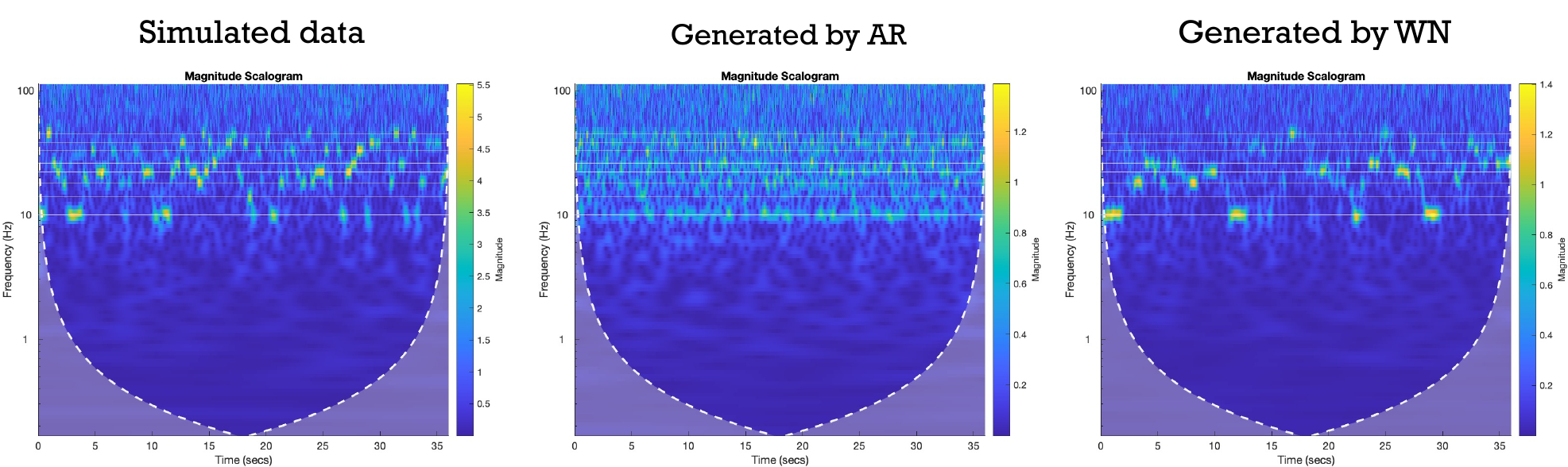}
    \caption{Wavelet transforms for simulated data (left), AR-generated data (middle), and Wavenet-generated data (WN, right). White horizontal lines indicate ground-truth state frequencies.}
    \label{fig:wavelet}
\end{figure}

To examine how frequencies evolve over time, we computed wavelet transforms \citep{mallat1999wavelet} for 36 seconds of generated data, from models trained on the simulation with 8 states. As shown in Figure~\ref{fig:wavelet}, qualitative differences emerged despite the similar power spectra. \texttt{SimpleWavenet} produced clear periods dominated by a single frequency, closely matching the true generative process. By contrast, the AR model blended frequencies. This demonstrates \texttt{SimpleWavenet}'s ability to capture greater signal complexity, likely due to nonlinearity.

\begin{figure}[!t]
    \centering
    \includegraphics[width=0.5\textwidth]{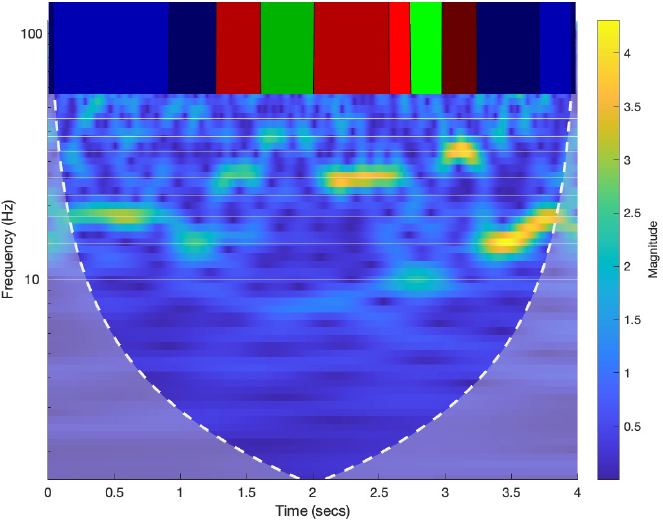}
    \caption{Wavelet transform for simulated data with the HMM-inferred state time course superimposed. States coincide with distinct frequencies. Each state is a different colour.}
    \label{fig:wavelet_hmm}
\end{figure}

\begin{figure}[!t]
    \centering
    \includegraphics[width=1.0\textwidth]{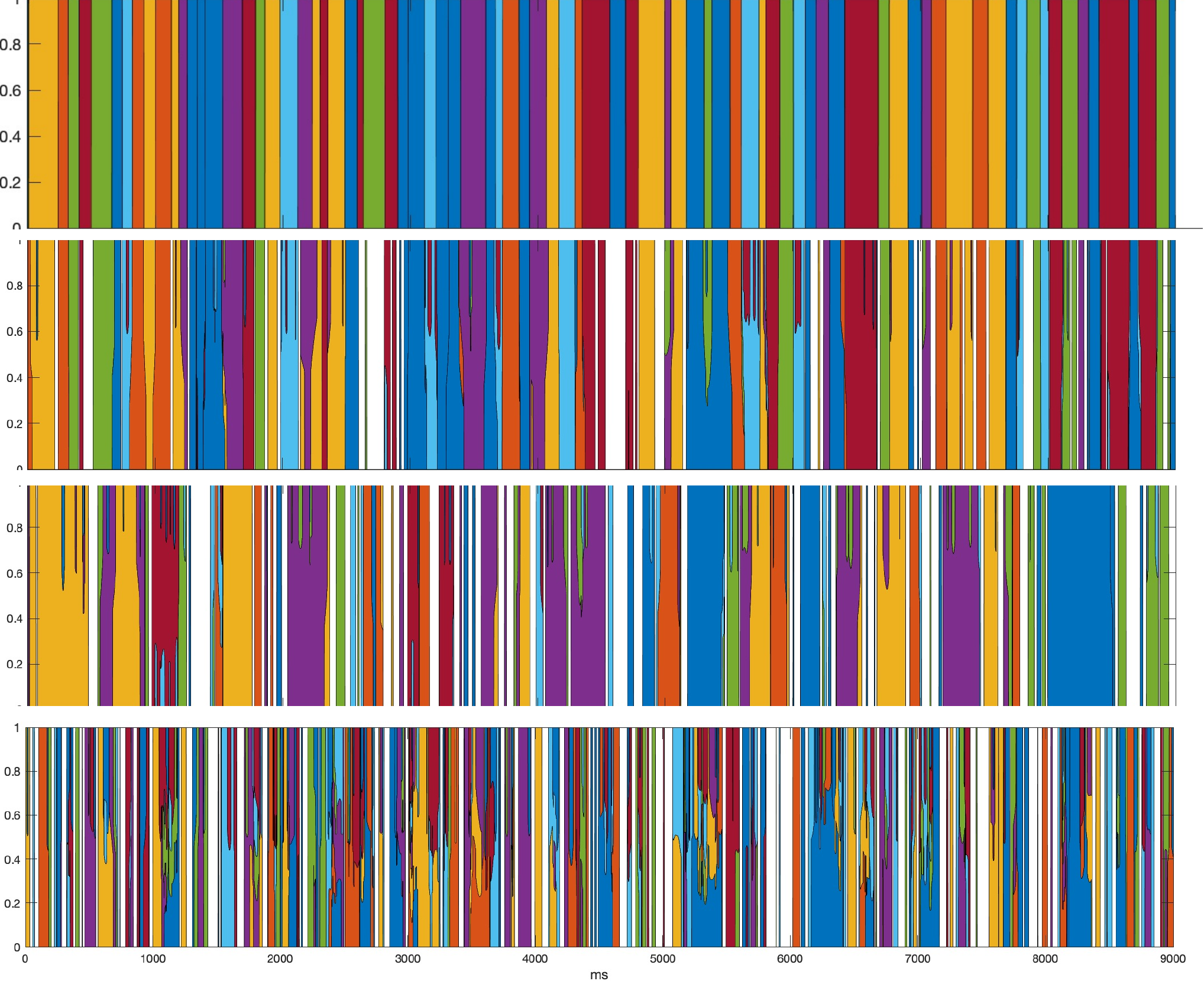}
    \caption{Comparing state probability time courses extracted by the naive method for simulated data (2nd row), \texttt{SimpleWavenet} generated data (3rd row), and AR generated data (bottom). The top row shows the ground-truth state time course used to generate the simulated data. \texttt{SimpleWavenet} and AR time courses do not line up with the simulated data, since data is generated from random noise. The horizontal axis shows time in milliseconds (ms). The vertical axis shows the probability distribution of states represented by different colours.}
    \label{fig:stc}
\end{figure}

To quantitatively evaluate how well \texttt{SimpleWavenet} reproduced the state-switching dynamics, we extracted state time courses from the generated data. The wavelet analysis clearly illustrated frequency switching, so we first extracted time courses for each of the 8 known frequencies. At each timestep, the frequency with maximum power was treated as the predicted state. For a more principled approach, we trained a HMM on these frequency time courses to infer states in an unsupervised manner \citep{vidaurre2018spontaneous}. Figure~\ref{fig:wavelet_hmm} shows the HMM-inferred state time course aligns closely with frequency switching in the wavelet transform. State time courses extracted by taking the most probable frequency (naive method) at each timestep are compared in Figure~\ref{fig:stc}. Visually, \texttt{SimpleWavenet} largely reproduced the switching structure, while the AR model did not.

\begin{figure}[!t]
    \centering
    \includegraphics[width=1.0\textwidth]{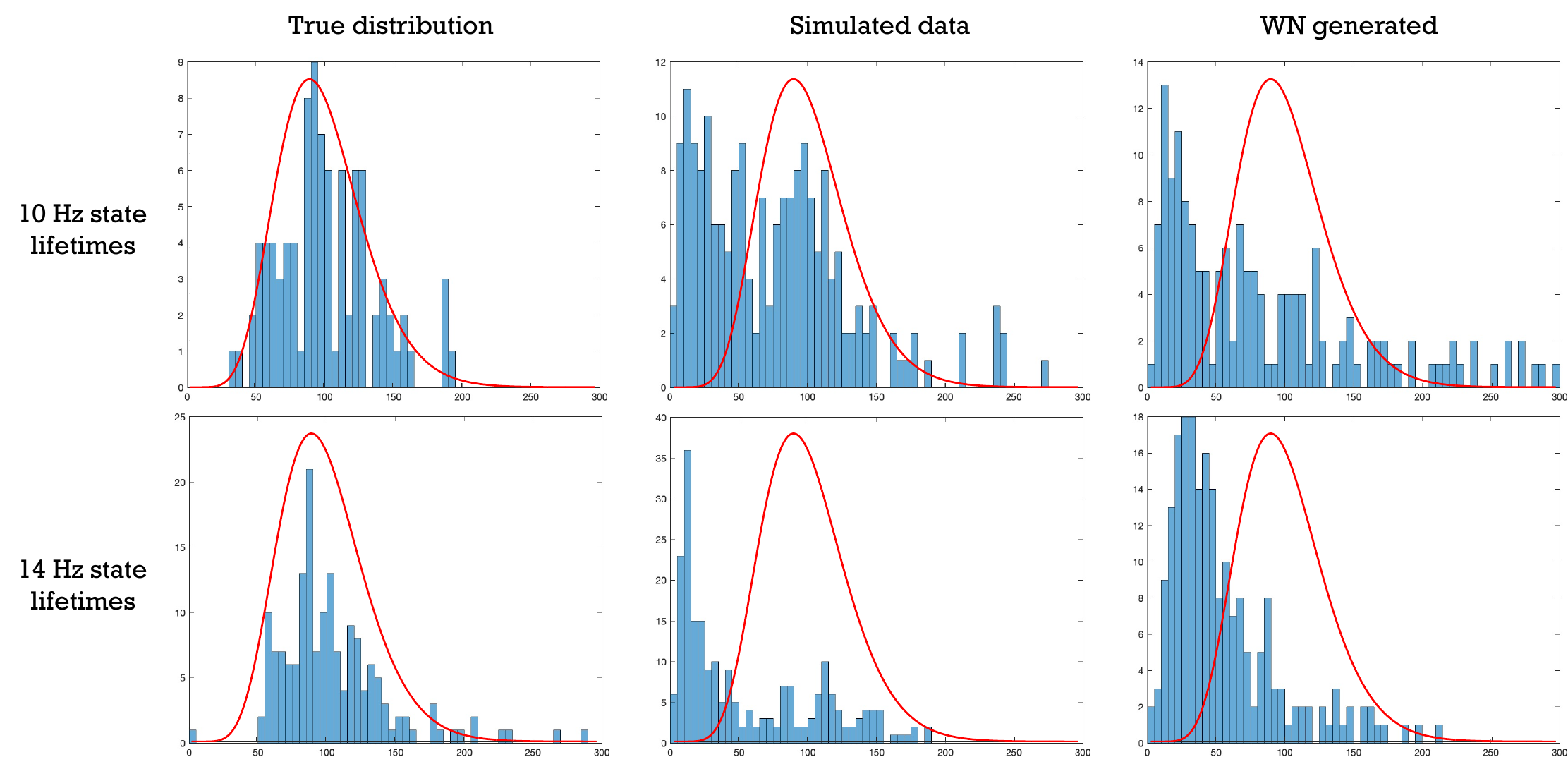}
    \caption{Lifetime distributions (in milliseconds) for the 10 and 14 Hz states. The first column is the true distribution originally sampled to generate the simulated data. The second and third columns are the state lifetime distributions based on the HMM state time courses inferred from the simulated and \texttt{SimpleWavenet} (WN) generated time series, respectively. The red curve shows the true gamma probability density function from which the state lifetimes were sampled for the simulated data.}
    \label{fig:lifetimes}
\end{figure}

Finally, we analysed state lifetime distributions, which characterise state persistence. Taking the $\mathrm{argmax}$ of the HMM time course gave a state sequence from which lifetimes were calculated. Figure~\ref{fig:lifetimes} compares lifetime distributions for simulated and \texttt{SimpleWavenet} data. The noisy state extraction meant that distributions for simulated data differed from the true gamma distribution. However, the \texttt{SimpleWavenet} lifetimes closely matched the simulated data, with slightly more short states due to noisier state switching.

\begin{figure}[!t]
    \centering
    \includegraphics[width=1.0\textwidth]{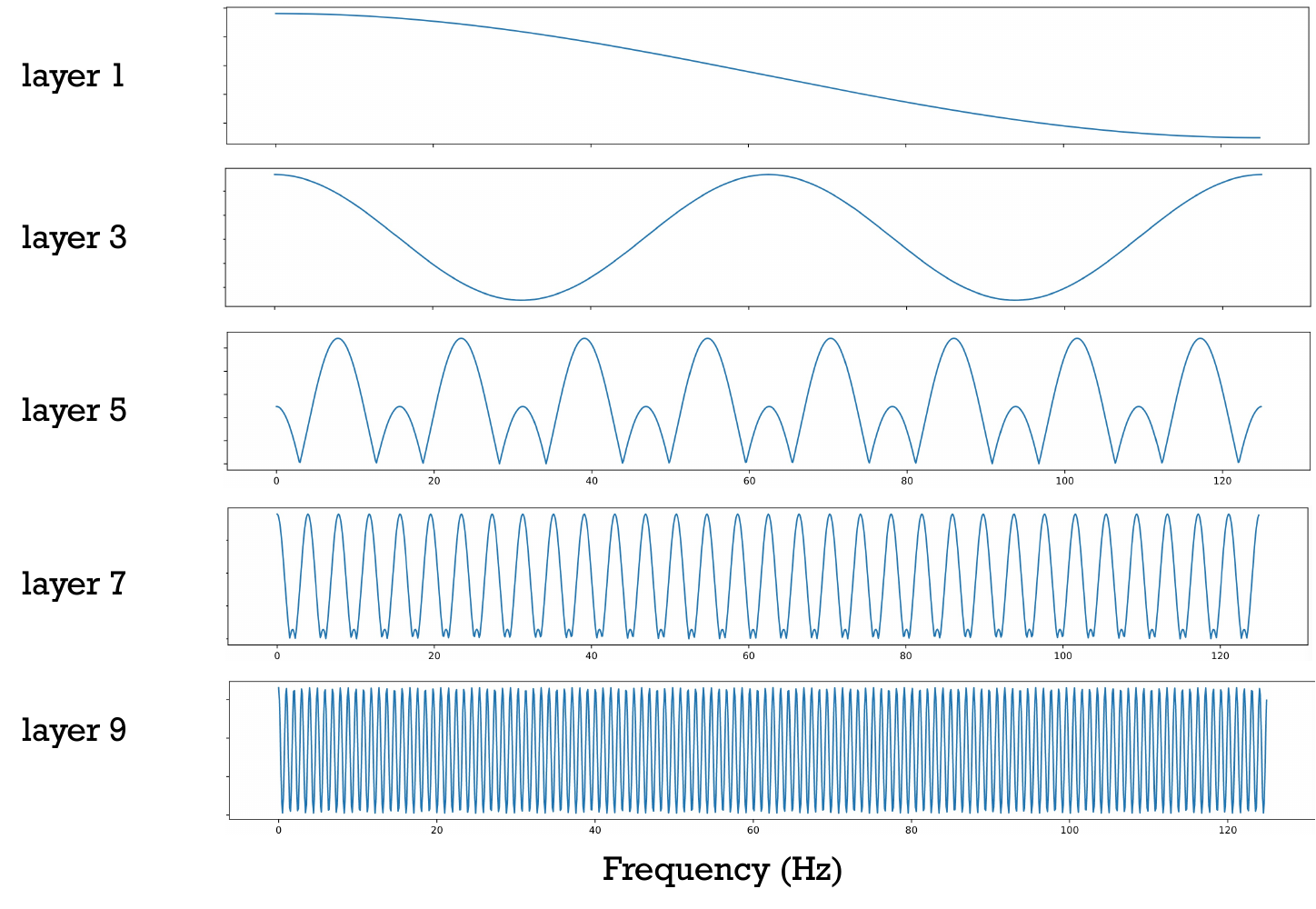}
    \caption{The power spectra of 5 random kernels (as FIR filters) across layers of \texttt{SimpleWavenet}.}
    \label{fig:kernel_filter}
\end{figure}

\begin{figure}[!t]
    \centering
    \includegraphics[width=1.0\textwidth]{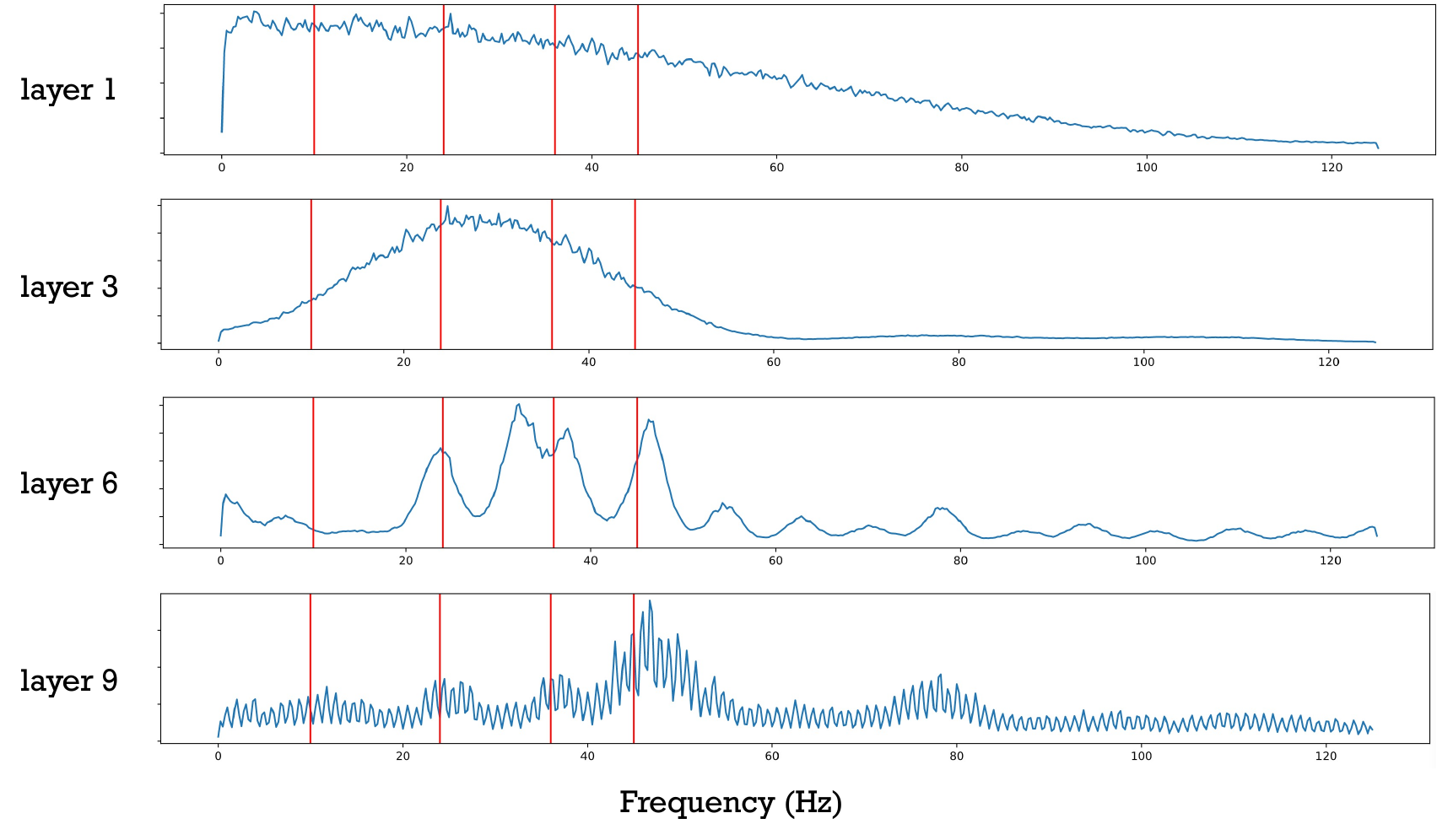}
    \caption{The power spectra of 5 random kernels from kernel FIR analysis of \texttt{SimpleWavenet}. Vertical lines indicate the 4 ground truth frequencies.}
    \label{fig:kernel_network_fir}
\end{figure}

Additionally, we analysed the power spectra of kernels across layers to understand how the network processes frequencies. The power spectra predominantly exhibits the effects of dilation seen in Figure~\ref{fig:kernel_filter} - sparser kernels have narrower periodic peaks. \textit{Kernel FIR analysis}, however, reveals that deeper kernels became more selective for the 4 ground truth frequencies used in this simulation, likely reflecting the effect of dilations enabling longer temporal receptive fields (Figure~\ref{fig:kernel_network_fir}). Early layers can only apply wide filters, while deeper layers can be more selective. The observed power spectra looks like a superposition of the dilation effect from Figure~\ref{fig:kernel_filter} and the ground truth frequency peaks.

Together these results on controlled simulated data provide promising evidence that \texttt{SimpleWavenet} can capture structure in simulated electrophysiological signals better than linear models. The model demonstrated an ability to reproduce complex temporal dynamics, such as state switching.

\subsection{Quantised simulated data}

From the previous section, we can conclude that a simplified version of Wavenet applied to continuous simulated data performs well. As our baseline model is a linear autoregressive (AR) model, this showed that using a multi-layer nonlinear architecture can better capture the dynamics of the data such as switching between events characterised by different oscillations. Next, we wanted to test whether the full Wavenet model is also able to produce this switching behaviour on the simulated data when it is quantised. This is primarily aimed at validating the quantised Wavenet approach, and whether training through cross-entropy loss and generating data through sampling from the output probability distribution works as well as the Wavenet trained on continuous data. Validating the GPT2 approach on simulated data and drawing comparisons with Wavenet is left for future work.

Specifically, we generated 1000 seconds of simulated data with 8 events, using the frequencies described in the previous section. The data was generated at 1000 Hz, then a 1-100 Hz infinite impulse response (IIR) bandpass filter was applied, and then downsampled to 200 Hz. This approach was used to match preprocessing steps of real data. Finally, the data was quantised using the mu-law companding transform to 256 bins. Data was split into train and validation sets with a 4:1 ratio.

We set the model order (receptive field) of the linear AR model to 255. As our Wavenet model we used \texttt{WavenetFullChannel} with a matched receptive field. We used two identical dilation blocks stacked on top of each other. A single block contained 7 layers with doubling of the dilation factors in successive layers. Dropout rate between layers was set to 0.2. The embedding for the quantised inputs was set to size 64, and the hidden channel size of the convolutions was 128. The channel dimension of the skip convolutions was set to 512. The linear AR model is applied directly to the quantised values and produces continuous outputs.

The accuracy and mean squared error (MSE) of predictions on the validation set is presented in Table~\ref{table:simulated_quantised}. Accuracy is computed over the 256 bins. For the linear AR model, the closest bin of the prediction is used to compute accuracy. For \texttt{WavenetFullChannel}, MSE is obtained by reconstructing the original signal from the quantised output, by applying the inverse of the mu-law transform, and comparing to the target values. The results clearly show that \texttt{WavenetFullChannel} is somewhat better at predicting future timesteps compared to the AR(255) model. We also tried an AR(10) model which showed the same performance as AR(255). This demonstrates that the linear AR model is not able to leverage longer receptive fields effectively. Interestingly, the next-timestep prediction accuracy of the AR model is worse than the repeat baseline, however the MSE is much lower. Because the AR model was optimised with MSE, it makes sense that accuracy might not reflect its performance as well.

\begin{table}[t]
\centering
\begin{tabular}{l|l|l}

& Accuracy  & MSE \\
\midrule
\texttt{WavenetFullChannel}& 4.3\% & 0.038 \\

AR(255)& 2.1\%  & 0.049 \\

Repeat baseline& 2.3\%  & 0.092
\end{tabular}

\caption{\label{table:simulated_quantised}Comparing the accuracy and MSE of a linear AR model with order 255 and \texttt{WavenetFullChannel}. The last row shows the performance achieved by a baseline model which always repeats the last timestep. Chance level is $100/256\%$.}
\end{table}

\begin{figure}[!t]
\begin{subfigure}{0.49\textwidth}
  \centering
  \includegraphics[width=1.0\linewidth]{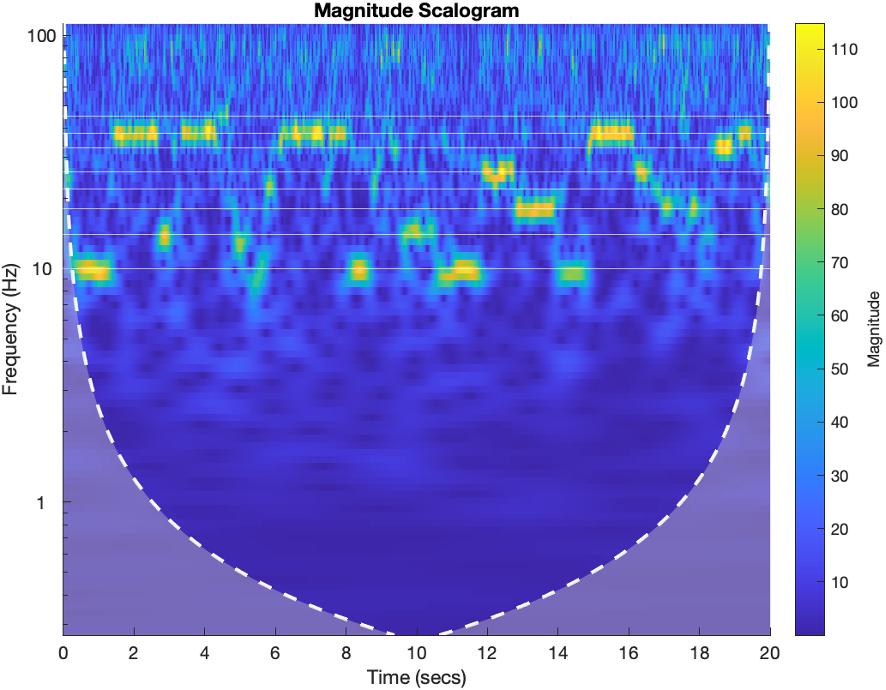}
  \caption{\texttt{WavenetFullChannel}}
  \label{fig:wavelet_wavenetfullchannel}
\end{subfigure}%
\begin{subfigure}{0.49\textwidth}
  \centering
  \includegraphics[width=1.0\linewidth]{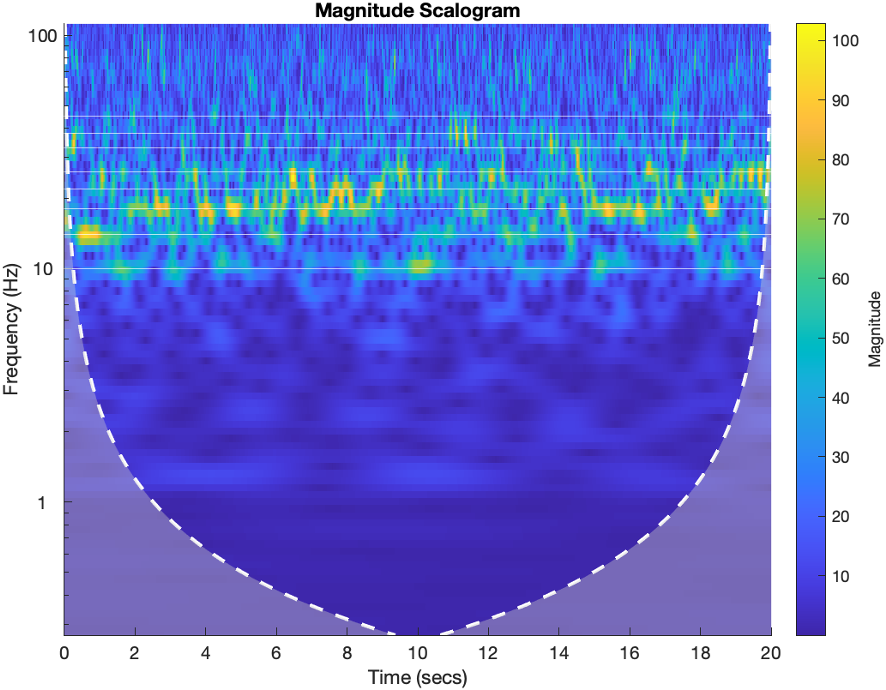}
  \caption{Linear \texttt{WavenetFullChannel}}
\end{subfigure}
\caption{Wavelet transform of the generated data from \texttt{WavenetFullChannel} (left) and the ablated (linear) version (right). White horizontal lines show the true frequencies used to create the simulated data.}
\label{fig:wavelet_linear_comparison}
\end{figure}

\begin{sloppypar}
Next, we tested whether \texttt{WavenetFullChannel} can generate the event-switching dynamics of the simulated data. We sampled from the full output distribution during generation. As shown in Figure~\ref{fig:wavelet_wavenetfullchannel}, it has similarly capabilities to \texttt{WavenetSimple}. Thus, the simulation now becomes too simplistic for these models, and we move to applying them to real data in the next section.
\end{sloppypar}

We also performed an ablation of the nonlinearity of \texttt{WavenetFullChannel}. In the ablated model, we replaced all activation functions with the identity function $y=x$. This ensures that the model is essentially linear, other than the softmax function at the output, and the nonlinear training dynamics caused by dropout and successive dilated convolutions. A comparison of generated data between the standard \texttt{WavenetFullChannel} and the ablated (linear) version is shown in Figure~\ref{fig:wavelet_linear_comparison}. This clearly shows that the linear version is much worse at capturing the event-switching structure of the simulated data, and its generation spectrum looks closer to the linear AR model.

In summary, we have demonstrated that the full Wavenet model with quantised inputs can successfully model the switching dynamics in simulated time series data, outperforming linear AR models. The nonlinear nature of Wavenet was shown to be critical through an ablation study.

\subsection{Next time-step prediction performance}
\label{ssec:flatgpt_results}

For \texttt{FlatGPT2} we set the (temporal) receptive field to be between 120 and 240 because of memory constraints. Note that the total (actual) receptive field of the model is the temporal receptive field multiplied by the number of buckets + 1. All embedding sizes were set to 96, and we used 8 GPT2 layers, with 8 attention heads. Dropout was set to 0 and we used early stopping on the validation set. The quantisation parameters are given in Table~\ref{table:flatgpt_params}.

Next-timestep forecasting accuracy for different models on a sample subject is shown in Table \ref{table:forecast_results}. Beyond standard accuracy  (the number of true positives divided by the number of all examples), we also evaluated top-5 accuracy, counting a prediction as correct if the true bin was within the 5 most probable bins. Surprisingly, all models performed only slightly better than a naive baseline of repeating the previous timestep's value. This suggests next-timestep metrics do not effectively capture model performance.

As expected, the linear AR model had lower MSE but worse accuracy than the nonlinear models. This can be because MSE measures the distance of the prediction to the target, while accuracy is only 1 if the prediction is in the target bin. Thus it can be that the AR model always predicts values that are slightly closer to the target, but never quite falling in the target bin. While \texttt{WavenetFullChannel} appears to be worse, \texttt{WavenetFullChannelMix} and \texttt{ChannelGPT2} have nearly identical performance.

All these observations are very likely a consequence of these metrics not capturing actual goodness of modelling the data. Specifically MSE and accuracy measure only how well models predict the next timestep. As we have seen in Chapter~\ref{Chap5}, these models have very different dynamics when generating data over longer temporal horizons. Perhaps looking at these metrics when recursively generating multiple timesteps in the future might be more informative.

\texttt{FlatGPT2} performance is not comparable to other models, since the output distribution is over 16384 tokens, and the prediction is done sequentially for channels as well. This latter point is probably the reason why we observe such high accuracies, since it is much easier to predict the same timestep of one channel at a time, while some others (possibly with high correlation) have already been predicted. In addition it is possible to have a skewed distribution of tokens and thus true chance level may be higher than $1/16384$.

\begin{table}[t!]
\centering
        \begin{tabular}{l|ccc}
        %\toprule
           \bf Model  &\bf MSE & \bf Top-1 Accuracy & \bf Top-5 Accuracy   \\ \midrule
            
            Repeat baseline & 0.024 & 1.5 & 7.6  \\ 
            AR(255) & 0.016 & 1.5 & 7.5  \\
            \texttt{WavenetFullChannel} & 0.026 & 2.0 & 9.8 \\
            \texttt{WavenetFullChannelMix} & 0.022 & 2.2 & 10.8 \\
            \texttt{ChannelGPT2} & 0.023 & 2.2 & 10.9 \\ \midrule
            \texttt{FlatGPT2} & - & 3.0 & 10.8 \\
            \texttt{FlatGPT2} recursive & - & 0.0015 & 0.0069

            %\bottomrule
        \end{tabular}
    \caption{\label{table:forecast_results} Test data next-timestep prediction performance across various models. Accuracy values are given in percentages. Note that \texttt{FlatGPT2} is not comparable to other models as the prediction is done for buckets with much larger vocabularies. Chance-level for \texttt{FlatGPT2} is $1/16384$, while for other models it is $1/256$. \texttt{FlatGPT2} \textit{recursive} refers to recursive prediction of all buckets within the same timestep.}
\end{table}

To allow for true next-timestep prediction, where the model can not access information from other channels in the same timestep, we also computed a recursive loss. This is done by recursively generating all of the channels/buckets within a timestep $t$. Thus the model has to rely on its own predictions and can not "cheat" by using the true value of buckets $1, ..., b$, when predicting bucket $b+1$ within timestep $t$. Importantly this metric is computed after the model is trained, so there is a discrepancy in how the model was trained (autoregressively over buckets) and tested. As we can see in Table~\ref{table:forecast_results} the accuracy of recursive predictions is much lower than in the normal operating mode. This kind of analysis ties into losses that optimise for multi-timestep future horizons. We believe this may be important to consider as future research for improving \texttt{FlatGPT2} and \texttt{ChannelGPT2}.

\begin{figure}[!t]
\begin{subfigure}{0.49\textwidth}
  \centering
  \includegraphics[width=1.0\linewidth]{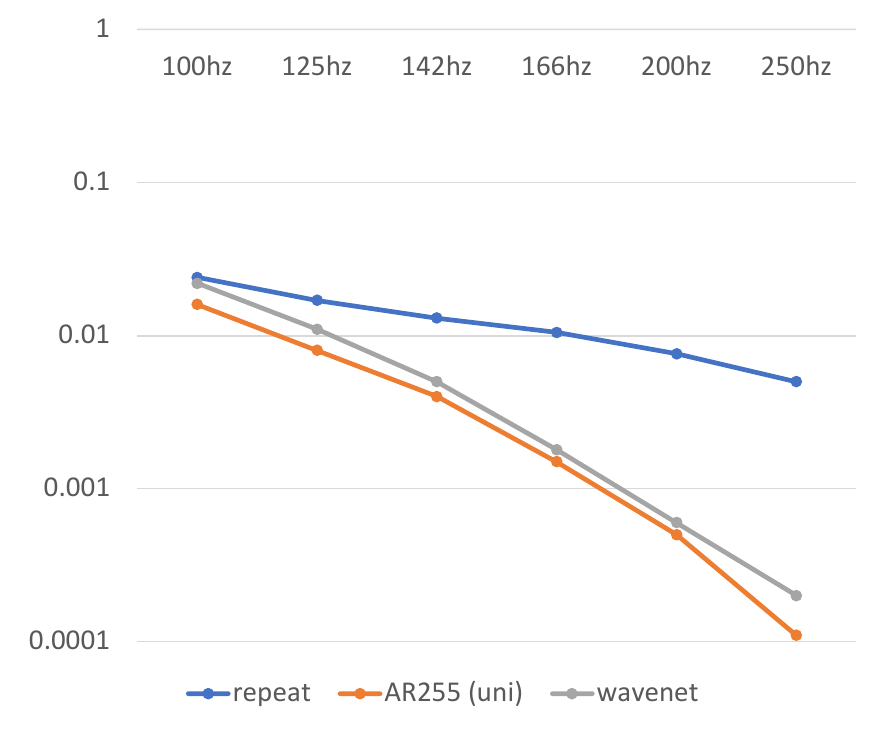}
  \caption{Test MSE}
\end{subfigure}%
\begin{subfigure}{0.49\textwidth}
  \centering
  \includegraphics[width=1.0\linewidth]{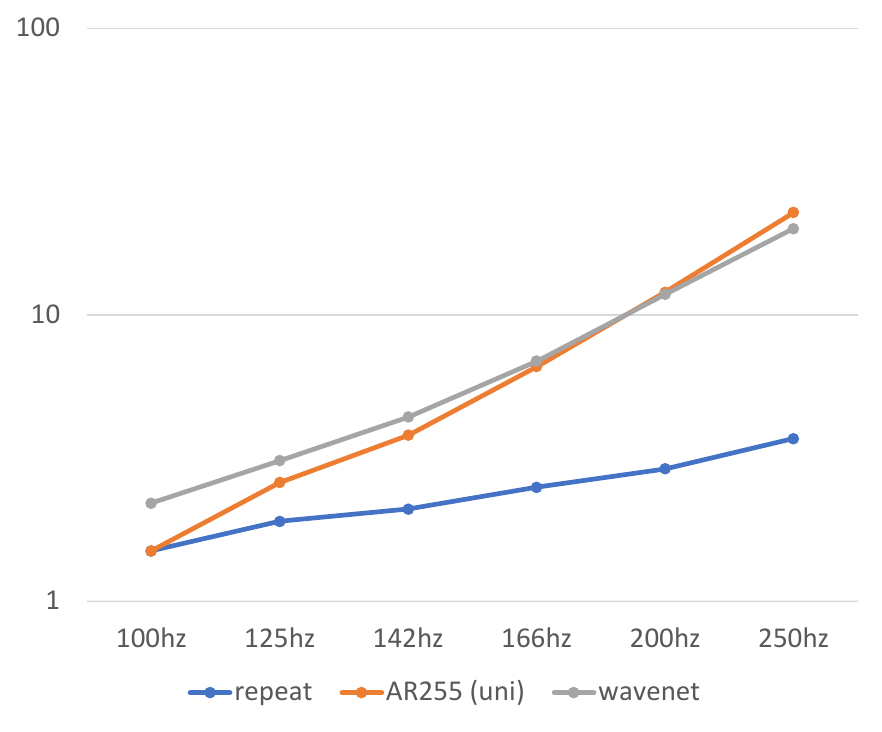}
  \caption{Test accuracy (\%)}
\end{subfigure}
\caption{Comparing AR(255) and \texttt{WavenetFullChannelMix} (wavenet) across increasing sampling rates of the data. \textit{repeat} refers to the repeat baseline. Accuracy is given in percentages.}
\label{fig:sr_comparison}
\end{figure}

We further analysed sampling rate effects on forecasting performance in Figure \ref{fig:sr_comparison}. We trained the AR(255) and \texttt{WavenetFullChannelMix} models on increasing sampling rates of the data from 100 Hz to 250 Hz. The lowpass filter was kept the same at 50 Hz. The receptive fields were kept the same in terms of timesteps, thus they decreased accordingly in terms of actual time in seconds. As expected, both AR and Wavenet models improved markedly with higher sampling rates, as the prediction task became easier when timesteps were closer together. The performance gap between models and the repeating baseline also grew with sampling rate. However, these trends are likely influenced by both the changing prediction interval and filtering artefacts. It is unlikely that such marked improvement would be caused by better modelling of higher-frequency content. Varying the low-pass cutoff with sampling rate reduced performance, suggesting filtering effects dominate. Removal of noise with lower lowpass filters is also a possible explanation. Overall, next-timestep prediction remains a problematic metric. Not only does it not differentiate between models of very different characteristics and dynamics it is also heavily dependent on arbitrary factors like sampling rate.

\subsection{FlatGPT2 on group data}

\begin{figure}[!t]
\centering
\begin{subfigure}{0.33\textwidth}
  \centering
  \includegraphics[width=1.0\linewidth]{forecast_figures/data_psd.pdf}
  \caption{Data}
  \label{fig:data_psd2}
\end{subfigure}%
\begin{subfigure}{0.33\textwidth}
  \centering
  \includegraphics[width=1.0\linewidth]{forecast_figures/gptflat_p95_psd.pdf}
  \caption{\texttt{FlatGPT2}}
  \label{fig:gptflat_p95}
\end{subfigure}%
\begin{subfigure}{0.33\textwidth}
  \centering
  \includegraphics[width=1.0\linewidth]{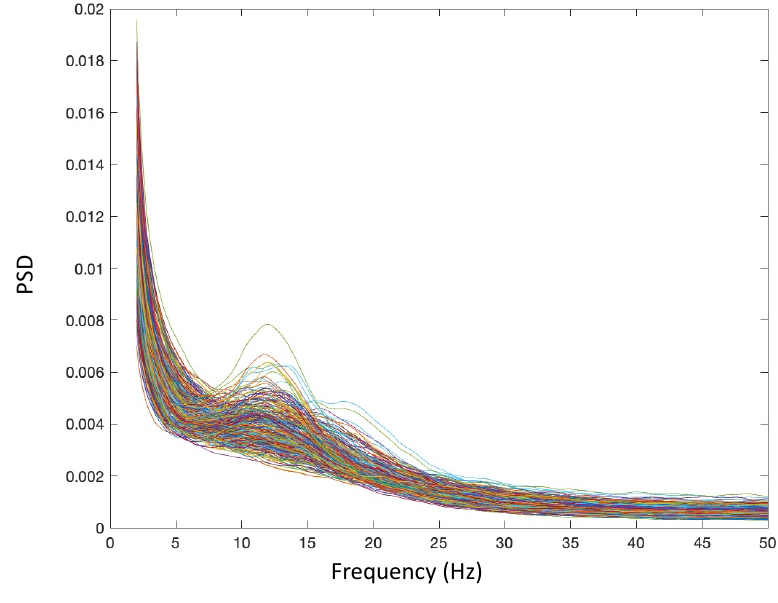}
  \caption{\texttt{FlatGPT2-group}}
  \label{fig:gptflat_group_psd}
\end{subfigure}
\caption{Comparison of generated data PSD across data single-subject \texttt{FlatGPT2} and \texttt{FlatGPT2-group}. Each line represents a different MEG channel.}
\label{fig:psd_compare3}
\end{figure}

Unfortunately, even scaling \texttt{FlatGPT2} did not improve evoked generation. However, we did find that the spectral content of the generated data matched the real data much better than the single-subject version of \texttt{FlatGPT2} (Figure~\ref{fig:psd_compare3}). \texttt{FlatGPT2-group} seemed to scale particularly well with model size as larger models achieved lower and lower loss, improving test accuracy by multiple folds (16.1\% top-1 and 40.1\% top-5 accuracy) over single-subject \texttt{FlatGPT2} (3\% top-1 and 10.8\% top-5 accuracy). This is interesting behaviour compared to \texttt{ChannelGPT2-group} which did not improve much on our forecasting metrics. It remains to be seen whether even more data and larger models are needed to make this type of architecture viable.

\subsection{Ablations}

\begin{figure}[!hb]
    \centering
    \includegraphics[width=0.7\textwidth]{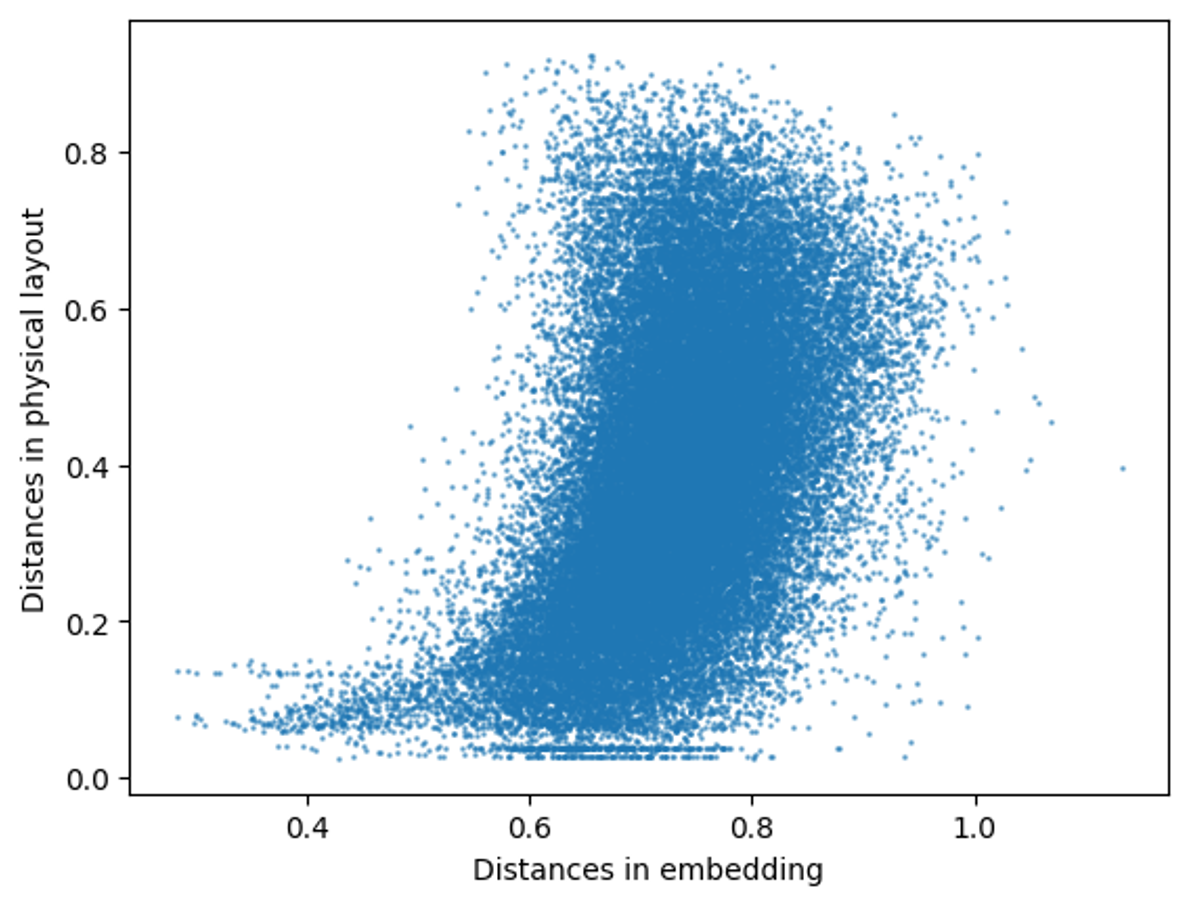}
    \caption{Plotting pairwise Euclidean distances of channels in real, physical space versus embedding space. Sensors that are near to each other in the real sensor montage tend to have more similar embeddings. Each point represents a different pair of channels. Correlation is 0.45.}
    \label{fig:emb_distances}
\end{figure}

\chapter{Decoding thoughts}

\section{Results}

\subsection{Evoked analysis}
\label{ssec:inner_speech_evoked}

\begin{figure}[!t]
    \centering
    \includegraphics[width=0.99\textwidth]{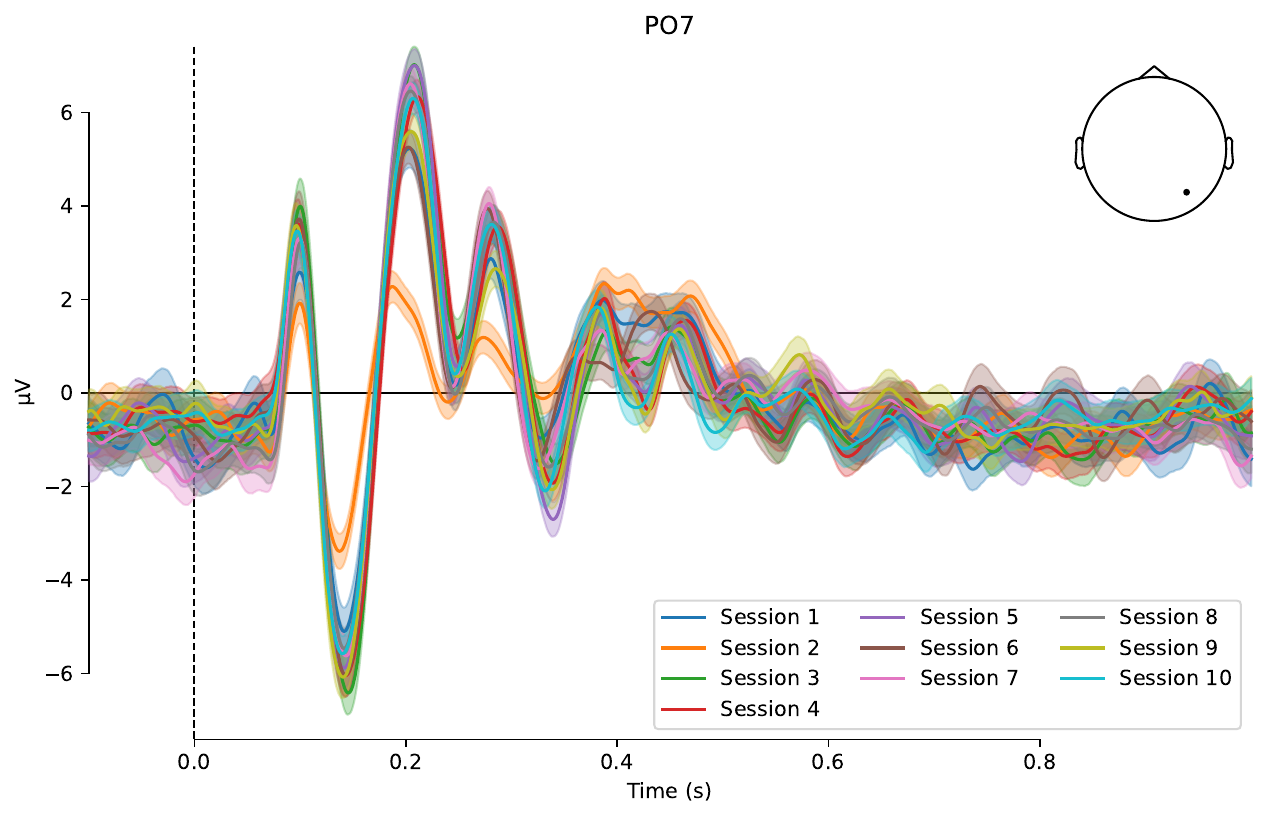}
    \caption{Evoked responses across the 10 EEG sessions of P4 for 1 electrode (PO7) in the visual area. Shading indicates 95\% confidence interval across trials. Timepoint 0 indicates stimulus (cross) onset.}
    \label{fig:po7_sessions}
\end{figure}

We computed evoked responses jointly across the two inner speech types without any baseline correction. We visualise these for each session for electrode PO7 (visual area) in Figure~\ref{fig:po7_sessions}. It is evident that evoked responses across sessions are very similar in the visual area, except for session 2 which appears to be an outlier. The plot also shows the expected response to visual stimulus, with the first peak as early as 100 ms post-stimulus (P100), followed by several peaks and troughs. This demonstrates the oscillatory nature of the evoked response in the visual area, likely due to the cross cue used in the inner speech task.

\begin{figure}[!t]
    \centering
    \includegraphics[width=0.99\textwidth]{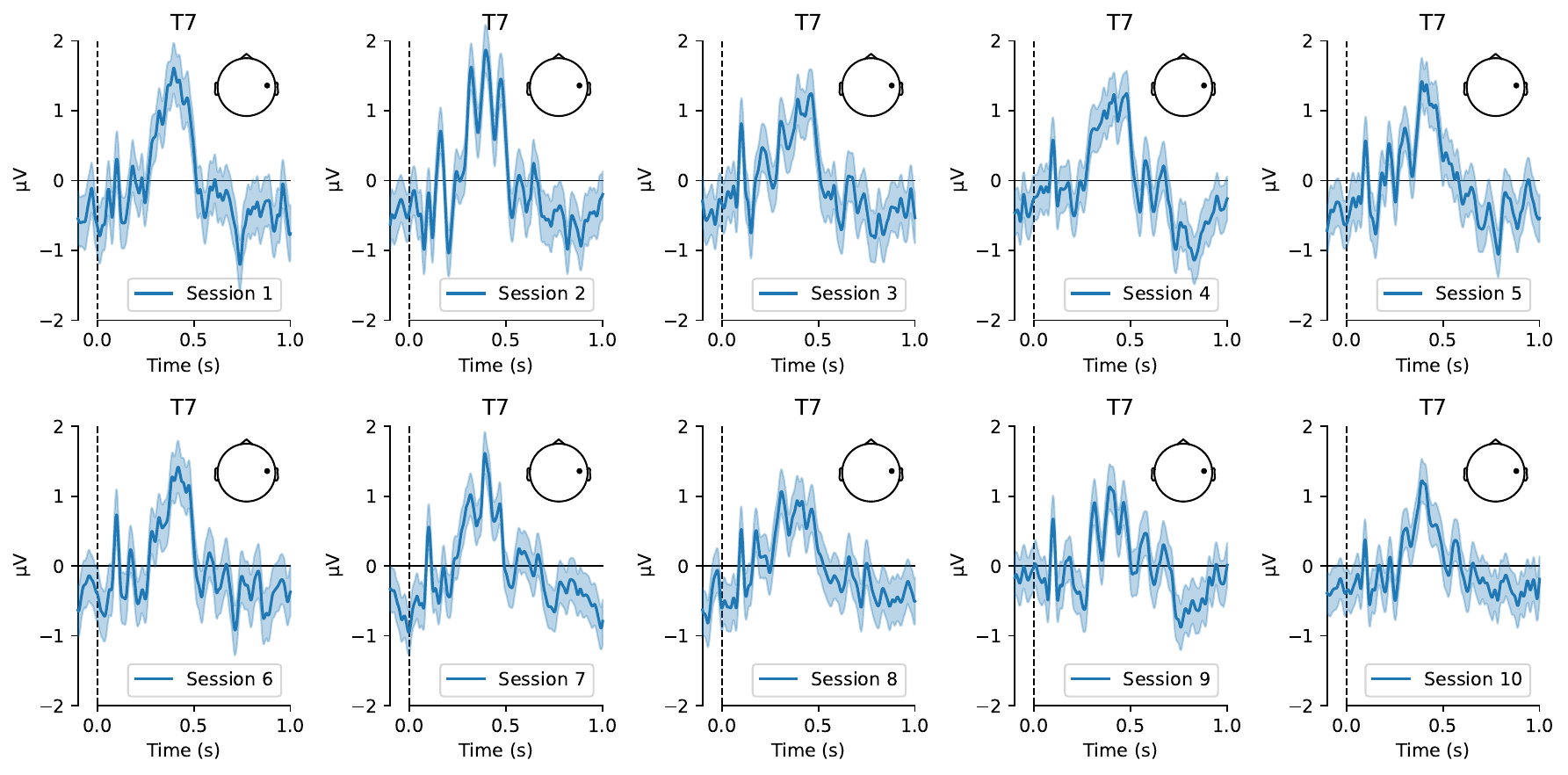}
    \caption{Evoked responses across the 10 EEG sessions of P4 for 1 electrode (T7) above the temporal lobe. Shading indicates 95\% confidence interval across trials. Timepoint 0 indicates stimulus (cross) onset.}
    \label{fig:t7_sessions}
\end{figure}

While we were also interested to compare other channels across sessions, non-visual channels exhibit more noise. Thus, we plot evoked responses in separate plots per-session. Figure~\ref{fig:t7_sessions} shows this for the T7 electrode, which is above the temporal lobe. Evoked responses in the temporal lobe are much more variable across sessions, but all display peak activity around 400 ms post-stimulus. However, it is questionable whether this reflects language-related activity (due to inner speech), or merely spreading/propagation of the visual response. The latter is more probable.

\begin{figure}[!t]
    \centering
    \includegraphics[width=0.99\textwidth]{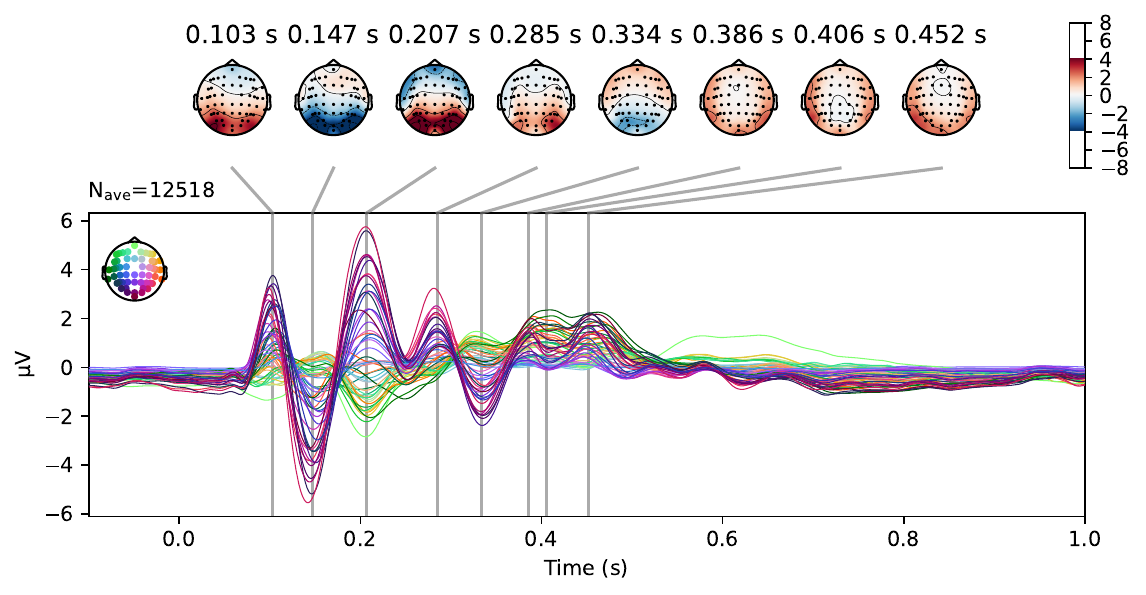}
    \caption{Joint evoked responses for each channel averaged across all 10 EEG sessions of P4. The spatial topography and timestamp of notable peaks is shown in the upper part.}
    \label{fig:evoked_spatiotemporal}
\end{figure}

It is important to note that since we utilise the Cz electrode for referencing, our evoked results are influenced by this. Any evoked response present at the reference is subtracted from all other channels. To better elucidate the spatiotemporal evolution of the evoked response, we plot responses averaged across sessions for all channels concurrently (Figure~\ref{fig:evoked_spatiotemporal}). This demonstrates that after the initial two visual peaks, at 200 ms a third positive visual peak emerges, accompanied by a smaller negative activity in the frontal area. Then, at 285 ms another smaller visual peak occurs, followed by a negative peak at 334 ms. At this time, some positive activity also arises in the frontal area. Finally, more visual positivity is observed around 386 ms, which shifts slightly to temporal/lateral areas at 406 ms, returning to the visual area at 452 ms. This plot provides a robust characterisation of the evoked response to inner speech across a substantial number of trials and sessions. However, we suspect the described activity remains principally due to the cross-cue presentation.

\begin{figure}[!t]
\begin{subfigure}{0.95\textwidth}
  \centering
  \includegraphics[width=1.0\linewidth]{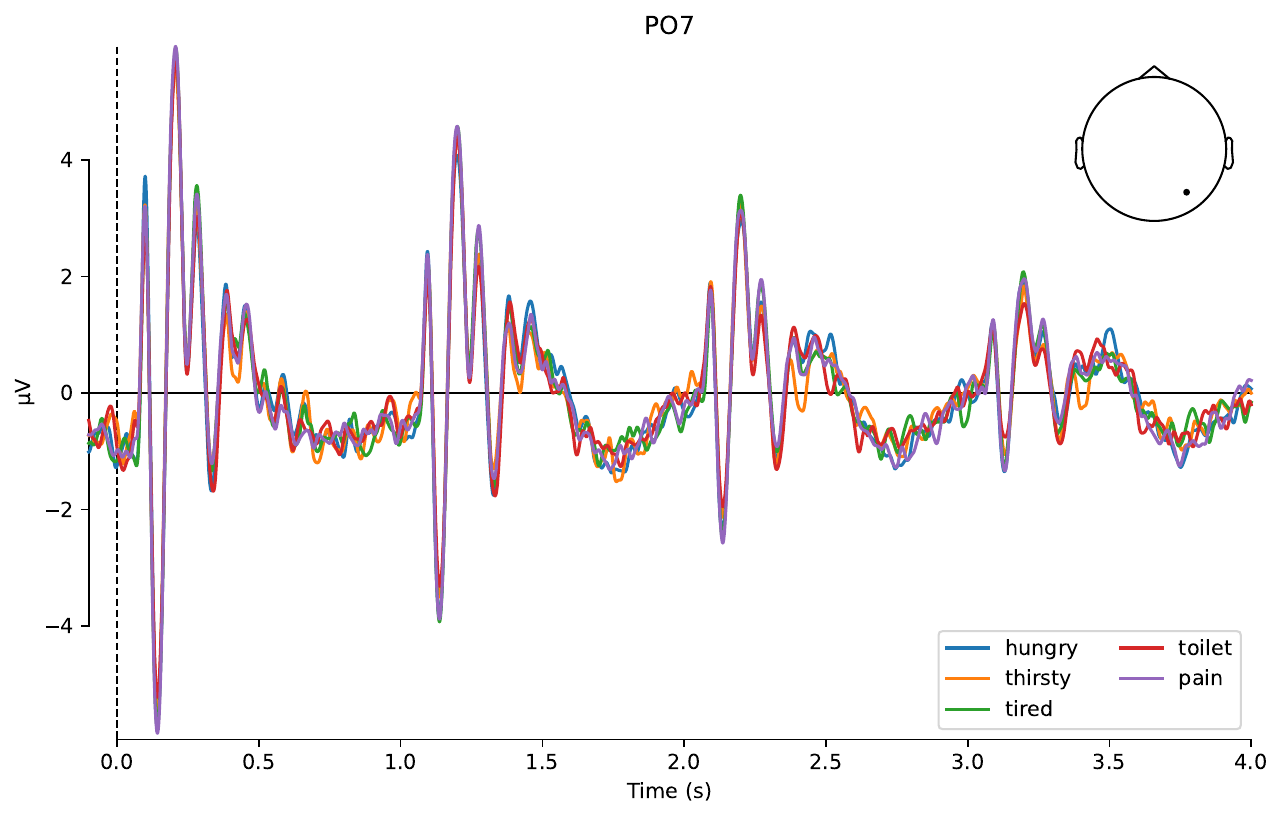}
  \caption{PO7}
  \label{fig:po7_words}
\end{subfigure}
\begin{subfigure}{0.95\textwidth}
  \centering
  \includegraphics[width=1.0\linewidth]{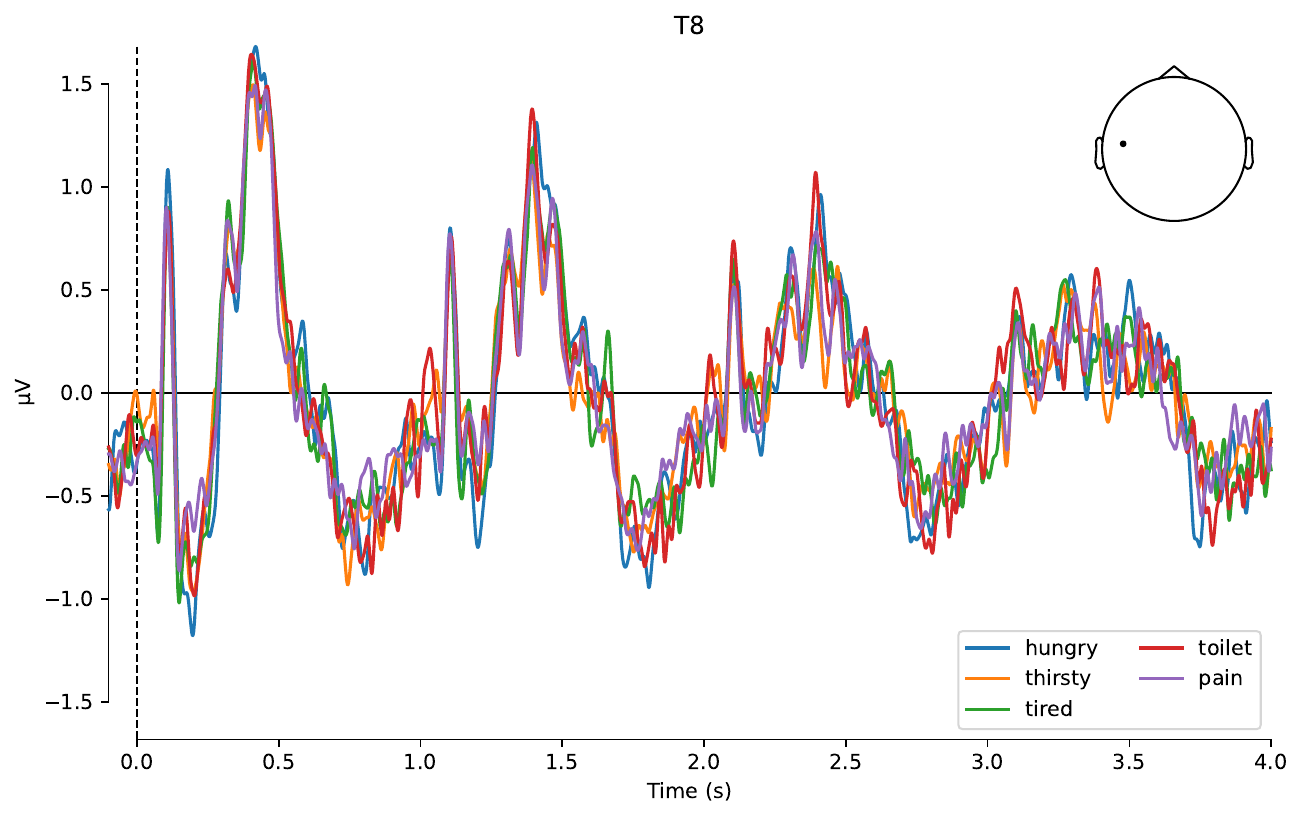}
  \caption{T8}
  \label{fig:t8_words}
\end{subfigure}
\caption{4-second evoked responses in 2 channels (PO7 and T8) across the 5 words averaged across all 10 EEG sessions of P4. Each line represents a different word.}
\label{fig:evoked_words_4s}
\end{figure}

Figure~\ref{fig:evoked_words_4s} displays the evoked response for each word across all sessions. Again, we only examine the two inner speech types here. We also opted to plot the entire 4-second trial with the 4 consecutive crosses, rather than averaging over these. We can discern that as the trial continues, the response to subsequent cross cues diminishes. This could reflect genuine activity changes and/or more noise across sessions later in the 4-second trial. There are no apparent differences between the evoked responses of words. This is anticipated since inner speech should elicit very subtle distinctions in EEG that would be nullified when averaging over many trials and sessions.

We wanted to verify that the evoked responses are purely visual in nature. However, there is no straightforward way to separate the visual and inner speech-related activity. The cross cue is essential to provide consistent timing for inner speech production; otherwise, variability in timing would impede decoding. Still, we conducted 1 EEG session with P4, where we implemented 3 tasks. First, we utilised the standard repetitive inner speech task with the 4 consecutive cues (\textit{cue+inner speech}). Then, we included a task where the visual stimuli were identical, but the participant was instructed not to think/internally vocalise the words, simply observe the crosses (\textit{cue-only} task). Finally, we incorporated a task with only 1 cross cue at the beginning of the 4-second trial, after which the participant attempted to repeat the inner speech 4 times as in the original task, but without timing alignment (\textit{inner speech-only} task).

\begin{figure}[!t]
\begin{subfigure}{0.99\textwidth}
  \centering
  \includegraphics[width=0.98\linewidth]{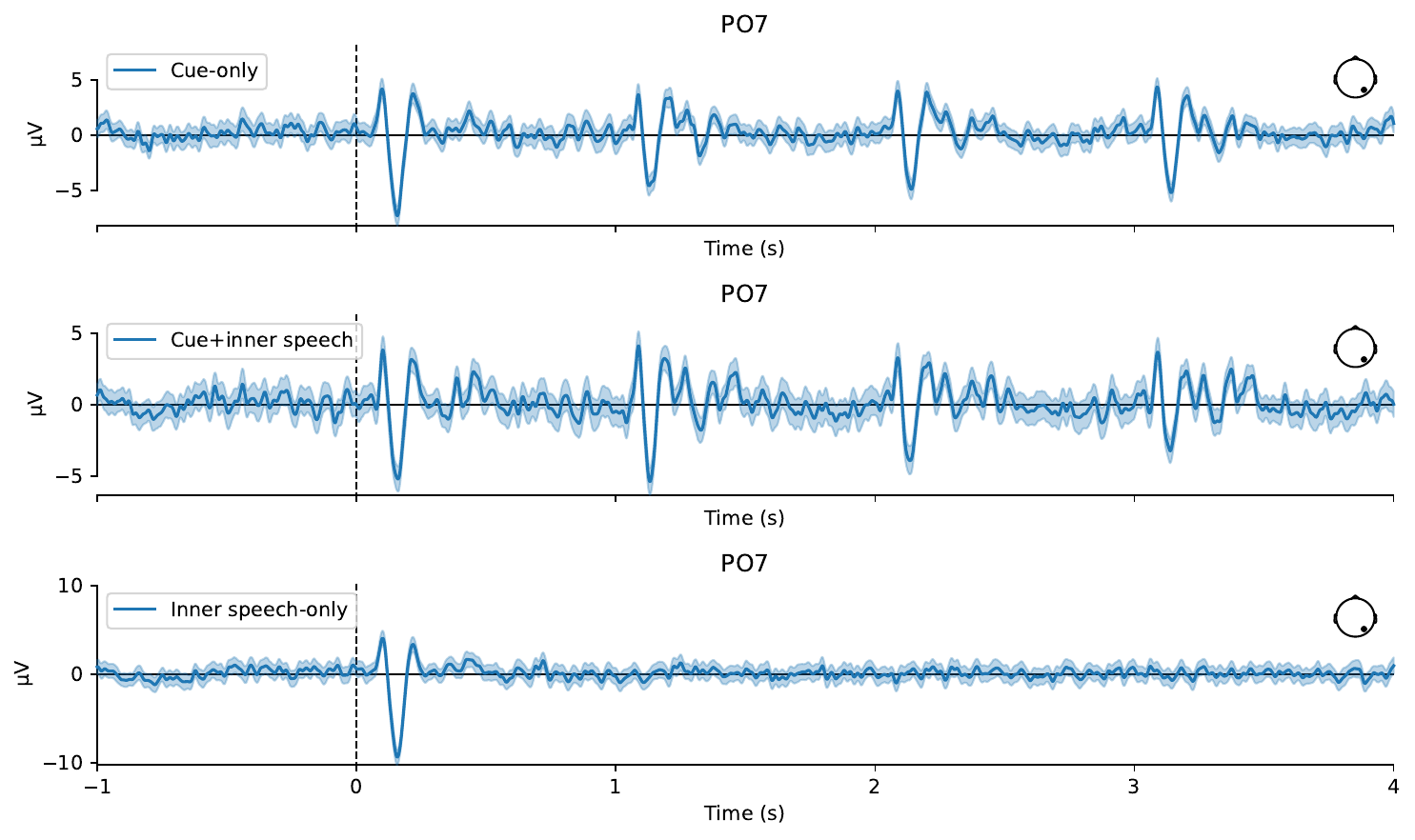}
  \caption{PO7}
  \label{fig:po7_cross_nocross}
\end{subfigure}
\begin{subfigure}{0.99\textwidth}
  \centering
  \includegraphics[width=0.98\linewidth]{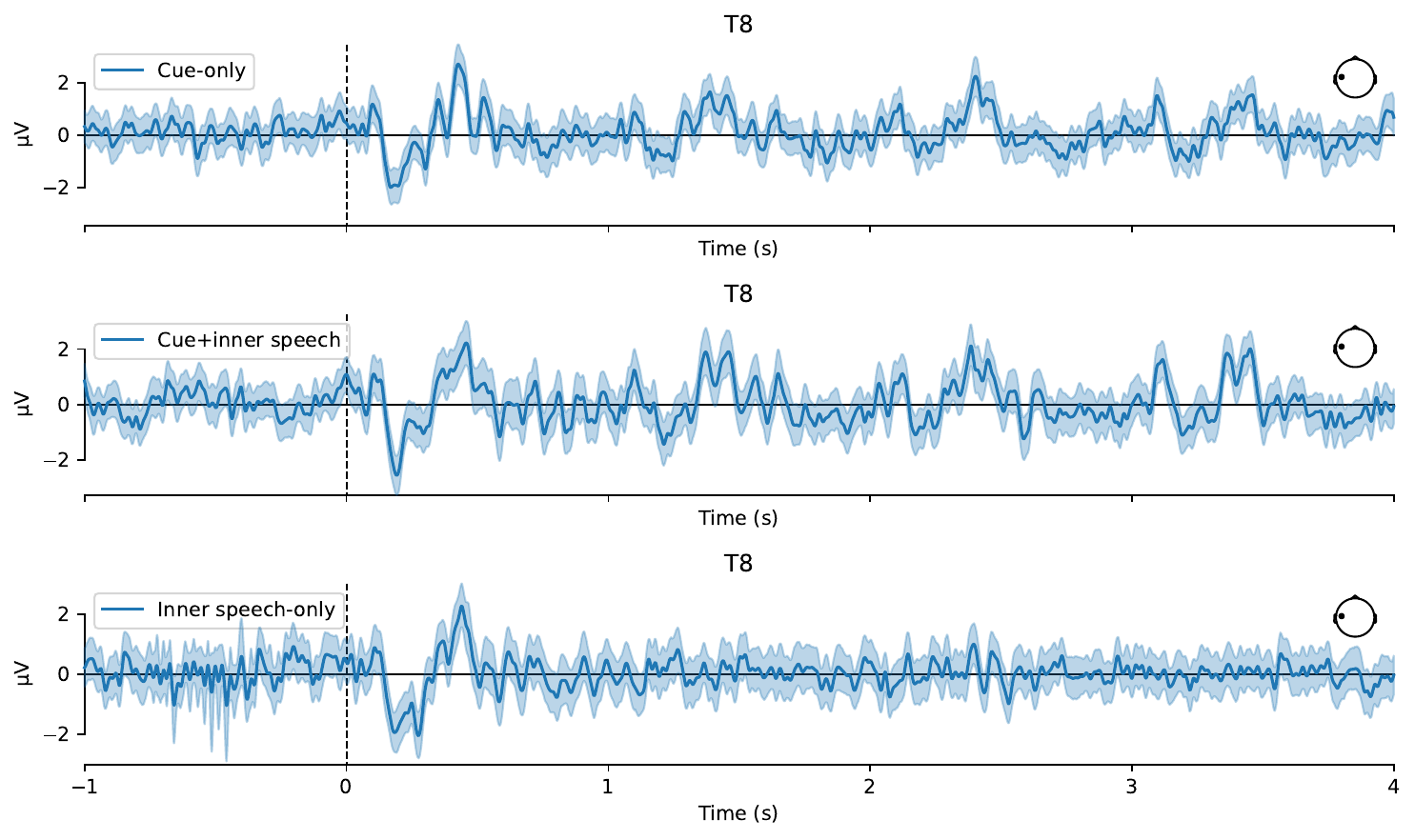}
  \caption{T8}
  \label{fig:t8_cross_nocross}
\end{subfigure}
\caption{4-second evoked responses in 2 channels, PO7 and T8, for the EEG session with 3 tasks. The evoked response across the 4-second trial is shown for the cue-only (top), cue+inner speech (middle), and inner speech-only (bottom) tasks, in both (a) and (b). Shading indicates 95\% confidence interval across trials.}
\label{fig:evoked_is_visual}
\end{figure}

Evoked responses across the 4-second trials for the three tasks from this single session are depicted in Figure~\ref{fig:evoked_is_visual}. This provides unambiguous evidence that the previously observed evoked responses are elicited by the cross cue. There are no discernible differences between the brain activity of solely observing the cues, compared to also engaging inner speech. The task where inner speech had to be repeated without visual cues shows that after the evoked response to the initial cue, there are no subsequent evoked responses attributable to inner speech. This could also stem from variability in timing. However, it is more likely that utilising inner speech is too subtle to generate brain signals exceeding baseline noise. These findings imply that decoding inner speech may be an equally challenging endeavour.

\end{appendices}

\end{document}

%% file: Packages.tex
\usepackage[T1]{fontenc}
\usepackage[utf8]{inputenc}
\usepackage[english]{babel}
\usepackage{siunitx}
\usepackage{graphicx}
\usepackage{tipa} % for the \ark{} command
\usepackage{graphics} % for pdf, bitmapped graphics files
\usepackage{times} % assumes new font selection scheme installed
\usepackage{amsmath}
\usepackage{latexsym}
\usepackage{amscd}% for commutative diagrams
\usepackage{mathrsfs} %this package is for the script font \mathscr
\usepackage{relsize}
\usepackage{ulem}
\usepackage{mathtools}
\usepackage{delarray}
\usepackage[round]{natbib}
\usepackage{pstricks}
\usepackage{theorem}
\usepackage{changepage}
\usepackage{euscript}
\usepackage{textcomp}
\usepackage{esvect}
\usepackage{parskip}
\usepackage{placeins}
\usepackage{subcaption}
\usepackage{array}
\usepackage{delarray}
\usepackage{stmaryrd}
\usepackage{fancyhdr}
\usepackage{graphpap}
\usepackage{makeidx}
\usepackage{enumerate}
\usepackage{esint}
\usepackage{datetime}
\usepackage{amsfonts} 
\usepackage{caption}
\usepackage{microtype}
\usepackage{nicefrac}
\usepackage{smartdiagram}
\usesmartdiagramlibrary{additions}
\usepackage{abstract}
\usepackage{color}
\definecolor{igreen}{rgb}{0.0, 0.56, 0.0}
\usepackage{xcolor, colortbl}
\colorlet{gred}{-red!75!green!65!}
\colorlet{mamber}{-red!75!green!15!blue!50!}
\colorlet{grown}{-red!75!blue!20!green}
\colorlet{bled}{-red!85!blue!40!green!45!}
\colorlet{waters}{cyan!25} % Define color for the water
\colorlet{water}{cyan!25!green!20!} % Define color for the water
\definecolor{grin}{HTML}{00F9DE}
\usepackage{rotating}

\usepackage{arydshln}
\usepackage{booktabs}
\usepackage{graphicx}
\usepackage{tikz}
\usepackage{tikz-3dplot}
\usetikzlibrary{
arrows,
arrows.meta,
automata,
backgrounds,
calc,
decorations,
decorations.pathmorphing,
decorations.pathreplacing,
decorations.fractals,
external,
fit,
matrix,
petri,
positioning,
shadows,
shapes,
shapes.multipart,
topaths,
intersections
}
\usepackage{eso-pic}
\def\ba{\begin{array}}
\def\ea{\end{array}}
\def\beann{\begin{eqnarray*}}
\def\eeann{\end{eqnarray*}}
\def\bea{\begin{eqnarray}}
\def\eea{\end{eqnarray}}

\DeclareMathOperator*{\argmin}{arg\!\min}

\makeatletter
\newlength\qvec@height
\newlength\qvec@depth
\newlength\qvec@width
\newcommand{\qvec}[2][]{
    \settoheight{\qvec@height}{$#2$}
    \settodepth{\qvec@depth}{$#2$}
    \settowidth{\qvec@width}{$#2$}
  \def\qvec@arg{#1}
  \raisebox{.2ex}{\raisebox{\qvec@height}{\rlap{% 
    \kern.05em
    \begin{tikzpicture}[scale=1,shorten >=-3pt,shorten <=-3pt]
    \pgfsetroundcap
    \coordinate (Stx) at (.05em,0) ;
		\coordinate (Arx) at (\qvec@width-.05em,0) ;
    \draw[->](Stx) to[bend left] (Arx);
    \end{tikzpicture}
  }}}
  #2
}
\makeatother
\makeatletter
\newlength\pvec@height
\newlength\pvec@depth
\newlength\pvec@width
\newcommand{\pvec}[2][]{
    \settoheight{\pvec@height}{$#2$}
    \settodepth{\pvec@depth}{$#2$}
    \settowidth{\pvec@width}{$#2$}
  \def\pvec@arg{#1}
  \raisebox{.2ex}{\raisebox{\pvec@height}{\rlap{% 
    \kern.05em
    \begin{tikzpicture}[scale=1,shorten >=-3pt,shorten <=-3pt]
    \pgfsetroundcap
    \coordinate (Stx) at (.05em,0) ;
		\coordinate (Arx) at (\pvec@width-.05em,0) ;
    \draw[->](Stx) to[bend right] (Arx);
    \end{tikzpicture}
  }}}
  #2
}
\makeatother
\makeatletter
\newlength\vvec@height%
\newlength\vvec@depth%
\newlength\vvec@width%
\newcommand{\vvec}[2][]{%
  \ifmmode%
    \settoheight{\vvec@height}{$#2$}%
    \settodepth{\vvec@depth}{$#2$}%
    \settowidth{\vvec@width}{$#2$}%
  \else 
    \settoheight{\vvec@height}{#2}%
    \settodepth{\vvec@depth}{#2}%
    \settowidth{\vvec@width}{#2}%
  \fi%
  \def\vvec@arg{#1}%
  \def\vvec@dd{:}%
  \def\vvec@d{.}%
  \raisebox{.2ex}{\raisebox{\vvec@height}{\rlap{%
    \kern.05em%
    \begin{tikzpicture}[scale=1]
    \pgfsetroundcap
    \draw (.05em,0)--(\vvec@width-.05em,0);
    \draw (\vvec@width-.05em,0)--(\vvec@width-.15em, .075em);
    \draw (\vvec@width-.05em,0)--(\vvec@width-.15em,-.075em);
    \ifx\vvec@arg\vvec@d%
      \fill(\vvec@width*.45,.5ex) circle (.5pt);%
    \else\ifx\vvec@arg\vvec@dd%
      \fill(\vvec@width*.30,.5ex) circle (.5pt);%
      \fill(\vvec@width*.65,.5ex) circle (.5pt);%
    \fi\fi%
    \end{tikzpicture}%
  }}}%
  #2%
}
\makeatother
\def\ba{\begin{array}}
\def\ea{\end{array}}
\def\beann{\begin{eqnarray*}}
\def\eeann{\end{eqnarray*}}
\def\bea{\begin{eqnarray}}
\def\eea{\end{eqnarray}}

\usepackage{titlesec}
\usepackage{multirow}
\usepackage[hidelinks]{hyperref} 
\usepackage{lipsum}
\usepackage{tikz-cd}
\usepackage{float}
\usepackage{titling}
\usepackage{epigraph}
\usepackage[title, titletoc]{appendix}
\titleformat{\chapter}{\normalfont\LARGE}{\thechapter\,$\vert$}{20pt}{\LARGE}{\setcounter{chapter}{0}}
\titlespacing*{\chapter}{0pt}{-70pt}{40pt} %Move chapter titles up
\makeatletter
\newcommand\BackgroundPicturea[3]{
	\setlength{\unitlength}{1pt}
	\put(0,\strip@pt\paperheight){
		\parbox[t]{\paperwidth}{
			\vspace{#2}\hspace{#3}
			\mbox{\includegraphics[scale=0.5]{#1}}
}}}
\newcommand\BackgroundPictureb[3]{
	\setlength{\unitlength}{1pt}
	\put(0,\strip@pt\paperheight){
		\parbox[t]{\paperwidth}{
			\vspace{#2}\hspace{#3}
			\mbox{\includegraphics[scale=0.3]{#1}}
}}}
\makeatother
\newenvironment{abbreviations}{\begin{list}{}{}}{\end{list}}
\usepackage{setspace}
\addto\captionsenglish{
	\renewcommand{\contentsname}%
	{Table of Contents}
	\setcounter{tocdepth}{3}% Include \subsubsection in ToC
	\setcounter{secnumdepth}{3}% Number \subsubsection in ToC
	}